\documentclass[a4paper, 12pt]{book}

\usepackage{phdthesis}
\usepackage[T3,T1]{fontenc}      		
\usepackage{geometry}        		
\usepackage[frenchb, english]{babel}
\usepackage[options]{titlepage_IRMES}	
\usepackage{fancybox}
\usepackage{psboxit}
\usepackage[avantgarde]{quotchap}
\usepackage[calcwidth]{titlesec}
\usepackage[table]{xcolor}
\usepackage{multicol}
\usepackage{float}
\usepackage{array}
\usepackage{url}
\usepackage{soul}
\usepackage{upgreek}
\usepackage{wrapfig}
\usepackage{graphicx}
\usepackage{scrtime}
\usepackage{pstricks}
\usepackage{amssymb}
\usepackage{amsmath}
\usepackage{infix-RPN,pst-math}
\usepackage{pst-node}
\usepackage{pst-tree}
\usepackage{pst-plot}
\usepackage{caption}
\usepackage{pst-bar}
\usepackage{pstricks-add}
\usepackage{pst-solides3d}
\usepackage{pst-3dplot} 

\makeatletter
\addto\extrasenglish{\bbl@frenchspacing}
\addto\extrasfrenchb{\bbl@frenchspacing}
\newcommand*{\rom}[1]{\expandafter\@slowromancap\romannumeral #1@}
\makeatother

\newcommand{\smallbox}{\psframe*(0,0)(0.4,0.4)}
\newcommand{\esmallbox}{\psframe[linewidth=0.4pt,fillstyle=solid,fillcolor=white](0,0)(0.4,0.4)}
\DeclareSymbolFont{tipa}{T3}{cmr}{m}{n}
\DeclareMathAccent{\invbreve}{\mathalpha}{tipa}{16}

\newpsbarstyle{light_blue}{fillcolor=blue!30,fillstyle=solid,framearc=0}
\newpsbarstyle{sepia_style}{fillcolor=sepia!30,fillstyle=solid,framearc=0}

\title{\begin{tabular}[t]{@{}c@{}}The phenotypic expansion\\and its boundaries\end{tabular}}

\author{Geoffroy Berthelot}
\begin{document}
\frenchspacing
\maketitle
\restoregeometry
\tableofcontents

\chapter*{Abstract}
{\small The development of sport performances in the future is a subject of myth and disagreement among experts. In particular, an article in 2004 \cite{tatem} gave rise to a lively debate in the academic field. It stated that linear models can be used to predict human performance in sprint races in a far future. As arguments favoring and opposing such methodology were discussed, other publications empirically showed that the past development of performances followed a non linear trend \cite{Blest1996, nevill1}. Other works, while deeply exploring the conditions leading to world records, highlighted that performance is tied to the economical and geopolitical context \cite{marion1}. Here we investigated the following human boundaries: development of performances with time in Olympic and non-Olympic events, development of sport performances with aging among humans and others species (greyhounds, thoroughbreds, mice). Development of performances from a broader point of view (demography \& lifespan) in a specific sub-system centered on primary energy were also investigated. We show that the physiological developments are limited with time \cite{berthelot2008, fdd1, nourTHESIS} and that previously introduced linear models are poor predictors of biological and physiological phenomena. Three major and direct determinants of sport performance are age \cite{moore1975, berthelot2011}, technology \cite{neptune, berthelot2010b} and climatic conditions (temperature) \cite{Nour2012}. However, all observed developments are related to the international context including the efficient use of primary energies. This last parameter is a major indirect propeller of performance development. We show that when physiological and societal performance indicators such as lifespan and population density depend on primary energies, the energy source, competition and mobility are key parameters for achieving long term sustainable trajectories. Otherwise, the vast majority (98.7\%) of the studied trajectories reaches 0 before 15 generations, due to the consumption of fossil energy and a low mobility rate. This led us to consider that in the present turbulent economical context and given the upcoming energy crisis, societal and physical performances are not expected to grow continuously.\\[.6cm]
\textbf{Keywords}: Sport, physiological limits, primary energy, limits in human sub-systems, multi-agents systems, physics.\\[0.6cm]
\textbf{Hosting laboratory:}\\
The work described in this document was performed at IRMES -Institut de recherche biom\'{e}dicale et d\textquoteright \'{e}pid\'{e}miologie du sport- situated in the INSEP -Institut national du sport, de l\textquoteright expertise et de la performance-.\\[.3cm]
IRMES is headed by Pr. Jean-Fran\c{c}ois Toussaint.\\[.3cm]
Adress: 11 avenue du Tremblay, 75012 Paris, France\\
Email: irmes$@$insep.fr\\
+33.1.41.74.41.29}

\chapter*{R\'{e}sum\'{e}}
Le d\'{e}veloppement futur des performances sportives est un sujet de mythe et de d\'{e}saccord entre les experts. Un article, publi\'{e} en 2004, a donn\'{e} lieu \`{a} un vif d\'{e}bat dans le domaine universitaire \cite{tatem}. Il sugg\`{e}re que les mod\`{e}les lin\'{e}aires peuvent \^{e}tre utilis\'{e}s pour pr\'{e}dire -sur le long terme- la performance humaine dans les courses de sprint. Des arguments en faveur et en d\'{e}faveur de cette m\'{e}thodologie ont \'{e}t\'{e} avanc\'{e}s par diff\'{e}rent scientifiques et d\textquoteright autres travaux ont montr\'{e} que le d\'{e}veloppement des performances est non lin\'{e}aire au cours du si\`{e}cle pass\'{e} \cite{Blest1996, nevill1}. Une autre \'{e}tude a \'{e}galement soulign\'{e} que la performance est li\'{e}e au contexte \'{e}conomique et g\'{e}opolitique \cite{marion1}. Dans ce travail, nous avons \'{e}tudi\'{e} les fronti\`{e}res suivantes: le d\'{e}veloppement temporel des performances dans des disciplines Olympiques et non Olympiques, avec le vieillissement chez les humains et d\textquoteright autres esp\`{e}ces (l\'{e}vriers, pur sangs, souris). Nous avons \'{e}galement \'{e}tudi\'{e} le d\'{e}veloppement des performances d\textquoteright un point de vue plus large en analysant la relation entre performance, dur\'{e}e de vie et consommation d\textquoteright \'{e}nergie primaire. Nous montrons que ces d\'{e}velopments physiologiques sont limités dans le temps \cite{berthelot2008, fdd1, nourTHESIS} et que les mod\`{e}les lin\'{e}aires introduits pr\'{e}c\'{e}demment sont de mauvais pr\'{e}dicteurs des ph\'{e}nom\`{e}nes biologiques et physiologiques \'{e}tudi\'{e}s. Trois facteurs principaux et directs de la performance sportive sont l\textquoteright \^{a}ge \cite{moore1975, berthelot2011}, la technologie \cite{neptune, berthelot2010b} et les conditions climatiques (temp\'{e}rature) \cite{Nour2012}. Cependant, toutes les \'{e}volutions observ\'{e}es sont li\'{e}es au contexte international et \`{a} l\textquoteright utilisation des \'{e}nergies primaires, ce dernier \'{e}tant un param\`{e}tre indirect du d\'{e}veloppement de la performance. Nous montrons que lorsque les indicateurs des performances physiologiques et soci\'{e}tales -tels que la dur\'{e}e de vie et la densit\'{e} de population- d\'{e}pendent des \'{e}nergies primaires, la source d\textquoteright \'{e}nergie, la comp\'{e}tition inter-individuelle et la mobilit\'{e} sont des param\`{e}tres favorisant la r\'{e}alisation de trajectoires durables sur le long terme. Dans le cas contraire, la grande majorit\'{e} (98,7\%) des trajectoires étudi\'{e}es atteint une densit\'{e} de population \'{e}gale \`{a} 0 avant 15 g\'{e}n\'{e}rations, en raison de la d\'{e}gradation des conditions environnementales et un faible taux de mobilit\'{e}. Ceci nous a conduit \`{a} consid\'{e}rer que, dans le contexte \'{e}conomique turbulent actuel et compte tenu de la crise \'{e}nerg\'{e}tique \`{a} venir, les performances soci\'{e}tales et physiques ne devraient pas cro\^{\i}tre continuellement.\\[0.6cm]
\textbf{Laboratory d\textquoteright accueil:}\\
IRMES -Institut de recherche biom\'{e}dicale et d\textquoteright \'{e}pid\'{e}miologie du sport- \`{a} l\textquoteright INSEP -Institut national du sport, de l\textquoteright expertise et de la performance-.\\[.3cm]
The IRMES est dirig\'{e} par Pr. Jean-Fran\c{c}ois Toussaint.\\
Adresse: 11 avenue du Tremblay, 75012 Paris, France\\
Email: irmes$@$insep.fr\\
+33.1.41.74.41.29

\begin{savequote}[10pc]
\sffamily
Soyons reconnaissants aux personnes qui nous donnent du bonheur; elles sont les charmants jardiniers par qui nos \^{a}mes sont fleuries.
\qauthor{Marcel Proust (1871-1922)}
\end{savequote}
\chapter*{Remerciements / Acknowledgments}
\vspace{3cm}
Je tiens, en tout premier lieu, \`{a} exprimer mes plus vifs remerciements \`{a} Jean-Fran\c{c}ois Toussaint qui fut pour moi un directeur de th\`{e}se attentif et disponible malgr\'{e} ses nombreuses charges. Sa comp\'{e}tence, sa rigueur scientifique et sa clairvoyance m\textquoteright ont beaucoup appris. Elles ont \'{e}t\'{e} et resteront des moteurs de mon travail de chercheur.\\[.3cm]
J\textquoteright exprime \'{e}galement mes remerciements \`{a} l\textquoteright ensemble des membres de mon jury, des sages parmis les sages, \`{a} cot\'{e} desquels je me sens minuscule: Madame Val\'{e}rie Masson-Delmotte, Messieurs Gilles Boeuf, Didier Sornette, Jean-Fran\c{c}ois Dhainaut, Denis Couvet, Vincent Bansaye et Jean Fran\c{c}ois Toussaint. Je tiens egalement \`{a} remercier Sidi Kaber, pour sa patience et son aide pr\'{e}cieuse.\\[.3cm]
Merci \`{a} mes nombreux et pr\'{e}cieux coll\`{e}gues de l\textquoteright institut pour leur bonne humeur, leurs moments de folie, en vrac et dans un ordre plus ou moins pseudo-al\'{e}atoire: H\'{e}l\`{e}ne, St\'{e}phane, Amal, Juliana, Laurie-Anne, Marion, Nour, Hala, Muriel, Val\'{e}rie, Fr\'{e}déric, Adrien (le po\`{e}te ou pouet ca d\'{e}pend des moments), Julien + Andy + Adrien (alias \textquoteleft le trio infernal\textquoteright), Karine, Julie pour son expertise des odom\`{e}tres, Jean pour sa bonne humeur et tous ceux que j\textquoteright oublie.\\[.3cm]
Un remerciement tout particulier \`{a} mes amis musiciens qui se sont accord\'{e}s à mon rythme: Fabien \textquoteleft Fabulous\textquoteright, Olivier \textquoteleft Olis\textquoteright, Renaud, Vincent, Nicolas \textquoteleft Nixx\textquoteright et Ludovic.\\[.3cm]
Mes amis doctorants, que j\textquoteright ai rencontr\'{e}s pendant cette th\`{e}se, et qui gagnent \`{a} \^{e}tre connus: Mlles Ana\"{\i}s, B\'{e}n\'{e}dicte, Mr Vivien et ce cher Cl\'{e}ment \textquoteleft +5\textquoteright~D., on se revoit bient\^{o}t \`{a} la taverne.\\[.3cm]
Enfin, les mots les plus simples \'{e}tant les plus forts, j\textquoteright adresse toute mon affection \`{a} ma famille dont mon petit R\'{e}mi, et en particulier \`{a} ma maman.\\[.3cm]
Last but not least, I\textquoteright d like to thank the people who took some time to read some pages of this document and correct my creepy english style: Ladies Liz (the beered/scientific/party girl) and her teammates Bugs, Puff and Alex (The Fantastic Dessert Maker), Katrine (aka \textquoteleft DJ Ginger Kreutz\textquoteright) and Maya.
\chapter{Introduction}
When Pierre de Coubertin revived the Olympic games in 1896, he expressed his ideals in the Olympic creed: ``\textit{The most important thing in life is not the triumph, but the fight; the essential thing is not to have won, but to have fought well}'' \cite{OSymbols}. The aim was to bring the Olympic games back to life in an international competition where athletes from various countries compete for the victory. The first Olympic games started with 9 disciplines and a focus on track and fields, cycling, fencing, gymnastics, shooting, swimming, tennis, weightlifting and wrestling. Fourteen countries and 241 athletes (only men) entered the competition. As time went, the participation increased and the Olympics became the field of competition of world wide nations with a higher number of represented disciplines. The physical performances improved in all events, and they became the major sporting event where athletes compete to gain fame and push back human limits. The Olympic Games thus celebrated the reflection of the human spirit toward a continuous expansion. After more than 100 years of competition, the processes leading to elite performances were sharply optimized. Competitions and disciplines, while being amateur at their early stages, gradually professionalized toward sports entertainment \cite{nourTHESIS}. At the same time, technologies dramatically improved and now allowed for the construction of huge databases, containing all best performances and records of elite athletes. The number of empirical studies based on this data progressively rose in recent times, as well as the debate on forecasting human limits in sports \cite{reinbound}. Linear \cite{tatem} and non linear \cite{Blest1996, nevill1} methodologies were used to describe the past development of performances with time and in an attempt to predict upcoming human capabilities in running or swimming.\\[.3cm]
Nevill et al. revealed that the progression of physical performances in track and field and swimming is limited \cite{nevill1, nevill2}. There have been a strong improvement in the performances in the early stage of the Olympics Games (1896-1939), but the 1960-2010 period showed a slowdown in the progression of the performances for the vast majority of events and disciplines. The causes of the initial increase are multiple.\\[.3cm]
Robert Fogel demonstrated that, during the past two centuries, there have been an strong development of human height, lifespan and bodymass of individuals \cite{fogel}. Such developments are pictured in Fig. \ref{FigIntro} (left panel), where the population strongly increased after the industrial revolution. Guillaume et al. replaced the sport performance in a broader frame and showed that sport performances are influenced by the economical and geopolitical context, strengthening the link between the social infrastructure and the development of physical performances. Similar decelerations were observed in over features of human development: the rate of progression of life expectancy decreased after 1960 and the infant mortality rates significantly declined between 1960 and 2001 \cite{UNICEF}. These features are tied to the societal infrastructure and synchronized in time \cite{fogel}. The environmental and societal changes resulting from the human expansion question the ability of man to further adapt upcoming alterations in the current paradigm \cite{JFT2012}. There is a need to assess the impact of policies in order to preserve current societal performance (ie. world records, lifespan, economical growth, industrial output, etc.), taking into account the environmental aspects. New models are progressively designed in this goal \cite{Sokolov2005} and R. Fogel previously embraced such a vision \cite{fogel, Fogel2004b}.\\[.3cm]
This thesis work aims at investigating the boundaries of human physical performances in sport with a focus on their development with aging. We also analyze the progression of athletic (greyhounds and thoroughbreds) and non athletic (mice) species with age. While sports performance is the starting point of our investigations, we also consider performance in a wide sense, with a particular interest on energy requirements and use in sport and non-sport environments. At the end of this work, we focus on exploring the sustainable solutions in a specific system, designed with cooperation/competition and environmental aspects. We are interested in sketching the boundaries in human sub-systems such as sport performance, lifespan and interactions between individuals. We focus in the last two centuries time frame (1800-2010) because of major impacts after the industrial revolution \cite{fogel}.\\[.3cm]
Figure \ref{FigIntro} (right panel) reveals that the energy use dramatically increased during the past century and has been a driving force for achieving current societal infrastructure and performances. Primary energy uses in the specific context of sports have not been widely investigated to our knowledge. Apart from the work of Boussabaine \cite{Boussabaine1999, Boussabaine2001} and Beuskera \cite{Beuskera2012}, focusing on the energy consumption of structures dedicated to sport or sporting events, no approach was intended to bind primary energy consumption and performance. In a broader perspective, we do not know any world-scale systemic approach using primary energy as the cornerstone. Energy is usually embedded as part of a global system. Sophisticated models, such as the MIT Integrated Global System Model (IGSM) \cite{Sokolov2005} and the World3 model \cite{meadows1, meadows2, meadows3}, aim at wider ambitious goals including assessing the impact of policies on climate change, ecosystems, industrial output, and so on. We rather focus on global dynamics resulting from local interactions between individuals, and their environment using typical methodologies issued from two-dimensional cellular automatas (CAs), Multi Agents Systems (MAS) and numerical simulations. We also choose to investigate societal performances: lifespan and population density using energy as the leading parameter. Moreover, we are interested in the specific scenario of energy scarcity and related short and long-term possible trajectories.\\[.3cm]
The document is divided in three sections: the first section aims at defining the determinants of sport performance and its limits. The middle section binds the development of human performances with the one of other species and at defining the \textquoteleft Phenotypic Expansion\textquoteright, a central theme of the document. Finally, the last section extends the concept of phenotypic expansion outside the sports environment and introduces a simple model to study the relationship between low primary energy flow and sustainable development of a population of agents.
\begin{center}
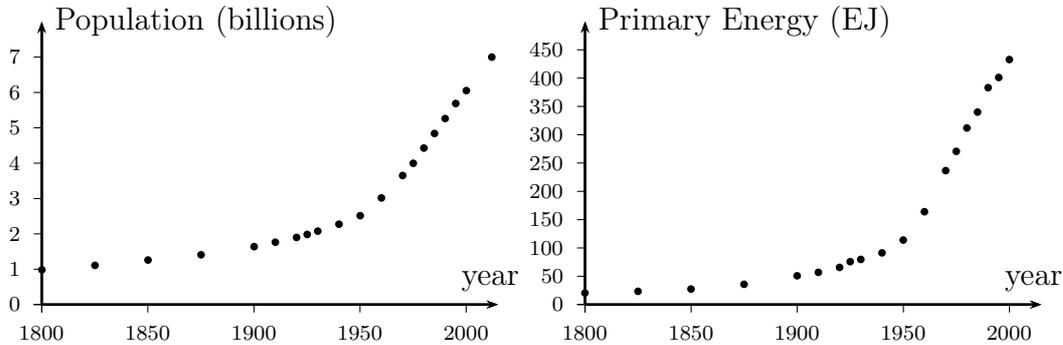

\begin{tabular}{ c c c }
\psset{xunit=0.027906977cm,yunit=0.46875cm}
\begin{pspicture}(1800,0)(2015,8)
\psaxes[Dx=50,Dy=1,Ox=1800,Oy=0,ticksize=-3pt,labelFontSize=\scriptstyle]{->}(1800,0)(2015,8)[year,100][Population (billions),0]
    \fileplot[plotstyle=dots, dotscale=0.8]{figure44-pop.prn}
\end{pspicture} & \hspace{0.3cm} &
\psset{xunit=0.027906977cm,yunit=0.0075cm}
\begin{pspicture}(1800,0)(2015,500)
\psaxes[Dx=50,Dy=50,Ox=1800,Oy=0,ticksize=-3pt,labelFontSize=\scriptstyle]{->}(1800,0)(2015,500)[year,100][Primary Energy (EJ),0]
    \fileplot[plotstyle=dots, dotscale=0.8]{figure44-Ener.prn}
\end{pspicture}
\end{tabular}
\captionof{figure}[One figure]{\label{FigIntro} {\footnotesize Development of world population (in billions) and primary energy use (EJ) on the left and right panels respectively (source: \cite{Grubler2008}). Both rates od development strongly increased after the industrial revolution (about 1760 to 1840). Population development is a societal performance and requires high energy inputs.}}
\end{center}
\chapter{Physiological boundaries}
This chapter introduces several key studies about the development of sport performances during the modern Olympic era, the determinants of sport performance and the lifespan of elite athletes compared with the values of the general population. Boundaries of the performances are studied with the data of elite athletes in order to demonstrate that the upper limits of the physical expansion are stagnating. The determinants of top performances are analyzed to show that they are related to environmental conditions and the expansion of societal infrastructure: the mobility (increased number of international competitions) and health parameters (average body size and mass, lifespan). R. Fogel previously demonstrated that they improved during this same period \cite{fogel}. We also compare the lifespan of Olympic athletes to a cohort of supercentenarians to study if they both follow the same time dynamics. Elite physical and physiological performances are here set in a global frame and compared with the values of the population.
\section{Boundaries in Olympic sports}
\label{sec:OlympicSports}
Since the introduction of the modern Olympic games in 1896, the development of sport performances have been investigated in many aspects in the scientific literature. Various mathematical models were introduced to describe and extrapolate the performances in various sport events. A recent study in the description of performance development in sport triggered controversy among the experts \cite{tatem, reinbound, rice}: a linear model was adjusted to the development of modern world records (WR) in the track and field 100m dash. By forecasting his model in a distant future, the author suggested that the WR would follow a constant progression rate, and eventually lead to an instant 100m in the future, and indeed establish negative marks. This approach sounded inappropriate in regards of known biological and physics theories. Other authors chose a more physiological approach by using different non-linear models \cite{Blest1996, nevill1, nevill2}. The models relied on exponential, logistic, or S-Shaped functions and included a limit (an asymptote). They can produce estimates of actual and futures performances that sounded more realistic in comparison to linear models. However, all these models were applied to only a reduced number of sport events.\\[0.3cm]
In order to gain insights on the physiological development of WR on an exhaustive basis, we investigate all the 147 quantifiable Olympic events in the modern Olympic era. Five disciplines are included in the study: track and field, cycling, speed skating, weight lifting and swimming and a total number of 3,263 WR are gathered over the 1896-2007 period. Two parameters are introduced in order to study \textit{i)} the annual frequency $\lambda$ and \textit{ii)} the relative improvement $\kappa$ of WR series. The first parameter $\lambda$ is computed as the average number of WR per year per Olympic event (eq. \ref{eq1-lambda}). The measure is adjusted to the effective number of Olympic events each year, as the number of official events change over the time (Fig. \ref{Fig. 2-lambda}). For each studied event, the parameter $\kappa$ shows the progression rate of a WR, comparatively to the previous mark (eq. \ref{eq2-kappa}). A high $\kappa$ value depicts a strong progression, while a low value means a weak development. The parameter is averaged over the 147 events to describe the mean improvement of WRs (Fig. \ref{Fig. 3-kappa}). Both parameters show a decrease with time: one starting in 1972 in $\lambda$ and following a strong increase after the second world war, and the other starting in 1925 in $\kappa$. These decreases in both the frequency and progression rate of WR reveal a major slow down in their development and suggest the characteristics of an exponential decrease. The pattern of WR development also unveils a step-wise progression related to various technological, pharmacological or sociological improvements \cite{nourTHESIS}. Each period of improvement is fitted using a simple exponential equation of the form:
\begin{equation}
  \label{eq0-modelGenericForm}
  y(t) = \Delta \exp^{-a \cdot t} + b
\end{equation}
with $\Delta$ the relative difference between the first and last data observed. The last period of improvement is forecasted toward infinity (ie. $t \rightarrow \infty$) and a fraction of the estimated asymptotic performance value is gathered. The year corresponding to this asymptotic value is then calculated for each of the 147 events. The distribution of the 147 years shows that 50\% of the events will reach their limits in 2027. This assumption is accurate as long as the performance\textquoteright s environment of the last period used to make predictions still prevail. Any alteration of this environment resulting from the introduction of a new factor (technology, sociology, \ldots) may trigger a new period. For instance, the development related to the introduction of Erythropoietin (EPO) in sport initiated a new period of performance\textquoteleft s improvement in cycling \cite{nour1}. This new period was also modeled using eq. \ref{eq0-modelGenericForm}.\\[0.3cm]
The model is based on an exponential equation (eq. \ref{eq0-modelGenericForm} with 2 parameters) and nevertheless produced good fitting statistics (average $R^2 = 0.91 \pm 0.08$). A summary of all the usual models used to describe the development of WR and appearing in the scientific literature could be made:
\begin{center}
     \begin{tabular}{|m{5cm}|m{6cm}|m{4.5cm}|} 
     \hline
     Type of model & General formulation & Authors\\ \hline
     linear & $y = \theta_1 \times x + \theta_2$ & Tatem \cite{tatem}, Blest \cite{Blest1996}\\ \hline
     exponential & $y = \theta_1 \times exp^{\theta_2 \cdot x} + \theta_3$ & Berthelot \cite{berthelot2008}, Desgorces \cite{fdd1}, El Helou \cite{nour1, nourTHESIS}, Blest \cite{Blest1996}\\ \hline
     generalized logistic function / Richards\textquoteright~curve & $y = \theta_1 + \dfrac{\theta_2 - \theta_3}{\left(1 + \theta_4 \cdot exp^{-\theta_5 \left( x - \theta_6\right)}\right)^{1/\theta_7}}$ & Blest \cite{Blest1996}, Nevill \cite{nevill1, nevill2}\\ \hline
     Gompertz & $y = \theta_1 \cdot exp^{\theta_2 \cdot exp^{\theta_3 x}} $ & Blest \cite{Blest1996}, Berthelot \cite{berthelot2010a}\\ \hline
     \end{tabular}
    \captionof{table}{\label{Table1-nonlinearmodels} {\footnotesize Listing of the various models used to describe the development of WR.}}
\end{center}
where $\theta_i$ denotes a parameter or constant of the equation. Additional models such as the extended Chapman-Richards (eq. \ref{ChapmanRichards} \cite{Blest1996, Ratkowsky1990}) or the antisymmetric exponential model (eq. \ref{antisymetric}), were also tested with significant results. However, according to Blest \cite{Blest1996} and other studies \cite{berthelot2008, nevill1} the popular models remain the piecewise exponential, Gompertz and logistic / sigmoidal.
\begin{equation}
  \label{ChapmanRichards}
  y = \theta_1 - \theta_2 \cdot \left[1 - exp^{-\theta_3 \cdot x} \right]^{\theta_4}
\end{equation}
\begin{equation}
\label{antisymetric}
    y =
    \begin{cases}
        \theta_1 + \theta_2 \cdot exp^{-\theta_3\left(x - \theta_4 \right)} & \text{if } x \geq \theta_4 \\
        \theta_1 + \theta_2 \left[2 - exp^{\theta_3 \left(x - \theta_4\right)} \right] & \text{if } x < \theta_4
    \end{cases}
\end{equation}
A firm conclusion is that the linear model can be dismissed to describe and forecast performances\textquoteright~development during the modern Olympic era. Another adequate statement is that all the proposed models in the literature reach a limit as time increases (Tab. \ref{Table1-nonlinearmodels} and eq. \ref{antisymetric}, \ref{ChapmanRichards}). Thus polynomial models are poor candidates for describing such a problem. Many of the presented models can produce a S-Shaped pattern suggesting that the overall performances\textquoteright~development is S-Shaped. As demonstrated in the article above, the development can also be described by a stepwise exponential model \cite{berthelot2008, fdd1, nour1, nourTHESIS}, suggesting that inside the overall S-Shaped pattern, multiple stages (or steps) of development occur suddenly. The exact causes of such steps of development are multifactorial, and will be quantified later on (see section \ref{sec:Technology}).\\[0.3cm]
The article \cite{berthelot2008} that introduced the exponential model and the methodology is presented in the following text:
\newpage
\noindent
\textbf{The Citius End: World Records Progression Announces the Completion of a Brief Ultra-Physiological Quest}\\
G. Berthelot, V. Thibault, M. Tafflet, S. Escolano, N. El Helou, X. Jouven, O. Hermine, J.-F. Toussaint
\\[0.6cm]
\noindent
\textbf{\textsc{Abstract}} World records in sports illustrate the ultimate expression of human integrated muscle biology, through speed or strength performances. Analysis and prediction of man's physiological boundaries in sports and impact of external (historical or environmental) conditions on WR occurrence are subject to scientific controversy. Based on the analysis of 3263 WR established for all quantifiable official contests since the first Olympic Games, we show here that WR progression rate follows a piecewise exponential decaying pattern with very high accuracy (mean adjusted $R^2$ values $= 0.91 \pm 0.08$ (s.d.)). Starting at 75\% of their estimated asymptotic values in 1896, WR have now reached 99\%, and, present conditions prevailing, half of all WR will not be improved by more than 0,05\% in 2027. Our model, which may be used to compare future athletic performances or assess the impact of international antidoping policies, forecasts that human species\textquoteright~physiological frontiers will be reached in one generation. This will have an impact on the future conditions of athlete training and on the organization of competitions. It may also alter the Olympic motto and spirit.
\begin{multicols}{2}
Olympic Games were reintroduced in 1896 by Pierre de Coubertin. One hundred and eleven years later, world record collection shows the progression of human performance as elite athletes periodically pushed back the frontiers of \textquoteleft ultra-physiology\textquoteright. This unplanned experiment could have been written as the phenotypic maximization of present human genotype under the pressure of regulated competition \cite{MacArthur2007}. This large scale investigation can now be appraised, but the best methodology to do it is a disputed scientific issue \cite{Whipp1992, reinbound, nevill1, nevill2}, with some literary perspectives \cite{Lovett2007, Miller2006}. Linear regression models \cite{Whipp1992, tatem} have been criticized \cite{reinbound} for their inaccuracy and non physiological relevance. A flattened S-shaped model has been elaborated by Nevill and Whyte \cite{nevill1, nevill2} on 8 running and 6 swimming events, but closer observation would suggest more detailed variations of the WR curves, adding historical or technical influences to biological parameters (Fig. \ref{Fig. 1.1 - WRevents}). Here we identify a common progression pattern for world records from all quantifiable Olympic events and propose a model that predicts the end of the quest.\\[0.3cm]
\textbf{\textsc{Material and methods}}
\\[0.3cm]
We conducted a qualitative and quantitative analysis of 3263 WR in all 147 measurable Olympic events from five disciplines \cite{IOC, SwimingUHome, FINA, LiftUp} in order to identify WR progression patterns. Data were gathered from 1896 to 2007 (modern Olympic era).\\[0.3cm]
\noindent
\textbf{Descriptive analysis: $\lambda$, $\kappa$ factors}\\
Two indicators were introduced in order to describe WR\textquoteright~development. Because the WR number established each year also depends on the number of events, we defined factor $\lambda$ as the annual ratio at year $t$ of the new WR number over the total number of official Olympic events:
\begin{equation}
  \label{eq1-lambda}
  \lambda_t = \frac{\sum \left( \text{newWR} \right)_t }{\sum \left( \text{events} \right)_t}
\end{equation}
WR evolution is also analyzed through the progression step $\kappa$, which measures the relative improvement of the $n^{th}$ best performance as compared to the $n-1^{th}$ value:
\begin{equation}
  \label{eq2-kappa}
  \kappa_n = \frac{\mid \text{WR}_n - \text{WR}_{n-1} \mid}{\text{WR}_{n-1}}
\end{equation}
with a mean $\overline{\kappa_t}$ annually calculated for all official events at year $t$.\\[0.3cm]
\noindent
\textbf{Description of the model}\\
WR series for each event were fitted by the function
\begin{equation}
  \label{eq3-model}
  y_j(t) = \Delta_{\text{WR}} \cdot \exp^{-a_j \cdot t'} + b_j
\end{equation}
where $\Delta_{\text{WR}} = \text{WR}_{i,j} - \text{WR}_{f,j}$ is an event indicator for the studied $j$ period; it is positive for the chronometric events (with decreasing WR values) and negative for the non-chronometric ones (increasing WR values); $\text{WR}_{i,j}$ and $\text{WR}_{f,j}$ are the initial and final WR values, respectively; $a_j$ is the positive curvature factor given by non linear regression; $b_j$ is the asymptotic limit. Normalization of $t$ in the $[0, 1]$ interval ensures the objective function (eq. \ref{eq3-model}) to be well-defined for all values of $t$. As a consequence, we used:
\begin{equation}
  \label{eq4-normalization}
  t'_j = \frac{t_j - t_{i,j}}{t_{f,j} - t_{i,j}}
\end{equation}
where $t'$ is the WR year after the linear transformation of $t$; $t_{i,j}$ and $t_{f,j}$ are the years of initial and final WR in the current $j$ period, respectively. Equation \ref{eq3-model} assumes that WR will achieve an asymptotic value within a given span starting at $\text{WR}_{i,j}$.\\[0.3cm]
\noindent
\textbf{Splitting WR series into periods}\\
A procedure based on the best adjusted $R^2$ is used to split WR series into periods. The algorithm is initiated by the first three WR values. The series is iteratively fitted by adding the next WR point using equation \ref{eq3-model}. For each fit, the adjusted $R^2$ is obtained; local maxima provide the changes of incline corresponding to the beginning of a new period. The minimum period duration is 6 years, the minimal WR number is three per period.\\[0.3cm]
For each event, this piecewise exponential decaying model provides successive periods. A period refers to a time slot defined by a group of consecutive WR, following a rupture of incline. During the period $j$, parameters $a_j$ and $b_j$ are estimated using the Levenberg-Marquardt algorithm (LMA) \cite{Levenberg1944, Marquardt1963, More1978} in a non linear least-squares regression to fit the model to WR. High values of the curvature coefficient a are seen in highly curved periods showing weak margin of final progression. Coefficient $b$ is the asymptotic value; the comparison of the initial ($\text{WR}_i$) and final ($\text{WR}_f$) records to $b$ are described through the $b'$ and $b''$ ratios respectively. The progression step over the Olympic era is equal to $b'' - b'$ and expressed as a percentage of the asymptotic value.\\[0.3cm]
\noindent
\textbf{Coefficients description}\\
The initial progression range is given by $b'$. In order to compare the predicted final progression, $b''$ are calculated for terminal periods of events. For chronometric events ($\text{WR}_i > \text{WR}_f$):
\begin{equation}
  \label{eq5-b1}
    \begin{array}{c c}
        b' = \dfrac{b_j}{\text{WR}_{i,j}} & \text{and \hspace{0.1cm}} b'' = \dfrac{b_j}{\text{WR}_{f,j}}
    \end{array}
\end{equation}
For non chronometric events ($\text{WR}_i < \text{WR}_f$):
\begin{equation}
  \label{eq6-b2}
        \begin{array}{cc}
            b' = \dfrac{\text{WR}_{i,j}}{b_j} & \text{and \hspace{0.1cm}} b'' = \dfrac{\text{WR}_{f,j}}{b_j}
        \end{array}
\end{equation}
This presentation also allows for a comparison of each record as a percentage of the estimated asymptotic value.\\[0.3cm]
\noindent
\textbf{Prediction}\\
Data set used for prediction was reduced to 125 events. From the 22 discarded events, two resulted from javelin weight change and 20 referred to weight lifting: 9 Clean and Press events were removed from official list in 1972 (Fig. S2, \cite{SupInfA1}) and 11 suffered major rule\textquoteright s alterations (weight categories). For prediction purpose, the inverse function of eq. \ref{eq3-model}, is given by:
\begin{equation}
  \label{eq7-inverseEq}
    t' = \frac{1}{-a} \cdot log\left( \dfrac{y-b}{\Delta_{\text{WR}}} \right)
\end{equation}
Coefficients of eq. \ref{eq7-inverseEq} are calculated for the last period of each event through LMA. The result is used to estimate the year $t$ when the 99.95\% (1/2000) limit is reached. This limit is set by $y = b + b/2000$ for chronometric events, and $y = b - b/2000$ for non chronometric records. The 1/2000 value was chosen in reference to the chronometric limits used on the quickest track race: it represents half of a $1/100^{th}$ of a second on the 100m. (and about 4s. on a marathon or 100g. in weight lifting disciplines).\\[0.3cm]
\noindent
\textbf{Estimating prediction variability}\\
Credibility interval is computed using a simulation method of Monte Carlo \cite{Mosegaard2002}. Previously estimated coefficients from eq. \ref{eq7-inverseEq} are used to draw $10 000$ new coefficients in a bi-dimensional normal distribution. Median was chosen as a robust measure of the center of the distribution in a non parametric approach.\\[0.3cm]
We used the 2.5th percentile, median and 97.5th percentile to produce the prediction errors for the estimated year at 99.95\% and the estimated WR asymptotic values. The credibility interval \cite{Willink2006} is given by the mean of the 2.5th and 97.5th percentiles for all 125 predictable events (Tab. S1 \cite{SupInfA1}).
\end{multicols}
\begin{center}
\begin{tabular}{l l l}
\psset{xunit=.05cm,yunit=.02cm}
\begin{pspicture}(1910,230)(2015,400)
    \psaxes[Dx=20,Dy=20,Ox=1910,Oy=230,ticksize=-3pt,labelFontSize=\scriptstyle]{->}(1910,230)(2015,400)[,-90][Performance (s.),0]
    \fileplot[plotstyle=dots, dotscale=0.8]{Figure16-A1.prn}
    \fileplot[plotstyle=line, linecolor=red]{Figure16-A2.prn}
    \fileplot[plotstyle=line, linecolor=red]{Figure16-A3.prn}
    \rput[b](2010,390){\textbf{(a)}}
\end{pspicture} & \hspace{0.5cm} &
\psset{xunit=.05cm,yunit=0.048cm}
\begin{pspicture}(1910,190)(2025,260)
    \psaxes[Dx=20,Dy=10,Ox=1910,Oy=190,ticksize=-3pt,labelFontSize=\scriptstyle]{->}(1910,190)(2025,260)[year,-90][Performance (s.),0]
    \fileplot[plotstyle=dots, dotscale=0.8]{Figure16-B1.prn}
    \fileplot[plotstyle=line, linecolor=red]{Figure16-B2.prn}
    \fileplot[plotstyle=line, linecolor=red]{Figure16-B3.prn}
    \rput[b](2015,255){\textbf{(c)}}
\end{pspicture} \\
\psset{xunit=.05cm,yunit=0.0006cm}
\begin{pspicture}(1910,12000)(2015,18000)
    \psaxes[Dx=20,Dy=2000,Ox=1910,Oy=12000,ticksize=-3pt,labelFontSize=\scriptstyle]{->}(1910,12000)(2015,18000)[,-90][Performance (s.),0]
    \fileplot[plotstyle=dots, dotscale=0.8]{Figure16-C1.prn}
    \fileplot[plotstyle=line, linecolor=red]{Figure16-C2.prn}
    \fileplot[plotstyle=line, linecolor=red]{Figure16-C3.prn}
    \rput[b](2010,17400){\textbf{(b)}}
\end{pspicture} & \hspace{0.5cm} &
\psset{xunit=.05cm,yunit=0.025cm}
\begin{pspicture}(1910,145)(2025,330)
    \psaxes[Dx=20,Dy=20,Ox=1910,Oy=145,ticksize=-3pt,labelFontSize=\scriptstyle]{->}(1910,145)(2025,285)[year,-90][Performance (kg.),0]
    \fileplot[plotstyle=dots, dotscale=0.8]{Figure16-D1.prn}
    \fileplot[plotstyle=line, linecolor=red]{Figure16-D2.prn}
    \fileplot[plotstyle=line, linecolor=red]{Figure16-D3.prn}
    \fileplot[plotstyle=line, linecolor=red]{Figure16-D4.prn}
    \fileplot[plotstyle=line, linecolor=red]{Figure16-D5.prn}
    \rput[b](2015,275){\textbf{(d)}}
\end{pspicture}
\end{tabular}

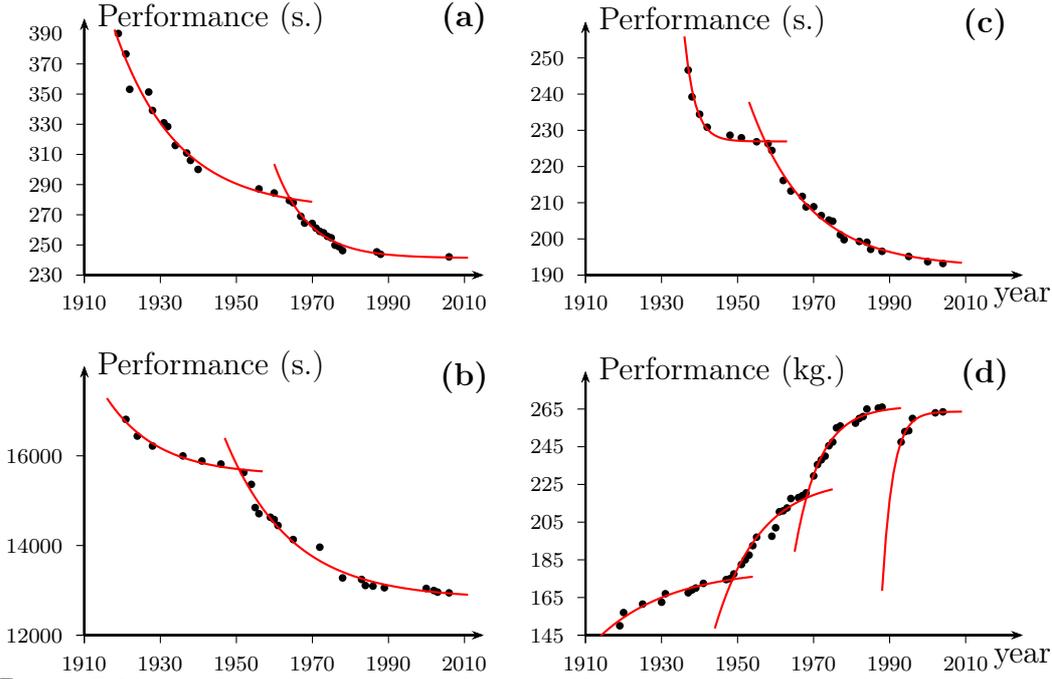
\captionof{figure}[One figure]{\label{Fig. 1.1 - WRevents} {\footnotesize \textbf{a}. Women 400m freestyle (swimming) with biexponential decaying curve, adjusted $R^2_{i} = 0.976$ and $R^2_{ii} = 0.966$; \textbf{b}. Men 50km walk (track), $R^2_{i} = 0.972$, $R^2_{ii} = 0.977$; \textbf{c}. Men $4\times100$m freestyle relay (swimming), $R^2_{i} = 0.985$, $R^2_{ii} = 0.988$; \textbf{d}. Clean \& Jerk Super Heavyweight (weight lifting), $R^2_{i} = 0.939$, $R^2_{ii} = 0.937$, $R^2_{iii} = 0.975$ and $R^2_{iv} = 0.946$. Weight categories were altered in 1948, 1968 and 1992 and control reinforced in 1988-1992 in weight lifting.}}
\end{center}
\begin{multicols}{2}
\textbf{\textsc{Results}}
\\[0.3cm]
Chronometric events represent 58\% of the data set (swimming, track, cycling, speed skating) with a decreasing tendency of WR values ; 42\% are non chronometric events (field, weight lifting) with increasing WR values.
Factor $\lambda$ evolution during the Olympic era (Fig. \ref{Fig. 2-lambda}) reveals three major phases of decline starting in 1913, 1938 and 1971 respectively. World Wars impact on $\lambda$ results in two major lag times, estimated by their width at mid-height: $\Delta_{\text{WWI}} = 6.4$ years for World War I and $\Delta_{\text{WWII}} = 13.4$ years for World War II. The calculated mean delay between each new WR is $2.62 \pm 3.05$ (s.d.) years.
$\overline{\kappa_t}$ significantly diminishes over the whole studied era (linear model: $F(1,102)=27.14$, $p<0.001$) (Fig. \ref{Fig. 3-kappa}) supporting the hypothesis of a constant reduction of WR progression possibilities.
Using the best adjusted $R^2$ iterative algorithm, 363 periods were obtained with mean $a = 3.00 \pm 2.87$, $b' = 0.75 \pm 0.15$ (Fig. S4 \cite{SupInfA1}), $b'' = 0.99 \pm 0.008$, an average progression step between initial and final records of 24\%, and mean adjusted $R^2 = 0.91 \pm 0.08$ (Fig. S3 \cite{SupInfA1}). The evolutionary profile of WR series over the Olympic era shows 2 to 3 periods for most of the events (2.47 periods $\pm 1.18$, range 1-6), with a mean of 8.98 WR per period and a mean duration of 25.8 years $\pm 14.8$ per period.
\begin{center}
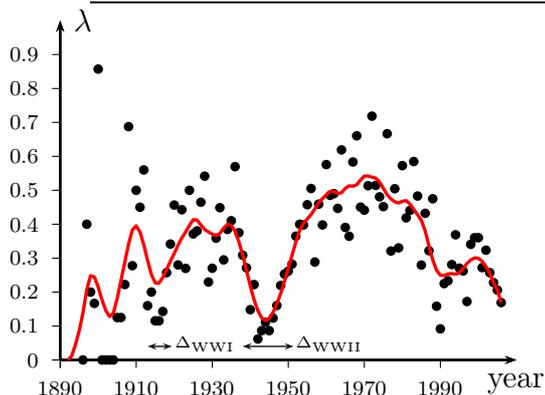

\psset{xunit=.05cm,yunit=4.5cm}
\begin{pspicture}(1890,-0.04)(2010,0.9)
    \psaxes[Dx=20,Dy=0.1,Ox=1890,Oy=0,ticksize=-3pt,labelFontSize=\scriptstyle]{->}(1890,0)(2010,1)[year,-90][$\lambda$,0]
    \fileplot[plotstyle=dots, dotscale=1]{figure14-1.prn}
    \fileplot[plotstyle=line, linecolor=red, linewidth=1.25pt]{figure14-2.prn}
    \rput[b](1928,0.025){\tiny $\Delta_{\text{WWI}}$}
    \rput[b](1960.8,0.025){\tiny $\Delta_{\text{WWII}}$}
    \psline[linewidth=0.5pt,linearc=0]{<->}(1913,0.04)(1919.4,0.04)
    \psline[linewidth=0.5pt,linearc=0]{<->}(1938,0.04)(1951.4,0.04)
\end{pspicture}
\captionof{figure}[One figure]{\label{Fig. 2-lambda} {\footnotesize Exact numbers (black dots) are filtered with a 60Hz second order low pass butterworth filter (red curve). World wars show major impact on $\lambda$: $\Delta_{\text{WWI}} = 6.4$ years; $\Delta_{\text{WWI}} = 13.4$ years.}}
\end{center}
We predicted the asymptotic value of each record, using the inverse function of eq. \ref{eq3-model} on the last period of 125 exploitable events (Tab. S1 \cite{SupInfA1}). A Monte-Carlo procedure was used to define the credibility interval of the prediction. The mean credibility interval of the asymptotic WR values is $[-2.28\%, +2.28\%]$. We also predicted the year when a record will be established at 99.95\% of its asymptotic value using the same method on the last period of the 125 exploitable events. The distribution of the 125 dates is expressed by decades (Fig. \ref{Fig. 4-distrib}): 12.8\% of these asymptotic WR have been reached in 2007. By 2027, half of the records will reach 99.95\% of their asymptotic value, within a [2002, 2120] credibility interval (Tab. S1 \cite{SupInfA1}) for each event prediction).\\[0.3cm]
\textbf{\textsc{Discussion}}
\\[0.3cm]
The proposed piecewise exponential decaying model, describing momentary expansion in a finite context, suggests a major global fading of WR progression. During an initial phase of rapid improvement, interrupted by two major events (Fig. S1 \cite{SupInfA1}), WR progression rate may have been described by a linear model. With a 40 years hindsight on the WR rate decline (Fig. \ref{Fig. 2-lambda}), debate on the limits now clearly emerges. As expected from biology, accurately fitted curves (high $R^2$ values) now refute the linear model. In all measurable Olympic contests from five different disciplines, involving either aerobic (10,000m. skating) or anaerobic (weight lifting) metabolic pathways, leg muscles mainly (cycling) or all muscles (decathlon), lasting seconds (shots) or hours (50km walk), either in men or women, small (Fly weight) or tall athletes (100m free style), individual or collective events (relays), all progression curves follow the same pattern, supporting the universality of the model.\\[.3cm]
Recently introduced events, such as women weight lifting starting in 1998 (Fig. S2 \cite{SupInfA1}), may require closer follow-up. Also final periods appearing in the last decade, with smaller data samples, may have a wider progression margin than estimated. Record measurement accuracy may be enhanced by using more precise technologies (times recorded in milliseconds, jumps in millimeters). Such decisions, however, are expected to have no effect on the WR progression rate $\lambda$, as it only alters the curve sampling and not the exponential shape nor the asymptotic value.
\begin{center}
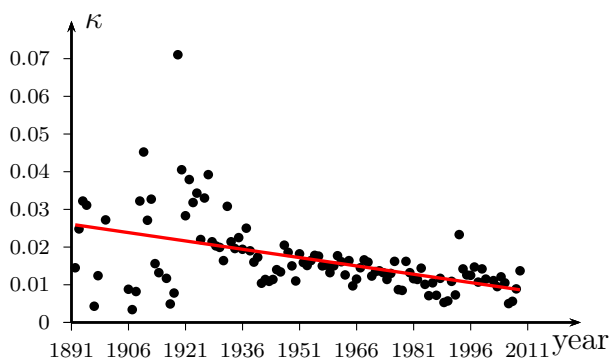

\psset{xunit=.05cm,yunit=50cm}
\begin{pspicture}(1890,-0.005)(2025,0.08) 
    \psaxes[Dx=15,Dy=0.01,Ox=1891,Oy=0,ticksize=-3pt,labelFontSize=\scriptstyle]{->}(1891,0)(2025,0.08)[year,-90][$\kappa$,0]
    \fileplot[plotstyle=dots, dotscale=1]{Figure19-1.prn}
    \fileplot[plotstyle=line, linecolor=red, linewidth=1.25pt]{Figure19-2.prn}
\end{pspicture}
\captionof{figure}[One figure]{\label{Fig. 3-kappa} {\footnotesize Annual evolution of WR relative improvement: $\kappa$ decreases from 0.024 in the first 30 years to 0.010 in the last 10 years (Linear model: $y = -1.46\times10^{-4} \cdot x + 0.301$, $F(1,102) = 27.14$, $p < 0.001$). This decrease is representative of the growing difficulty to improve previously established WR values.}}
\end{center}
Historical circumstances and WR evolution are closely linked: the impact of world wars results in two delays (Fig. \ref{Fig. 2-lambda}) with $\Delta_{\text{WWII}}$ being twice as large as $\Delta_{\text{WWI}}$. Starting in 1971, a much larger $\lambda$ reduction is observed (from 0.72 to 0.17), in the absence of major conflict \cite{Harbom2005} and despite the Cold War, which boosted sport competition among east European and western nations. In addition economic development between 1950 and 1980, with major technological, nutritional and medical advances, offered a constant elevation of life resources in the few countries (USA, Russia, Australia, Canada, Japan and European countries) that provide 95\% of WR. This last WR rate decrease also happened despite a considerable expansion of participating countries (and athletes): from 14 nations in 1896 in Athens (240 competitors), 69 nations in Helsinki\textquoteright s 1952 games (4950 athletes) and 202 nations for the last summer games in Athens again ($11100$ athletes). Rule modifications and anti-doping control reinforcements may have generated specific WR evolutions, as in Clean \& Jerk super heavyweight (Fig. \ref{Fig. 1.1 - WRevents}.D), where weight categories changed in 1948, 1968 and 1992. Finally, the decrease of $\lambda$ still appeared despite improvement of selection and training processes (time allotted to practice, new jumping or race-starting techniques, recruitment of taller athletes \cite{Norton2001}). All of these may have triggered new periods, but did not alter the global pattern, which is common to sport events as different as marathon, $4 \times 100$m. medley relay or pole vault.\\[.3cm]
Individual or team doping strategies have been used throughout the Olympic era, and state controlled protocols were developed since 1970 \cite{Kalinski2003, Geipel2001}: both may have contributed to slow down the $\lambda$ slope. Such practices however did not prevent the decline observed after the Mexico Games. Situations where doping would be legalized or not properly prevented \cite{Eichner2007, Kayser2007, Editorial2007, Mitchell2007} may again partially alter the record course in the future. The fact that suspected or belatedly convicted athletes hold some of the final records may exaggerate our predictive model, such that the year when half of WR will reach the 99.95\% limit may even be closer. In fact recent data show no progression of the 10 best performers in the last 20 years for the 100m track women or men high jump \cite{ReinboundWebsite} suggesting these WR may not be challenged anymore, especially when anti-doping agencies increase their actions and penalties. This is also observed when comparing the sprint events in running, swimming and speed skating over the second half of the XX$^{th}$ century \cite{Seiler2007}.
\begin{center}
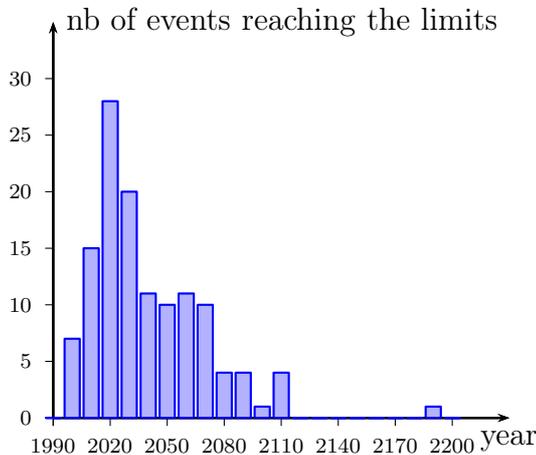

\psset{xunit=.025cm,yunit=0.15cm}
\begin{pspicture}(1990,-1.2)(2230,35) 
\psaxes[Dx=30,Dy=5,Ox=1990,Oy=0,ticksize=-3pt,labelFontSize=\scriptstyle]{->}(1990,0)(2230,35)[year,-90][nb of events reaching the limits,0]
\readdata{\data}{Figure15-1.prn}
\listplot[linecolor=blue,plotstyle=bar,barwidth=0.2cm,fillcolor=blue!30,fillstyle=solid,opacity=1]{\data}
\end{pspicture}
\captionof{figure}[One figure]{\label{Fig. 4-distrib} {\footnotesize Distribution of estimated limits at 99.95\% of the asymptotic value. Results are sampled by decades. Half of the asymptotic records will be established in 2027, and 90\% in 2068.}}
\end{center}
Major international rule changes, new technologies or gene profiling have also been tested in sport \cite{Buckley2000, Dennis2005}. However 50\% of asymptotic WR values will be obtained in one generation: sport organizations may then try to create new events, drop the WR quest, favor sports less directly associated with pure performance or promote health benefits of physical activity \cite{Warburton2006}. The \textquoteleft citius, altius, fortius\textquoteright~motto may be reworded within this century. Toward a \textquoteleft sanius?\textquoteright~remains an open question.\\[.3cm]
In summary, an epidemiological analysis of sport performances demonstrates that WR progression follows a piecewise exponential decaying pattern, altered by historical events. Results point out that in 2007, WR have reached 99\% of their asymptotic value. Present conditions prevailing for the next 20 years, half of all WR won\textquoteright t be improved by more than 0.05\%. As compared to the positivism triumphing at the time Coubertin inspired Olympic renewal, the present analysis emphasizes the ineluctable rarefaction of the quantifiable proofs of human physiological progression.
\end{multicols}
\section{The influence of environmental conditions}
\label{sec:NonOlympicSports}
\subsection{Performance development in outdoor sports}
The previous approach was only tested in Olympic events. We here aim to demonstrate that the development of performances established in outdoor and non-standardized environments, follow the same exponential pattern. Therefore, we analyze the development of 10 non-Olympics outdoors events including boat races, speed skating races or country ski races. Some of these events -such as the Oxford-Cambridge boat race- have performances starting in 1836 and providing exceptional historical trends. All the performances recorded in these 10 events are influenced by the environmental conditions, material or technical constraints. We decide to include both the technological and non technological development of the hour speed record in cycling. \textquoteleft Artificial\textquoteright~performances were recorded by the International Human Powered Vehicle Association (I.H.P.V.A.) and reached the ultimate speed of 91.56 km.h$^{-1}$ using streamlined recumbent bicycles. In comparison, the official record of the ICU (International Cyclist Union) is 49.7km.h$^{-1}$. The piecewise exponential model developed in section \ref{sec:OlympicSports}) is applied to the best performances in each events. The model fits the different periods with accuracy (mean $R^2 = 0.95 \pm .07$) and reveals that the development of performances is identical in standardized and non-standardized events. Whereas the studied events differ in longevity, duration or environment (ice, snow, water, ground, air), the piecewise decaying exponential model describes a common evolutive pattern. Older events (such as the Oxford Cambridge boat race) exhibit a sharp development in their early phase.\\[0.3cm]
Technological improvements in the early century also strongly participated in the development of performance and numerous aspects of sport environment (physical training, nutrition, medicine,\ldots) were impacted by technology. Robert Fogel previously introduced the term \textquoteleft techno-physiological evolution\textquoteright~to describe the anthropometric gains over the last three centuries \cite{fogel}. Our results show that the model remains robust to the introduction of new technologies that eventually led to a brutal development of performance\textquoteright~progression. This is best demonstrated in the IHPVA hour cycling record where all new periods of progression are fitted with accuracy.\\[0.3cm]
The model also demonstrates that the physiological limits in the studied outdoor events may become more and more evident in the XXI\textsuperscript{th} century. The model show that the most recent event (the Hawaii Ironman) almost reaches its asymptotic value after only three decades of competition. This event may have benefited from the advances previously made in the three sports involved in triathlon (swimming, cycling, running). Women weight lifting events present as well a fast and unique progression pattern and may have also benefited from the advances achieved in men weight lifting. In this study, we also investigate the delay for achieving limits regarding the technological degree of an event. It appears that this delay may be larger for events highly influenced by technology: estimated limits are of 2030 for the \textquoteleft physiological\textquoteright~group and of 2075 for the \textquoteleft technological\textquoteright~ group). One reason is that technology keeps modifying the condition of performance (boats in transatlantic records, skis and bicycles in those disciplines). However, the strength of the new periods of progression triggered by the introduction of new technologies appear to progressively weaken with time \cite{berthelot2008, nour1}.\\[0.3cm]
The article \cite{fdd1} and the complete discussion are presented in the following text:
\newpage
\noindent
\textbf{From Oxford to Hawaii Ecophysiological Barriers Limit Human Progression in Ten Sport Monuments}\\
F.-D. Desgorces, G. Berthelot, N. El Helou, V. Thibault, M. Guillaume, M. Tafflet, O. Hermine, J.-F. Toussaint
\\[0.6cm]
\noindent
\textbf{\textsc{Abstract}} In order to understand the determinants and trends of human performance evolution, we analyzed ten outdoor events among the oldest and most popular in sports history. Best performances of the Oxford-Cambridge boat race (since 1836), the channel crossing in swimming (1875), the hour cycling record (1893), the Elfstedentocht speed skating race (1909), the cross country ski Vasaloppet (1922), the speed ski record (1930), the Streif down-hill in Kitzbühel (1947), the eastward and westward sailing transatlantic records (1960) and the triathlon Hawaii ironman (1978) all follow a similar evolutive pattern, best described through a piecewise exponential decaying model ($R^2 = 0.95 \pm 0.07$). The oldest events present highest progression curvature during their early phase. Performance asymptotic limits predicted from the model may be achieved in forty years ($2049 \pm 32$ years). Prolonged progression may be anticipated in disciplines which further rely on technology such as sailing and cycling. Human progression in outdoor sports tends to asymptotic limits depending on physiological and environmental parameters and may temporarily benefit from further technological progresses.
\begin{multicols}{2}
World records highlight the progression of human performance. Our group recently analyzed WR from 5 measurable Olympic disciplines and demonstrated a major global fading in WR progression following a piecewise exponential decaying pattern [1]. Track and field, weight lifting or swimming competitions take place in a finite context with standardization of competitive fields, controlled environmental factors and major influence of physiological capacity on performance whereas outdoor sports are usually considered to be more influenced by environmental conditions, material or technical constraints. Therefore, outdoor events among the oldest and most popular have been ignored by previous studies that have analyzed human performance evolution \cite{berthelot2008, nevill1}.\\[0.3cm]
Performance depends on trainable variables (related to physiology, psychology, biomechanics, tactics) and other factors beyond the athlete\textquoteright s control (genetics, environment, climatic conditions) \cite{Smith2003}. Environmental factors may modify results according to the discipline rules, mobility milieu (snow, ice, water, air), motion type (running, cycling, skiing, swimming) or duration of the event. As a result, International Federation of Rowing Associations, as well as other federations do not provide world records, though events \textquoteleft best times\textquoteright~are available \cite{RowingF}. We hypothesized that performance evolution of outdoor sports events would also follow a piecewise exponential decaying pattern.\\[0.3cm]
\textbf{\textsc{Material and methods}}
\\[0.3cm]
Ten outdoor events among the most popular in the world, performed in non motorized sports and presenting large variations in duration, longevity, competitive circumstances and frequency were analyzed (Table \ref{TabFDD1}). The Oxford-Cambridge rowing race is challenged since 1829, with an unchanged course against the streams of the Thames from \textquoteleft Putney to Mortlake\textquoteright~since 1845 \cite{OxfordCambridge, Wikipedia}. The Vasaloppet is a Swedish long distance cross-country skiing race held on the first Sunday of March between the village of S\"{a}len and town of Mora for 90km \cite{Wikipedia, VasaloppetWeb}. Since 1909 the speed skating race \textquoteleft Elfstedentocht\textquoteright~takes place in the canals of the Dutch Friesland when ice freezes over the 200km route around Leeuwarden \cite{Wikipedia, Elfstedentocht}. The record of transatlantic crew sailing eastward from New-York \textquoteleft Ambrose Light\textquoteright~tower to English \textquoteleft Lizard Point\textquoteright~(5417km) was first set by Charlie Barr\textquoteright s crew in 1905 but the record has only been challenged for a few decades starting with Eric Tabarly \cite{Wikipedia, WSSRC}. The oldest solo ocean race is challenged sailing transatlantic westward every four years from Plymouth to New England (Newport or Boston, orthodromic track: 5185km) starting with Sir Francis Chichester in 1960 \cite{Wikipedia, WSSRC}. The speed ski record is challenged since 1930 in varied places (Portillo, Chile; Silverton, USA; Les Arcs, France) and determined in a timing zone 100 meters long \cite{Wikipedia, FSVWeb}. The hour cycling record is accounted as firstly challenged outdoor in Paris since 1876 and registered by the International Cyclist Union (ICU) and the International Human Powered Vehicle Association (IHPVA) with different acceptance in the bicycle type used for the record \cite{Wikipedia, UCIWeb, IHPVAWeb}. The Streif (Kitzbühel, Austria) down-hill ski is 3312 m long with an 27\% average gradient, held since 1930 and part of the ski world cup since 1967 \cite{Wikipedia, KitzbuelWeb}. The Hawaiian Ironman is the first modern long-distance triathlon (3.86km swimming, 180.2km cycling and 42.2km running). Starting in 1978, each year it is considered to be the World championship in long-distance triathlon \cite{Wikipedia, Ironman}. The channel crossing swim is usually challenged between Shakespeare beach (Dover, England) to \textquoteleft Cap gris nez\textquoteright~(France) over 33.8km; firstly held in 1875 it has been regulated since 1927 \cite{Wikipedia, CSAweb}. For all events, best performances (BP) only (equivalent to the race record) are accounted for into the analysis \cite{OxfordCambridge, Wikipedia, VasaloppetWeb, Elfstedentocht, WSSRC, FSVWeb, UCIWeb, IHPVAWeb, KitzbuelWeb, Ironman, CSAweb}.
\begin{center}
 \begin{table*}[ht]
  \footnotesize
      \begin{tabular}{ | l | l | l | l | p{2cm} | p{2cm} |}
        \hline
        Event & Sport & Dates & Frequency & Performance number (total) & Performance number (BP)\\ \hline
        Oxford-Cambridge & Rowing & 1829 & Annually & 153 & 19 \\
        Transatlantic record & Crew sailing & 1980 & Free & 11* & 11 \\
        Transatlantic record & Solo sailing & 1960 & Every 4 years & 13 & 9 \\
        Channel crossing & Swimming & 1875 & Free & 772 & 16\\
        Ironman Hawaii & Triathlon & 1978 & Annually & 30 & 12\\
        Hour cycling record UCI & Cycling & 1876 & Free & 29* & 29\\
        Hour cycling record IHPVA & Cycling & 1933 & Free & 20* & 20\\
        Elfstedentocht & Speed ice skating & 1909 & Annually & 16 & 9\\
        Vasaloppet & Cross-country ski & 1922 & Annually & 86 & 17\\
        Speed ski record & Down-hill ski & 1930 & Free & 32* & 32\\
        Streif down-hill & Down-hill ski & 1930 & Annually & 74 & 20\\
        \hline
    \end{tabular}
     \captionof{table}{\label{TabFDD1} {\footnotesize Event parameters: longevity, occurrence and available performance number. *Performance is only registered when the best performance is improved.}}
 \end{table*}
\end{center}
\textbf{Function description and prediction}\\
We performed BP modeling as previously reported \cite{berthelot2008} (sec. \ref{sec:OlympicSports}). The ratios $\beta$ (progression range since the beginning of the event) and $\beta'$ (present progression range) were calculated to describe the improvement over the final time frame in each event by using the following equation: $\beta = \dfrac{\text{BP}_i}{b}\times 100$ and $\beta' = \dfrac{\text{BP}_f}{b}\times 100$. Note that data are expressed as mean $\pm$ standard deviation.
\newpage
\textbf{\textsc{Results}}
\\[0.3cm]
The collected data provided a mean of $17.6 \pm 7.1$ BP per event. Events had large differences in their initial progression range ($\beta = 46.27 \pm 13.5\%$). Present mean achievement of the asymptotic performance, $\beta'$, was $94.6 \pm 7.2 \%$ with seven events over 97\%. Events with the lowest $\beta$ and $\beta'$ coefficients were: sailing transatlantic records, Streif and speed ski record and IHPVA hour cycling record (gathered in group 1: $\beta_1 = 37.5 \pm 9.5 \%$; $\beta'_1 = 90.4 \pm 8.8 \%$) whereas group 2 sport events: Oxford-Cambridge, Vasaloppet, Elfstedentocht, Ironman, Channel crossing and ICU Hour cycling record had higher values $\beta_2 = 53.5 \pm 12.1 \%$; $\beta'_2 = 98.0 \pm 3.2 \%$).\\[.3cm]
The model fits progression periods, depending on the events\textquoteright~longevity, with high accuracy ($R^2 = 0.95 \pm 0.07$). A first progression period (XIX\textsuperscript{th} and beginning of XX\textsuperscript{th} century) was modeled for the older events (Oxford-Cambridge, Vasaloppet, speed ski record, ICU cycling: $R^2 = 0.97 \pm 0.03$, $a = 1.93 \pm 0.96$ and $\beta' = 57.6\pm8.9\%$; Fig. \ref{FigFDD1}). A second progression period starts in the early XX\textsuperscript{th} century for seven events (the four previous ones plus Channel crossing, Elfstedentocht (Fig. \ref{FigFDD2}) and IHPVA cycling: $R^2 = 0.95 \pm 0.03$, $a = 0.97 \pm 0.45$, $\beta' = 73.15 \pm 16.8\%$). Modeled curves for the last period of all 10 events start in the middle of the XX\textsuperscript{th} century ($R^2 = 0.95\pm0.04$, $a = 1.28 \pm 0.51$, $\beta' = 94.5\pm7.2\%$). The progression of the transatlantic records, the Hawaii Ironman and Streif down-hill are modeled by a mono-exponential curve (Fig. \ref{FigFDD3}). The curves for the hour cycling record progression according to the specific associations\textquoteright~rules (ICU and IHPVA) are compared in Fig. \ref{FigFDD4}.\\[.3cm]
The year when mean BP will be established at 99.95\% of their asymptotic value is predicted to be $2049\pm32.1$ years (Tab. \ref{TabFDD2}).
\begin{center}
\begin{tabular}{l}
\psset{xunit=.033cm,yunit=.0023cm}
\begin{pspicture}(1805,900)(2000,2500) 
    \psaxes[Dx=20,Dy=200,Ox=1820,Oy=900,ticksize=-3pt,labelFontSize=\scriptstyle]{->}(1820,900)(2015,2500)[,-90][Performance (s.),0]
    \fileplot[plotstyle=dots, dotscale=0.8]{Figure20-1.prn}
    \fileplot[plotstyle=line, linecolor=red]{Figure20-2.prn}
    \fileplot[plotstyle=line, linecolor=red]{Figure20-3.prn}
    \fileplot[plotstyle=line, linecolor=red]{Figure20-4.prn}
    \rput[b](1920,2200){\textbf{Oxford-Cambridge}}
\end{pspicture} \\
\psset{xunit=.065cm,yunit=0.25cm} 
\begin{pspicture}(1907,10)(2000,30) 
    \psaxes[Dx=10,Dy=2.5,Ox=1915,Oy=13,ticksize=-3pt,labelFontSize=\scriptstyle, ylabelFactor=\cdot10^{3}]{->}(1915,13)(2015,28)[,-90][,0]
    \fileplot[plotstyle=dots, dotscale=0.8]{Figure20-B1.prn}
    \fileplot[plotstyle=line, linecolor=red]{Figure20-B2.prn}
    \fileplot[plotstyle=line, linecolor=red]{Figure20-B3.prn}
    \fileplot[plotstyle=line, linecolor=red]{Figure20-B4.prn}
    \rput[b](1965,26){\textbf{Vasaloppet}}
\end{pspicture} \\
\psset{xunit=.065cm,yunit=1.7cm}
\begin{pspicture}(1907,1.25)(2000,3.35) 
    \psaxes[Dx=10,Dy=0.4,Ox=1915,Oy=1.4,ticksize=-3pt,labelFontSize=\scriptstyle]{->}(1915,1.4)(2015,3.5)[,-90][,0]
    \fileplot[plotstyle=dots, dotscale=0.8]{Figure20-C1.prn}
    \fileplot[plotstyle=line, linecolor=red]{Figure20-C2.prn}
    \fileplot[plotstyle=line, linecolor=red]{Figure20-C3.prn}
    \fileplot[plotstyle=line, linecolor=red]{Figure20-C4.prn}
    \rput[b](1965,3.2){\textbf{Speed Ski Record}}
\end{pspicture}
\end{tabular}

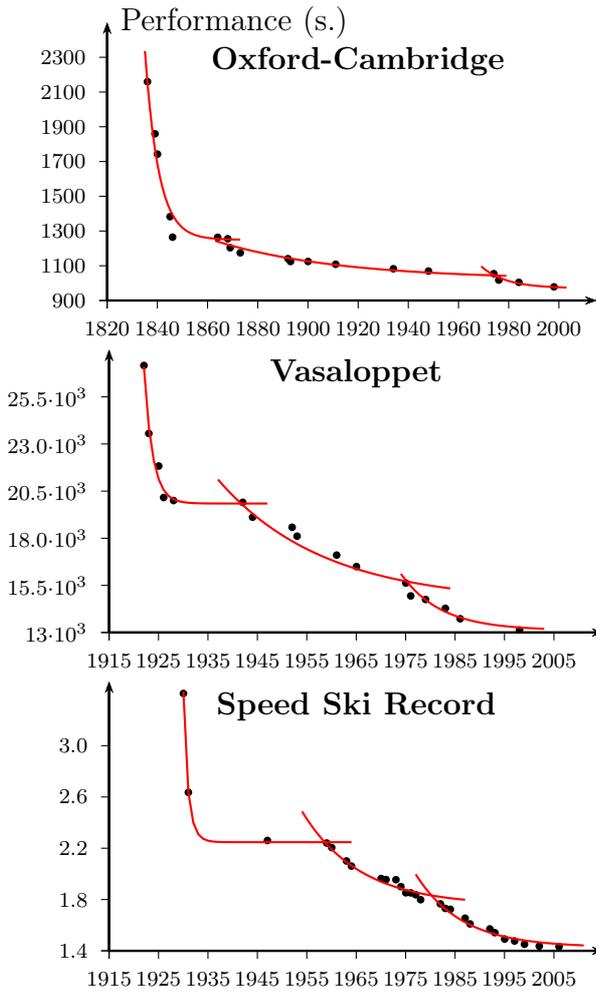
\captionof{figure}[One figure]{\label{FigFDD1} {\footnotesize Model fitting for events with 3 progression periods (x-axis: year, y-axis: speed). Performances of the Oxford-Cambridge boat race, Vasaloppet cross-country ski and speed ski record (for covering a 100m distance) in seconds.}}
\end{center}
\textbf{\textsc{Discussion}}
\\[0.3cm]
Our study is the first to model the performance evolution of outdoor sport events influenced by physiological, technological and environmental factors. Although the studied events largely differ in longevity, duration and physical environment (ice, snow, water, ground, air), the piecewise decaying exponential model describes a common evolutive pattern. Older events BPs follow a progression with high curvature coefficients suggesting the same rapid improvement in the early phase and a profile made up of two or three periods over the XIX\textsuperscript{th} and XX\textsuperscript{th} centuries. In addition, the initial ß state that we estimated here is lower than the value previously calculated for Olympic sports (46\% vs 60\%), which may be due to the fact that here we took into account all records starting at day 0, whereas national Track \& Field or Speed Skating competitions had already started long before the first Olympic games \cite{berthelot2008}. This emphasizes the particular status of these sport monuments in their own discipline.\\[.3cm]
Robert Fogel \cite{fogel} previously used the term \textquoteleft techno-physiological evolution\textquoteright~to describe human health and anthropometric gains over the last three centuries. Similar improvements occurred over the XX\textsuperscript{th} century (enhanced physical training, higher number of participants, nutritional practice, biological knowledge and medical advances) in sports like rowing \cite{Fiskestrand2004}, cross-country skiing \cite{Eisenman1989}, \cite{Rusko1992}, speed skating \cite{Schenau1994} or cycling \cite{Lucia2001}, which are sports with high aerobic requirements. Indeed, these events present similar progression patterns ($a$ and $\beta$ coefficients) than those obtained in Olympic disciplines performed in a controlled environment and quantified by world records \cite{berthelot2008}.
\begin{center}
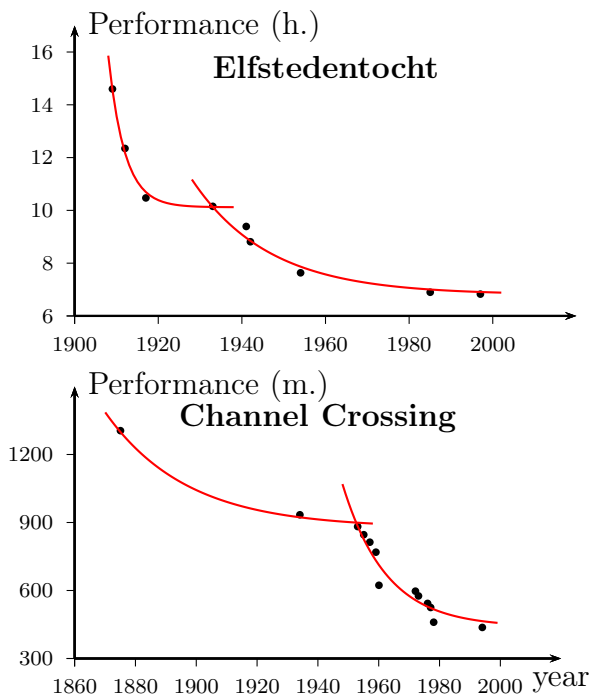

\begin{tabular}{l}
\psset{xunit=.055cm,yunit=.35cm}
\begin{pspicture}(1900,6)(2015,17)
    \psaxes[Dx=20,Dy=2,Ox=1900,Oy=6,ticksize=-3pt,labelFontSize=\scriptstyle]{->}(1900,6)(2020,17)[,-90][Performance (h.),0]
    \fileplot[plotstyle=dots, dotscale=0.8]{Figure21-A1.prn}
    \fileplot[plotstyle=line, linecolor=red]{Figure21-A2.prn}
    \fileplot[plotstyle=line, linecolor=red]{Figure21-A3.prn}
    \rput[b](1960,15){\textbf{Elfstedentocht}}
\end{pspicture} \\
\psset{xunit=.04cm,yunit=0.003cm}
\begin{pspicture}(1860,250)(2015,1760)
    \psaxes[Dx=20,Dy=300,Ox=1860,Oy=300,ticksize=-3pt,labelFontSize=\scriptstyle]{->}(1860,300)(2020,1500)[year,-90][Performance (m.),0]
    \fileplot[plotstyle=dots, dotscale=0.8]{Figure21-B1.prn}
    \fileplot[plotstyle=line, linecolor=red]{Figure21-B2.prn}
    \fileplot[plotstyle=line, linecolor=red]{Figure21-B3.prn}
    \rput[b](1940,1300){\textbf{Channel Crossing}}
\end{pspicture}
\end{tabular}
\captionof{figure}[One figure]{\label{FigFDD2} {\footnotesize Model fitting for events with two progression periods. Performances of the Elfstedentocht speed skating race (normalized distance of 200km) in hours and of the Channel crossing swimming in minutes.}}
\end{center}
Predictions of the BP asymptotic limits from the piecewise exponential model suggest that the achievement of human limits may occur during the XXI\textsuperscript{th} century. Furthermore, our model suggests that the limits will be rapidly reached in events highly dependent on physiological capabilities, e.g. expected performance improvements over the next century in Oxford-Cambridge, Vasaloppet and Elfstedentocht are lower than 2\%. The most recent of these events, the Hawaii Ironman, may have already achieved its asymptotic value ($\beta'$ value: 99.64\%) although it has been challenged since three decades only. The performance evolution for this event may be based on its rapidly growing popularity and benefits from the advances previously made in the three sports involved in triathlon (swimming, cycling, running). These results compared to those published for world records \cite{berthelot2008} suggest that environmental and climatic conditions, though influencing a year to year performance progression, do not modify the secular trend toward ultimate performance achievement.\\[.3cm]
Maximal boat speed in transatlantic records largely depends on environmental conditions (north Atlantic depression speed and numbers) rather than on individuals\textquoteright~physical power. Tools have been developed to control, at least partly, environmental factors for the athlete\textquoteright s benefits: e.g. navigation equipments allow for information to be transmitted to, or from, the boat in order to better take advantage of a major depression as well as of the lightest wind. Also, an increased knowledge and online computation now allows for a better use of sea streams. In addition, some events (speed ski, eastward transatlantic sailing, hour cycling) can be organized when optimal spatial and meteorological conditions are met in order to enhance probabilities to break a record. Choice for an event frequency and location may possibly influence BP occurrence and allow for a tight control of physical and climatic parameters. On the other side, the inability to organize the Elfstedentocht race due to the lack of thick ice between 1986 and 1997 and until now also demonstrates the major dependence of these competitions on climatic conditions \cite{Elfstedentocht}. Therefore, the progression of the Elfstedentocht performances could not benefit from the new skate technology that improved world skating records since 1998 \cite{berthelot2008, DeKoning2000}. In a pre-determined place and date, Oxford defeated Cambridge in 2008 with the slowest time since 1947 and because of blustery and rough conditions the Cambridge boat sank in 1978 \cite{OxfordCambridge}. In 1987, the finishing time of the coldest Vasaloppet (-30$\,^{\circ}\mathrm{C}$) was 16 minutes longer than the record stated the year before. On another hand, due to extremely mild weather, the race was canceled in 1990. However, as calculations integrate performances under-constraints, the particular environmental and climatic influences do not change the common pattern of performance evolution.
\end{multicols}
\begin{center}
\begin{tabular}{l l l}
\psset{xunit=.1444cm,yunit=.0125cm}
\begin{pspicture}(1970,50)(2015,350)
    \psaxes[Dx=10,Dy=50,Ox=1970,Oy=50,ticksize=-3pt,labelFontSize=\scriptstyle]{->}(1970,50)(2015,350)[,-90][Performance (h.),0]
    \fileplot[plotstyle=dots, dotscale=0.8]{Figure22-A1.prn}
    \fileplot[plotstyle=line, linecolor=red]{Figure22-A2.prn}
    \rput[b](2000,290){{\footnotesize \textbf{Eastward Crew Transatlantic}}}
\end{pspicture} & \hspace{0.5cm} &
\psset{xunit=.0866cm,yunit=0.00375cm}
\begin{pspicture}(1950,150)(2025,1150)
    \psaxes[Dx=10,Dy=200,Ox=1950,Oy=150,ticksize=-3pt,labelFontSize=\scriptstyle]{->}(1950,150)(2030,1150)[year,-90][Performance (h.),0]
    \fileplot[plotstyle=dots, dotscale=0.8]{Figure22-B1.prn}
    \fileplot[plotstyle=line, linecolor=red]{Figure22-B2.prn}
    \rput[b](2000,950){{\footnotesize \textbf{Westward Solo Transatlantic}}}
\end{pspicture} \\
\psset{xunit=.21cm,yunit=0.009375cm}
\begin{pspicture}(1975,400)(2005,875)
    \psaxes[Dx=5,Dy=50,Ox=1975,Oy=400,ticksize=-3pt,labelFontSize=\scriptstyle]{->}(1975,400)(2005,800)[,-90][Performance (m.),0]
    \fileplot[plotstyle=dots, dotscale=0.8]{Figure22-C1.prn}
    \fileplot[plotstyle=line, linecolor=red]{Figure22-C2.prn}
    \rput[b](1995,740){{\footnotesize \textbf{Hawaii Ironman}}}
\end{pspicture} & \hspace{0.5cm} &
\psset{xunit=.0764cm,yunit=0.025cm}
\begin{pspicture}(1940,100)(2025,250)
    \psaxes[Dx=10,Dy=20,Ox=1940,Oy=100,ticksize=-3pt,labelFontSize=\scriptstyle]{->}(1940,100)(2030,250)[year,-90][Performance (s.),0]
    \fileplot[plotstyle=dots, dotscale=0.8]{Figure22-D1.prn}
    \fileplot[plotstyle=line, linecolor=red]{Figure22-D2.prn}
    \rput[b](1997,227){{\footnotesize \textbf{Streif Down-Hill}}}
\end{pspicture}
\end{tabular}

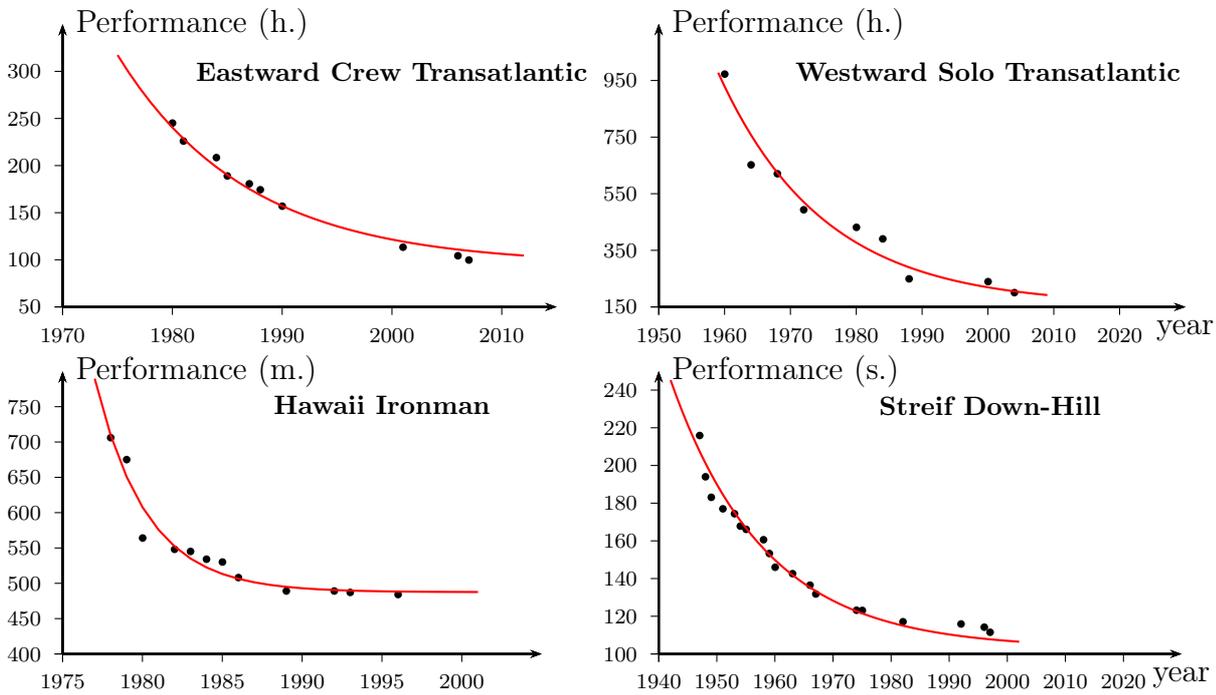
\captionof{figure}[One figure]{\label{FigFDD3} {\footnotesize Model fitting for three events with a single progression period. Performances of the Eastward Crew and Westward Solo transatlantic records in hours, of the Hawaii Ironman in minutes and of the Streif down-hill in seconds.}}
\end{center}
\begin{multicols}{2}
Our results demonstrate that higher progression rates remain plausible when rules allow for greater impact from technology. Sailing records present the lowest $\beta$ and $\beta'$ coefficients compared with rowing, cross-country skiing, triathlon or ICU cycling (group 2). The influence of technology is best demonstrated in the hour cycling record: the ICU \textquoteleft physiological\textquoteright~record limit is almost achieved at about 50km/h, when using a vehicle similar to the Eddy Merckx bicycle but the predicted limit is much higher for the \textquoteleft technological\textquoteright~IHPVA value ($91.94 \pm 3.58 \text{km/h}$). In addition, the delay for achieving limits may be larger than for events highly influenced by physiological capabilities (2030 for group 2 vs. 2075 for group 1), as technology keeps modifying boats in transatlantic records (e.g. gauge, size and class, hull and keel types, sail surface), equipment for ski records (e.g. ski length and composition, aerodynamic helmets, latex or polyurethane suits) or aerodynamic surface in torpedo-like bikes \cite{WSSRC, FSVWeb}. Although techno-physiological improvements allow for the enhancement of propulsive efficiency they remain under sport rules control: conception of rowing boats introduced carbon fiber use in 1972 and larger blades increased propulsive efficiency after 1991, but FISA banned sliding-rigger boats, which largely increased boat speed in order to limit technology influence and too expensive materials \cite{RowingF}.\\[.3cm]
Anti-doping policies are elaborated to restrain pharmacological impacts within health perspectives \cite{Kayser2007} but technological advances are accepted within sport rules and identity (e.g. minimum weights for each class boat in rowing; Vasaloppet performed in classic style; Channel crossing with swim suits and hat without thermal protection or buoyancy capabilities) \cite{RowingF, VasaloppetWeb, CSAweb}. Conversely, the speed ski record, one of the fastest non motorized sport on land, seems to have reached its limits after 70 years of technological improvements under major environmental constraints (e.g. air penetration coefficient, snow quality, slope). Thus, if international federations, pushed by public demand and media coverage, want to develop sport on a pure performance basis with the fascination of newly established records, their need for technology will constantly increase. Swimming is the recent demonstration of such evolution with the introduction of swimming suits in 1998 and of second generation models in 2008 that allow for a 1 to 2\% increase in speed \cite{Len2009}. However, close future and popularity of these sport monuments depend more on their history, identity and scenery than on continuous performance increase.
\begin{center}
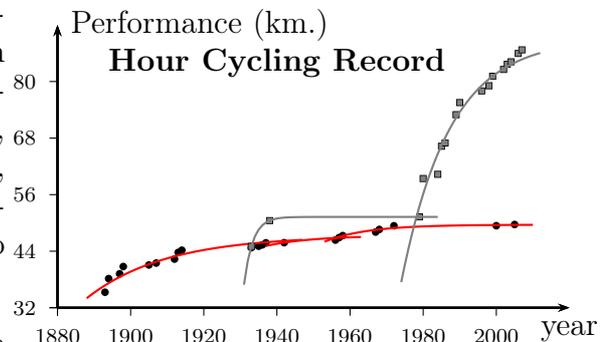

\psset{xunit=.0481cm,yunit=.0625cm}
\begin{pspicture}(1880,29)(2015,92)
    \psaxes[Dx=20,Dy=12,Ox=1880,Oy=32,ticksize=-3pt,labelFontSize=\scriptstyle]{->}(1880,32)(2020,92)[year,-90][Performance (km.),0]
    \fileplot[plotstyle=dots, dotscale=0.8]{Figure23-A1.prn}
    \fileplot[plotstyle=line, linecolor=red]{Figure23-A2.prn}
    \fileplot[plotstyle=line, linecolor=red]{Figure23-A3.prn}
    \fileplot[plotstyle=line, linecolor=red]{Figure23-A4.prn}
    \fileplot[plotstyle=dots, dotstyle=Bsquare, fillcolor=gray, dotscale=0.8]{Figure23-A5.prn}
    \fileplot[plotstyle=line, linecolor=gray]{Figure23-A6.prn}
    \fileplot[plotstyle=line, linecolor=gray]{Figure23-A7.prn}
    \rput[b](1940,81){\textbf{Hour Cycling Record}}
\end{pspicture}
\captionof{figure}[One figure]{\label{FigFDD4} {\footnotesize Model fitting for the hour cycling record. Hour cycling records in kilometers. Red lines and black dots for record stated by the International Cyclist Union. Grey lines and squares for the record stated by the International Human Powered Vehicle Association.}}
\end{center}
The present study suggests: \textit{i)} the universality of the piecewise exponential model to describe the evolution of very short (1.4s) to very long sport performances (400 hours), six orders of magnitude apart; \textit{ii)} a common law of progression in most sports toward eco-physiological limits; \textit{iii)} major physiological constraints in the progression pattern of rowing, cross-country skiing, swimming, triathlon and ICU cycling record; \textit{iv)} technology as a unique way to push back human physiological limits under environmental constraints, but with a major drawback: a constantly growing dependence on it.\\[.3cm]
In an accelerated process occurring over the XIX\textsuperscript{th} and XX\textsuperscript{th} centuries, sport performances took advantage of a common techno-physiological progression pathway \cite{fogel} but soon reach their limits. Environmental constraints now add to these development barriers. Human performance progression during the XXI\textsuperscript{th} century may essentially rely on technological improvements with its own reliance on energy and economy.
\end{multicols}
\begin{center}
 \begin{table}[htbp!]
      \begin{tabular}{ | l | l | l | l | l | l | l |}
        \hline
        \cellcolor{gray!10}Event & \multicolumn{2}{|p{3cm}|}{\cellcolor{gray!10}Date for asymptote achievement} & \multicolumn{2}{|l|}{\cellcolor{gray!10}$b$} & \cellcolor{gray!10}$\beta$ & \cellcolor{gray!10}$\beta'$ \\
        \hline
         & & \textit{CI} & & \textit{CI} & & \\
        \hline
        Eastward Crew Transatlantic & 2084 & 2050-2134 & 84.7 & 70.2-98.7 & 34.53 & 84.77 \\
        Westward Solo Transatlantic & 2107 & 2050-2206 & 155.8 & 66-246 & 23.89 & 77.71 \\
        Streif Down-hill & 2058 & 2045-2072 & 108.7 & 106-110 & 49.89 & 97.56 \\
        Speed ski-record & 2043 & 2021-2077 & 1.39 & 1.36-1.43 & 41.07 & 97.73 \\
        Hour cycling record IHPVA & 2086 & 2057-2126 & 91.94 & 88.3-95.5 & 38.45 & 94.35 \\
        Oxford-Cambridge & 2020 & 1990-2107 & 973.1 & 957-989 & 45.05 & 99.40 \\
        Vasaloppet & 2022 & 1994-2096 & 12998 & 12650-13357 & 47.84 & 98.95 \\
        Elfstedentocht & 2045 & 2027-2065 & 6.45 & 6.40-6.50 & 46.33 & 99.03 \\
        Ironman Hawaii & 2003 & 2000-2006 & 482.2 & 480-483 & 68.31 & 99.64 \\
        Channel crossing & 2051 & 2017-2109 & 399.4 & 325-460 & 42.76 & 91.41 \\
        Hour cycling record ICU & 2026 & 1984-2134 & 49.80 & 49.4-51.2 & 70.89 & 99.79 \\
        \hline
    \end{tabular}
 \caption{ \label{TabFDD2} {\footnotesize Coefficients and function fit for the asymptotic limit prediction. Year calculated for achieving 99.95\% of the asymptote with confidence interval (CI); $b$ is the asymptotic limit; $\beta$ is the progression range from the beginning of the event expressed as a percentage of the predicted asymptotic limit; $\beta'$ is the level of the actual BP expressed as a percentage of the asymptotic limit. Asymptotic limit $b$ expressed in seconds for Oxford-Cambridge, Vasaloppet, Streif and speed ski record (for 100m); in minutes for the Hawaii Ironman and channel crossing; in hours for the Eastward, Westward transatlantic records and Elfstedentocht; in kilometers for the hour cycling records.}}
 \end{table}
\end{center}
\subsection{Assessing the role of environmental and climatic conditions}
Beside the fact that outdoor sports are dependent in environmental conditions, we demonstrated that top performances in outdoor et indoor sports share the same progression pattern. In order to focus on the relationship between climatic and performance, we analyzed the performances in the marathon, a competition with a significant number of competitors. This approach was intended to be a comprehensive survey of all the performances of the 6 largest marathons worldwide: Paris, London, Berlin (European marathons), Boston, Chicago, New York (American marathons) from 2001 to 2010. The relationship between $1.7\times 10^6$ performances and environmental factors such as temperature, humidity, dew point and atmospheric temperature was analyzed. Major pollutants were also included in the study: nitrogen dioxide (NO$_2$), sulfur dioxide (SO$_2$), Ozone (O$_3$) and particulate matter (PM$_{10}$). All performances per year and race were found to be normally distributed with distribution parameters varying accordingly with the environmental factors. Air temperature was the most significant environmental parameter: it was significantly correlated with all performance levels in both male and female runners. Humidity was the second parameter with a high impact on performance; it was significantly correlated with women\textquoteright s and men\textquoteright s performance levels. Finally, the dew point and atmospheric pressure only had a slight influence in both genders, and did not affect the other performance levels. The effect of pollutant with performance was not trivial: pollutant combine with other components of air. In addition most marathons are held on Sunday mornings, when urban transport activity and its associated emissions are low, and photochemical reactions driven by solar radiation have not yet produced secondary pollutants such as ozone. Taking into account theses limitations, we found that NO$_2$ had the most significant correlation with performance: it was significantly correlated with the first quartile, the interquartile range and the median for both genders. Other pollutants only had a slight influence on performances, except for the O$_3$ who had a significant impact in several marathons (Berlin, Boston, Chicago and New York). However, this may be linked to temperature.\\[0.3cm]
When temperature increased above an optimum, performance decreased, such that the relationship between air temperature and performance was described by a simple quadratic polynomial function of the form:
\begin{equation}
  \label{eQuad_Withdraw}
    \%\text{withdrawals} = -a T + b T^2 + c
\end{equation}
with $T$ the temperature. The eq. \ref{eQuad_Withdraw} was adjusted to the percentage of runners withdrawals ($R^2 = 0.36$; $p < 0.0001$) and to the runner\textquoteright s speeds (women: $R^2 = 0.27$; $p<0.001$, estimated peak at $9.9\,^{\circ}\mathrm{C}$; men: $R^2 = 0.24$; $p<0.001$, estimated peak at $6\,^{\circ}\mathrm{C}$). The maximal average speeds were performed at an optimal temperature comprised between 3.8 and $9.9\,^{\circ}\mathrm{C}$ depending on the performance level. The thermal optimum of the model was close to the one where mortality rates are minimal suggesting that both sports performance and mortality are thermodynamically regulated. The work was published in \cite{Nour2012} and is presented below:\\[0.6cm]
\noindent
\textbf{Impact of Environmental Parameters on Marathon Running Performance}\\
N. El Helou, M. Tafflet, G. Berthelot, J. Tolaini, A. Marc, M. Guillaume, C. Hausswirth, J.-F. Toussaint
\\[0.6cm]
\noindent
\textbf{\textsc{Abstract}} The objectives of this study were to describe the distribution of all runners\textquoteright~performances in the largest marathons worldwide and to determine which environmental parameters have the maximal impact. We analyzed the results of six European (Paris, London, Berlin) and American (Boston, Chicago, New York) marathon races from 2001 to 2010 through 1791972 participants\textquoteright~performances (all finishers per year and race). Four environmental factors were gathered for each of the 60 races: temperature ($^{\circ}\mathrm{C}$), humidity (\%), dew point ($^{\circ}\mathrm{C}$), and the atmospheric pressure at sea level (hPA); as well as the concentrations of four atmospheric pollutants: NO$_2$ - SO$_2$ - O$_3$ and PM$_{10}$ ($\upmu$g.m$^-3$). All performances per year and race are normally distributed with distribution parameters (mean and standard deviation) that differ according to environmental factors. Air temperature and performance are significantly correlated through a quadratic model. The optimal temperatures for maximal mean speed of all runners vary depending on the performance level. When temperature increases above these optima, running speed decreases and withdrawal rates increase. Ozone also impacts performance but its effect might be linked to temperature. The other environmental parameters do not have any significant impact. The large amount of data analyzed and the model developed in this study highlight the major influence of air temperature above all other climatic parameter on human running capacity and adaptation to race conditions.
\begin{multicols}{2}
Like most phenotypic traits, athletic performance is multifactorial and influenced by genetic and environmental factors: exogenous factors contribute to the expression of the predisposing characteristics among best athletes \cite{Lippi2008a, McArthur2005}. The marathon is one of the most challenging endurance competitions; it is a mass participation race held under variable environmental conditions and temperatures sometimes vary widely from start to finish \cite{Cheuvront2001, Kenefick2007, WUW2011}. Warm weather during a marathon is detrimental for runners and is commonly referenced as limiting for thermoregulatory control \cite{Cheuvront2001, Vihma2010}. More medical complaints of hyperthermia (internal temperature = 39$\,^{\circ}\mathrm{C}$) occur in warm weather events, while hypothermia (internal temperature = 35$\,^{\circ}\mathrm{C}$) sometimes occurs during cool weather events \cite{Cheuvront2001}. In addition, participating in an outdoor urban event exposes athletes to air pollution which raises concerns for both performance and health \cite{Shephard1984}. Runners could be at risk during competitions as they are subject to elevated ventilation rate and increased airflow velocity amplifying the dose of inhaled pollutants and carrying them deeper into the lungs \cite{Shephard1984, Chimenti2009, Marr2010}. They switch from nasal to mouth breathing, bypassing nasal filtration mechanisms for large particles. Both might increase the deleterious effects of pollutants on health and athletic performance \cite{Chimenti2009, Lippi2008b}. Exposure to air pollution during exercise might be expected to impair an athlete\textquoteright s performance in endurance events lasting one hour or more \cite{Shephard1984, Lippi2008b}. The relationship between marathon performance decline and warmer air temperature has been well established. Vihma \cite{Vihma2010} and Ely et al. \cite{Ely2007a, Ely2007b} found a progressive and quantifiable slowing of marathon performance as WBGT (Wet Bulb Globe Temperature) increases, for men and women of wide ranging abilities. Ely et al. \cite{Ely2008} as well as Montain et al. \cite{Montain2007} also found that cooler weather (5-10$\,^{\circ}\mathrm{C}$) was associated with better ability to maintain running velocity through a marathon race compared to warmer conditions especially by fastest runners; weather impacted pacing and the impact was dependent on finishing position. Marr and Ely \cite{Marr2010} found significant correlations between the increase of WBGT and PM$_{10}$, and slower marathon performance of both men and women; but they did not find significant correlations with any other pollutant. Previous studies have mostly analyzed the performances of the top 3 males and females finishers as well as the 25th-, 100th-, and 300th- place finishers \cite{Lippi2008b, Ely2008, Montain2007, Martin1999, Trapasso1989}. Here we targeted exhaustiveness and analyzed the total number of finishers in order to quantify the effect of climate on the full range of runners. The objectives of this study were 1) to analyze all levels of running performance by describing the distribution of all marathons finishers by race, year and gender; 2) to determine the impact of environmental parameters: on the distribution of all marathon runners\textquoteright~performance in men and women (first and last finishers, quantiles of distribution); and on the percentage of runners withdrawals. We then modeled the relation between running speed and air temperature to determine the optimal environmental conditions for achieving the best running performances, and to help, based on known environmental parameters, to predict the distribution and inform runners on possible outcomes of running at different ambient temperatures. We tested the hypothesis that all runners\textquoteright~performances distributions may be similar in all races, and may be similarly affected by temperature.\\[0.3cm]
\textbf{\textsc{Material and methods}}
\\[0.3cm]
\noindent
\textbf{Data collection}\\
Marathon race results were obtained from six marathons included in the \textquoteleft IAAF Gold Labeled Road Races\textquoteright~and \textquoteleft World Marathon Majors\textquoteright: Berlin, Boston, Chicago, London, New York and Paris. From 2001 to 2010 (available data are limited before 2001) the arrival times in hours: minutes: seconds, of all finishers were gathered for each race. These data are available in the public domain on the official internet website of each city marathon, and on marathon archives websites \cite{OWARDW} and complementary data when needed from official sites of each race. Written and informed consent was therefore not required from individual athletes. The total number of collected performances was 1,791,972 for the 60 races (10 years $\times$ 6 marathons), including 1,791,071 performances for which the gender was known. We also gathered the total number of starters in order to calculate the number and the percentage of non-finishers (runner withdrawal) per race.\\[0.3cm]
Hourly weather data corresponding to the race day, time span and location of the marathon were obtained from \textquoteleft weather underground website\textquoteright~\cite{WUW2011}. Four climatic data were gathered for each of the 60 races: air temperature ($^{\circ}\mathrm{C}$), air humidity (\%), dew-point temperatures ($\,^{\circ}\mathrm{C}$), and atmospheric pressure at sea level (hPA). Each of these parameters was averaged for the first 4 hours after the start of each race. Hourly air pollution data for the day, time span and location of each race were also obtained through the concentrations of three atmospheric pollutants: NO$_2$ - SO$_2$ - O$_3$ ($\upmu$g.m$^-3$) from the Environmental Agency in each state (the Illinois Environmental Protection Agency for Chicago marathon, the Massachusetts Department of environmental Protection for Boston marathon and the New York State Department of Environmental Conservation for New York marathon), and the Environmental agency websites of the three European cities \cite{AirParif, SDEA, LAW}. All pollutants values were averaged for the first 4 hours after the start of each race.\\[0.3cm]
Concurrent measurements of air pollution for all ten race years (2001-2010) were only available for 3 pollutants, because air pollution monitoring sites typically measure only a subset of pollutants and may not have been operational in all years. In addition, particulate matters PM$_{10}$ were collected in Paris and Berlin, but there were not enough measurements in the other four cities races days.\\[0.3cm]
\noindent
\textbf{Data Analysis and selection}\\
Men and women performances were analyzed separately. For each race and each gender every year, we fitted the Normal and log-Normal distributions to the performances and tested the normality and log normality using the Kolmogorov-Smirnov $D$ statistic. We rejected the null hypothesis that the sample is normally or log-normally distributed when $p$ values $<0.01$. The following statistics (performance levels) were determined for all runners\textquoteright~performances distribution of each race, every year and for each gender:
\begin{itemize}
   \item the first percentile of the distribution (P1), representing the elite of each race.
   \item the winner.
   \item the last finisher.
   \item the first quartile of the distribution (Q1), representing the 25th percentile of best performers of the studied race.
   \item the median.
   \item the inter quartile range (IQR), representing the statistical dispersion, being equal to the difference between the third and first quartiles.
\end{itemize}
\begin{center}
 \begin{table*}[htbp!]
 \scriptsize
    \begin{tabular}{ | l | l | l | l | l | l | l |}
        \hline
        Marathon & Parameter & $N$ & Mean & Std Dev & Minimum & Maximum \\
        \hline
        Berlin & Temperature ($^{\circ}\mathrm{C}$) & 10 & 14.9 & 3.2 & 11.3 & 21.3 \\
        Run in September  & Dew Point ($^{\circ}\mathrm{C}$) & 10 & 10.6 & 1.8 & 5.8 & 12.3 \\
        Starts 9am  & Humidity (\%) & 10 & 78.0 & 14.5 & 55.0 & 98.5 \\
         & Atmospheric pressure (hPA) & 10 & 1017.0 & 6.3 & 1003.0 & 1029.0 \\
         & NO$_2$ ($\upmu$g.m$^-3$) & 10 & 26.5 & 4.0 & 20.8 & 33.2 \\
         & O$_3$ ($\upmu$g.m$^-3$) & 10 & 41.0 & 17.3 & 21.2 & 81.8 \\
         & PM$_{10}$ ($\upmu$g.m$^-3$) & 8 & 25.1 & 11.4 & 7.6 & 46.5 \\
         & SO$_{2}$ ($\upmu$g.m$^-3$) & 10 & 5.0 & 3.1 & 1.1 & 10.7 \\
         \hline
        Boston & Temperature ($^{\circ}\mathrm{C}$) & 10 & 11.8 & 5.1 & 8.0 & 25.2 \\
        Run in April & Dew Point ($^{\circ}\mathrm{C}$) & 10 & 3.9 & 3.8 & -2.1 & 10.2 \\
        Starts 10am  & Humidity (\%) & 10 & 62.6 & 19.9 & 28.3 & 91.0 \\
         & Atmospheric pressure (hPA) & 10 & 1013.0 & 12.4 & 981.6 & 1029.0 \\
         & NO$_2$ ($\upmu$g.m$^-3$) & 10 & 29.3 & 10.3 & 14.6 & 50.5 \\
         & O$_3$ ($\upmu$g.m$^-3$) & 10 & 73.5 & 25.7 & 18.5 & 122.7 \\
         & PM$_{10}$ ($\upmu$g.m$^-3$) & 0 &  &  &  &  \\
         & SO$_{2}$ ($\upmu$g.m$^-3$) & 10 & 7.0 & 2.9 & 1.6 & 12.1 \\
         \hline
        Chicago & Temperature ($^{\circ}\mathrm{C}$) & 10 & 12.1 & 7.5 & 1.7 & 25.0 \\
        Run in October & Dew Point ($^{\circ}\mathrm{C}$) & 10 & 4.9 & 7.6 & -5.9 & 19.0 \\
        Starts 7:30am  & Humidity (\%) & 10 & 62.8 & 8.1 & 52.3 & 79.2 \\
         & Atmospheric pressure (hPA) & 10 & 1022.0 & 6.4 & 1012.0 & 1031.0 \\
         & NO$_2$ ($\upmu$g.m$^-3$) & 10 & 27.9 & 13.0 & 9.7 & 52.0 \\
         & O$_3$ ($\upmu$g.m$^-3$) & 10 & 57.1 & 15.1 & 35.9 & 84.0 \\
         & PM$_{10}$ ($\upmu$g.m$^-3$) & 2 & 26.7 & 11.6 & 15.3 & 38.0 \\
         & SO$_{2}$ ($\upmu$g.m$^-3$) & 9 & 6.5 & 3.1 & 2.1 & 12.4 \\
         \hline
        London & Temperature ($^{\circ}\mathrm{C}$) & 10 & 12.4 & 3.2 & 9.5 & 19.1 \\
        Run in April & Dew Point ($^{\circ}\mathrm{C}$) & 10 & 6.0 & 2.9 & 0.8 & 10.7 \\
        Starts 9:30am  & Humidity (\%) & 10 & 66.9 & 16.7 & 42.9 & 86.1 \\
         & Atmospheric pressure (hPA) & 10 & 1010.0 & 12.5 & 976.4 & 1020.0 \\
         & NO$_2$ ($\upmu$g.m$^-3$) & 10 & 44.8 & 14.5 & 22.8 & 72.2 \\
         & O$_3$ ($\upmu$g.m$^-3$) & 9 & 51.4 & 17.1 & 35.0 & 92.3 \\
         & PM$_{10}$ ($\upmu$g.m$^-3$) & 2 & 27.8 & 14.5 & 13.7 & 41.9 \\
         & SO$_{2}$ ($\upmu$g.m$^-3$) & 10 & 4.5 & 2.8 & 0.0 & 8.8 \\
         \hline
        New York & Temperature ($^{\circ}\mathrm{C}$) & 10 & 12.5 & 4.1 & 7.1 & 18.4 \\
        Run in November & Dew Point ($^{\circ}\mathrm{C}$) & 10 & 2.3 & 6.4 & -5.6 & 12.8 \\
        Starts 10am  & Humidity (\%) & 10 & 51.1 & 12.1 & 36.5 & 79.8 \\
         & Atmospheric pressure (hPA) & 10 & 1020.0 & 7.8 & 1009.0 & 1034.0 \\
         & NO$_2$ ($\upmu$g.m$^-3$) & 9 & 55.1 & 17.2 & 21.9 & 77.3 \\
         & O$_3$ ($\upmu$g.m$^-3$) & 10 & 32.6 & 12.3 & 11.1 & 53.8 \\
         & PM$_{10}$ ($\upmu$g.m$^-3$) & 10 & 5.0 & 0.0 & 5.0 & 5.0 \\
         & SO$_{2}$ ($\upmu$g.m$^-3$) & 9 & 19.7 & 12.2 & 4.8 & 42.4 \\
         \hline
        Paris & Temperature ($^{\circ}\mathrm{C}$) & 10 & 9.2 & 3.2 & 4.8 & 17.4 \\
        Run in April & Dew Point ($^{\circ}\mathrm{C}$) & 10 & 4.2 & 4.1 & -3.6 & 13.4 \\
        Starts 8:45am  & Humidity (\%) & 10 & 72.4 & 10.1 & 45.9 & 85.4 \\
         & Atmospheric pressure (hPA) & 10 & 1019.0 & 6.2 & 1005.0 & 1026.0 \\
         & NO$_2$ ($\upmu$g.m$^-3$) & 10 & 43.0 & 13.7 & 23.4 & 73.1 \\
         & O$_3$ ($\upmu$g.m$^-3$) & 10 & 66.9 & 9.8 & 55.2 & 82.1 \\
         & PM$_{10}$ ($\upmu$g.m$^-3$) & 10 & 37.9 & 32.6 & 16.6 & 132.7 \\
         & SO$_{2}$ ($\upmu$g.m$^-3$) & 10 & 6.4 & 3.7 & 1.5 & 12.2 \\
        \hline
    \end{tabular}
 \caption{\label{TabNOUR1} {\footnotesize Average and range values of all weather and pollution parameters for the six marathons.}}
 \end{table*}
\end{center}
A Spearman correlation test was performed between each performance level and climate and air pollution parameters, in order to quantify the impact of weather and pollution on marathon performances. Spearman correlation tests were also performed between each environmental parameter. The year factor was not included because we previously demonstrated that for the past ten years, marathon performances were now progressing at a slower rate \cite{berthelot2010a}.\\[0.3cm]
\noindent
\textbf{Temperature and running speed}\\
We modeled the relation between running speed of each performance level for each gender and air temperature, using a second degree polynomial quadratic model, which seems appropriate to depict such physiological relations \cite{Zeng2009, Eliason2011, Kirschbaum2011}. The second degree polynomial equation was applied to determine the optimal temperature at which maximal running speed is achieved for each level of performance for each gender, and then used to calculate the speed decrease associated with temperature increase and decrease above the optimum. We similarly modeled the relation between air temperature and the percentage of runners\textquoteright~withdrawal. All analysis were performed using the MATLAB and SAS software.\\[0.3cm]
\textbf{\textsc{Results}}
\\[0.3cm]
The total numbers of starters and finishers of the 6 marathons increased over the 10 studied years (Fig. \ref{FigNOUR1}). Marathons characteristics are described in supplementary data (Tab. S1 \cite{SupInfA3}). The race with the least number of finishers was Boston 2001 with 13381 finishers and the highest number was seen in New York 2010 with 44763 finishers. Three marathons were held in April, the other three during fall. Air temperatures ranged from $1.7\,^{\circ}\mathrm{C}$ (Chicago 2009) to $25.2\,^{\circ}\mathrm{C}$ (Boston 2004) (Tab. \ref{TabNOUR1}).\\[0.3cm]
\noindent
\textbf{Performance distribution}\\
For all 60 studied races, the women and men\textquoteright s performance distributions were a good approximation of the log-normal and normal distributions ($p$-values of Kolmogorov-Smirnov statistics $=0.01$). Fig. \ref{FigNOUR2} illustrates examples of 4 races\textquoteright~performances distribution fit: men\textquoteright s performances distribution of two races in Paris (2002: $T\,^{\circ} = 7.6\,^{\circ}\mathrm{C}$; and 2007: $T\,^{\circ} = 17.4\,^{\circ}\mathrm{C}$) and Chicago (2002: $T\,^{\circ} = 5.4\,^{\circ}\mathrm{C}$; and 2007: $T\,^{\circ} = 25\,^{\circ}\mathrm{C}$).\\[0.3cm]
We notice a stable gap between male and female performances at all levels in all marathons, women being on average $10.3\%\pm1.6\%$ (mean $\pm$ standard deviation) slower than men (Tab. S1 \cite{SupInfA3}); mean female winners are $9.9\%\pm1.5\%$ slower than male winners, mean female median is $9.9\%\pm1.6\%$ than male median, and mean female Q1 are $11.1\%\pm1.5\%$ slower that male Q1.
\end{multicols}
\begin{center}
 \includegraphics[scale=0.8]{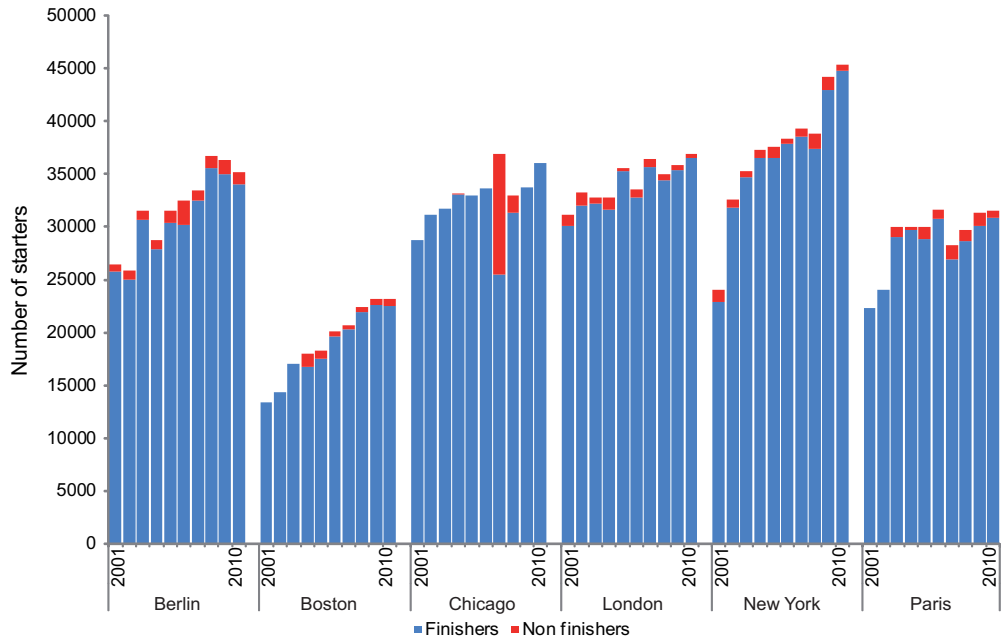}
\end{center}
\captionof{figure}[One figure]{\label{FigNOUR1} {\footnotesize Number of starters and finishers by marathon and year (missing data points for Boston, Chicago and Paris marathons).}}
\begin{multicols}{2}
\noindent
\textbf{Correlations}\\
Spearman correlations results are displayed in Tab. \ref{TabNOUR2}, detailed correlations by marathon are available in supplementary data (Tab. S2 \cite{SupInfA3}). The environmental parameter that had the most significant correlations with marathons performances was air temperature: it was significantly correlated with all performance levels in both male and female runners. Humidity was the second parameter with a high impact on performance; it was significantly correlated with women\textquoteright s P1 and men\textquoteright s all performance levels. The dew point and atmospheric pressure only had a slight influence ($p<0.1$) in men\textquoteright s P1 and women\textquoteright s P1 respectively, and did not affect the other performance levels.\\[0.3cm]
Concerning the atmospheric pollutants, NO$_2$ had the most significant correlation with performance: it was significantly correlated with Q1, IQR and the median for both genders. Sulfur dioxide (SO$_2$) was correlated with men\textquoteright s P1 ($p<0.01$) and had a slight influence ($p<0.1$) on men\textquoteright s Q1. Finally ozone (O$_3$) only had a slight influence ($p<0.1$) on men\textquoteright s Q1. In the marathon by marathon analysis, ozone ($O_3$) had the most significant correlation with performance (Tab. S2 \cite{SupInfA3}): it was significantly correlated with all performance levels (P1, Q1, IQR and the median) of the Berlin and Boston (except men\textquoteright s IQR) marathon for both genders. It also affected Chicago (men\textquoteright s P1, Q1, and men\textquoteright s median), and New York (women\textquoteright s Q1) marathons.\\[0.3cm]
\noindent
\textbf{Temperature and running speed}\\
When temperature increased above an optimum, performance decreased. Fig. \ref{FigNOUR3} describes the relationship between marathons running speeds and air temperature, fit through a quadratic second degree polynomial curve for women\textquoteright s P1 and men\textquoteright s Q1 of all 60 races. For each performance level the speed decrease associated with temperature increase and decrease is presented in supplementary data (Tab. S3 \cite{SupInfA3}). For example the optimal temperature at which women\textquoteright s P1 maximal running speed was attained was $9.9\,^{\circ}\mathrm{C}$, and an increase of $1\,^{\circ}\mathrm{C}$ from this optimal temperature will result in a speed loss of 0.03\%. The optimal temperatures to run at maximal speed for men and women, varied from $3.8\,^{\circ}\mathrm{C}$ to $9.9\,^{\circ}\mathrm{C}$ according to each level of performance (Tab. S3 \cite{SupInfA3}).\\[0.3cm]
Warmer air temperatures were associated with higher percentages of runners\textquoteright~withdrawal during a race (Fig. \ref{FigNOUR4}). After testing linear, quadratic, exponential and logarithmic fits, the quadratic equation was the best fit ($R^2 = 0.36$; $p<0.0001$) for modeling the percentage of runners withdrawals $\%\text{W}$ associated with air temperature (Fig. \ref{FigNOUR4}):
\begin{equation}
  \label{eQuad_WithdrawC}
    \%\text{W} = -0.59 T + 0.02 T^2 + 5.75
\end{equation}
\\[.3cm]
\textbf{\textsc{Discussion}}
\\[0.3cm]
Our study is the first to our knowledge to analyze the exhaustiveness of all marathon finishers\textquoteright~performances in the three major European (Berlin, Paris and London, which were not previously analyzed) and three American marathons. Previous studies have mostly analyzed American marathons including Chicago, Boston and New York that are analyzed in the present paper \cite{Marr2010, Ely2007a, Ely2007b, Ely2008, Montain2007, Martin1999}, but they have only included the performances of the top 3 males and females finishers as well as the 25th-, 100th-, and 300th- place finishers \cite{Ely2007a, Ely2008, Montain2007, Martin1999}. In the present study we analyzed the total number of finishers in order to exhaustively quantify the effect of climate on runners from all performance levels. Updating and extending earlier results, this study still concludes that the main environmental factor influencing marathon performance remains temperature. The pattern of performance reduction with increasing temperature is analogous in men and women, suggesting no apparent gender differences. In addition the mean gap between male and female performances is the same across all marathons and all performance levels (Tab. \ref{TabNOUR1}). This is consistent with our previous work that showed that the gender gap in athletic performance has been stable for more than 25 years, whatever the environmental conditions \cite{Thibault2010}.
\end{multicols}
\begin{center}
 \includegraphics[scale=0.8]{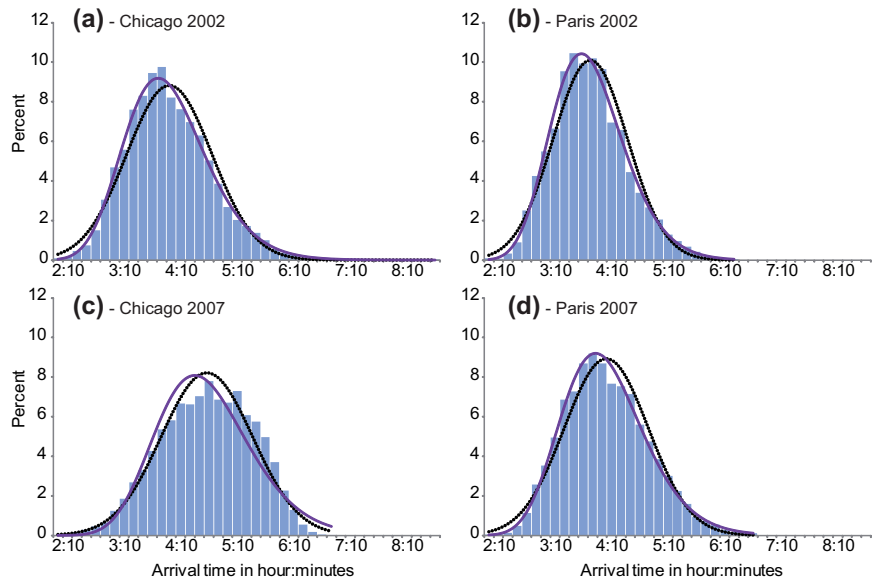}
\end{center}
    \captionof{figure}[One figure]{\label{FigNOUR2} {\footnotesize Distribution of performances: example of men\textquoteright s performances distribution for Chicago (in 2002: $T\,^{\circ} = 5.4\,^{\circ}\mathrm{C}$; and in 2007: $T\,^{\circ} = 25\,^{\circ}\mathrm{C}$); and Paris (in 2002: $T\,^{\circ} = 7.6\,^{\circ}\mathrm{C}$; and in 2007: $T\,^{\circ} = 17.4\,^{\circ}\mathrm{C}$).}}
\begin{multicols}{2}
The more the temperature increases, the larger the decreases in running speeds (Tab. S3 \cite{SupInfA3}). This is supported by the increased percentage of runners\textquoteright~withdrawals when races were contested in very hot weather (Fig. \ref{FigNOUR4}), and by the significant shift of the race\textquoteright s results through the whole range of performance distribution (Fig. \ref{FigNOUR2}). The significant effect of air temperature on the median values (Tab. \ref{TabNOUR2}) also suggests that all runners\textquoteright~performances are similarly affected by an increase in air temperature, as seen in Fig. \ref{FigNOUR2} showing performances distribution of races in Paris and Chicago with different air temperatures: the significant shift of performance towards the right concerns all runners categories, from the elite to the less trained competitors. In addition the percentage of runner\textquoteright s withdrawals in Chicago 2007 was the highest (30.74\%) among all 60 studied races (Fig. \ref{FigNOUR1} and \ref{FigNOUR4}). Roberts \cite{Roberts2010} reported that organizers tried to interrupt the race 3.5h after the start. This was not successful as most of the finishers crossed the finish line much later (up to 7h after the start); 66 runners were admitted to the hospital (12 intensive care cases with hydration disorders, heat shock syndromes and 1 death). During the 2004 Boston Marathon ($T\,^{\circ} = 22.5\,^{\circ}\mathrm{C}$) more than 300 emergency medical calls were observed, consequently the race\textquoteright s start time changed from noon to 10 am in order to decrease heat stress and related casualties \cite{Roberts2010}. The 2007 London Marathon was hot by London standards (air $T\,^{\circ} = 19.1\,^{\circ}\mathrm{C}$ vs. an average of $11.6\,^{\circ}\mathrm{C}$ for the nine other years analyzed in our study), 73 hospitalizations were recorded with 6 cases of severe electrolyte imbalance and one death, the total average time (all participants\textquoteright~average) was 17 min slower than usual. In contrast, the number of people treated in London 2008 in cool and rainy conditions ($T\,^{\circ} = 9.9\,^{\circ}\mathrm{C}$), was 20\% lower \cite{Roberts2010}. Our results showed that the percentage of runners\textquoteright~withdrawals from races significantly increases with increasing temperature (Fig. \ref{FigNOUR4}). The acceptable upper limit for competition judged by the American College of Sports Medicine (ACSM) is a WBGT of $28\,^{\circ}\mathrm{C}$, but it may not reflect the safety profile of unacclimatized, non-elite marathon runners \cite{Cheuvront2001, Roberts2010, Zhang1992, Armstrong1996}. Roberts \cite{Roberts2010} stated that marathons should not be allowed to start for non-elite racers at a WBGT of $20.5\,^{\circ}\mathrm{C}$. Our results suggest that there is no threshold but a continuous process on both side of an optimum: the larger the gap from the optimal temperature, the lower the tolerance and the higher the risk. In fact, in environments with high heat and humidity, not only is performance potentially compromised, but health is also at risk \cite{Maughan2007}; both are similarly affected. As soon as WBGT is higher than $13\,^{\circ}\mathrm{C}$ the rate of finish line medical encounters and on-course marathon dropouts begin to rise \cite{Roberts2010} as similarly seen in our study in Fig. \ref{FigNOUR4}.
\begin{center}
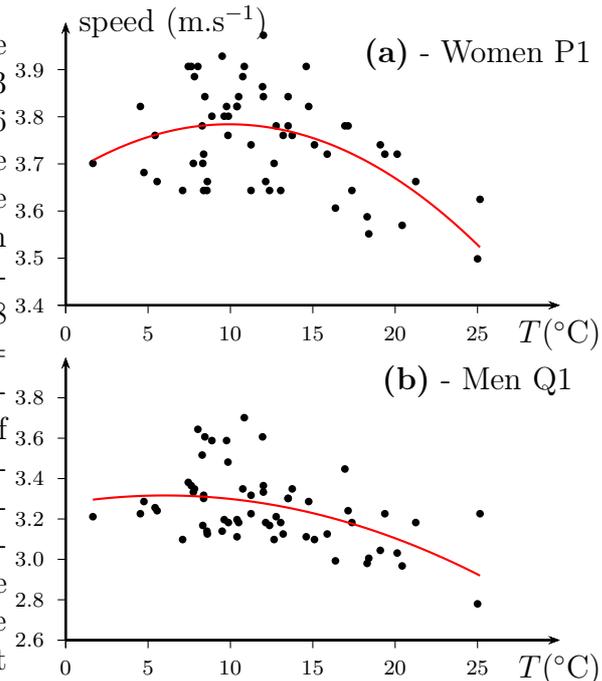

    \begin{tabular}{ l }
\psset{xunit=.2166667cm,yunit=6.25cm}
\begin{pspicture}(-0.75,3.4)(30,4.0)
    \psaxes[Dx=5,Dy=0.1,Ox=0,Oy=3.4,ticksize=-3pt,labelFontSize=\scriptstyle]{->}(0,3.4)(30,4.0)[$T (^{\circ}\mathrm{C})$,-90][speed (m.s$^{-1}$),0]
    \fileplot[plotstyle=dots, dotscale=0.8]{Figure31-B1.prn}
    \fileplot[plotstyle=line, linecolor=red]{Figure31-B2.prn}
    \rput[b](25,3.9){\textbf{(a)} - Women P1}
\end{pspicture} \\
\psset{xunit=.2166667cm,yunit=2.678571429cm}
\begin{pspicture}(-0.75,2.6)(30,4.2) 
    \psaxes[Dx=5,Dy=0.2,Ox=0,Oy=2.6,ticksize=-3pt,labelFontSize=\scriptstyle]{->}(0,2.6)(30,4.0)[$T (^{\circ}\mathrm{C})$,-90][,0]
    \fileplot[plotstyle=dots, dotscale=0.8]{Figure31-A1.prn}
    \fileplot[plotstyle=line, linecolor=red]{Figure31-A2.prn}
    \rput[b](25,3.8){\textbf{(b)} - Men Q1}
\end{pspicture}
    \end{tabular}
    \captionof{figure}[One figure]{\label{FigNOUR3} \textbf{a}. {\footnotesize Quadratic second degree polynomial fit for Women\textquoteright s P1 running speeds vs. air temperature ($T$), $R^2 = 0.27$; $p<0.001$; max = $9.9\,^{\circ}\mathrm{C}$ \textbf{b}. Men\textquoteright s Q1 running speeds vs. air temperature, $R^2 = 0.24$; $p<0.001$; max = $6\,^{\circ}\mathrm{C}$.}}
\end{center}
Warm weather enhances the risk of exercise induced hyperthermia; its first measurable impact is the reduction of physical performance \cite{Kenefick2007, Montain2007, Maughan2007, Hargreaves2008, Walters2000} as it is detrimental for the cardiovascular, muscular and central nervous systems \cite{Coyle2007, Gonzalez2007}. More recent work suggested that central fatigue develops before any elevation in body temperature occurs: evidence supported that subjects would subconsciously reduce their velocity earlier after the start of an exercise in hot environment, when internal temperatures are still lower than levels associated with bodily harm. Exercise is thus homeostatically regulated by the decrease of exercise intensity (decrease of running performance and heat production) in order to prevent hyperthermia and related catastrophic failures \cite{Tucker2004, Tucker2006}. On the other hand, cool weather is associated with an improved ability to maintain running velocity and power output as compared to warmer conditions, but very cold conditions also tend to reduce performance \cite{Maughan2007, Nimmo2004, Weller1997}.\\[0.3cm]
Among the studied races\textquoteright~winners, men\textquoteright s marathon world record was beaten in Berlin in 2007 and 2008 (Haile Gebrselassie in 02:03:59), as well as women\textquoteright s marathon world record, beaten in London 2003 (Paula Radcliffe in 02:15:25). The winners\textquoteright~speeds couldn\textquoteright t be affected in the same way than the other runners by air temperature and the other environmental parameters, because top performances can fluctuate from year to year due to numerous factors, such as prize money, race strategies, or overall competition \cite{Ely2007a}. Another explanation is that, in all of our 60 studied races, 89.5\% of male winners were of African origin (57.9\% from Kenya; 21.1\% from Ethiopia; and 10.5\% from Eritrea, Morocco and South Africa); as well as 54.5\% of female winners (27.3\% from Kenya and 27.3\% from Ethiopia- data not shown). African runners might have an advantage over Caucasian athletes, possibly due to a unique combination of the main endurance factors such as maximal oxygen uptake, fractional utilization of VO$_{2max}$ and running economy \cite{Larsen2003}. They might also perform better in warm environments as they are usually thinner than Caucasian runners (smaller size and body mass index) producing less heat with lower rates of heat storage \cite{Larsen2003, Marino2000, Marino2004}. Psychological factors may also play a role; some hypothesis suggested that regardless of the possible existence of physiological advantages in East African runners, belief that such differences exist may create a background that can have significant positive consequences on performance \cite{Hamilton2000, Baker2003b}.\\[0.3cm]
Genetics and training influence the tolerance for hyperthermia \cite{Kenefick2007, Larsen2003, Sawka2006}. Acclimatization involving repeated exposures to exercise in the heat also results in large improvements in the time to fatigue. Optimal thermoregulatory responses are observed in runners who have been acclimatized to heat and who avoid thirst before and during the race. Their best performances might be less influenced by temperature as winners had been more acclimatized to it \cite{Kenefick2007, Maughan2007, Hargreaves2008, Zouhal2009}. The avoidance of thirst sensation rather than optimum hydration prevents the decline in running performance \cite{Goulet2011}; contradicting the idea that dehydration associated with a body weight loss of 2\% during an exercise will impair performance, recent studies reported that Haile Gebrselassie lost 10\% of his body weight when he established his world record \cite{Goulet2011, Zouhal2011, Beis2012}.\\[.3cm]
Previous studies suggested that the impact of weather on speed might depend on running ability, with faster runners being less limited than slower ones \cite{Vihma2010, Ely2008, Montain2007, Maughan2007}. This could be attributable to a longer time of exposition to the environmental conditions of slower runners during the race \cite{Ely2007a}. Also, slower runners tend to run in closer proximity to other runners with clustering formation \cite{Alvarez-Ramirez2006, Alvarez-Ramirez2007}, which may cause more heat stress as compared with running solo \cite{Dawson1987}. These elements, however, are not supported after analyzing the full range of finisher\textquoteright s data; at a population level, temperature causes its full effect whatever the initial capacity. Differences in fitness relative to physiological potential may also contribute to differences in performance times and ability to cope with increasing heat stress \cite{Ely2007a, Alvarez-Ramirez2006, Alvarez-Ramirez2007}.\\[0.3cm]
There was a strong correlation of running speed with air temperature (Fig. \ref{FigNOUR3}). The maximal average speeds were performed at an optimal temperature comprised between $3.8\,^{\circ}\mathrm{C}$ and $9.9\,^{\circ}\mathrm{C}$ depending on the performance level (Tab. S3 \cite{SupInfA3}); small increases in air temperatures caused marathon performances to decline in a predictable and quantifiable manner. On the other hand, large decreases in air temperatures under the optimum also reduce performances. These optimal temperatures found in the present study are comprised in the optimal temperature range of $5-10\,^{\circ}\mathrm{C}$ WBGT found in previous studies \cite{Montain2007}; other studies stated that a weather of $10-12\,^{\circ}\mathrm{C}$ WBGT is the norm for fast field performance and reported a decrease of performance with increasing WBGT \cite{Ely2007b, Zhang1992, Galloway1997, Buoncristiani1983}. Best marathon times and most marathon world records were achieved in cool environmental temperatures ($10-15\,^{\circ}\mathrm{C}$) and have been run in the early morning during spring and fall \cite{Ely2007b}. Analyzing Gebrselassie\textquoteright s performances in Berlin reveals that they follow the same trend, with both World Records obtained at the lowest temperatures ($14\,^{\circ}\mathrm{C}$ in 2007 and $13\,^{\circ}\mathrm{C}$ in 2008, vs. $18\,^{\circ}\mathrm{C}$ in 2009 and $22\,^{\circ}\mathrm{C}$ in 2006 when he also won these two races without beating the world record).
\begin{center}
 \includegraphics[scale=0.8]{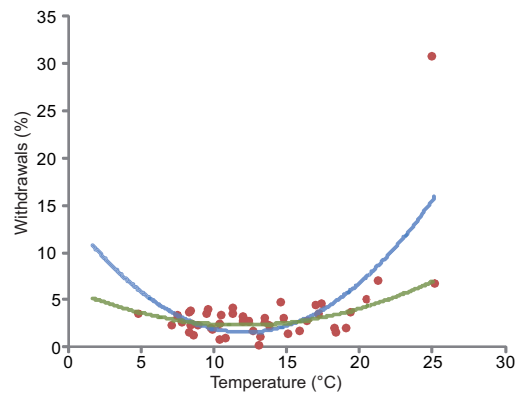}
\end{center}
\captionof{figure}[One figure]{\label{FigNOUR4} {\footnotesize Relationship between air temperature and the percentage of runners\textquoteright~withdrawals, modeled with a quadratic fit (blue curve, $R^2 = 0.36$; $p<0.0001$). The green curve represents the quadratic fit without the maxima (Chicago 2007: 30.74\% withdrawals at a race temperature of $25\,^{\circ}\mathrm{C}$).\\[.1cm]}}
The relationship between running speed and air temperature defined in our study (Fig. \ref{FigNOUR3}) is similar to the relationship found between mortality and air temperature (asymmetrical U-like pattern) in France defined by Laaidi et al \cite{Laaidi2006}, where mortality rates increase with the lowest and the highest temperatures. A \textquoteleft thermal optimum\textquoteright~occurs in between, where mortality rates are minimal \cite{Laaidi2006}. The great influence that temperature has on performance is comparable to the influence it has on mortality, suggesting that both sports performance and mortality are thermodynamically regulated. This also emphasizes the utility of prevention programs, the assessment of public health impacts and acclimatization before participating in hot marathons \cite{Laaidi2006}. Similar correlations were also found between temperature and swimming performance in juvenile southern catfish \cite{Zeng2009}, and between increases in summer water temperature and elevated mortality rates of adult sockeye salmon \cite{Eliason2011}; suggesting that physiological adaptations to temperature, similarly occur in various taxons, but vary within specific limits that depend on species and will modify performances.
\begin{center}
 \begin{table*}[htbp!]
 \begin{center}
 \small
    \begin{tabular}{ | l | l | l | l | l | l |}
        \hline
            Parameter & Gender & P1 & Median & Q1 & IQR \\
            \hline
            Temperature & Women & 0.31* & 0.30* & 0.35** & 0.15\\
            & Men & 0.48*** & 0.40*** & 0.44*** & 0.25\$\\
            Dew Point & Women & 0.14 & 0.18 & 0.21 & 0.01\\
            & Men & 0.25\$ & 0.19 & 0.20 & 0.10\\
            Humidity & Women & -0.3* & -0.16 & -0.19 & -0.21\\
            & Men & -0.34** & -0.28* & -0.32* & -0.19\\
            Atm. Pressure & Women & 0.22\$ & 0.06 & 0.07 & 0.06\\
            & Men & 0.13 & 0.04 & 0.06 & 0.06\\
            NO$_2$ & Women & 0.11 & 0.40** & 0.43*** & 0.33*\\
            & Men & 0.25\$ & 0.38** & 0.35** & 0.27*\\
            O$_3$ & Women & 0.01 & -0.15 & -0.11 & -0.20\\
            & Men & -0.05 & -0.21 & -0.24\$ & -0.11\\
            PM$_{10}$ & Women & 0.08 & 0.15 & -0.25 & 0.03\\
            & Men & 0.10 & 0.10 & 0.09 & 0.16\\
            SO$_{2}$ & Women & 0.21 & 0.13 & 0.21 & 0.02\\
            & Men & 0.37** & 0.20 & 0.25\$ & 0.04\\
        \hline
    \end{tabular}
 \caption{\label{TabNOUR2} {\footnotesize Spearman correlations results between all marathons performance levels and environmental parameters: \$ = $p<0.1$; * = $p<0.05$; ** = $p<0.01$; *** = $p<0.001$.}}
  \end{center}
 \end{table*}
\end{center}
\noindent
\textbf{Air pollution and performance}\\
The measured levels of pollution had no impact on performance, except for ozone (Tab. S2 \cite{SupInfA3}) and NO2 (Tab. \ref{TabNOUR2}). Assessing the effect of any single air pollutant separately is not simple; it is not isolated in the inhaled air, but rather combined with other parameters. Therefore any possible influence might probably be due to a combination of components. In addition most marathons are held on Sunday mornings, when urban transport activity and its associated emissions are low, and photochemical reactions driven by solar radiation have not yet produced secondary pollutants such as ozone \cite{Marr2010}. This is the most probable explanation to our results, confirming previous studies. Among the air pollutants analyzed in the present study, ozone and NO$_2$ had the greatest effect on decreasing marathon performances (Tab. S2 \cite{SupInfA3}). Ozone concentrations on the ground increase linearly with air temperature \cite{Shephard1984, Chimenti2009, Lippi2008b}; thus the effect of ozone in our study may be mainly associated with the temperature effect, as seen in Berlin and Chicago. However ozone and other pollutants effects are known to be detrimental to exercise performance only when exposure is sufficiently high. Many studies showed no effect of air pollutants on sports performance \cite{Marr2010}. Some of them showed that PM$_{2.5}$ and aerosol acidity were associated with acute decrements in pulmonary function, but these changes in pulmonary function were unlikely to result in clinical symptoms \cite{Korrick1998}. Others showed that chronic exposure to mixed pollutants during exercise may result in decreased lung function, or vascular dysfunction, and may compromise performance \cite{Rundell2012}. During the marathons studied here, concentrations of air pollutants never exceeded the limits set forth by national environmental agencies (US Environmental Protection Agency- EPA; AirParif; European Environmental Agency- EEA) or the levels known to alter lung function in laboratory situations \cite{Marr2010}.\\[0.3cm]
\textbf{\textsc{Conclusion}}
\\[0.3cm]
Air temperature is the most important factor influencing marathon running performance for runners of all levels. It greatly influences the entire distribution of runners\textquoteright~performances as well as the percentage of withdrawals. Running speed at all levels is linked to temperature through a quadratic model. Any increase or decrease from the optimal temperature range will result in running speed decrease. Ozone also has an influence on performance but its effect might be linked to the temperature impact. The model developed in this study could be used for further predictions, in order to evaluate expected performance variations with changing weather conditions.
\end{multicols}
\section{Performance development of the best performers}
\label{sec:Performance}
\begin{wrapfigure}{l}{0.5\textwidth}
   \begin{center}
\psset{xunit=.12cm,yunit=.05cm}
\begin{pspicture}(1972,480)(2018,560)
    \psaxes[Dx=10,Dy=10,Ox=1972,Oy=480,ticksize=-3pt,labelFontSize=\scriptstyle]{->}(1972,480)(2018,560)[year,-90][time (s.),0]
    \fileplot[plotstyle=dots, dotscale=0.5]{Figure17.prn}
    \psframe[linecolor=red, linewidth=0.5pt](1992.4,505)(1993.6,483)
    \rput[b](2004,483){\scriptsize CHN performances}
\end{pspicture}
   \end{center}
   \caption{\label{Fig.2.1: 1993 in TF} {\footnotesize Development of performances in the 3000m women (T\&F). The red frame encapsulates the outstanding performances established by the Chinese women athletes in 1993. They were established in the same competition in Beijing on the 12\textsuperscript{th} and 13\textsuperscript{th} September and 4 of them on the same race, a singular moment in track and field history.}}
\end{wrapfigure}
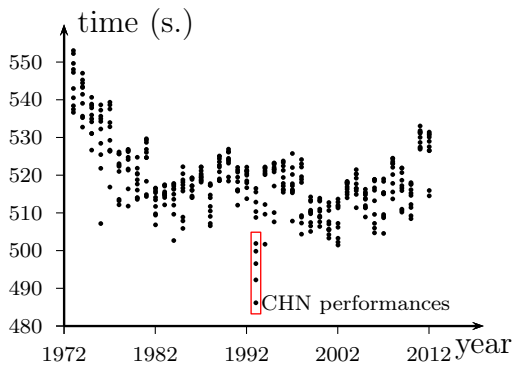

The two previous sections were only based on the WR metric \cite{berthelot2008, fdd1}. Although it remains a good representation of the development of ultimate physiology during the past century, it did not provide insights into the development of performance of athletes, who are not record holders. WR are a monotonic measure of physiology and the analysis of both their frequency and relative improvement can shred to light historical and geopolitical connections \cite{marion1, berthelot2008, fdd1}. However, in order to investigate a finest measure of physiology and its fluctuations during the past 120 years, we gather the best performance of the 10 best performers each year. This data is available in 36 T\&F events and 34 swimming events and represents a data-set of more than $40,000$ performances. Performances show a S-Shaped development, in line with the one appearing in WR. They also reveal that performance development is synchronized with historical events. Performance considerably decreases during both world wars: -0.44\% during 6 years (1913 to 1918 for WWI and -0.45\% in a 6-year time span from 1940 to 1945 for WWII. Performance also shows a strong increase during the sports globalization period, starting in 1950 \cite{marion1}. These different phases of fluctuation are synchronized for all the 70 events and are modeled using a Gompertz function:
\begin{equation}
  \label{eqGompertz}
  y(t) = a \exp^{b \exp^{c \cdot t}} + d
\end{equation}
It is a special case of the generalized logistic function (or Richards\textquoteright~curve) \cite{Richards1959}. This function is used for time series and presents an asymmetrical shape: the right-hand (or future value) of the function is approached much more gradually by the curve than the left-hand or lower valued asymptote. In contrast to the simple logistic function in which both asymptotes are approached by the curve symmetrically. The model shows good statistics on the 147 periods fitted (mean $R^2 = 0.68\pm0.22$). As many S-Shaped models, it admits a limit as time increases. We compute the estimated dates of the limit in each of the 70 events based on the approach developed in section \ref{sec:OlympicSports} and \ref{sec:NonOlympicSports} (also see \cite{berthelot2008, fdd1}). We find that 64\% of T\&F events no longer progress in 2010 (the first event stops its progression in 1981). On another hand, 47\% of the swimming events stagnate after 1990 before the introduction of a new generation swimsuits: the LZR racer. Since then, 100\% of the swimming events show a new period of progression in 2008 \& 2009. This is yet another example of a new period triggered by the diffusion of a new technology among elite athletes. We also focus on the development of performances between each Olympic games, and find an Olympic seasonality with an increase of performances at each Olympic year of $1\%\pm0.6$ and a decrease of $-0.3\%\pm0.5$ the following year.\\[0.3cm]
In these sets, we deal with samples of 10 performances every year. This data-set allows for the additional quantification of the \textquoteleft atypicity\textquoteright~of the yearly first performer. We introduce three statistical descriptors $d_1$, $d_2$ and $d_3$ that provide a measurement of atypicity. The first descriptor is the distance of the first performer to the other 9 performers. The second descriptor is the \textquoteleft durability\textquoteright~of the performance (ie. how long it will last until being beaten by another performer). The last descriptor is the variability introduced by the single studied performance over the remaining performance of the event. The standardized sum of the three descriptors defines the atypicity \textbf{A} of a given first performer. The value of \textbf{A} peaks at 4 dates in T\&F: 1916, 1943, 1968 and 1988 and the highest values occurs in 1943 and 1988. No peak is found in swimming. Additionally, we study the outlying values of each descriptor\textquoteright~distribution and find three particular dates in T\&F: 1943, 1988 and 1993 and one date in swimming (1994). Sport\textquoteright s environment, training volume (and conditions), competitions are not standardized between the performers before the before the WWII \cite{marion1}. Therefore it is difficult to explain the peak observed in \textbf{A} and in the different descriptors. On the other hand, the peaks observed in 1968, 1988 and 1993 partly correspond to historical events: the Olympic games of Mexico (1968), the Olympic games of Seoul and the occurrence of a number of atypical performances in 1988. The year 1993 is a post Olympic year (1992 Barcelona), and a decrease of top performances is excepted. However, Chinese women athletes achieve exceptional performances and hold 33\% of the first performances, 33\% of the second performances and 39\% of the third performances. These ratios have never been equalled by China since then. The same year at the Chinese National Games in Beijing, 5 Chinese women athletes beat the 3000m WR, a singular moment in T\&F history (Fig. \ref{Fig.2.1: 1993 in TF}). In swimming, the year 1994 also corresponds to a doping affair, where 7 Chinese athletes were caught by a surprise test at the 1994 Asian Games in Hiroshima, Japan \cite{NYT1}. Fish, Hersh, Whitten and Yesalis also questioned the role of East German coaches in China\textquoteright s sport programs after the fall of Communism in Europe \cite{Fish1994, Hersh1993, Whitten1994, Yesalis2002}.\\[0.3cm]
During the London Olympic games in 2012, Ye Shiwen, a 16-year-old Chinese athlete, shattered the world record in the swimming women\textquoteright s 400m individual medley. It bore numerous interrogations, especially in the journal Nature, where a press article depicted Ye\textquoteright s performance as \textquoteleft anomalous\textquoteright~\cite{Callaway2012}.
According to the editor: ``[the article] \textit{draw an extraordinary level of outraged response. The volume of comments has been so great that our online commenting system is unable to cope: it deletes earlier posts as new ones arrive}''. In fact, criticism about the methodology rapidly arose. The article is using \textquoteleft performance profiling\textquoteright~ as a robust method to nab atypical athletes: it uses the relative performance improvement between two consecutive years to state that the athlete may have used doping substances. Using only the previous year\textquoteright s progression rate as a reference appears limited and the whole career seems necessary to better understand typical and abnormal trajectories. However, it is not clear if such transitions between typical and atypical states during the career is a memoryless process. As an example some athletes pop out from the top performers and suddenly establish exceptional marks (a famous case was Bob Beamon in the 1968 Olympics), while some others always exhibit exceptional performances with normal progression rates since their younger age (such as Usain Bolt). Nonetheless, it reveals the appealing of the scientific and non-scientific community for such statistical approach, complementary to traditional biological analyzes. A number of different methods would help classify and identify atypical trajectories: supervised (provided the data allow for the construction of a robust learning data-set) or unsupervised learning algorithms, LMS method \cite{Cole1998} and analyzing the distributions of performances variations. Interesting careers to start the study on are those of Shelly-Ann Fraser, Marion Jones or Florence Griffith-Joyner.\\[0.3cm]
Another example is given by the recent development in road cycling. El Helou reported a distinctive 6.38\% improvement in speed in European professional road racing from 1993 onwards \cite{nour1}. A period which coincides with the years of intense use of EPO in professional cycling. In January 2013, Lance Armstrong, the rider with most wins (seven victories between 1999 and 2005) admitted doping in a television interview, despite having made denials throughout his career \cite{ESPN1}. This testimony strengthen the observations of El Helou and reveals that, despite confounding variables that may distort evaluations of riders\textquoteright~accomplishments in professional road races over the years, the last progression period in road cycling may have been strongly influenced by the use of pharmacological means \cite{nour1}.\\[0.3cm]
The article \cite{berthelot2010a} is presented in the following text:\\[0.6cm]
\noindent
\textbf{Athlete Atypicity on the Edge of Human Achievement: Performances Stagnate after the Last Peak, in 1988}\\
G. Berthelot, M. Tafflet, N. El Helou, S. Len, S. Escolano, M. Guillaume, H. Nassif, J. Tolaïni, V. Thibault, F.-D. Desgorces, O. Hermine, J.-F. Toussaint
\\[0.6cm]
\noindent
\textbf{\textsc{Abstract}}
The growth law for the development of top athletes performances remains unknown in quantifiable sport events. Here we present a growth model for 41351 best performers from 70 track and field (T\&F) and swimming events and detail their characteristics over the modern Olympic era. We show that 64\% of T\&F events no longer improved since 1993, while 47\% of swimming events stagnated after 1990, prior to a second progression step starting in 2000. Since then, 100\% of swimming events continued to progress. We also provide a measurement of the atypicity for the 3919 best performances (BP) of each year in every event. The secular evolution of this parameter for T\&F reveals four peaks; the most recent (1988) followed by a major stagnation. This last peak may correspond to the most recent successful attempt to push forward human physiological limits. No atypicity trend is detected in swimming. The upcoming rarefaction of new records in sport may be delayed by technological innovations, themselves depending upon economical constraints.
\begin{multicols}{2}
Sport performances may cease to improve during the XXI\textsuperscript{th} century, possibly due to physiological limits \cite{berthelot2008} and interactions between genomic \cite{MacArthur2007} and environmental parameters \cite{fdd1}. The progression of top athletes\textquoteright~performances remains unknown for quantifiable sport events. We hypothesize that such an evolution also mirrors our social and historical development \cite{marion1} and symbolize our quest for the Citius. Modeling this progression enables us to identify the underlying trends and behaviors in sports and history on a world scale. After the publication of initial mathematical models \cite{nevill1, nevill2}, world records (WR) were shown to follow a piecewise exponential model \cite{berthelot2008} over the modern Olympic era (1896-2007). The present study is based on the analysis of a large scope of sport performances as an indicator of our species\textquoteright~physiological maxima. It encloses the human physical potential and may be seen as a complement to direct laboratory measurements on a sample of elite athletes \cite{Schumacher2009}.\\[.3cm]
In this study we analyze the single best result of the top 10 world performers every year from two major olympic disciplines (track \& field - T\&F - and swimming) in order to establish their law of progression after one century of sport development. We also introduce the concept of \textquoteleft atypicity\textquoteright, defined as the singularity trait of a given performance, and we measure its trend in the sport law of progression. Atypicity is seen in all systems. A previous study ranked T\&F WR using extreme value theory \cite{Einmahl2008} but did not investigate their temporal tendencies or the relationship between the best performer and the other athletes. A closer observation allows for scoring all performances using 3 statistical descriptors, characterizing atypicity, and providing a reliable trend of the avant-garde of human performers.\\[0.3cm]
\textbf{\textsc{Material and methods}}
\\[0.3cm]
We collected the single best result of the top 10 world performers every year in 70 events from the major two quantifiable Olympic disciplines: 36 T\&F events over the modern Olympic era (1891-2008) and 34 swimming events over the 1963-2008 period \cite{Swimnews} \cite{FINA} \cite{Rabinovich2010}. A total number of 41351 performances including 3919 Best Performances (BP) were gathered.\\[0.3cm]
\noindent
\textbf{Growth law}\\
The law of progression after one century of sport development is modeled using a Gompertz function, widely used in biology \cite{Rossi2003}, economic dynamics \cite{Jarne2005} or technology diffusion \cite{Michalakelis2008, Cardenas2004}. The physiological limit for each event was given by computing the year corresponding to 99.95\% (1/2000\textsuperscript{th}) of the estimated asymptotic value (Tab S1 \cite{SupInfA4}). Events presenting a limit before 2008 were considered as \textquoteleft halted\textquoteright~events. In addition, we conducted a residual analysis (Materials and Methods S1 \cite{SupInfA4}) and determined the Most Recent Change Of Incline (MRCOI, Materials and Methods S1 \cite{SupInfA4}) for swimming events.\\[0.3cm]
\noindent
\textbf{Descriptive Analysis}\\
Descriptive statistics were also conducted to assess the impact of the Olympic Games on performances every four years and to measure the yearly variation between each performer using the yearly mean relative performance improvement $\overline{\kappa}$ and the coefficient of variation $\overline{cv(t)}$.\\[1cm]
\noindent
\textbf{Measurement of the Atypicity}\\
We finally focused on the \textit{atypicity} of each BP using a set of specific descriptors: $d_1$ measured its relative distance to all other performances during the year, $d_2$ was the \textquoteleft durability\textquoteright~of a BP over the years before it is beaten by another performance and $d_3$ characterized the weight of each BP over all other performances for each event during the Olympic era. The highest 5\% values of each descriptor are selected and studied for both disciplines. We define \textquoteleft \textit{atypicity}\textquoteright~\textbf{A} as the distance from each BP to the origin of the descriptors\textquoteright~uniformized Euclidian space.
\end{multicols}

\begin{center}
\begin{tabular}{l l l}
\psset{xunit=.05cm,yunit=.14cm}
\begin{pspicture}(1890,45)(2015,70)
    \psaxes[Dx=20,Dy=5,Ox=1890,Oy=45,ticksize=-3pt,labelFontSize=\scriptstyle]{->}(1890,45)(2015,70)[,-90][Performance (s.),0]
    \fileplot[plotstyle=dots, dotscale=0.5]{Figure18-A1.prn}
    \fileplot[plotstyle=line, linecolor=red]{Figure18-A2.prn}
    \fileplot[plotstyle=line, linecolor=red]{Figure18-A3.prn}
    \rput[b](2010,65){\textbf{(a)}}
\end{pspicture} & \hspace{0.5cm} &
\psset{xunit=.1cm,yunit=.25cm}
\begin{pspicture}(1960,58)(2020,72)
    \psaxes[Dx=10,Dy=2,Ox=1960,Oy=58,ticksize=-3pt,labelFontSize=\scriptstyle]{->}(1960,58)(2020,72)[year,-90][Performance (s.),0]
    \fileplot[plotstyle=dots, dotscale=0.5]{Figure18-C1.prn}
    \fileplot[plotstyle=line, linecolor=red]{Figure18-C2.prn}
    \fileplot[plotstyle=line, linecolor=red]{Figure18-C3.prn}
    \rput[b](2015,69){\textbf{(c)}}
\end{pspicture} \\
\psset{xunit=.05cm,yunit=1.2cm}
\begin{pspicture}(1890,6.5)(2015,10) 
    \psaxes[Dx=20,Dy=0.5,Ox=1890,Oy=6.5,ticksize=-3pt,labelFontSize=\scriptstyle]{->}(1890,6.5)(2015,9.5)[,-90][Performance (m.),0]
    \fileplot[plotstyle=dots, dotscale=0.5]{Figure18-B1.prn}
    \fileplot[plotstyle=line, linecolor=red]{Figure18-B2.prn}
    \fileplot[plotstyle=line, linecolor=red]{Figure18-B3.prn}
    \fileplot[plotstyle=line, linecolor=red]{Figure18-B4.prn}
    \rput[b](2010,9.0){\textbf{(b)}}
\end{pspicture} & \hspace{0.5cm} &
\psset{xunit=.1cm,yunit=0.0144cm}
\begin{pspicture}(1960,850)(2020,1100)
    \psaxes[Dx=10,Dy=50,Ox=1960,Oy=850,ticksize=-3pt,labelFontSize=\scriptstyle]{->}(1960,850)(2020,1100)[year,-90][Performance (s.),0]
    \fileplot[plotstyle=dots, dotscale=0.5]{Figure18-D1.prn}
    \fileplot[plotstyle=line, linecolor=red]{Figure18-D2.prn}
    \fileplot[plotstyle=line, linecolor=red]{Figure18-D3.prn}
    \rput[b](2015,1055){\textbf{(d)}}
\end{pspicture}
\end{tabular}

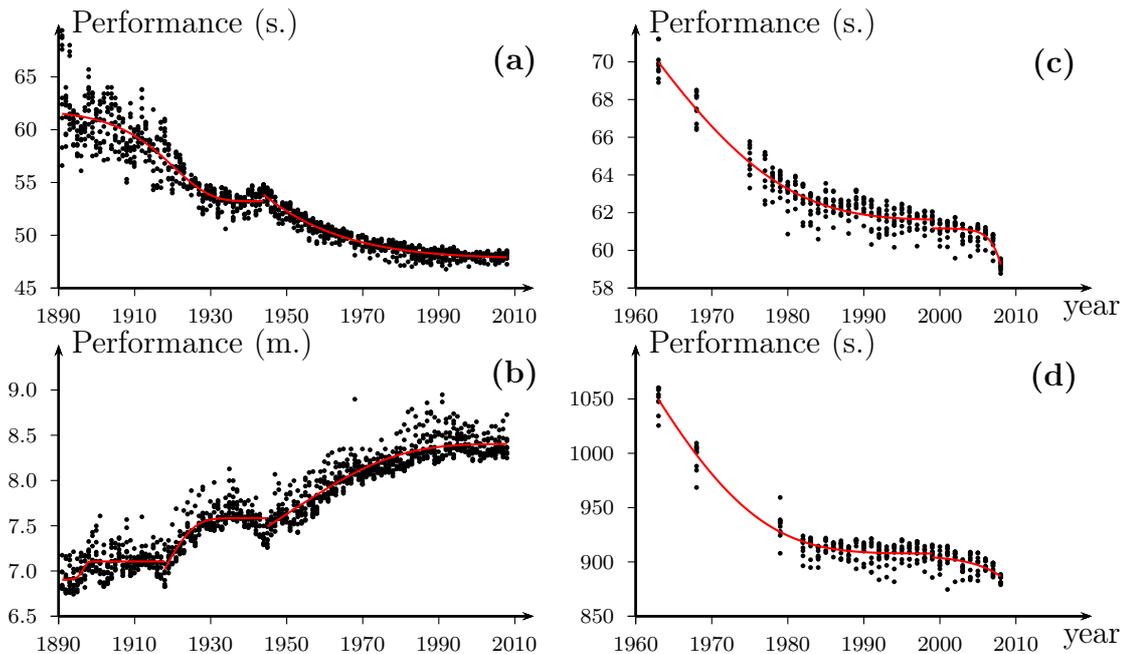
\captionof{figure}[One figure]{\label{FigGB1} {\footnotesize \textbf{a}. Men 400m hurdles (T\&F) fitting ($R^2_i = 0.77$; $R^2_{ii} = 0.91$), progressing event. \textbf{b}. Men long jump (T\&F) ($R^2_i = 0.22$; $R^2_{ii} = 0.55$; $R^2_{iii} = 0.81$), \textquoteleft halted\textquoteright~event since 2001.5 (Tab. S1 \cite{SupInfA4}). \textbf{c}. Women 100m back (swimming) ($R^2_i = 0.90$; $R^2_{ii} = 0.63$). \textbf{d}. Men 1500m freestyle (swimming) ($R^2_i = 0.92$; $R^2{ii} = 0.31$). Fewer data are available in swimming (\textbf{c}, \textbf{d}) between 1963 and 1977. Prior to the MRCOI, C was in progression and \textbf{d} was a \textquoteleft halted\textquoteright~event since 1991.9. A new progression trend appears after the MRCOI (2000 (\textbf{c})/1999 (\textbf{d})) for swimming events, as a result of the introduction of swimsuits.}}
\end{center}
\begin{multicols}{2}
\textbf{\textsc{Results}}
\\[0.3cm]
\noindent
\textbf{Growth law}\\
The average adjusted $R^2$ for all 147 historical curves is $0.68\pm0.22$ (mean $\pm$standard deviation).

Among T\&F events, 63.9\% no longer progress (77.8\% of the 18 women events; 50\% of the 18 men events). The average year for the detected dates of halt in performance is $1992.8\pm7.9$ ($1991.8\pm8.0$ for women; $1994.8\pm7.9$ for men). Dates of halt range from 1980.9 (1500m women) to 2007.1 (triple jump men, Tab. S1 \cite{SupInfA4}).

The changes of incline in the T\&F curve are mostly related to World Wars I and II (WWI, WWII) or changes in event timing methods, improving measurement accuracy, which is especially important over short events. However, in the last twenty years no rule alterations or technological improvements were made that could have resulted in a new curve.

Thirteen T\&F events still progress (4 women\textquoteright s and 9 men\textquoteright s), nine of which are middle and long distance races.

Among swimming events, 100\% still progress. The average MRCOI in swimming is $1997.2\pm2.7$ and the peak appears in 2000. Specifically, the mean MRCOI for fly and breast swim styles is $1994.4\pm2.5$ while the mean MRCOI for freestyle and back is $1998.1\pm2.1$. Prior to the MRCOI, progression in 16 (47.1\%) of the 34 events had halted: 8 women and 8 men events. Average year of halt is $1990\pm4.4$ ($1988.6\pm4.8$ for women, $1991.5\pm3.6$ for men).\\[0.3cm]
\noindent
\textbf{Descriptive Analysis}\\
The impact of WWI and WWII for T\&F events over $\overline{\kappa_t}$ are computed: a regression of -0.44\% of the mean performances within a six-year time span from 1913 to 1918 is observed for WWI; and -0.45\% in a 6-year time span from 1940 to 1945 for WWII.

Average Olympic periodicity is measured with a mean increase of T\&F performances of $0.99\%\pm0.56$ for Olympic year $t$, $-0.32\%\pm0.49$ for $t+1$, $0.48\%\pm0.45$ for $t+2$ and $0.37\%\pm0.41$ for $t+3$.

The mean yearly coefficient of variation $\overline{cv(t)}$ regresses over the century and range from $3 \times 10^{-2}$ (1891) to $1.08 \times 10^{-2}$ (2008) for T\&F, and $0.81 \times 10^{-2}$ (2008) for swimming.\\[0.3cm]
\noindent
\textbf{Atypical Performances Measured through A}\\
All of the descriptors\textquoteright~distributions for T\&F and swimming reveal a right-skewed profile: outlying performances are located in right tails.

Secular evolution of the highest 5\% values for each descriptor reveals 4 historical peaks in T\&F: 1943 (14 cases), 1988 (13 cases), 1993 (7 cases) and 1998 (16 cases). Evolution of the highest 5\% values remains steady in swimming over the period 1982-2008 except in 1994 where 17 cases are spotted.\\[.3cm]
The secular trend is given for each descriptor: descriptor $d_1$ decreases throughout the century, reaching a stable value in the 1980\textquoteright s, showing that the distance from the best performer to the others is decreasing, except in the WWII era\textquoteright s peak in 1943. Life expectancy or \textquoteleft durability\textquoteright~of a BP ($d_2$) is stable over the century, with high values appearing in the 1980\textquoteright s in T\&F. The most \textquoteleft durable\textquoteright~BPs were established during this period. The deviation produced by each BP over all other performances of the same event ($d_3$) is stable over the century. High values are found during the following years: 1943 ($d_1$, $d_3$), 1988 ($d_2$), 1993 ($d_3$) for T\&F and 1994 ($d_1$, $d_4$) for swimming.\\[.3cm]
Historical evolution of yearly \textquoteleft Atypicity\textquoteright~(\textbf{A}) shows 4 peaks in T\&F: 1916, 1943, 1968 and 1988 of amplitudes 0.39, 0.46, 0.31 and 0.46 respectively. Four corresponding cycles were found with durations of: 21, 28, 23 and 27 years accordingly.\\[.3cm]
There are no significant variations of \textbf{A} in swimming between 1982 and 2008.\\[0.3cm]
\textbf{\textsc{Discussion}}
\\[0.3cm]
\noindent
\textbf{Progression Law of the \textquoteleft 10 Best\textquoteright~and Descriptive Statistics}\\
Analysis of top athletes\textquoteright~performances suggests that the progression of human performances may reach its limit soon \cite{berthelot2008, fdd1, nevill1, nevill2, Einmahl2008}. WR were shown to follow a similar development \cite{berthelot2008, fdd1} that also highlighted that $1/4$ of T\&F WR had reached their limit in 2008. We here introduce a new tool to describe the physiological dynamics of elite sport performances by modeling growth curves over a large data sample in 34 events since 1963 and 36 events over 118 years.\\[.3cm]
The two studied disciplines show different progression schemes: most (63.9\%) of T\&F events have stopped progressing since $1993\pm8$ years while all swimming events were progressing until 2009 (Fig. \ref{FigGB1}). This halt occurred 34 years earlier than the estimated stagnation of half of the WR in 5 Olympic disciplines \cite{berthelot2008}; it may reveal that most of T\&F athletes are already beyond the edge of stagnation. Both genders present a slightly different evolution in T\&F events, suggesting male events still have some potential reserve whereas the majority of women events (77.8\%) has stopped progressing since $1992\pm8$ years. Women may have reached their limits before men, despite a later entry into Olympic competition \cite{Thibault2010}. Positive values of the mean relative performance improvement are also attributable to the recent introduction of women events (pole vault, marathon, triple jump and 1500m).
\end{multicols}

\begin{center}
\includegraphics[scale=0.55]{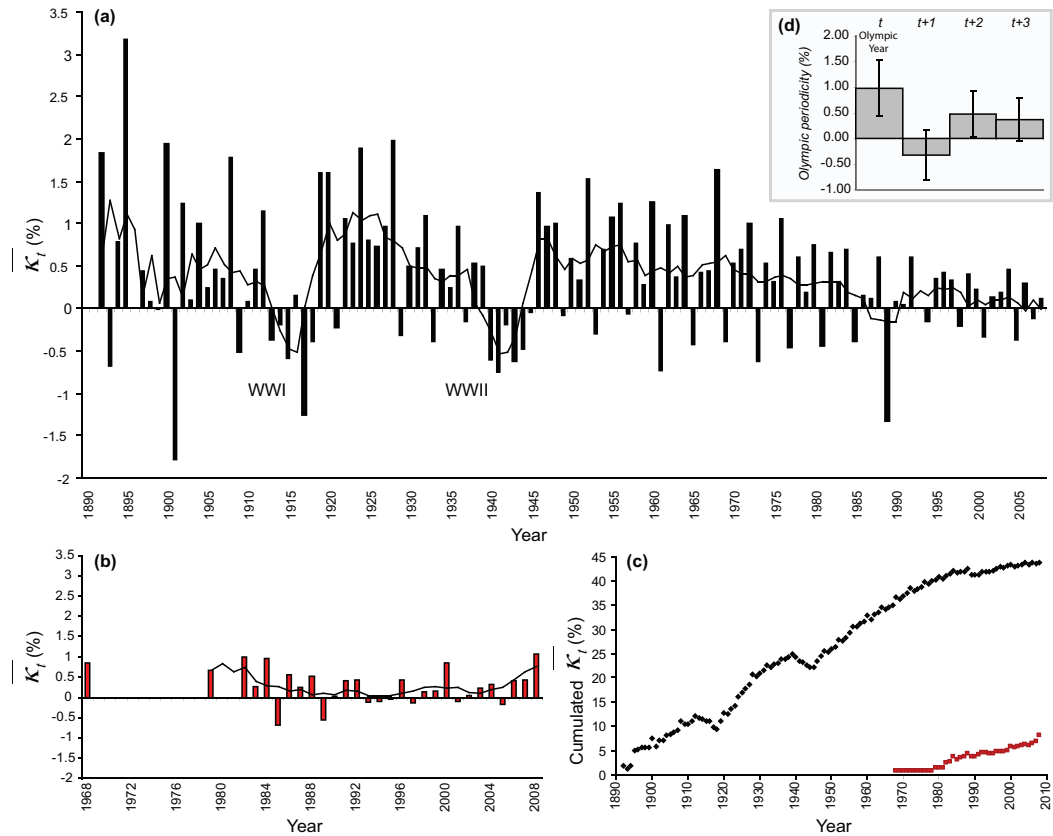}
\captionof{figure}[One figure]{\label{FigGB2} {\footnotesize Evolution of performance improvement $\overline{\kappa_t}$ and Olympic periodicity. \textbf{a}. Evolution of $\overline{\kappa_t}$ in T\&F. \textbf{b}. Evolution of $\overline{\kappa_t}$ in swimming. \textbf{C}. Cumulative evolution of $\overline{\kappa_t}$ in T\&F (black dots) and swimming (red squares). \textbf{d}. Average Olympic periodicity with standard deviation in T\&F and swimming. Both $\overline{\kappa_t}$ evolutions in T\&F and swimming (\textbf{a}, \textbf{b}) are given with a 4 year smoothing average (black line). After a large period of performance improvement, hindered by the two World Wars, the development of performance slow down since 1960. A larger regression is observed in 1989. At this time, random anti-doping tests were established. The impact of the Olympic games, aka Olympic periodicity reveals that performances increase by 1\% during an Olympic year ($t$), while they regress in the following year ($t+1$).}}
\end{center}
\begin{multicols}{2}
The analysis of residuals ($\overline{\text{YADR}}$) in T\&F events reveals the impact of both world wars on performance development (Fig. S1 \cite{SupInfA4}). Following the Cold War period, a large regression (3.07\%) is noticed in 4 events (shot put women, discus throw women, high jump men, long jump men, (Fig. S2 \cite{SupInfA4}). The fact that the major two powers were in a dense competition \cite{marion1} may have lead to a transitory extra-improvement which disappeared shortly after the war. From that point on, physiological progression may be limited in a majority of T\&F events, and performance will not increase until international instances or federations allow for major technological improvement \cite{Hood2005, Miodownik2007}.\\[.3cm]
The recent progression period in swimming results owes much to the introduction of swimsuits (Fig. \ref{FigGB2}). This new technology, allowed by FINA in 1999, enhances hydrodynamic penetration and largely reduces drag forces \cite{Swimnews, ToussaintHM2002}. Results show a divergence in recent performance development between \textquoteleft profiled\textquoteright~styles (freestyle, back) and \textquoteleft turbulent\textquoteright~swim styles (fly, breast). Previous studies support this observation as they show that swimsuits may have more impact in breaststroke, the most turbulent style with the largest drag resistance to flow \cite{Holmer1974, Kolmogorov1997}. The next step of performance development is seen in 2008 in relation to the introduction of second-generation swimsuits. This major performance increase culminated in the last Olympic Games (Beijing, 2008), where a single swimsuit was the common determinant of 96\% of the 22 new WR, 95\% of gold medals and 90\% of all medals \cite{Wood2008, Len2009}.\\[.3cm]
Sport performance is an important indicator of the optimization of physiological functions, enzyme isoforms (actin, myosin, alpha-actinin \cite{MacArthur2007}), energy use in aerobic and anaerobic metabolisms, oxygen transport and psychological resources that mobilize them \cite{Joyner2008}. Genes encoding each of these proteins or functions and their co-segregation or clustering are theoretically limited \cite{William2008}. Applying the Gompertz model to the top performances of 70 sport events illustrates this concept through a description of performances evolving toward their limits. In fact, for a majority of T\&F events, best performances show a secular progression hindered by the two WWs before reaching a phase of slower progression at the end of the XX\textsuperscript{th} century (Fig. \ref{FigGB2}). The evolution difference between the two disciplines is mostly based on technological improvement and rules alterations. In Beijing, T\&F records were scarce compared to swimming: only 5 new WR were established with 3 assigned to the same exceptional athlete. The Berlin (2009 T\&F World Championship: 3 WR) vs. Rome (2009 Swimming World Championship: 43 WR) competition confirmed the demonstration.\\[0.3cm]
\noindent
\textbf{Atypicity of the BP}\\
The Gompertz model describes the evolution of the first 10 world performers in the past century (Fig. \ref{FigGB1}). However, some top performers stand apart from this group, suggesting their performances are remarkable. Such singular marks are in contrast with the tight evolution of the whole group. The stability of the yearly mean coefficient of variation (Fig. S3 \cite{SupInfA4}) is noticeable after 1960, suggesting that all athletes are benefiting from similar conditions for competitions (training facilities, technological and medical advances). The initial era (1891-1930) was a period during which social or economical conditions may have greatly differed between athletes. Later on, sport\textquoteright s internationalization (1960-2008) has lead to the implementation of similar organizations, competition calendars and regulations among nations \cite{marion1}. The continuous technological advances aiming to enhance performance were previously described by Robert Fogel as a \textquoteleft Techno-physiological evolution\textquoteright~\cite{fogel}.\\[.3cm]
The study of the highest 5\% values of descriptors reveals 4 peaks for T\&F and 1 peak for swimming (Fig. \ref{FigGB3}). Considering the post 1930 era in T\&F, the 1943 peak is related to WWII. In this context, exceptional athletes are even more unique. The case of Fanny Blankers-Koen highlights that brilliant performers can score a maximum number of highest performances in such a period (4 gold medals in London Olympic games, 5 European titles and 16 WR in 8 events: 100 yards, 100m, 200m, 100m hurdles, high jump, long jump, pentathlon and $4\times100$m relay). This may be due to the presence of fewer skilled opponents.
\end{multicols}

\begin{center}
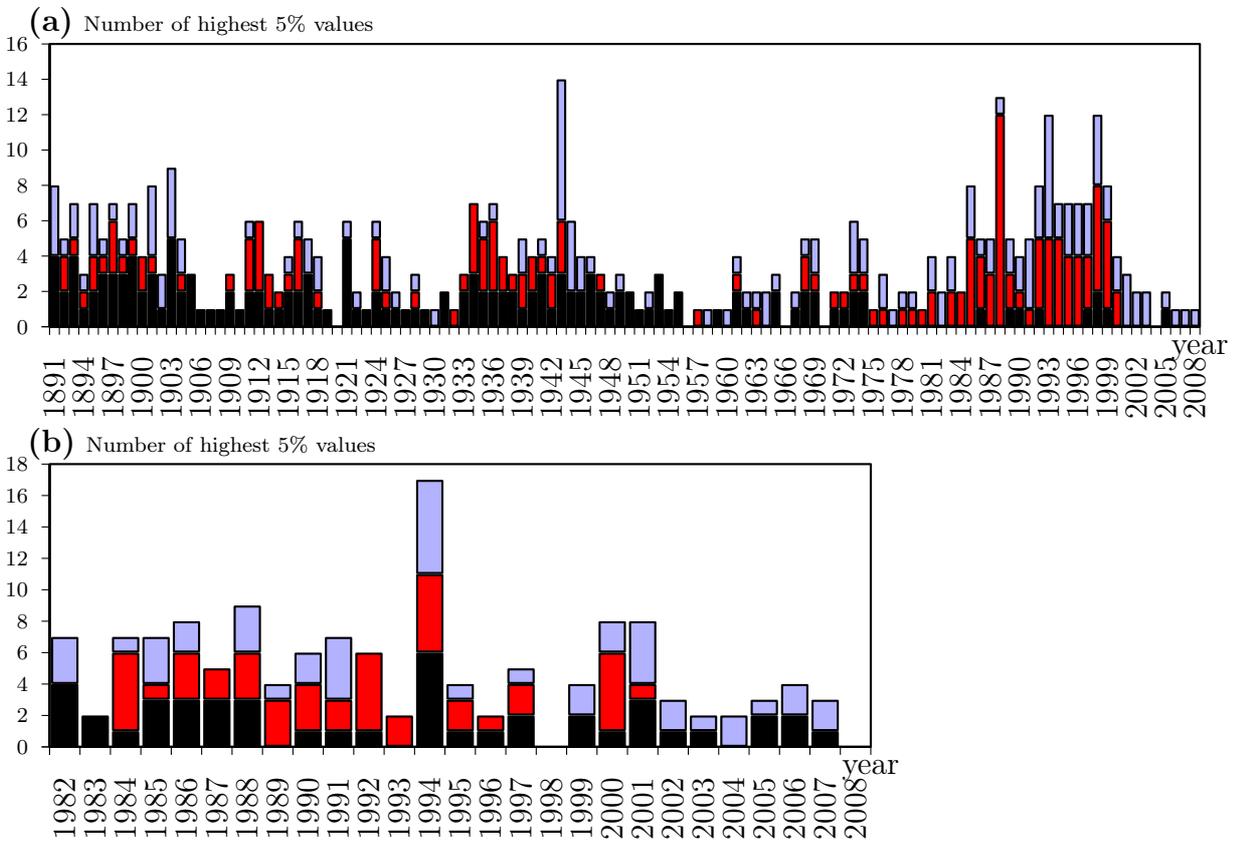

\begin{tabular}{l}
\psset{xunit=0.128205128cm, yunit=0.234375cm}
\begin{pspicture}(0,0)(118,16)
    \readpsbardata{\data}{Figure29-A1.prn}
    \psaxes[axesstyle=frame,Ox=0,Oy=0,Dx=1,Dy=2,labels=y,ticksize=-3pt,labelFontSize=\scriptstyle](0,0)(118,16)[year,-90][,0]
    \psbarchart[barsep=0,barcolsep=0,barlabelrot=90,barstyle={black,red,light_blue},chartstyle=stack]{\data}
    \rput[b](15.5,16.3){\textbf{(a)} \scriptsize{Number of highest 5\% values}}
\end{pspicture} \\
\psset{xunit=0.4cm, yunit=0.208333333cm}
\begin{pspicture}(0,-4)(27,26) 
    \readpsbardata{\data}{Figure29-A2.prn}
    \psaxes[axesstyle=frame,Ox=0,Oy=0,Dx=1,Dy=2,labels=y,ticksize=-3pt,labelFontSize=\scriptstyle](0,0)(27,18)[year,-90][,0]
    \psbarchart[barsep=0,barcolsep=0.1,barlabelrot=90,barstyle={black,red,light_blue},chartstyle=stack]{\data}
    \rput[b](5,18.3){\textbf{(b)} \scriptsize{Number of highest 5\% values}}
\end{pspicture} \\
\end{tabular}
\captionof{figure}[One figure]{\label{FigGB3} {\footnotesize Number of highest 5\% values by year and descriptor. \textbf{a}. T\&F. \textbf{b}. swimming. The highest 5\% values are gathered from each descriptor distribution in both disciplines: $d_1$ (black), $d_2$ (red), $d_3$ (light blue).}} 
\end{center}
\begin{multicols}{2}
The 1988 peak matches an Olympic year (Fig. \ref{FigGB3}). Eleven outstanding WR were beaten this year and 7 of them, in women events exclusively, still remain today. Only 3 were established in Seoul (200m women, $4\times400$m relay women, heptathlon). The others were established a few weeks earlier in Stara Zagora (BUL, 100m hurdles), Indianapolis (USA, 100m), Neubrandenburg (GDR, discus throw) and Leningrad (USSR, long jump). In this period, local competitions may also have had fewer monitoring procedures favoring illegal enhancement behaviors. Procedures for unexpected, out-of-competition anti-doping controls were officially approved one year later and 1988 can be considered as the T\&F golden year of exceptional marks, which subsequently resulted in a large stagnation in the women\textquoteright s events.\\[.3cm]
Outstanding BP values of the 1993 peak do not correspond to an Olympic year, but may be linked to a generation of exceptional performers. This year, Chinese women athletes achieved exceptional performances in T\&F with 33\% of the BP, 33\% of the second performances and 39\% of the third performances. These ratios have never been equalled by China since then. On 1993 at the Chinese National Games in Beijing, 5 Chinese women athletes have beaten the 3000m WR, a singular moment in T\&F history. A similar occurrence was only observed on the 26\textsuperscript{th} of July 1976 in Montreal, when 6 women athletes have beaten the WR in the same 800m race. All were East European athletes (Union of Soviet Socialist Republics (USRR), German Democratic Republic (GDR) and Bulgaria).\\[.3cm]
In swimming, evolution of atypical performances peaked in 1994 (Fig. S4 \cite{SupInfA4}). This year, Chinese swimmers achieved exceptional performances, obtaining 64.7\% of the women\textquoteright s BP. In the last 30 years, only the GDR has established such a supremacy, with a 70.6\% ratio in 1983, and 64.7\% in 1987.
\begin{center}
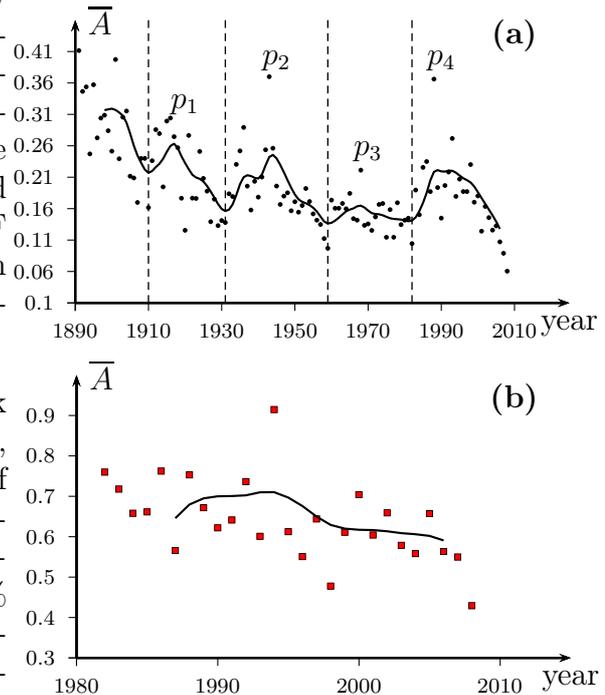

\begin{tabular}{l}
\psset{xunit=0.048148148cm,yunit=8.3333333cm}
\begin{pspicture}(1883,0.1)(2025,0.55)
    \psaxes[Dx=20,Dy=0.05,Ox=1890,Oy=0.1,ticksize=-3pt,labelFontSize=\scriptstyle]{->}(1890,0.1)(2025,0.55)[year,-90][$\overline{A}$,0]
    \fileplot[plotstyle=dots, dotscale=0.5]{Figure28-A1.prn}
    \fileplot[plotstyle=line, linecolor=black]{Figure28-A2.prn}
    \psline[linewidth=0.5pt, linestyle=dashed, dash=3pt 2pt](1910,0.1)(1910,0.55)
    \psline[linewidth=0.5pt, linestyle=dashed, dash=3pt 2pt](1931,0.1)(1931,0.55)
    \psline[linewidth=0.5pt, linestyle=dashed, dash=3pt 2pt](1959,0.1)(1959,0.55)
    \psline[linewidth=0.5pt, linestyle=dashed, dash=3pt 2pt](1982,0.1)(1982,0.55)
    \rput[b](1920,0.4){$p_1$}
    \rput[b](1945,0.47){$p_2$}
    \rput[b](1970,0.325){$p_3$}
    \rput[b](1990,0.47){$p_4$}
    \rput[b](2010,0.50){\textbf{(a)}}
\end{pspicture} \\
\psset{xunit=0.185714286cm,yunit=5.357142857cm}
\begin{pspicture}(1978.1,0.3)(2015,1.15)
    \psaxes[Dx=10,Dy=0.1,Ox=1980,Oy=0.3,ticksize=-3pt,labelFontSize=\scriptstyle]{->}(1980,0.3)(2015,1)[year,-90][$\overline{A}$,0]
    \fileplot[plotstyle=dots, dotstyle=Bsquare, fillcolor=red, dotscale=0.8]{Figure28-A3.prn}
    \fileplot[plotstyle=line, linecolor=black]{Figure28-A4.prn}
    \rput[b](2011,0.9){\textbf{(b)}}
\end{pspicture}
\end{tabular}
\captionof{figure}[One figure]{\label{FigGB4} {\footnotesize Average yearly  over time for each discipline. \textbf{a}. Secular evolution of  for T\&F (black dots); \textbf{b}. Evolution of \textbf{A} for swimming (red squares). Both evolutions are given with a 60 Hertz, second order low pass Butterworth filter. Besides the initial era (1890-1910), 4 peaks ($p_1$-$p_4$) appear in T\&F. Swimming curve (\textbf{b}) does not show any trend over the years. The 1994 peak is mainly related to Chinese swimmers, who performed very high performances (also observable in 1993 ($p_4$) in T\&F).}}
\end{center}
Both disciplines present different distributions and spatial repartitions of descriptors (Fig. S4, S5 \cite{SupInfA4}). As a result, measure \textbf{A} shows two different evolutions (Fig. \ref{FigGB4}). In T\&F the first 2 peaks of \textbf{A} have the same significance as the first 2 peaks of the highest 5\% values, corresponding to the world wars (Fig. \ref{FigGB4}). The third peak corresponds to the year of the Mexico Olympic Games (1968), during which numerous outstanding performances were established. This corresponds to the year with the highest number of WR \cite{berthelot2008}. The largest and final peak (1988) is followed by a major regression of performances and a \textquoteleft halt\textquoteright~of T\&F progression. This peak may correspond to the last successful attempt of our species to push its physiological limits forward.\\[.3cm]
The observed discrepancy between athletes of different countries may be related to training protocols improvements. Russian training loads were increased three fold between 1968 and 1998, in speed and strength sports \cite{Silber1998}. Most countries have followed such a trend in training protocols and volumes. Between 1950 and 1990, East European countries such as the GDR were involved in doping protocols \cite{franke} and may be responsible for many performances by undetected doped athletes \cite{franke, Riordan1993, Riordan1999, Donald2008, nour1, Dilger2007}. As a more recent example, recombinant erythropoietin (EPO) was marketed in 1989 and the International Olympic Committee prohibited its use in sports in 1990 \cite{IOC, Diamanti-Kandarakis2005}. Several studies quantified EPO\textquoteright s effect on performance by measuring maximal oxygen uptake (VO2max) and showed a 6.3 to 6.9\% increase \cite{Ashenden2001, Parisotto2000}. The introduction of out-of-competition controls in 1988 may have led to a reduction of drugs use among athletes. The later introduction of World Doping Anti-Agency (WADA) in 1999 has led to the harmonization of rules and regulations governing the anti-doping struggle in 2004. These recent efforts in the fight against doping may have had an effect on the proportion of doped athletes in competition. This may have been the case in T\&F where we see a general decline of performance after 1988 except in middle and long distance running races \cite{Schumacher2009}. The same effect is observable in cycling \cite{nour1}. In the present study, we cannot quantify the proportion of doped athletes so far. However, the analysis of this proportion among BP, using available lists from official sources \cite{Swimnews, IAAF} might allow an estimation of the ratio of physiological vs. pharmacologically enhanced performances.\\[.3cm]
In swimming, \textbf{A} remains stable over the last quarter of century as swimsuits benefited most of the best performers (Fig. \ref{FigGB4}). Major advancements in sports heavily rely on materials developments. Duralumin, carbon, polyurethane (tartan track, first used in the 1968 Olympic Games, or swimsuits) granted temporary improvements in Olympic disciplines. Innovation is a driving force of performance development and technology will need continuous research development in order to produce materials with higher energetic efficiency and to defy performances stagnation. These new programs will also depend on economical incentives and balances \cite{Lu2008}.\\[0.3cm]
\noindent
\textbf{Conclusion}\\
The Gompertz model of the first 10 world performers reveals that performances now stagnate in T\&F while swimming still progressed until today. Atypicity quantifies the irregularity of the yearly best performer. The pinnacle of atypical T\&F performances occurred in 1988, hastening the race toward human limits. This present halt of performances and the previously demonstrated stagnation of WR \cite{berthelot2008, fdd1} emphasize that our physiological evolution will remain limited in a majority of Olympic events. Present performances may now be enhanced through extremely exceptional individuals at the frontier of our genomic condition or with the artificial help of technology. However, the recent decision of FINA regarding the ban of specific swimsuits in 2010 will impact the future performance progression. If such a decision is confirmed, we may observe a rapid convergence toward the previously estimated physiological asymptotes. The limitation of artificial enhancements may drive performances back down to the physiological frontiers, which in turn depend upon growing economical or environmental constraints.
\end{multicols}
\section{Technology and physiology}
\label{sec:Technology}
In the previous sections we repeatedly stated that performance development was and still is strongly related to technological innovation and rules alteration. We also showed that the studied disciplines present different steps of progression and a majority of performances development in mature events slow down in the 90\textquoteright s. In some cases, disciplines such as swimming or cycling, rely on technological innovations to overcome physiological limits \cite{berthelot2010b, neptune, nourTHESIS}. It results in a brutal development of performances, though at high costs: the latest generation swimsuit in polyurethane costs about \$400 for a single race in swimming in 2010. Performances quickly reached a new asymptote since then and question the ability of technology to durably help athletes. R. Neptune published a review on the influence of advanced technology on physiology and concluded that ``\textit{recent technological innovations have produced faster running tracks, bicycles, speed skates, swimsuits, and pools to expand the limits of human performance}'' \cite{neptune}. It supports the idea that without such technologies, actual performances may not have reached the actual level. However, their use remain under sport instances control.\\[0.3cm]
The technological enhancement can roughly come into three categories: reducing drag forces, enhancing physiological power and indirect environmental parameters (training volumes, etc.). Here we list some crucial technological improvements in various disciplines:
\subsection{Cycling}
Cycling is one of the most technology dependant discipline. At high speed, drag forces (pressure drag in particular) are responsible for the majority of the resistances encountered \cite{Wilson2004, Bartlett1997}. The IHPVA decides to focus on technology and introduces specifically enhanced bikes. These \textquoteleft technological\textquoteright~cycles with a pilot in supine position provides a protection toward usual environmental stresses such as the wind, rain, etc. Due to their streamlined body shape, they also reduce the overall aerodynamic drag that usually affects cyclists using upright cycles. The aerodynamical drag coefficient on frontal area of a racing bike (rider crouched with tight clothing) is about 0.88 and around 0.10-0.15 for vector-faired recumbent cycles or tricycles (\cite{Wilson2004}, p.188).\\[0.3cm]
Another technological improvement focus on the weight of cycles. In 1869, the average weight of a standard cycle is 40kg. It decreases to 18kg in the 1930\textquoteright~and the introduction of the duralumin in 1934 reduces the weight to 12kg \cite{nour1}. Actual cycles approximatively weight 7kg and are made of titanium, aluminium or carbon fibers (introduced in the 80\textquoteright s) \cite{nour1}. Other technological improvements include handlebars \cite{Faria2005}, derailleur gears (1937 in professional tracks) and in-ear monitors for optimizing race strategies.
\subsection{Track and Field}
Again, materials play an important role in the development of T\&F performances. In pole vault, the pole progressively shifts from bamboo to aluminium and actual poles are made of fiberglass or carbon fibers. Track surface (known as Tartan track) also improves from \textquoteleft ashes\textquoteright~, rubber to asphalt and polyurethane. Track shoes evolves with the introduction of spikes and became lighter. Specific shoes are actually developed for long running distances.
\subsection{Swimming}
Swimming is another technological discipline and numerous technological improvements are made, especially to reduce turbulence through \textit{(i)} alteration of the swimmer\textquoteright s environment \textit{(ii)} alteration of the skin property of the swimmer.
\subsubsection{\textit{(i)} Swimming pools}
Waves generated by swimmers usually bounce off of what they hit, including walls, floors, and swimmers. Technological innovations are introduce to minimize turbulence and generally alter the boundary conditions of the pool. Competitive swimming pools now include wave-crushing lane ropes that diminish and deflect waves and make each lane in the swimming pool less turbulent. This means swimmers battle less against waves - theirs or others - which helps them swim faster. They are introduced in the 60\textquoteright s and have been improved over time. Other modifications include extra-wide swimming pools that have extra space outside lanes 1 and 8 in order to prevent additional turbulence. Some pools are built with 10 or more lanes, and sometimes only the center 8 are used for competition. The 2008 Beijing Olympics\textquoteright~National Aquatic Center - the Water Cube - is built that way \cite{NS1, NS2, AboutSwimming}.\\[0.3cm]
Other swimming pools are extra-deep. The hypothesis is that once a swimming pool gets deeper than about 1.5m, the waves generated by a swimmer that go down and hit the bottom of the pool tend to bounce back up after the swimmer passage. The idea is not new but building a pool much over 1.5m deep is not usually done, even for the Olympics, due to extra expenses after the Games. The higher the volume of water in a pool, the higher the operating expenses, and a 10-lane, 3m deep pool holds a lot of water. Deeper also means less use of the pool for other purposes.\\[0.3cm]
Other reports said that the designers of the Beijing pool planned to use a special porous wall material that would absorb speed-killing waves \cite{NS1, NS2, AboutSwimming}. The report says the idea was dropped due to its cost. We can see that such technological improvements comes at high costs.\\[0.3cm]
Since the 1\textsuperscript{st} of January 2010, the FINA allows for the introduction of angled starting blocks, that favor start.
\subsubsection{\textit{(ii)} Swimsuits}
However, the major technological improvement in swimming remains the swimsuits. Three generations of swimsuits are used during the modern Olympic era. The first generation is officially introduced and allowed by the FINA in 1999. The second generation is introduced in 2008 and is elaborated with the help of NASA\textquoteright~engineers. Part of the swimsuit is made of polyurethane and the seams are ultrasonically welded to further reduce drag. The Beijing Olympics highlights the suit, with 94\% of all swimming races being won with it. Moreover, 23 out of the 25 newly established world records, are achieved by swimmers competing in the suit. In 2009, the last generation of swimsuits is derived from the previous one and included even more polyurethane. It allows for trapping air for buoyancy and lead competitors to wear two or more suits for an increased effect. As of 24 August 2009, 93 world records are broken by swimmers wearing the suit \cite{berthelot2010b}. We investigate the gains provided by the 3 swimsuit generations by studying the performance development of the 10 best in 34 events from 1990 to 2009 (for a total of 6790 performances gathered). Three peaks of progression are identified in 2000, 2008 and 2009. They correspond to the year of introduction of each swimsuit. Average gains are of $0.74\% \pm0.26\%$ (2000); $1.16\% \pm0.48\%$ (2008); $0.68\% \pm0.55\%$ (2009) for men, and $1.00\% \pm0.37\%$ (2000); $0.97\% \pm0.57\%$ (2008); $0.27\% \pm0.70\%$ (2009) for women. The total gain in 3 years (2000,2008,2009) is of $2.58\% \pm1.29\%$ for men and $2.24 \pm1.64\%$ for women.
\subsection{Speed skating}
Speed skating is approved as an Olympic sport in 1924 and encounters several technical and technological improvements. Due to the high speed of skaters, the main resistive force to overcome are air resistance (80\%) and friction drag (due to the friction between skate and ice, account for 20\%) \cite{Schenau1982}. Again, various modifications are performed to improve both \textit{(i)} skater\textquoteright s environment and \textit{(ii)} skates.
\subsubsection{\textit{(i)} Ice rinks}
The introduction of 400m wide artificially refrigerated ice rinks allows for diminishing the friction \cite{Koning2010}. These rinks are practicable regardless of the season, and allow for a longer training time. In 1950, skaters are able to skate only 3 months per year on average ($\sim60$ days), today\textquoteright s training time is around 200 days per year \cite{Koning2010}.
\subsubsection{\textit{(ii)} Suits \& skates}
One piece spandex suits are introduced in 1974 to reduce air friction. They are improved in 2002 \cite{Schenau1982, Koning2010}. One major improvement in speed skating is the introduction of specific skates known as clap skates. Although the concept is introduced in the early 1900\textquoteright s, it only appeared in pro tours in the late 90\textquoteright s. These skates allow for the blade to longer remain in contact with the ice, as the ankle can be extended toward the end of the stroke, thereby distributing the energy of the leg more efficiently \cite{DeKoning2000, Houdijk2000}. They allow for an improved velocity and stroke frequency \cite{Houdijk2000b}. After their introduction in 1997, all world records are beaten during the Nagano\textquoteright s Olympics and the additional speed provided by these skates is estimated to 5\% \cite{Koning2010, Houdijk2000}. De Koning suggested that the observed improvement in speed skating in the recent years is, for the first half, due to the technological innovations and for the second half, to physiological improvements.
\subsection{Sailing, rowing}
As mentioned in \cite{fdd1}, technology keeps modifying boats in transatlantic records. Numerous alteration of the ship such as the boat\textquoteright s gauge, size, hull, sail surface of keel types, allowed for reducing turbulent frictions forces. They account for 80 to 90\% of the overall drag \cite{Greidanus2012}. Other modifications such as the use of composite materials (carbon fibers, \ldots) allowed for an reducing the overall weight of the ship. In rowing, carbon fiber was introduced in 1972 and the use of larger blades increased propulsive efficiency after 1991.
\subsection{Speed Skiing}
Equipment for speed ski records improved: ski length and composition, aerodynamic helmets or polyurethane suits. Downhill skiers also use wind tunnels extensively, not only to test clothing and equipment, but also to refine their moves so they minimize atmospheric drag on the slopes.
\subsection{Banned technologies}
A well known banned technology are performance-enhancing drugs composed of pharmacological products. Doping substances such as EPO make history in endurance sports such as cycling, rowing, distance running, race walking, cross country skiing, biathlon, and triathlons. Other doping drugs such as anabolic steroids have been banned in all sports in the seventies. The World Anti-Doping Agency (WADA) annually updates the list of performance-enhancing substances used in all sports. It includes anabolic steroids and precursors as well as hormones and related substances. Genetic modification, or \textquoteleft gene doping\textquoteright, is today\textquoteright s doping grail. It is supposed to alter the genetic material in order to improve specific factors of athletic performance. It may enhance or directly alter phenotypic expression by inducing muscle hypertrophy through overexpression of insulin-like growth factor-1, or increase oxygen delivery by raising the hematocrit through the use of EPO \cite{Wells2008, Baoulina2007, Baoulina2010}. However, after 30 years of development and clinical trials, the developments of genetic treatments for human disease show that the prospects for gene doping is still essentially theoretical so far \cite{Wells2008, Baoulina2007, Baoulina2010}.\\[0.3cm]
Other banned technologies may arise from innovations that come against rules and regulations. As a famous example, Oscar Pistorius, a disabled athlete without fibula nor feet, enters the T\&F able-bodied competition in 2007 using carbon fiber transtibial prostheses. Criticism rapidly arise as his prostheses seems to give him an unfair advantage over non handicapped runners. The IAAF then amends its competition rules to ban the use of ``\textit{any technical device that incorporates springs, wheels or any other element that provides a user with an advantage over another athlete not using such a device}''. After monitoring his track performances and carrying out tests, scientists take the view that O. Pistorius enjoys considerable advantages over athletes without such prosthetic limbs. It is estimated that Pistorius\textquoteright s limbs plus protheses use 25\% less energy than runners with complete natural legs at the same speed, and that they lead to less vertical motion combined with 30\% less mechanical work for lifting the body \cite{BBCS12008}. On the 14 of January 2008 the IAAF ruled him ineligible for competitions, including the 2008 Summer Olympics. This decision was reversed by the Court of Arbitration for Sport on 16 May 2008. The Court ruled that since he was slower out of the blocks than an able-bodied athlete, there was insufficient evidence that Pistorius had an overall net advantage over able-bodied athletes. He is lately admitted up to the 400m semifinal in the London 2012 Olympic Games, before collapsing into a gloomy criminal affair.\\[0.3cm]
History provides us with other examples: polyurethane swimsuits in swimming are banned on the January 1\textsuperscript{th} 2010. The FINA Congress voted almost unanimously to ban all body-length swimsuits, including polyurethane ones. However, all other records set by a swimmer wearing the suit stood as valid. In the same way, the IFRA (International Federation of Rowing Associations) bans sliding-rigger boats, which largely increases boat speed in order to limit technology influence and too expensive materials \cite{fdd1}.\\[0.3cm]
\subsection{Importance of technological designs in sport}
The athlete outfit plays a vital role in high speed sports including cycling, skiing, bobsleigh, sprint, speed skating and swimming as tiny time margins in those sports makes a difference between the winner and loser. This assertion is illustrated in the following example, with the example of polyurethane swimsuits and their impact in the swimming performances:\\[0.6cm]
\noindent
\textbf{Technology \& swimming: 3 steps beyond physiology}\\
G. Berthelot \& S. Len, P. Hellard, M. Tafflet, N. El Helou, S. Escolano, M. Guillaume, K. Schaal, H. Nassif, F.-D. Desgorces, J.-F. Toussaint
\\[0.6cm]
\noindent
\textbf{\textsc{Abstract}}
We focus here on the impact of material science in swimming by measuring the impact of the three successive generations of swimsuits on human performance and estimate the upcoming performance drop consecutive to the decision of the FINA to suspend their use. We investigate the recent evolutions of the best performers over the 1990 - 2009 period and demonstrate that three bursts of performances occurred in 2000, 2008 and 2009. The overall observed gains of these bursts exceed 2.0\% for both sexes. The drop in performance that may result from this rule change may return to similar levels as seen in 1999.
\begin{multicols}{2}
The science of engineering materials and the development of materials science during human history have strongly evolved over the past two centuries \cite{Shackelford2009, Ashby2005}. Other new technological fields such as particle physics, computer science, nanoscience also flourished \cite{Bhushan2009}, all leading to innovations that impacted sport. Polymers and metal alloys such as carbon fibres are exemplars of materials now widely used in various disciplines \cite{Brown2001}. In 2008, polyurethane made its first appearance in swimming with the use of a new swimsuit generation. The result was a sudden improvement of performances, allowing athletes to go beyond physiological limits that have been nearly reached \cite{berthelot2008, fdd1}. This study aimed to quantify the gain provided by the three generations of swimsuits introduced in 1999, 2008, 2009 and to estimate the upcoming performance drop in 2010. Using a recently published methodology \cite{berthelot2010a}, we analyzed the single best result each year for the world\textquoteright s top ten swimmers from 1990 to 2009 in order to assess the sudden progression trends and quantify the total performance gain.\\[0.3cm]
\textbf{\textsc{Material and methods}}
\\[0.3cm]
We collected the best performance of the world\textquoteright s top ten swimmers every year in 34 swimming events from 1963 to 2009 \cite{FINA, Swimnews, SwimingUHome}. A total of 6790 individual performances were selected from the data spanning the 1990 - 2009 period as they present a complete measure each year. The mean time and standard deviation of the 10 best performances were computed for each year in all events (Fig. \ref{Fig. 2.11}).\\[0.3cm]
\noindent
\textbf{Anova test}\\
An ANOVA test was conducted to estimate the variance of the mean of the 10 best swimmers. This test was intended to study the yearly evolution of performances and to identify any significant sudden development of performances. The test was performed each year between the means from 1990 to 2009 for all events. The number of significant values ($p < 0.05$) was then summed each year for both sexes (Fig. \ref{Fig. 2.12}).\\[0.3cm]
\noindent
\textbf{Performance gains}\\
The relative improvement or \textquoteleft gain\textquoteright~percentage between the mean $m$ of the 10 best from the year $t$ and $t+1$ was defined as:
\begin{equation}
  \label{eqPerfGains}
  g(t) = \dfrac{m_{t-1} - m_t}{m_{t-1}} \times 100
\end{equation}
and was computed between all years of all events from 1990 to 2009. To study the overall impact of the three generations of swimsuits in their year of introduction and use, gains obtained in 2000, 2008 and 2009 were summed in one measure by year for all events (overall summed gains) and for each event (summed gains). The \textquoteleft negative gains\textquoteright~correspond to a performance drop of the mean of the 10 best between the year $t$ and $t-1$. We analyzed whether swim distance and styles were predictors of summed gain with a linear model (distance and styles separately first and together second). The analysis was performed by gender. Distributions were assumed to be normal. The linear model was performed using MATLAB statistical toolbox.
\end{multicols}
\begin{center}
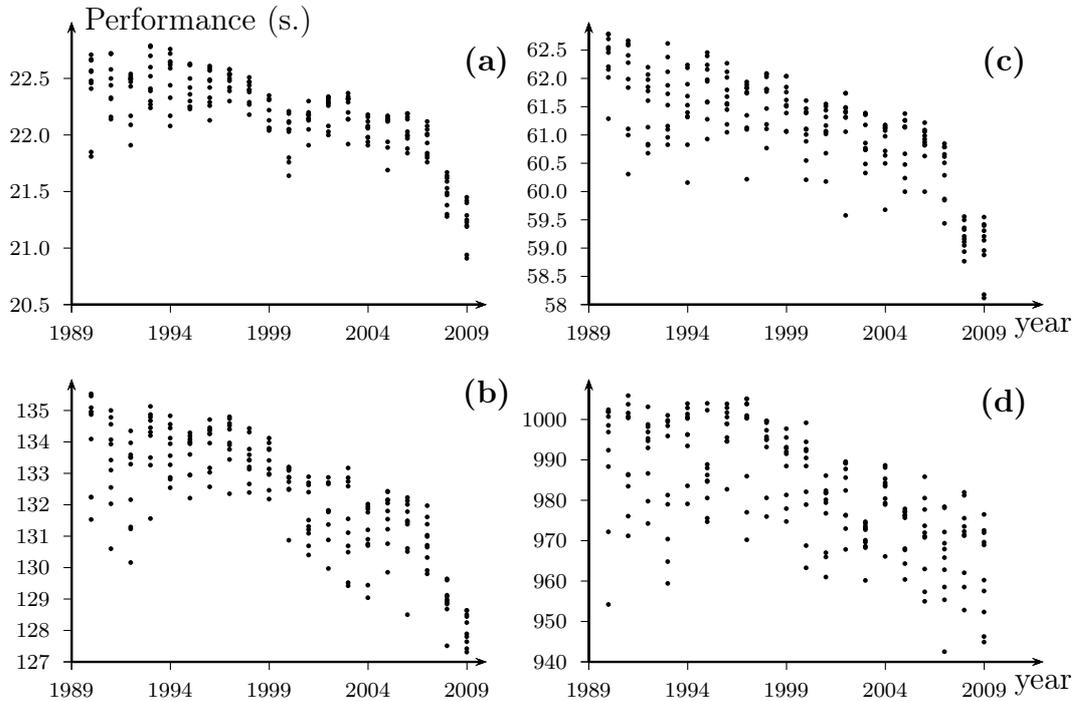

\begin{tabular}{l l l}
\psset{xunit=0.26cm,yunit=1.5cm}
\begin{pspicture}(1989,20.5)(2010,23)
    \psaxes[Dx=5,Dy=0.5,Ox=1989,Oy=20.5,ticksize=-3pt,labelFontSize=\scriptstyle]{->}(1989,20.5)(2010,23)[,-90][Performance (s.),0]
    \fileplot[plotstyle=dots, dotscale=0.5]{Figure27-A2.prn}
    \rput[b](2010,22.5){\textbf{(a)}}
\end{pspicture} & \hspace{0.5cm} &
\psset{xunit=0.26cm,yunit=.75cm}
\begin{pspicture}(1989,58)(2012,63)
    \psaxes[Dx=5,Dy=0.5,Ox=1989,Oy=58,ticksize=-3pt,labelFontSize=\scriptstyle]{->}(1989,58)(2012,63)[year,-90][,0]
    \fileplot[plotstyle=dots, dotscale=0.5]{Figure27-A3.prn}
    \rput[b](2010,62){\textbf{(c)}}
\end{pspicture} \\
\psset{xunit=0.26cm,yunit=0.41666667cm}
\begin{pspicture}(1989,127)(2010,138) 
    \psaxes[Dx=5,Dy=1,Ox=1989,Oy=127,ticksize=-3pt,labelFontSize=\scriptstyle]{->}(1989,127)(2010,136)[,-90][,0]
    \fileplot[plotstyle=dots, dotscale=0.5]{Figure27-A1.prn}
    \rput[b](2010,135){\textbf{(b)}}
\end{pspicture} & \hspace{0.5cm} &
\psset{xunit=0.26cm,yunit=0.05357143cm}
\begin{pspicture}(1989,940)(2012,1010)
    \psaxes[Dx=5,Dy=10,Ox=1989,Oy=940,ticksize=-3pt,labelFontSize=\scriptstyle]{->}(1989,940)(2012,1010)[year,-90][,0]
    \fileplot[plotstyle=dots, dotscale=0.5]{Figure27-A4.prn}
    \rput[b](2010,1000){\textbf{(d)}}
\end{pspicture}
\end{tabular}
\captionof{figure}[One figure]{\label{Fig. 2.11} {\footnotesize Four swimming events. \textbf{a}. Men 50m freestyle \textbf{b}. Men 200m breast \textbf{c}. Women 100m back \textbf{d}. Women 1500m freestyle (Swimming).}}
\end{center}
\begin{multicols}{2}
\noindent
\textbf{Predicted performance drop}\\
The prediction of performances drop $l_t$ corresponding to the present ban can be estimated using the relative difference between the mean of top performances in 2009 and the modeling of the previously developed model for the 10 best performances \cite{berthelot2010a}. The model is adjusted to the performances in each event from 1963 to 1999 and extrapolated in 2010. Thus the predicted values are estimated from the physiological period, before the introduction of the swimsuits:
\begin{equation}
  \label{eqPerfdrop}
  l_{2010} = \dfrac{\hat{m}_{2010} - m_{2009}}{m_{2009}} \times 100
\end{equation}
Where $\hat{m}_{2010}$ is the value estimated by the model of the 10 best mean in 2010, $m_{2009}$ is the mean value of the 10 best in 2009.\\[0.3cm]
\textbf{\textsc{Results}}
\\[0.3cm]
\textbf{Anova test}\\
Three peaks of variations are measured in 2000, 2008 and 2009 between the means corresponding to the year of introduction of each swimsuit. The first peak (2000) has a higher number of significant variations recorded for women vs. men (13 vs. 9). The second peak (2008) presents the highest number of variations, with an equal repartition of significant variations between men and women (13). The third peak (2009) shows a high number of variations for men (9) and few variations for women (3).\\[0.3cm]
\noindent
\textbf{Performance gains}\\
The gains are given for each event during the year of introduction of each new swimsuit generation (Table \ref{Table2.3}, Fig. \ref{Fig. 2.13}). The mean values of gains by year are: $0.74\% \pm0.26\%$ (2000); $1.16\% \pm0.48\%$ (2008); $0.68\% \pm0.55\%$ (2009) for men\textquoteright s events and $1.00\% \pm0.37\%$ (2000); $0.97\% \pm0.57\%$ (2008); $0.27\% \pm0.70\%$ (2009) for women\textquoteright s events. The cumulative mean values for the three years are: $2.58\% \pm1.29\%$ for men and $2.24\% \pm1.64\%$.

Overall summed gains of each events for the three specifics periods are: $12.66\%$ (men, 2001), $19.71\%$ (men, 2008), $11.48\%$ (men, 2009) and $17.06\%$ (women, 2001), $16.50\%$ (women, 2008), $4.59\%$ (women, 2009).

The number of negatives gains are: 0 (2000); 1 (2008); 4 (2009) for men and 0 (2000); 1 (2008); 3 (2009) for women. The nine events concerned are: 800m freestyle men in 2008, 200m fly men (2009), 400m medley men (2009), 400m freestyle men in 2009, 1500m freestyle men (2009), 1500m freestyle women (2008), 200m fly women (2009), 400m medley women (2009) and the $4 \times 100$m medley relay women (2009). Gains are also given by gender and distances (Table 1 \ref{Table2.3}).
\end{multicols}
\begin{center}
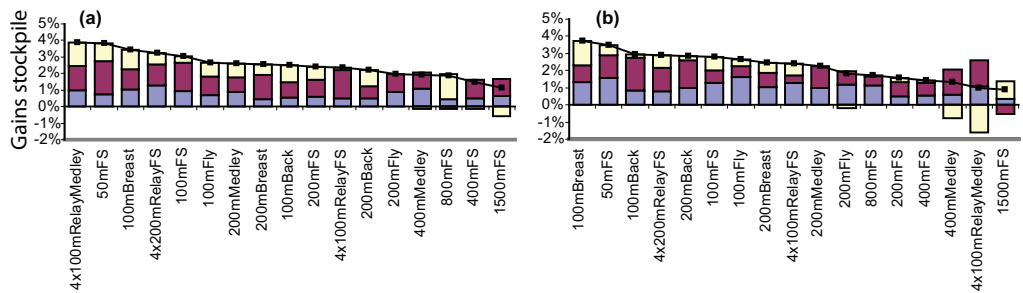

\begin{tabular}{l l l}
\psset{xunit=.285cm,yunit=.18cm}
\begin{pspicture}(1990,0)(2012,17)
\psaxes[Dx=3,Dy=2,Ox=1990,Oy=0,ticksize=-3pt,labelFontSize=\scriptstyle]{->}(1990,0)(2012,17)[,-90][Nb of significant values ($p < 0.05$),0]
\readdata{\data}{Figure25-A1.prn}
\listplot[linecolor=black,plotstyle=bar,barwidth=0.2cm,fillcolor=gray,fillstyle=solid,opacity=1]{\data}
\rput[b](2011,12){\textbf{(a)}}
\end{pspicture} & \hspace{0.3cm} &
\psset{xunit=.285cm,yunit=.18cm}
\begin{pspicture}(1990,0)(2014,17)
\psaxes[Dx=3,Dy=2,Ox=1990,Oy=0,ticksize=-3pt,labelFontSize=\scriptstyle]{->}(1990,0)(2014,17)[year,-90][Nb of significant values ($p < 0.05$),0]
\readdata{\data}{Figure25-A2.prn}
\listplot[linecolor=black,plotstyle=bar,barwidth=0.2cm,fillcolor=gray,fillstyle=solid,opacity=1]{\data}
\rput[b](2011,12){\textbf{(b)}}
\end{pspicture}
\end{tabular}
\captionof{figure}[One figure]{\label{Fig. 2.12} {\footnotesize Number of significant values for men (\textbf{a}) for women (\textbf{b}). The number of significant values ($p < 0.05$) of the ANOVA test are plotted by year. Three peaks are visible for men (2000, 2008 and 2009), while only two peaks are visible for women (2000, 2008).}}
\end{center}
\begin{multicols}{2}
The linear model reveals that a relation exists between summed gains and distance ($p=0.0048$ for men and $p=0.0144$ for women). Swimming styles show no relations with gains ($p=0.6245$ for men and $p=0.0985$ for women). A sensitivity analysis was then performed without the 1500 meter event and show that the distance effect is less significant in women events ($p = 0.08$).\\[0.3cm]
\noindent
\textbf{Predicted performance drop}\\
The calculated difference with the 1999 asymptote provides a potential drop of $3.65\% \pm0.78$ for all 34 events. The potential drop for men is $3.83\% \pm0.84$; and for women: $3.47\% \pm0.70$ (Fig. \ref{Fig. 2.14}, Table \ref{Table2.4}).\\[0.3cm]
\textbf{\textsc{Discussion}}
\\[0.3cm]
The evolution of numerous research fields and the development of technology in the recent era essentially benefited industrialized countries \cite{Brewer2005}. Technological improvement also helped increase the internationalization of sports meetings, enabling increased competition between nations, and increasing the importance and meaning of sport in society \cite{marion1}. Today, material science has become a major determinant of victory \cite{Miodownik2007, neptune}. Here we propose an analysis of the most recent technological enhancements in swimming, i.e. swimsuits, based on the measurement of swimmers\textquoteright~best performances. The analysis of this Olympic discipline was a convenient starting point for the analysis of the impact of major technological innovations on human performances; the highly controlled conditions of swimming competitions facilitated the measure of this impact.\\[0.3cm]
Swimming speed is limited by the resistive forces of water, also known as drag. They include skin friction, wave, and pressure forces \cite{Vorontsov2000, Mollendorf2004, Pendergast2006}, the most relevant force being pressure resistance that generates turbulent flow along the swimmer\textquoteright s body \cite{neptune, Vorontsov2000}. Several studies previously showed that using a swimsuit or wetsuit during effort gave its user an advantage by reducing drag resistance to water flow \cite{Toussaint2002, Toussaint2002b, Tomikawa2008, Cordain1991, Montagna2009}. The first generation swimsuit (1GS) was officially authorized by the FINA in October, 8\textsuperscript{th} 19998. The second generation swimsuit (2GS) was introduced in February, 13\textsuperscript{th} 2008. The third generation swimsuit (3GS) mostly used in 2009 was a derivative of the previous version with an enhancement of the preceding features. The very high number of new World Records (WR) set in a very short time span created controversy as it followed the introduction of the last two generations. FINA decided to suspend the use of swimsuits on January, 1\textsuperscript{st} 2010. On the other hand, it authorized the use of the new angled starting blocks.
\end{multicols}
\begin{center}
\includegraphics[scale=0.7]{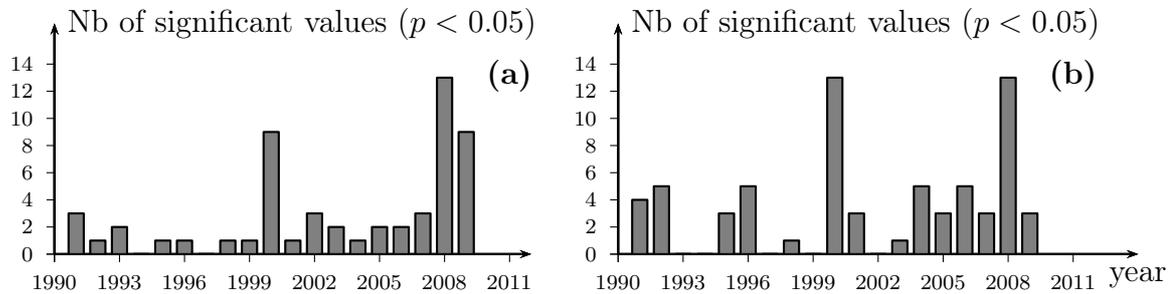}
\captionof{figure}[One figure]{\label{Fig. 2.13} {\footnotesize Number of significant values for men (\textbf{a}) for women (\textbf{b}). The number of significant values ($p < 0.05$) of the ANOVA test are plotted by year. Three peaks are visible for men (2000, 2008 and 2009), while only two peaks are visible for women (2000, 2008).}}
\end{center}
\begin{multicols}{2}
According to the analysis of the 10 best performers\textquoteright~swimming times between 1990 and 2009 (see Results), three bursts of rapid evolution in a number of swimming events stand out. The evolution of performances was expected to be a smooth progression, as the discipline was part of the first Olympic Games of 1896 along with track and field. To our knowledge, there was no other new ground breaking technique introduced during the studied era, strongly suggesting that each of these bursts corresponds to the introduction of each new suit generation in 2000, 2008 and 2009 respectively.
\begin{center}
\begin{table*}[htbp!]
\begin{tabular}{|l|r|r|r|r|}
  \hline
  Men event & 2000 (1GS) & 2008 (2GS) & 2009 (3GS) & sum \\
  \hline
  $4\times100$m relay medley & 0.99\% & 1.48\% & 1.38\% & 3.83\%\\
  50m freestyle & 0.74\% & 2.02\% & 1.05\% & 3.81\%\\
  100m breast & 1.01\% & 1.25\% & 1.16\% & 3.43\%\\
  $4\times200$m relay freestyle & 1.30\% & 1.25\% & 0.69\% & 3.24\%\\
  100m freestyle & 0.93\% & 1.72\% & 0.38\% & 3.02\%\\
  100m fly & 0.70\% & 1.12\% & 0.84\% & 2.65\%\\
  200m medley & 0.90\% & 0.85\% & 0.87\% & 2.62\%\\
  200m breast & 0.44\% & 1.46\% & 0.67\% & 2.57\%\\
  100m back & 0.56\% & 0.91\% & 1.03\% & 2.51\%\\
  200m freestyle & 0.60\% & 1.03\% & 0.77\% & 2.39\%\\
  $4\times100$m relay freestyle & 0.48\% & 1.73\% & 0.16\% & 2.37\%\\
  400m freestyle & 0.51\% & 1.14\% & 0.71\% & 2.35\%\\
  200m back & 0.48\% & 0.74\% & 0.98\% & 2.19\%\\
  200m fly & 0.88\% & 1.09\% & -0.01\% & 1.95\%\\
  400m medley & 1.06\% & 0.99\% & -0.14\% & 1.91\%\\
  800m freestyle & 0.45\% & -0.12\% & 1.52\% & 1.85\%\\
  1500m freestyle & 0.64\% & 1.04\% & -0.57\% & 1.11\%\\
  \hline
  Women event & 2000 (1GS) & 2008 (2GS) & 2009 (3GS) & sum \\
  \hline
  100m breast & 1.32\% & 0.97\% & 1.44\% & 3.74\%\\
  50m freestyle & 1.57\% & 1.35\% & 0.56\% & 3.48\%\\
  100m back & 0.83\% & 1.93\% & 0.18\% & 2.95\%\\
  $4\times200$m relay freestyle & 0.78\% & 1.37\% & 0.73\% & 2.88\%\\
  200m back & 1.00\% & 1.61\% & 0.25\% & 2.85\%\\
  100m freestyle & 1.28\% & 0.71\% & 0.80\% & 2.79\%\\
  100m fly & 1.63\% & 0.63\% & 0.39\% & 2.65\%\\
  200m breast & 1.05\% & 0.80\% & 0.60\% & 2.45\%\\
  $4\times100$m relay freestyle & 1.26\% & 0.46\% & 0.68\% & 2.40\%\\
  200m medley & 0.97\% & 1.31\% & 0.01\% & 2.28\%\\
  200m fly & 1.19\% & 0.79\% & -0.18\% & 1.80\%\\
  800m freestyle & 1.12\% & 0.56\% & 0.06\% & 1.74\%\\
  200m freestyle & 0.49\% & 0.84\% & 0.24\% & 1.56\%\\
  400m freestyle & 0.53\% & 0.73\% & 0.18\% & 1.44\%\\
  400m medley & 0.59\% & 1.49\% & -0.77\% & 1.30\%\\
  $4\times100$m relay medley & 1.10\% & 1.48\% & -1.61\% & 0.97\%\\
  1500m freestyle & 0.35\% & -0.51\% & 1.03\% & 0.87\%\\
  \hline
\end{tabular}
\caption{\label{Table2.3} {\footnotesize Table of Gains. Gains are given for each event and each swimsuit in descending order for men and women.}}
\end{table*}
\end{center}
A higher number of women events were affected by the 1GS than men events (Fig. \ref{Fig. 2.12}, \ref{Fig. 2.14}). Furthermore the peak of gains for women occurred in 2000 (Fig. \ref{Fig. 2.12}). This suggests that the compression of women\textquoteright s body parts (such as the breast) by the swimsuit may have been the key factor reducing drag resistance as early as 2000. Body shape is one of the factors altering drag \cite{Vorontsov2000}. For women, this reshaping of the body imparted by the swimsuits may have been the predominant factor, in the technological advancements made since 1999, leading to improved performances.\\[0.3cm]
The introduction of the 2GS in 2008 brought polyurethane to the realm of high level swimming. With an innovative seamless technology, this suit was made of polyurethane woven fabric with a texture based on shark scales \cite{Bhushan2009}, resulting in a large reduction of skin friction \cite{Bhushan2009, neptune, Toussaint2002b, Bechert1997, Bechert2000}. The 2GS provided an increased number of gains in men and women events (Fig. \ref{Fig. 2.12}, \ref{Fig. 2.13}). Its mean impact was $1.2\% \pm0.5\%$ for men\textquoteright s events and $1.0\% \pm0.6\%$ for women\textquoteright s events, and had a large effect on all distances and styles (Fig. \ref{Fig. 2.13}), except on the women\textquoteright s 1500m freestyle and the men\textquoteright s 800m freestyle.\\[0.3cm]
On May 19th 2009, FINA issued a list of 202 swimsuits approved for competition. Some full polyurethane swimsuits, the 3GS, were authorized until January 2010. However the impact of the 3GS on the 10 best performers is less homogeneous than the 2GS. Women showed a lower performance progression (Fig. \ref{Fig. 2.12}, \ref{Fig. 2.13}), while men experienced improvement.\\[0.3cm]
The impact of the three technological leaps due to the introduction of each swimsuit was measured in all events on the chosen years that presented significant variations: 2000, 2008 and 2009. We did not take into account the learning curve following the introduction of these new technologies. Thus, the given technological advances are measured by \textquoteleft default\textquoteright~and may exceed the 2.6\% (men) and 2.2\% (women) measured here if we consider the entire period 1990 - 2009. Furthermore, the proportion of the swimmers who competed without the swimsuits was not known. This proportion tends to be small but might constitute a bias in our analysis.\\[0.3cm]
The three measured impacts resulted in a larger improvement on short than long distances for both sexes ($p < 0.05$). It suggests the gain provided by the swimsuits may become marginal in long races. The resistance of polyurethane to tension and stretch might be limited on long distances, such as the 1500m freestyle (Fig. \ref{Fig. 2.13}). Another point is that the number of turns increases with the distance and turns generate hydrodynamic turbulence. It is possible that the swimsuit was not designed to improve speed during tumble turns. The relative improvement may therefore be reduced as distance increases.
\begin{center}
\begin{tabular}{l}
\psset{xunit=.1083333cm,yunit=.015cm}
\begin{pspicture}(1956,810)(2020,1100)
    \psaxes[Dx=10,Dy=50,Ox=1960,Oy=850,ticksize=-3pt,labelFontSize=\scriptstyle]{->}(1960,850)(2020,1100)[,-90][Performance (s.),0]
    \fileplot[plotstyle=dots, dotscale=0.5]{Figure24-A1.prn}
    \fileplot[plotstyle=line, linecolor=red]{Figure24-A2.prn}
    \psline[linestyle=dashed, dash=3pt 2pt](2010,907.76)(2014,907.76)
    \psline[linestyle=dashed, dash=3pt 2pt](2009,889.301)(2014,889.301)
    \psline{<->}(2012,889.301)(2012,907.76)
    \psdot[dotstyle=Bsquare, fillcolor=red, dotscale=1](2010,907.76)
    \psdot[dotstyle=Bsquare, fillcolor=red, dotscale=1](2009,889.301)
    \rput[b](2016,890){$l_t$}
    \rput[b](2015,1060){\textbf{(a)}}
\end{pspicture} \\
\psset{xunit=.1083333cm,yunit=.0625cm}
\begin{pspicture}(1956,215)(2020,275)
    \psaxes[Dx=10,Dy=10,Ox=1960,Oy=215,ticksize=-3pt,labelFontSize=\scriptstyle]{->}(1960,215)(2020,275)[year,-90][Performance (s.),0]
    \fileplot[plotstyle=dots, dotscale=0.5]{Figure24-B1.prn}
    \fileplot[plotstyle=line, linecolor=red]{Figure24-B2.prn}
    \psline[linestyle=dashed, dash=3pt 2pt](2010.5,223.653)(2014,223.653)
    \psline[linestyle=dashed, dash=3pt 2pt](2009,228.36)(2014,228.36)
    \psline{<->}(2012,223.653)(2012,228.36)
    \psdot[dotstyle=Bsquare, fillcolor=red, dotscale=1](2010,228.36)
    \psdot[dotstyle=Bsquare, fillcolor=red, dotscale=1](2009,223.653)
    \rput[b](2016,224){$l_t$}
    \rput[b](2015,265){\textbf{(b)}}
\end{pspicture}
\end{tabular}

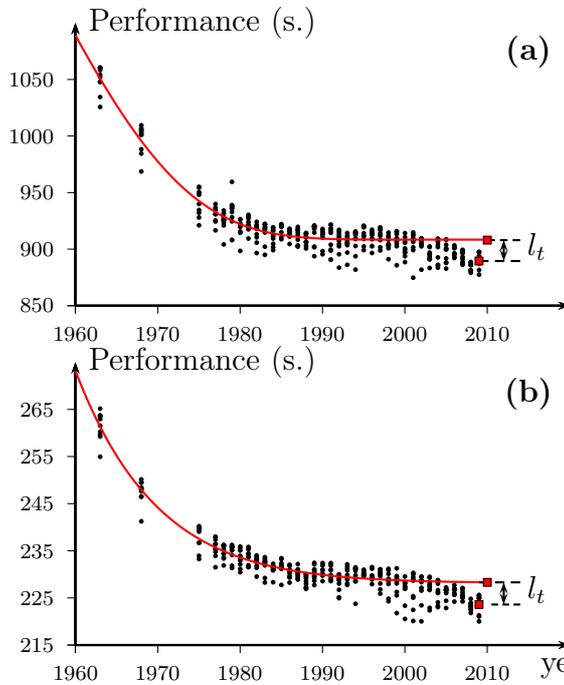
\captionof{figure}[One figure]{\label{Fig. 2.14} {\footnotesize Predicted performance drop in (\textbf{a}) 1500m freestyle men (\textbf{b}) 400m freestyle men. The model (red line) is adjusted to the performances (black dots) from 1963 to 1999 and extrapolated in 2010. The estimated performance drop $l_t$ is 2.12\% for the 1500m and 2.10\% for the 400m.}}
\end{center}
Compiling the following year of introduction of the three swimsuits, the average gain in performance improvement ranged from 1.11\% to 3.86\% for men and from 0.87\% to 3.74\% for women. Such considerable bursts of performance improvements in 12 months were not observed in any other Olympic discipline in the 1990 - 2009 period. However, similar improvements were observed at the introduction of new Olympic disciplines during the 1896 - 1914 period5, which may be considered as the athletes\textquoteright~learning curve. In cycling, similar bursts were observed when parallelogram systems (1930\textquoteright s), duralumin or carbon fiber cycles (1980\textquoteright s) were introduced \cite{nour1}. These bursts are now attributed to technological improvement rather than a learning curve or a physiological development.\\[0.3cm]
The estimated potential drop on the 1990 - 2009 period of $3.7\% \pm0.8$ for the year 2010 was based on the conditions prevailing from 1963 to 1999 (Fig. \ref{Fig. 2.11}). This value was obtained by modeling the evolution of the 10 best performances from this period7. The model was based on a Gompertz function and did not admit any brutal evolution. As the effect induced by the introduction of the swimsuits presented such a profile, we decided to model only the evolution of performances from 1963 to 1999, which follows a more typical physiological curve.
\begin{center}
\begin{table*}[htbp!]
\begin{center}
\begin{tabular}{|l|r|r|r|r|}
  \hline
  Event & 2010 pred. value & $l_{2010}$ (\%) \\
  \hline
  $4\times100$m medley relay men & 220.58 & 5.14 \\
  50m freestyle men & 22.30 & 4.85 \\
  100m back men & 54.94 & 4.60 \\
  $4\times100$m freestyle men & 199.92 & 4.49 \\
  100m breast women & 68.81 & 4.45 \\
  $4\times100$m freestyle relay women & 224.83 & 4.40 \\
  50m freestyle women & 25.22 & 4.35 \\
  200m medley men & 121.34 & 4.32 \\
  100m back women & 61.59 & 4.22 \\
  $4\times200$m freestyle relay men & 441.68 & 4.17 \\
  100m breast men & 61.38 & 4.15 \\
  100m freestyle men & 49.29 & 4.02 \\
  200m breast men & 133.02 & 3.87 \\
  100m fly men & 52.64 & 3.87 \\
  100m freestyle women & 55.12 & 3.84 \\
  200m back men & 118.71 & 3.80 \\
  100m fly women & 59.45 & 3.73 \\
  200m fly men & 118.01 & 3.72 \\
  $4\times200$m freestyle relay women & 484.84 & 3.70 \\
  200m back women & 131.52 & 3.64 \\
  800m freestyle men & 480.16 & 3.61 \\
  200m breast women & 147.52 & 3.56 \\
  200m freestyle women & 119.59 & 3.42 \\
  200m freestyle men & 108.12 & 3.35 \\
  200m medley women & 134.67 & 3.33 \\
  200m fly women & 130.54 & 3.27 \\
  1500m freestyle women & 990.28 & 3.14 \\
  400m medley men & 257.37 & 2.96 \\
  400m freestyle women & 250.10 & 2.88 \\
  $4\times100$m medley relay women & 248.54 & 2.54 \\
  400m medley women & 282.22 & 2.35\\
  800m freestyle women & 511.98 & 2.15 \\
  1500m freestyle men & 908.18 & 2.12 \\
  400m freestyle men & 228.36 & 2.10 \\
  \hline
\end{tabular}
\caption{ \label{Table2.4} {\footnotesize Performances drop per event. The performance drop $l_{2010}$ (\%) is estimated for each swimming event. The drop is more important on short distances. The given values may correspond to the average expected performance drop of the 10 best swimmers in each event if the FINA decide to draw back all technological enhancements provided since 2000 (including jammers, bodyskins, etc.).}}
\end{center}
\end{table*}
\end{center}
Since 2000, the three successive generations of swimsuits provided high gains along with new world records. At each competition, the media, the audience, the coaches and the athletes were expecting a new world record. Today, with the new FINA regulations, new records will be much more difficult to establish and raises some questions as to how the performance drop might affect the athletes psychologically. However, the estimated potential drops (Table \ref{Table2.4}) do not take into account the introduction of the new angled starting blocks or the technological innovations that may appear in the new swimsuits. In fact, this computed estimation corresponds to a return in traditional swim briefs, before the introduction of more elaborated swimsuits in the 2000 period. Since the ban of polyurethane swimsuits in 2010, the athletes compete in bodyskins, kneeskins or jammers that still provide a gain as compared to swim briefs. Along with the introduction of the new angled starting blocks, the predicted performance loss may be overestimated. The swimmers may therefore be expected to return to the 1999 - 2007 values. We can make the assumption that the new starting blocks may only increase the initial speed of the swimmers for the three styles in which they are used. The effect may fade with distance, resulting in little impact on performances over long distances. However the use of this new technology and any future technological innovations -whose implementation might be permitted by the FINA- may reduce the expected performance drop after the ban of these three generations of swimsuits in 2010.\\[0.3cm]
The enhancement of performances brought about by technological advances was previously described by Robert Fogel as a \textquoteleft Technophysiological evolution\textquoteright~\cite{fogel}. Swimsuits illustrate this process. The cost
of one swimsuit is about \$400 for one race. With the removal of high tech swimsuits from the sport, the financial investment of each team in each event will likely decrease. The trend observed between 2000 and 2009 could be used to demonstrate the close relationship between funds, technology and performance in this sport.\\[0.3cm]
\textbf{\textsc{Conclusion}}
\\[0.3cm]
To our knowledge, no previous study has been published with a precise and comparable quantification of the performance gain provided by the three swimsuit generations. These resulted in bursts of performance including new world records as soon as these technologies were introduced. The present analysis demonstrates the fact that technology has become a major factor increasing human sports performance. It may be our best hope to perform beyond our physiological limits5,6 while maintaining audience interest. However, the technological enhancement of performance may become limited by the financial costs that are needed to develop and maintain such technology. This questions the ability of sports official authorities to subsidize innovations in a moving economical context. Now that FINA has prohibited these swimsuits, we expect a return to the previous thresholds in 2010, that may be set between the level of the 1999 and 2007 asymptotes, except for the appearance of new technology on authorized jammers and kneeskins swimsuits or around the pools.
\end{multicols}
\section{Performances and lifespan}
\label{sec:Lifespan}
The expansion of human lifespan in the last two centuries suffers no controversy. It has increased after the industrial revolution, due to higher amount of nutrition (higher yields and production), breakthroughs in medicine, biotechnologies and a better hygiene \cite{Wilmoth20001111}. The Population Division of the United Nations estimates a growing proportion of living centenarians and supercentenarians: 23,000 in 1950, 110,000 in 1990, 150,000 in 1995, 209,000 in 2000, 324,000 in 2005 and 455,000 in 2009 \cite{UN1}. In particular, the development of human lifespan and life expectancy is widely investigated in the scientific literature and its limit are a subject of myth, speculation and debate \cite{Vaupel2010, Carnes2013a, Wilmoth20001111, Oeppen10052002, Tosato2007}. Again, similar debate arise between experts proposing a linear vs. non linear development in the future. L. Hayflick stated: ``\textit{To imagine that the current rate of life expectancy increase will continue indefinitely is as absurd as extrapolating the diminishing time taken to run a mile and concluding that it will sooner or later be done in one second}'' \cite{Watts2011}, picturing the models published and discussed in the earlier section (see sec. \ref{sec:OlympicSports} and \cite{Whipp1992, tatem}). The fact are that records of life expectancy follow a linear trend in the last few decades \cite{Oeppen10052002}. A linear forecast is only supported by extrapolating the previous trend, as in Tatem et al. \cite{tatem}. Some other studies refute this demonstration based on biological and thermo-dynamical constraints \cite{Watts2011, Carnes2013a, Olshansky2013a, Hayflick2007b, Hayflick2000, Tosato2007}. As of today, there is still a strong debate between experts \cite{Olshansky2013a}.\\[.3cm]
We here investigate the question of lifespan with a unique cohort of Olympic participants. It is exhaustive and contains the lifespan of all deceased Olympians having participated of the Olympic Games (OG) from 1896 up to 2012. It is also significant: Philip Clarke demonstrated that Olympic medallists live longer than the general population \cite{Clarke2012}. Finally, they can be compared with the world cohort of all validated supercentenarians deceased by now. This work is elaborated in collaboration with J. Antero-Jacquemin and A. Latouche.
\subsection{Material \& methods}
\subsubsection{Data}
\textbf{Olympians}\\
Day, month and year of birth and death were collected for 17,815 male and 1,197 female who have participated at least at one summer or winter OG since the first edition in 1896 up to 2012 and deceased by now. Data came from the most authoritative source of Olympians biography \cite{Clarke2012} carefully described elsewhere \cite{Zwiers2012}. The population was composed of 19,012 Olympians born between 1828 and 1991. The Olympians complete cohort (the cohort in which all subjects have entirely died out) ranges from 1828 to 1907. Date of death ranges from 1900 up to 2012 as detailed in table 1. Deaths were distributed among the world as follow: 1.7\% are originated from Africa, 19.8\% from America, 2.6\% from Asia, 73.7\% from Europe, 2.1\% from Oceania. The end of follow up is July 1\textsuperscript{st} 2013.\\[.3cm]
\textbf{Supercentenarians}\\
A verified and validated complete cohort of deceased supercentenarians born since 1850 was collected from the Gerontology Research Group \cite{Coles2013}. It totalized 1,419 supercentenarians (136 men and 1,283 women) born until 1897, last year without registered proof of a supercentenarian alive having the follow up end at July 1\textsuperscript{st} 2013.
\subsubsection{Lifespan distribution in time}
In order to investigate the distribution of lifespan with time, the density of the birth-dates and lifespan is estimated over a two-dimensional mesh.\\[.3cm]
\textbf{Mesh properties}\\
Let $X$, $Y$ be the date of birth and lifespan respectively, such that the data of an individual is expressed as $X_i$, $Y_i$ with $i=1,\ldots,N$. The density of lifespan is estimated over the nodes of a mesh $M$. The boundaries of $M$ are designed to encapsulate all the $X_i$ and $Y_i$: lower boundaries of $M$ $[L_X; L_Y]$ are defined as the largest integers that does not exceed $[\min(X_i); \min(Y_i)]$. Upper boundaries $[U_X; U_Y]$ are defined as the smallest integers that is not less than $[\max(X_i); \max(Y_i)]$. Note that in our case, the boundaries difference of the birth-date dimension $X$: $\Delta_X = U_X - L_X$ is always greater than the one of the lifespan dimension Y: $\Delta_Y = U_Y - L_Y$ , such that $\Delta_X > \Delta_Y$. The numbers of nodes in the $X$,$Y$ dimensions are respectively denoted $n_x$ and $n_y$, with respect to: $n_X,n_Y \in \mathcal{N}^*$ and $n_x, n_y \ge 2$. $M$ is set as a homogeneous mesh, such that each node is separated by the value $a$ in both dimensions, with:
\begin{equation}
  \label{eqJul1}
  a \in \left]0; \Delta_Y\right]
\end{equation}
Such that the maximum possible distance between two nodes does not exceed $U_Y$. The resolution $r$ of $M$ is given by:
\begin{equation}
  \label{eqJul2}
  r = n_X \times n_Y = \left[ \dfrac{U_X - L_X}{a} +1 \right] \times \left[ \dfrac{U_Y - L_Y}{a} +1 \right]
\end{equation}\\[.3cm]
\textbf{Estimate of the density}\\
The number of lifespan $d_j$ falling into the area of a $j$ ($j=1,\ldots,n$) node is summed, such that:
\begin{equation}
  \label{eqJul3}
  d_j = \# \left( X_i \subseteq \left[ X_j - \dfrac{a}{2}; X_j + \dfrac{a}{2} \right], Y_i \subseteq \left[ Y_j - \dfrac{a}{2}; Y_j + \dfrac{a}{2} \right] \right)
\end{equation}
is an estimate of the local continuous density and $\#()$ is the notation for the cardinal number.\\[.3cm]
\textbf{Representation of the density}\\
In order to avoid information loss due to an inadequate resolution of the mesh, we set the value of $a$ regarding the quantity of information provided by $M$ at a given resolution $r$. Shannon entropy was previously demonstrated as a good proxy for maximizing the information quantity in a viewpoint \cite{Vazquez2001}. In our approach we use the relative frequency of $d_j$ in $M$ of a given resolution provided the value of $a$ (eq. \ref{eqJul1}). In other words, we search for the values of $a$ that maximized entropy:
\begin{equation}
  \label{eqEntropy}
  H(M) = - \sum_{j=1}^n p(d_j) \log_2 p(d_j)
\end{equation}
Where $H(M)$ is the entropy of $M$, measured in bits and:
\begin{equation}
  \label{eqpdj}
  p(d_j) = \dfrac{\#(d_j)}{\#(d)}
\end{equation}
with $d_j \neq 0$. We browse the values of $H(M)$ for the values of $a$ in the interval defined in eq. \ref{eqJul1} using a step size of 0.1. We look for the values near the maximum of $H(M)$. As a consequence, we set $a=2$ (leading to $r=4067$ nodes) as a good representation of $d$ given the data set (Fig. \ref{Fig.Entropy}).
\subsubsection{Analysis of the instantaneous dynamics in a specific time frame}
In order to assess the development of lifespan among Olympians, we analyze the dynamics of the density in a specific time window, focusing on a frame defined by the coordinates of birth-date = $[1880; 1907]$ and lifespan = $[85; 106]$ years old, without time constraint (no Olympians born during this period are still alive \cite{Clarke2012}). The two dimensional gradient of the density is:
\begin{equation}
  \label{eqGradient}
  \nabla d = \dfrac{\partial d}{\partial X}\hat{i} + \dfrac{\partial d}{\partial Y}\hat{j}
\end{equation}
Where $\hat{i}$,$\hat{j}$ are the standard unit vectors. For each birth-date in the selected window, we collect the differences between adjacent elements of the densities in the $Y$ (lifespan) direction: $\alpha = [d_2-d_1, d_3-d_2, \ldots, d_n-d_{n-1}]$, as an estimate of $\dfrac{\partial d}{\partial Y} \hat{j}$. The total velocity per birth-dates $\bar{\alpha}$ is then estimated, for a given birthdate.
\begin{center}
\includegraphics[scale=0.8]{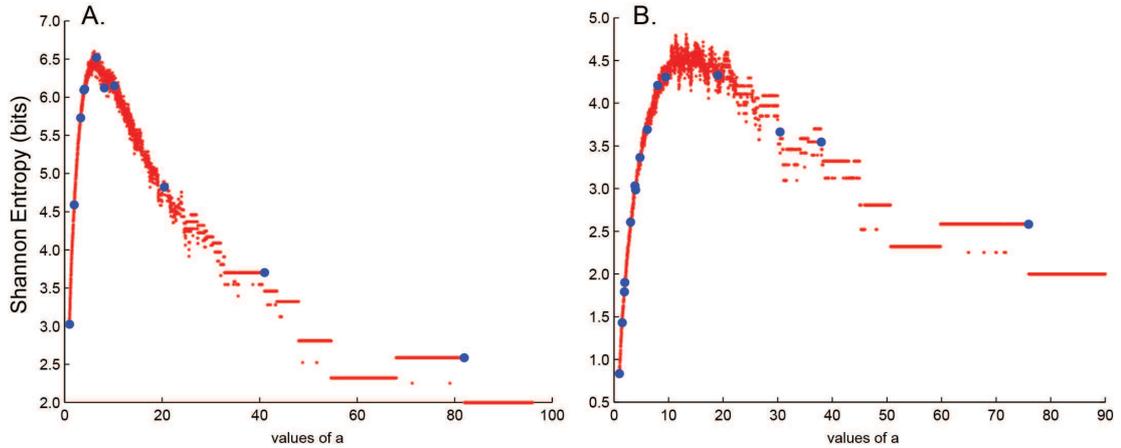}
\captionof{figure}[One figure]{\label{Fig.Entropy} {\footnotesize Quantity of information represented by different meshes of different resolutions, given different values of $a$. \textquoteleft Exotic\textquoteright~values of $a$ that lead to a mesh with $n_X,n_Y \notin \mathcal{N}^*$ , in contradiction to eq. \ref{eqJul1} are plotted in red (red dots). Correct values are shown in blue. \textbf{A.} Men. \textbf{B.} Women.}}
\end{center}
\subsection{Results}
The Olympians population is mostly composed of adults (mean age at entry in the cohort = $26.7 \pm 6.7$ years old for male and $23.9 \pm 7.1$ years old for women) from Western Europe and high income North America. The lifespan among them ranges from 15 to 106 years. The majority of supercentenarians also come from high income countries. The world record is 122 years old for a French woman (Jeanne Calment) born in 1875.

The lifespan density in function of birth\textquoteright s date is presented in Fig. \ref{Fig.Density}. The resolution was set to $a=2$ as one of the most expressive resolution, providing the best trade off between sensitivity and maximized entropy (Fig. \ref{Fig.Entropy}). The density scale ranges from dark blue illustrating the lowest density\textquoteright s values with fewer subjects to dark red corresponding to the highest density\textquoteright s values: these years concentrate the highest number of deaths. The vertical white line delineates the complete cohort (some Olympians are alive after this period). The horizontal white line separates the subjects with lifespan superior to 75 years old in order to delimit the analysis of superior lifespan.

The first games started in 1896, thus, the Olympians born before 1870 have participated at the games at an advanced age relatively to the mean age at first participation. Comparing with men the participation of women is minor, especially for those born before 1900.
\newpage
\textbf{General findings}\\[.15cm]
\textit{Men}. The maximum lifespan density among Olympians reaches 106 years. A gap may be visualized between the supercentenarians and the Olympians during the entire period. Among Olympians under 40 years old, two periods concentrate most of premature deaths and is delineated by two tracks, concerning subjects born between 1880-1895 and 1900-1920. The denser area, corresponding to the lifespan which concentrates the highest number of subjects, is formed by the Olympians born between 1900 and 1915 and died around 85 years old. The denser area move upward in function of the date of birth, until the time constraint start to limit this progression. The denser areas among the supercentenarians concentrate the doyens at 114 years.\\[.15cm]
\textit{Women}. Despite a restrained sample, women Olympians demonstrate a similar pattern of density layers. The maximum lifespan density reaches 15 years. Differently of men the two dense tracks delineating the premature deaths are not observed. The denser areas among female Olympians are seen between 85 and 90 years for those born between 1900 and 1915.
\begin{center}
\includegraphics[scale=0.9]{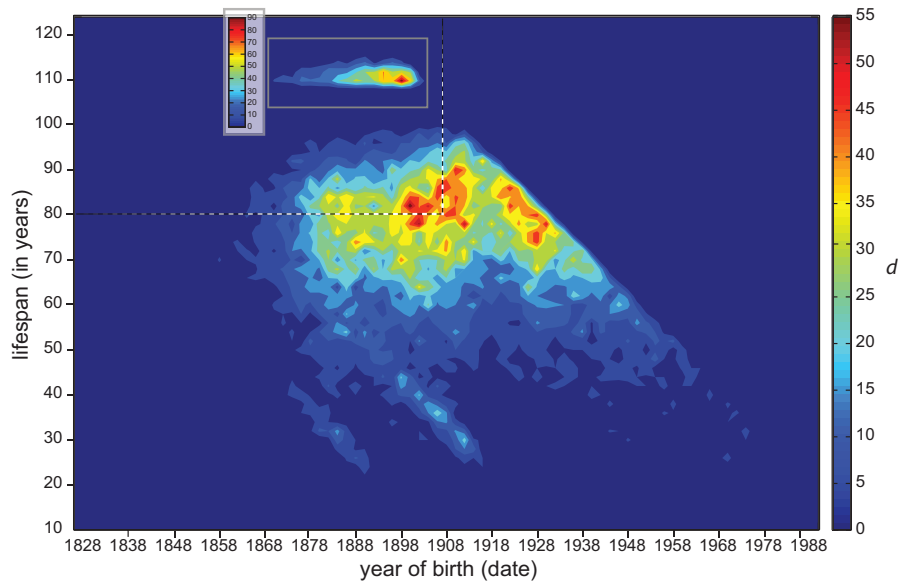}
\captionof{figure}[One figure]{\label{Fig.Density} {\footnotesize \textbf{Olympians and supercentenarians lifespan density\textquoteright s distribution}. Lifespan (in years) of all deceased Olympians (men and women) that have participated at the summer or winter games from 1896 until 2012 in function of their date of birth (in years). The resolution was defined as $a=2$ years. The density scale ranges from dark red representing the highest values of density to dark blue representing the lowest ones. The vertical white line delimitates the complete cohort with former Olympians having all deceased; and the horizontal white line separates the lifespan values superior to 80 years old. The top-left corner (gray rectangle) is the supercentenarians lifespan density.}}
\end{center}
\newpage
\textit{Lifespan upper limits}\\
The top-left corner of Fig. \ref{Fig.Density}, is the focus of the lifespan upper limit analysis for both male Olympians and supercentenarians. The development of different levels of densities, forming lifespan layers may be visualized in this corner. The superior layer, formed by the greatest values of lifespan creates a convex envelope of lifespan at approximately 100 years. This convex envelope points out the Olympians lifespan upper limit. The densities layers, below this convex envelope, move upward in function of birth date. The densities layers increase in lifespan values, and level off with the upper limit envelope. This shows increasing life duration of different age-structure evolving with time and leveling off with the upper limits.
A similar pattern is verified among the supercentenarians, where the inferior densities layers move upward following the date of birth and also leveling off with the convex envelope.\\[.3cm]
\textit{The compression phenomenon}\\
The development of the densities layers at the top-left corner on the Fig. \ref{Fig.Density} of both populations is closely studied in the \ref{Fig.Compression}. The superior contour of the different densities layers was smoothed and plotted. Both ancillary graphs attest an increased concentration of subjects in the direction of the superiors\textquoteright~layers in function of time, suggesting a phenomenon of compression. The steady slope among the supercentenarians, independent of the density level of subjects, suggests a limit to lifespan.
\subsection{Discussion}
The fittest of their country \cite{Clarke2012} composed the worldwide Olympians cohort. This selective population, is formed by subjects from regions that have historically dominated the Olympics as well as the world life expectancy record \cite{Engelman2010, Oeppen10052002}. To compare them with the supercentenarians kept our analysis focused on the lifespan records.\\[.3cm]
\textbf{Male Olympians}. The lifespan densities trends of the Olympians\textquoteright~cohort follow the general population trends of advanced countries but shift forward in relation to their period. The Olympians composing the complete cohort were born in the period when the greatest progress on life expectancy of developed countries took place \cite{Mackenbach2013}. That is due to a major reduction of mortality rates from infectious diseases among the younger people. The declined mortality at advanced ages occurs later on in the general population of developed countries, mostly due a cardiovascular mortality reduction \cite{Kannisto1994}. The denser areas upward displacement in function of time corresponds to the increasing lifespan of the majority of Olympians at advanced ages. These findings reveal that the Olympians, ahead of their time, are already taking benefit from a reduced mortality at older ages. They benefit from higher life expectation \cite{Teramoto2010} specially due to maintained aerobic expenditure \cite{Clarke2012} in its turn being associated with reduced cardiovascular diseases \cite{Warburton14032006}.\\[.3cm]
The dynamics of lifespan in function of time, observed among the majority of Olympians at advanced ages, no longer contribute to an expansion of the maximal lifespan values. In order to conceive the scenario which most part of the population surpasses the 100 years \cite{Christensen2009}, it is to be assumed that people will live each time longer, more often surpassing the 114, 120 year, 125 years, and so on.
\begin{center}
\includegraphics[scale=0.45]{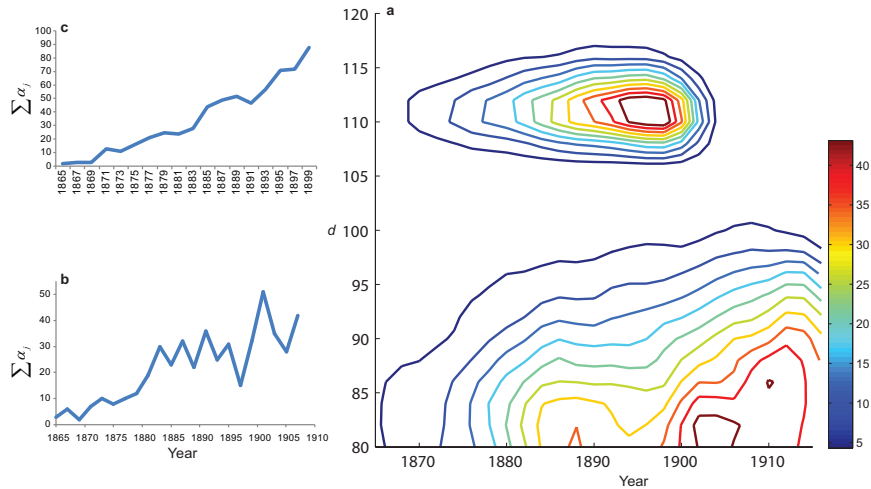}
\captionof{figure}[One figure]{\label{Fig.Compression} {\footnotesize \textbf{Slope of the Olympians and Supercentenarians lifespan densities layers}. Selected window for Supercentenarians and for Olympians (refers to the last year with no registry of subjects from these population still alive). A two-dimensional convolution kernel is applied to the density layers in order to smooth irregularities. $\bar{\alpha}$ is given in insets (\textbf{b} and \textbf{c}) for both populations. It quantifies the increase of the density\textquoteright s layers with time in the direction of the upper lifespan values. The left one (\textbf{b}) corresponds to the compression values of male Olympians population, the right one (\textbf{c}) refers to the compression of supercentenarians.}}
\end{center}
Conversely, if there is a fixed limit to human lifespan, the increasing populations\textquoteright~life expectancy will entail an accumulation of individuals living longer but dying closer to this limit. This scenario where the majority of lifespan is compressed into a narrow age range -close to the limit- is illustrated here. The slope of male Olympians densities decreases as the lifespan increases, in such slow pace that the maximal lifespan values remain nearly steady along the last decade. The steady level of maximum lifespan observed among the Olympians illustrates the invisible wall evoked by \cite{Kannisto2001}. Their lifespan superiors\textquoteright~density layers level off with the lifespan maximum values, while this one remains steady. It seems that we are here facing a limit for the Olympians lifespan. This suggests the existence of a limit to lifespan.\\[.3cm]
The gap between Olympians and supercentenarians attests the exceptional character of reaching 110 years old. Even among a highly selected population of Olympians surpassing the average age of death, none reached the status of a supercentenarian. Becoming a supercentenarian takes a complex sequence of rare and specific circumstances, involving constant favorable interactions between genetics and the environment. Our analyzes reinforce the heterogeneity and complexity of the selection that predominates among the supercentenarians. A favored population in terms of lifespan may surpass life expectations by benefiting of a reduced mortality rate at older ages, but this does not implicate reaching a supercentenarian status. It underlines the importance to distinguish the actuarial aging trends from biological aging possibilities. Moreover, in the case of additional stresses such as a pandemic or to a degraded economical context my alter the limits. Recent changes in the development of the life expectancy of countries such as Swaziland, Kenya or South Africa due to the AIDS pandemic provides approximations of the average dynamic of lifespan through time (Fig. \ref{FigJulAdd}). Another example is the Russian life expectancy that started to plateau in 1960\textquoteright s, before oscillating in 1980\textquoteright s. All the changes exhibited smooth or abrupt S-Shaped patterns and underline the sensibility of estimated life expectancy to a perturbed environment.
\begin{center}
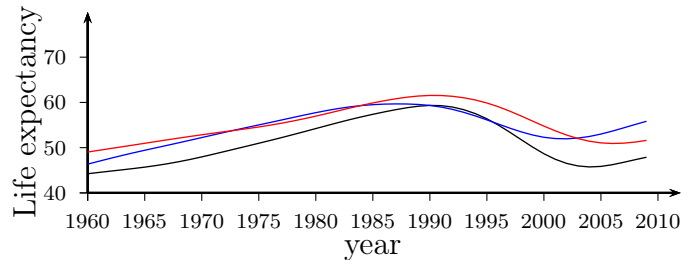

\psset{xunit=.15cm,yunit=.06cm}
\begin{pspicture}(1960,30)(2015,80)
    \psaxes[Dx=5,Dy=10,Ox=1960,Oy=40,ticksize=-3pt,labelFontSize=\scriptstyle]{->}(1960,40)(2012,80)
    \fileplot[linewidth=0.5pt, linecolor=black]{Fig55.txt} 
    \fileplot[linewidth=0.5pt, linecolor=blue]{Fig55b.txt} 
    \fileplot[linewidth=0.5pt, linecolor=red]{Fig55c.txt} 
    \rput[b]{90}(1956,58){Life expectancy}
    \rput[b](1985,25){year}
\end{pspicture}
\captionof{figure}[One figure]{\label{FigJulAdd} {\footnotesize Examples of altered developments of life expectancy (source: The World Bank) for three different countries (Swaziland: black curve, Kenya: blue curve and South Africa: red curve). The increasing and decreasing patterns are S-Shaped.\\[0.3cm]}}
\end{center}
\textbf{Olympians and supercentenarians: a similar compression pattern}. A common pattern of density compression is observed among Olympians and supercentenarians: this provides a second suggestion for a limited lifespan. As they compose the humanity\textquoteright s doyens, the observation of limits within this group allude to limited lifespan for mankind as previously suggested \cite{Carnes2013a}.\\[.3cm]
We kept our focus on supercentenarians lifespan\textquoteright~densities rather than on data from one single person, the one who lived longest \cite{Wilmoth1997}. Observing supercentenarians densities -the accumulation of longer lived subjects- increases in function of time. But the steady layers do not show an expansion response to this; on the contrary, they show a compression phenomenon. The Jeanne Calment world record could have been beaten for people born from 1875 until 1890. After this record, one single person (born in 1880) has lived 119 years, and after that, a plateau has been recorded up to now. We attest an increased accumulation of longer lived people without any progression trend, suggesting a limited lifespan. The more stable pattern of the densities slope comparing to Olympians\textquoteright~is our hint that the confrontation with this limit is not a matter for the future.\\[.3cm]
This study does not provide evidence to support the prolongivists claim \cite{DeGrey2004, Vaupel2010, Christensen2009}. However, the period of observation remains restrained, especially in comparison to populations studies dating from 1860s until 2008 \cite{Vaupel2010}. Nevertheless, analyzes of the most recent centenarians\textquoteright~birth cohort are limited in time to 1913. After this period all demographic forecasts are based on period life table (death rates from a calendar year is applied to people still alive) \cite{Oeppen10052002, Vaupel2010, Christensen2009} which remains speculative \cite{Carnes2013a}. Period life tables have constantly underestimated real population life expectancy \cite{DeGrey2007}. Extreme life extension forecast, in return, has largely overestimate life expectancy, as the gap between cohort and period life expectancy has fallen \cite{Carnes2013a}. Hence, the importance of analyzing a concrete cohort, that have shown to be ahead of their time in terms of life expectancy advantage. Finally comparing Olympians with the real lifespan record holders provides an objective overview of the present trends.\\[.3cm]
It is reasonable to suppose that if life expectancy keeps increasing, more people will constantly accumulate closer to the limit. This, however, may only happen if nutritional, climatic, social or economic conditions continuously improve. The word \textquoteleft continuously\textquoteright~is carrying great weight here because to maintain the life expectancy\textquoteright~progression the environmental conditions needs to improve in even faster pace. Because is easier to improve life expectancy by avoiding deaths among younger\textquoteright s than making older people live longer. The East Germany illustrates how life expectancy may progress among the frailer people: after the German reunification the Eastern improved on a much faster pace than the Western \cite{Vaupel2003}.\\[.3cm]
Prolongivists may call for important medical and technologies advances leading to life extension in a near future \cite{Goldman26092005}. On the other hand, major health determinants such as global warming (ref), air pollution \cite{Smith2013} economical recession or epidemic obesity \cite{Olshansky26092005} are already part of our present.\\[.3cm]
\textit{Premature death}\\
Deaths were observed in every age group for the whole period. The two diagonal tracks corresponding to the male Olympians born between 1880-1895 and 1900-1920 and deceased under 40 years old demonstrate that they were not spared from world wars consequences. The subjects born during these periods were 19 to 38 and 19 to 45 years old during the first and second war respectively. It corresponded to the most affected ages classes during both wars. This war effect was not observable among women.\\[.3cm]
\textbf{Women Olympians}. The women cohort is smaller than the male one because women participation at the first games was very limited, with a number of women per games inferior to 100. The limited sample of women born before 1900 restrains the analysis of upper limits before this date. Regarding lifespan of women Olympians born after 1900, they however demonstrate the same pattern of longevity, superior to men, that is observable in the general population \cite{Wilmoth20001111}. For the same time frame, denser areas of women Olympian\textquoteright s lifespan are, at least, 5 years superior to men. But the similar pattern of density slope allows us to consider transposable the results observed in the men\textquoteright s cohort.\\[.3cm]
\textbf{Conclusion}\\
A phenomenon of compression can be seen among the longest lived Olympians with a gap between them and the supercentenarians. In both cohorts, the developed compression does not support an unlimited increasing life expectancy. Conversely, the stabilized progression of the humanity\textquoteright~oldest provides arguments to claim a limited lifespan. In view of the scenario here demonstrated, we propose to start to consider a new forecast regarding life duration and anticipate on the increased constraint that may further limit it.
\chapter{Physiological boundaries in other species}
\label{sec:OtherSpeciesFDD}
In this chapter we investigate the development of physiological performances of other species related to mankind with the hypothesis that similar models can be used to describe both human and non-human performances progression patterns.
\section{Horse and greyhound}
Man domesticates several species such as goats, steers, sled dogs or draft horses for their capacity to produce food, fibers for clothing and their ability to tow material. He also organizes competitions for some of them. Horses and greyhounds are such examples of species selected for their athletic capacities: they both can reach around 64 km.h$^{-1}$ at maximum speed \cite{EEPTS2008}. In comparison, other domesticated and non athletics species have significant lower speeds but are domesticated for another type of performance: work energy, food,\ldots). We previously demonstrated that human physiological performances could be modeled by a simple Gompertz function \cite{berthelot2010a} (sec. \ref{sec:Performance}). We also showed that human sport performances plateaued in the recent times \cite{berthelot2008, berthelot2010a}, thus a driving rational would be \textquoteleft do the performances of other athletic species also plateau?\textquoteright. Did they benefited of the same enhancements? Is there a common pattern?. At first glance, it seems reasonable to think that this is the case.\\[0.3cm]
In order to test our hypothesis, we gather data for the 10 best performers (BP) in the 450-500m in dogs, 2200-2800m in horses and 200-1500m in humans. The data is collected over a span time of more than 110 years. The same model is applied to the 3 species \cite{berthelot2010a} (and eq. \ref{eqGompertz}). The model fits speed progression with a good accuracy (horses: $R^2 = 0.56$, $\text{MSE} = 0.169$; greyhounds: $R^2 = 0.81$, $\text{MSE} = 0.55$; and humans: 800-1500m. $R^2 = 0.97$, $\text{MSE} = 0.006$, 200-400m. $R^2 = 0.94$, $\text{MSE} = 0.008$). Overall speed progression over the studied period seems similar in the three species (11.1\% of initial value in horses, 9.4\% in greyhounds, 11.1\% in 200-400m. and 14.2\% in 800-1500m. in humans). We also compute the asymptotic values (speeds) and find that all the species are approaching their limit. As a conclusion, it seems that the performances progression of any species exhibit a logistic pattern, provided it is placed in a competitive environment. We assume that the performances of species in their natural landscapes maybe completely different, due to the predator-prey relationships and various hazards surrounding their daily activities.\\[0.3cm]
\noindent
\textbf{Similar slow down in running speed progression in species under human pressure}\\
F.-D. Desgorces, G. Berthelot, A. Charmantier, M. Tafflet, K. Schaal, P. Jarne and J.-F. Toussaint\\[0.6cm]
\noindent
\textbf{\textsc{Abstract}}
Running speed in animals depends on both genetic and environmental conditions. Maximal speeds were here analyzed in horses, dogs and humans using datasets on the 10 best performers covering more than a century of races. This includes a variety of distances in humans (200 to 1500m.). Speed has been progressing fast in the three species and this has been followed by a plateau. Based on a Gompertz model the current best performances reach 97.4\% of maximal velocity in greyhounds to 100.3 in humans. Further analysis based on a subset of individuals and using an \textquoteleft animal model\textquoteright~shows that running speed is heritable in horses ($h^2 = 0.438$, $p = 0.01$), and almost so in dogs ($h^2 = 0.183$, $p = 0.08$), suggesting the involvement of genetic factors. Speed progression in humans is more likely due to an enlarged population of runners, associated with improved training practices. The analysis of a data subset (40 last years in 800 and 1500m.) further showed that East Africans have strikingly improved their speed, now reaching the upper part of the human distribution while that of Nordic runners stagnated in the 800m. and even declined in the 1500m. Although speed progression in dogs and horses on one side and humans on the other have not been affected by the same genetic / environmental balance of forces, it is likely that further progress will be extremely limited.
\begin{multicols}{2}
Locomotor performance is a key-trait in mobile species that is closely associated to fitness \cite{Irschick1999}. In particular, improving running speed might allow for reaching prey or avoiding predator more efficiently \cite{husak2006}. It is therefore of importance to better evaluate those factors, whether genetic or environmental, affecting maximum speed. However, experimentally evaluating how far maximum speed can improve is difficult, because assessing it over meaningful time periods is not always feasible. We propose a way to alleviate this issue by using race records in species in which fastest speeds have been monitored over a long time-span. Such data are available in horses, dogs (greyhounds) and humans.\\[0.3cm]
Speed progression, i.e. the increase in maximal speed over time, in these three species results from both genetic and environmental factors, though to various extends according to species \cite{nevill1, Davids2007, Niemi2005}. Moreover some variables acting on speed progression depend on training (relating to physiology, psychology, biomechanics, technology or tactics improvement) while others are outside the athletes\textquoteright~control (genetics, anthropometric characteristics, climatic conditions) \cite{Brutsaert2006, Smith2003}. In domesticated animals, sustained selection by breeders has long been known to cause large and accelerated phenotypic changes contributing to short-term evolutionary processes \cite{Darwin1868, Falconer1996}. Racing horses have indeed been selected from the 16th century, based on closed populations and a very small number of founders \cite{Willett1975}. Beginning later, similar selection processes have been imposed on greyhounds for dog racing. This means that phenotypic expression (such as running speed) relies on a narrow genetic basis in comparison to the variation available in these species. However, genetic variance for performance has recently been detected in thoroughbred horses \cite{Gu2009, Hill2010}.\\[0.3cm]
The improvement of fast-running performances in humans does not (of course) rely on selective breeding, but on the detection of fast-running athletes generally during adolescence or later through national systems that were optimized after World War II \cite{marion1}. Importantly the athlete population has increased in size proportionally to the development of modern sport throughout the last century. For example, 241 athletes from 14 national Olympic committees participated to the first modern Olympic games (Athens, 1896) while 10942 athletes coming from 204 nations attended the 29th games in Beijing (2008). Thus, the best human runners are now selected from a much larger number of countries \cite{marion1}, presumably over a larger genetic basis for performance \cite{William2008, Yang2003}. This is to be opposed to the limited variation for running capacity observed in dogs and horses \cite{Denny2008}. A striking tendency in humans is also related to geographic origins of runners with e.g. the massive rise of African runners among best performers in middle-to long distance over the recent decades \cite{Onywera2006}.\\[0.3cm]
Other factors relating to runner physiology (e.g., training, running style, nutrition and, sometimes, doping activities), or to environment sensu lato (e.g., climatic conditions, riding style, rules, betting activity, reward) have also played a significant role in improving running performance \cite{Norton2001, Noakes2004, Pfau2009, Toutain2010}. The respective influence of genetic and non-genetic factors may therefore largely differ in domesticated animals and in humans \cite{Brutsaert2006}. Whatsoever recent studies have demonstrated that the maximal running speed may soon reach its limits in dogs, horses and humans \cite{berthelot2008, berthelot2010a, Denny2008}.\\[0.3cm]
Best running performances in humans have been under scrutiny \cite{berthelot2008, nevill1}, but a full analysis of speed progression in greyhounds and horses is not available. The comparison of speed progression in animals (horses and dogs) versus humans requires long-term data over comparable periods and distances.\\[0.3cm]
We collected data built up over more than a century (up to 118 years) for the 10 best performers (BP) in races (450-500m. in dogs, 2200-2800m. in horses and 200-1500m. in humans); the larger range in humans allowing direct comparison with the two other species over similar running times. We also tested the heritability of best performers using pedigrees of dogs and horses while, in humans, performance analysis from different geographical regions allowed us to test the progression of maximum running speed according to the geographic origin of runners. Our objectives here are (\textit{i}) to compare the progression of speed performances in humans, horses and dogs, and (\textit{ii}) to evaluate the potential influence of genetic factors in horses and dogs, based on a so-called \textquoteleft animal model\textquoteright~approach \cite{Kruuk2004}, and of geographic origin in humans. Our approach is comparative (not experimental) in essence since it is based on observational data over the long term in three species, though allows to dissect the respective influences of genetic and environmental factors on patterns of speed progression based on fairly large data-sets.\\[0.6cm]
\textbf{\textsc{Material and methods}}
\\[0.6cm]
\textbf{Data}\\
The performance progression of the 10 best performers (10 BP) was recorded yearly throughout the world in flat horses thoroughbred races (years: 1898-2009; distances: 2200-2800m.; effort duration from 130 to 170 seconds) and greyhound races (years: 1929-2009; distances: 450-500m.; effort duration from 24 to 28 seconds). Men\textquoteright s track and field races (years: 1891-2009; distances: 200, 400, 800 and 1500m.; effort duration from 20 to 230 seconds) were used to express human 10 BP. The means of 10 BP from sprint (200 and 400m.) and short-middle distances (800 and 1500m.) were considered separately for a best correspondence to thoroughbred and greyhounds BP times respectively. For each species, only the best yearly performance of a single athlete or animal was kept, so that any given athlete appears only once per year in the data-sets. The yearly BP records were obtained in thoroughbreds from 30 events at the highest competitive level over the world (\textquoteleft Group 1 and 2\textquoteright~in Europe, Oceania and Asia and \textquoteleft Stakes\textquoteright~for North America). In greyhounds, we collected the yearly 10 best performances recorded in short race\textquoteright s rings (25 events in Great Britain, Ireland, Spain, United States, Australia and New Zealand). In humans, track and field outdoor best performances were collected. Speed data was collected with similar methodology as recently reported \cite{Denny2008, fdd1} from various sources in humans (Associations of Track and Field Statisticians; \cite{Rabinovich2010}, horses \cite{GalopCourse2011, Galoppsieger2011} and greyhounds \cite{Greyhoundsdata2011}. A total of 6620 performances were gathered (4720 in humans, 1120 in horses and 810 in dogs). The BP in dogs and humans were male only while the horse data-set included some females as well. However as the fraction of female BP was limited, we did not account for a sex effect when analyzing the data. Speed data were expressed in m.s$^{-1}$.
\\[0.3cm]
\textbf{Progression pattern of maximal running speeds}\\
Previous studies reported that best performances progression over the last century was best described by logistics and/or multi-exponential curves (\cite{berthelot2008, nevill1, Denny2008}. Blest, analyzing a set of world records in athletics, showed that Gompertz models provided lower standard errors of estimates compared to other models (i.e., antisymmetric exponential and logistic models) \cite{Blest1996}. They also described well best performances in humans \cite{berthelot2010a}. We therefore opted for such a model using the following function:
\begin{equation}
  \label{eqGompertz1}
  y(t) = a \cdot \exp^{b \cdot \exp^{c \cdot t}} + d
\end{equation}
where a gives the upper asymptote of $y$, $b$ sets the value of $t$ displacement, $c$ the curve steepness, and $d$ accounts for the fact that the minimum $y$ value is not 0. $a$ and $d$ takes here positive values while $b$ and $c$ are negative. The model parameters were estimated using a non-linear least-squares method on the uniformized ($[0,1]$ range) vector of recorded values. The physiological limit for each species was given by computing the year corresponding to 99.95\% (1/2000th) of the estimated asymptotic value. Curve-fitting was performed using Matlab (Version 7.11.0.584, USA).
\\[0.3cm]
The yearly progression of human speed was compared to that of greyhounds (200-400m. in humans) and thoroughbreds (800-1500m. in humans) and expressed as the percentage of a human / animal ratio. The annual speed variation was also observed in each species through coefficient of variation (CV) changes.
\\[0.3cm]
\textbf{Speed heritability in dogs and horses}\\
For the recent period, genealogical data were available in both dogs and horses allowing for a quantitative genetics approach. The speeds of current 10 BP (2007-2009) were collected and compared with those of their ancestors (individual best performance) \cite{GalopCourse2011, Galoppsieger2011, Greyhoundsdata2011}. This ancestor data-sets bears on 67 dogs (1964 to 2009), and 64 horses (1960 to 2009) including 6 females in the current 10 BP (2007-2009). All dogs and horses are part of a pedigree (7 generations in dogs and 6 in horses) with only male links known.\\
In order to decompose the phenotypic variance expressed in speed of greyhounds and horses lineages, and to estimate the heritability ($h^2$) of this character, we used two methods: (\textit{i}) a classic father-midsons regression, and (\textit{ii}) a restricted estimate maximum likelihood procedure to run a mixed model with the software ASReml \cite{Gilmour2006}. An \textquoteleft animal model\textquoteright~\cite{Kruuk2004} was used in which the year was considered as a fixed effect and the total phenotypic variance ($V_P$) was broken into two components of variance as follows: $V_P = V_A + V_R$, where $V_A$ is the additive genetic variance, and $V_R$ is the residual variance, consisting of environmental effects, non-additive genetic effects, and error variance. A single speed record was available in ten years in dog races and 20 years in horses; hence a year could not be entered as a random effect in the model. The full model outlined above was compared to a simpler model where $V_A$ was removed using a $\chi^2$ test, in order to determine whether speed displayed significant additive genetic variance.
\\[0.3cm]
\textbf{Geographical origin of human best runners}\\
The full human data-set (200 to 1500m.) was split according to the geographical origin (Africa, America, Asia, Europe and Oceania) of runners. Africa, Asia and Oceania are poorly represented in the 10 BPs over the period considered (1891-2009; Supplementary file S.1, \cite{SupInfA5}). The analysis was further refined for short middle-distance (800-1500m.; year range: 1970-2009) by comparing the 5 BP from North European countries (Nordic: Denmark, Norway, Sweden, Finland) and from East Africa (EAf; Djibouti, Eritrea, Ethiopia, Somalia, Tanzania, Kenya, Uganda, Burundi), for which large data-sets are available, to runners from rest of world (ROW). An analysis of variance was used to determine the regional (Nordic, EAf, ROW) and time (data arranged per decade; 1970-1979; 1980-1989; 1990-1999; 2000-2009) influence on speed progression. When speed differences were detected, Tukey post-hoc tests were used to identify significant differences according to region and decade. The software package STATISTICA (Version 6.1, Statsoft, France) was used for statistical analyses. The level of significance for all analyses was set at $p < 0.05$. Data are expressed as mean $\pm$SD.
\\[2cm]
\textbf{\textsc{Results}}
\\[0.6cm]
The Gompertz model fits speed progression with good accuracy in animals (horses: $R^2 = 0.56$, $MSE = 0.169$); greyhounds: $R^2 = 0.81$, $MSE = 0.55$; Fig. \ref{Fig. 4.1}A, \ref{Fig. 4.1}B) and in humans (800-1500m. 10 BP: $R^2 = 0.97$, $MSE = 0.006$, Fig. \ref{Fig. 4.1}C; 200-400m. 10 BP: $R^2 = 0.94$, $MSE = 0.008$, Fig. \ref{Fig. 4.1}D). The asymptotic limit in horses and dogs was 16.59m.s$^{-1}$ and 17.76m.s$^{-1}$ respectively, i.e. 99.0 and 97.4\% of the respective Gompertz asymptotes. In humans, the speeds of sprint and middle distances have already reached their estimated maximum based on the Gompertz curves (sprint: 9.54m.s$^{-1}$ in 1996, 100.3\% of asymptotic speed; middle distance: 7.45m.s$^{-1}$ in 2001, 100.1\% of asymptotic speed). Speed progression  was evaluated by comparing the first 10 BP recorded to the 10 BP ever observed, it appears similar in the three species (11.1\% of initial value in horses, 9.4\% in greyhounds, 11.1\% in 200-400m. and 14.2\% in 800-1500m. in humans) (Fig. \ref{Fig. 4.1}). Even less difference in speed progression were observed in the two animal species when using the asymptotic speed limit of the Gompertz curves rather than the 10 BP ever observed (12.4\% in horses, 11.9\% in greyhounds, and 11.1 and 14.2\% in humans).\\[0.3cm]
Although the overall progression was similar, the curves differed in shape among species ($b$ and $c$ coefficients). The $\frac{b}{c}$ ratios revealed a mono-exponential shape ($\frac{b}{c} > 3$) in both horses and greyhounds, while performances of humans follow an S-shaped development ($\frac{b}{c} < 0.1$) (Fig. \ref{Fig. 4.1}). The CVs of dog and human speed were of the order of 0.02 (or less) over the period considered Fig. \ref{Fig. 4.2}. In horses, the CV was also 0.02 for the 1940-2009 period, but the 1898-1939 period had larger variance with CV up to 0.05. Humans have increased their mean speed compared to dogs over a comparable distance with a current human / dog ratio downward plateauing towards 53.2\% Fig. \ref{Fig. 4.3}. The comparison with horses shows a current human / horse ratio of 45.2\%. Both ratios have been stable for 18 years (dogs) and 42 years (horses) respectively.
\end{multicols}
\begin{center}
\begin{tabular}{l l l}
\psset{xunit=.05cm,yunit=.8cm}
\begin{pspicture}(1895,13)(2015,17) 
    \psaxes[Dx=20,Dy=0.5,Ox=1895,Oy=13,ticksize=-3pt,labelFontSize=\scriptstyle]{->}(1895,13)(2015,17)[,-90][speed (m.s$^{-1}$),0]
    \fileplot[plotstyle=dots, dotscale=0.3]{figure10a1.prn}
    \fileplot[plotstyle=line, linecolor=red]{figure10a2.prn}
    \rput[b](1903,16.3){\textbf{(a)}}
\end{pspicture} & \hspace{0.5cm} &
\psset{xunit=.045cm,yunit=2cm}
\begin{pspicture}(1890,6)(2028,7.8)
    \psaxes[Dx=20,Dy=0.2,Ox=1890,Oy=6,ticksize=-3pt,labelFontSize=\scriptstyle]{->}(1890,6)(2028,7.6)[year,-90][speed (m.s$^{-1}$),0]
    \fileplot[plotstyle=dots, dotscale=0.3]{figure10b1.prn}
    \fileplot[plotstyle=line, linecolor=red]{figure10b2.prn}
    \rput[b](1900,7.3){\textbf{(c)}}
\end{pspicture} \\
\psset{xunit=.05cm,yunit=1cm}
\begin{pspicture}(1895,14.3)(2015,18.5) 
    \psaxes[Dx=20,Dy=0.5,Ox=1895,Oy=14.5,ticksize=-3pt,labelFontSize=\scriptstyle]{->}(1895,14.5)(2015,18)
    \fileplot[plotstyle=dots, dotscale=0.3]{figure10c1.prn}
    \fileplot[plotstyle=line, linecolor=red]{figure10c2.prn}
    \rput[b](1903,17.5){\textbf{(b)}}
\end{pspicture} & \hspace{0.5cm} &
\psset{xunit=.045cm,yunit=2cm}
\begin{pspicture}(1890,8.1)(2028,10.5) 
    \psaxes[Dx=20,Dy=0.2,Ox=1890,Oy=8.2,ticksize=-3pt,labelFontSize=\scriptstyle]{->}(1890,8.2)(2028,10)[year,-90][,0]
    \fileplot[plotstyle=dots, dotscale=0.3]{figure10d1.prn}
    \fileplot[plotstyle=line, linecolor=red]{figure10d2.prn}
    \rput[b](1900,9.7){\textbf{(d)}}
\end{pspicture}
\end{tabular}
\end{center}
\begin{scriptsize}

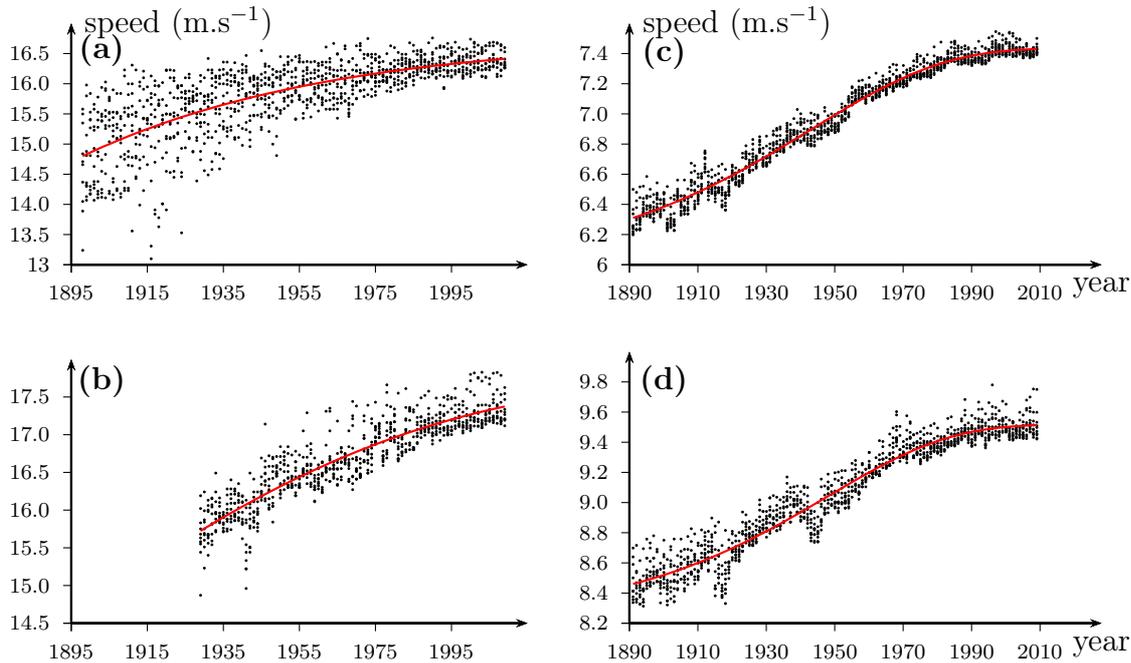
\captionof{figure}[One figure]{\label{Fig. 4.1} {\footnotesize Models fitting for speed progression (m.s$^{-1}$) between 1891 and 2009, in \textbf{(a)} 10 best thoroughbred performers ($R^2 = 0.56$); \textbf{(b)} 10 best greyhound performers on short ring races ($R^2 = 0.81$); \textbf{(c)} Mean speed of the 10 best human performers on 800 to 1500m. track races ($R^2 = 0.97$); \textbf{(d)} Mean speed of the 10 best human performers on 200 to 400m. track races ($R^2 = 0.94$). The calculated a, b, c, d parameters of the Gompertz function were for thoroughbred (respectively, 77.8, 3.77, 0.46, 16.59), greyhound (respectively, 14.99, 1.99, 0.60, 17.7), 800 to 1500m. human races (respectively, 1.45, 0.26, 3.06, 7.44) and for 200 to 400m. human races (respectively, 1.29, 0.20, 3.31, 9.52).}}
\end{scriptsize}
\begin{multicols}{2}
Heritability estimates based on father-midsons comparisons were neither significant in greyhounds (30 pairs; $h^2 = 0.042$, SE = 0.34, $p = 0.90$), nor in horses (34 pairs; $h^2 = 0.272$, SE = 0.23, $p = 0.25$). The \textquoteleft animal model\textquoteright~confirmed this trend in greyhounds ($N = 67$ data points): the additive variance was low (0.028, SE = 0.039), and model comparison (with and without $V_A$) showed that the heritability ($h^2 = 0.183$, SE = 0.240) did not differ from 0 ($\chi^2 = 3.08$, $p = 0.08$). In horses though ($N = 64$ data points), more additive variance was detected (0.085, SE = 0.059), and the heritability differed from 0 ($h^2 = 0.438$, SE = 0.267; $\chi^2 = 6.14$, $p = 0.013$). We note here that our approach, based on BP, underestimates the available phenotypic variance in both greyhounds ($V_P = 0.156$, SE = 0.003) and horses ($V_P = 0.193$, SE = 0.037).\\[0.3cm]
Human speeds achieved in the 800 and 1500m. races by the 5 BP over the 1970-2009 period differ according to performance years and geographical origin of runners (800m. races, $F_{1,39} = 5.89$, $p < 0.001$ and 1500m. races $F_{1,39} = 6.66$; Fig. \ref{Fig. 4.4}). EAf and ROW runners clearly increased their speed in both distances, compared to Nordic runners. The Nordic vs. EAf difference was not significant over the 1970-1989 decade, and both groups performed less well than the ROW group (all $p < 0.001$). Pairwise comparisons also showed a Nordic / EAf / ROW increasing hierarchy for 800m. races. In the most recent decade (2000-2009), speeds on the 800 and 1500m. races are lower in Nordic than in EAf runners (both $p < 0.001$), but there was no difference between  EAf and ROW ($p = 0.48$ in 800m. and $p = 0.59$ in 1500m. races). While the speed of ROW and EAf runners increased over both distances since 1970, that of Nordic runners actually stagnated (1500m.: $6.83 \pm 0.05$ in 1970-1979 vs. $6.73 \pm 0.04 \text{m.s}^{-1}$ in 2000-2009; 800m.: $7.42 \pm 0.08 \text{m.s}^{-1}$ in 1970-1979 vs. $7.44 \pm 0.09 \text{m.s}^{-1}$ in 2000-2009).\\[.1cm]
\begin{center}
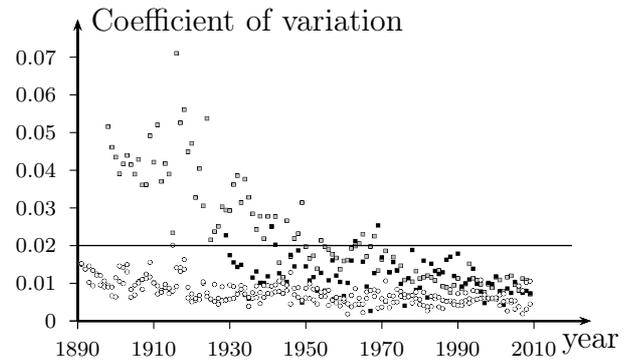

\psset{xunit=.05cm,yunit=50cm}
\begin{pspicture}(1870,-0.005)(2025,0.08) 
    \psaxes[Dx=20,Dy=0.01,Ox=1890,Oy=0,ticks=all,ticksize=-3pt,labelFontSize=\scriptstyle]{->}(1890,0)(2025,0.08)[year,-90][Coefficient of variation,0]
    \fileplot[plotstyle=dots, dotstyle=Bsquare, fillcolor=lightgray, dotscale=0.5]{figure11d1.prn}
    \fileplot[plotstyle=dots, dotstyle=square*, dotscale=0.5]{figure11d2.prn}
    \fileplot[plotstyle=dots, dotstyle=o, fillcolor=white, dotscale=0.5]{figure11d3.prn}
    \fileplot[plotstyle=dots, dotstyle=Bo, fillcolor=white, dotscale=0.5]{figure11d4.prn}
    \psline[linewidth=0.5pt,linearc=0]{-}(1890,0.02)(2020,0.02)
\end{pspicture}
\end{center}
\captionof{figure}[One figure]{\label{Fig. 4.2} {\footnotesize Coefficient of variations of 10 BP in thoroughbred (1899-2009; grey squares), greyhounds (1929-2009; black squares), human 200-400m. races (1891-2009; grey open circles) and humans 800-1500m. races (1891-2009; black open circles)\\[.3cm]}}
\textbf{\textsc{Discussion}}
\\[0.3cm]
Our study shows that the maximum running speed over short to middle distances increased over the last century in dogs, horses and humans, but is currently reaching its asymptotic value. This is a striking result because of the difference in genetic and environmental conditions leading to speed improvement in the three species. The fitted Gompertz curves do not exhibit the same shape (see values of $\frac{b}{c}$ for the three curves in Results section) -the initial plateau in humans is not detected in horses and greyhounds-, but this might simply be due to the larger variance in dogs and horses in the first decades of records or the facts that data were collected earlier in the speed progression in humans than in dogs and horses. Human speed has indeed already reached its asymptote, while horses and greyhounds speeds are 1.0 and 2.6\% down from their respective values. These results are in agreement with recent studies hypothesizing that current speeds are reaching the species\textquoteright~locomotory limits and explains their conclusions \cite{berthelot2008, Denny2008}.\\[.1cm]
\begin{center}
\psset{xunit=.05cm,yunit=0.225cm}
\begin{pspicture}(1890,39)(2025,60) 
    \psaxes[Dx=20,Dy=3,Ox=1890,Oy=40,ticks=all,ticksize=-3pt,labelFontSize=\scriptstyle]{->}(1890,40)(2025,60)[year,-90][Human vs. dog and horse speeds (\%),0]
    \fileplot[plotstyle=dots, dotstyle=square*, fillcolor=black, dotscale=0.5]{figure12d1.prn}
    \fileplot[plotstyle=dots, dotstyle=Bo, fillcolor=lightgray, dotscale=0.5]{figure12d2.prn}
    \psline[linewidth=0.5pt,linestyle=dashed, dash=2pt, linearc=0]{-}(1890,53)(2020,53)
    \psline[linewidth=0.5pt,linestyle=dashed, dash=2pt, linearc=0]{-}(1890,45)(2020,45)
\end{pspicture}
\end{center}

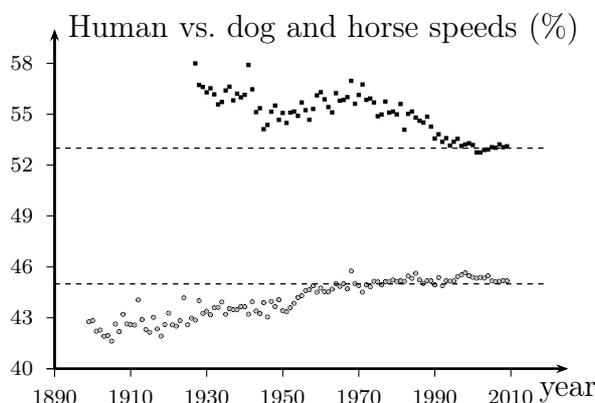
\captionof{figure}[One figure]{\label{Fig. 4.3} {\footnotesize Ratios of human vs. greyhound (black squares, comparison with human 200-400m. mean speed) and human vs. thoroughbred (grey circles, comparison with human 800-1500m. mean speed). Dashed lines for mean speed gap between humans and animals when stabilized.\\[0.1cm]}}
However, the similarities do not appear only in the progression patterns over the last century, but also in the progression ranges which were very close, especially when using the asymptotic speed limits to estimate the maximal speed progression per species. Over the last decades, we also noted that the speed gap between species pairs remained remarkably stable (Fig. \ref{Fig. 4.3}). The speed CVs among the ten best performers are low as well (Fig. \ref{Fig. 4.2}) and very similar across species. The large CV recorded in the earlier decades for horses and dogs might be accounted for by the fact that rank, rather than speed, was often recorded in races at the end of the 19th century and beginning of the 20th century. It is also likely that the homogenization in training methods and the selection of similar genetic backgrounds also lead to reduced CVs.
\begin{center}
\begin{tabular}{c}
\psset{xunit=.15cm,yunit=6cm}
\begin{pspicture}(1967,7.15)(2014,8.0) 
    \psaxes[Dx=5,Dy=0.1,Ox=1969,Oy=7.2,ticks=all,ticksize=-3pt,labelFontSize=\scriptstyle]{->}(1969,7.2)(2014,8.0)[,-90][Speed (m.s$^{-1}$),0]
    \fileplot[plotstyle=dots, dotstyle=*, fillcolor=black, dotscale=0.5]{figure13a1.prn}
    \fileplot[plotstyle=dots, dotstyle=o, fillcolor=lightgray, dotscale=0.5]{figure13a2.prn}
    \fileplot[plotstyle=dots, dotstyle=Bo, fillcolor=white, dotscale=0.5]{figure13a3.prn}
    \rput[b](1972,7.9){\textbf{(a)}}
\end{pspicture} \\
\psset{xunit=.15cm,yunit=6cm}
\begin{pspicture}(1967,6.55)(2014,7.4) 
    \psaxes[Dx=5,Dy=0.1,Ox=1969,Oy=6.6,ticks=all,ticksize=-3pt,labelFontSize=\scriptstyle]{->}(1969,6.6)(2014,7.3)[year,-90][,0]
    \fileplot[plotstyle=dots, dotstyle=*, fillcolor=black, dotscale=0.5]{figure13b1.prn}
    \fileplot[plotstyle=dots, dotstyle=o, fillcolor=lightgray, dotscale=0.5]{figure13b2.prn}
    \fileplot[plotstyle=dots, dotstyle=Bo, fillcolor=white, dotscale=0.5]{figure13b3.prn}
    \rput[b](1972,7.2){\textbf{(b)}}
\end{pspicture}
\end{tabular}
\end{center}

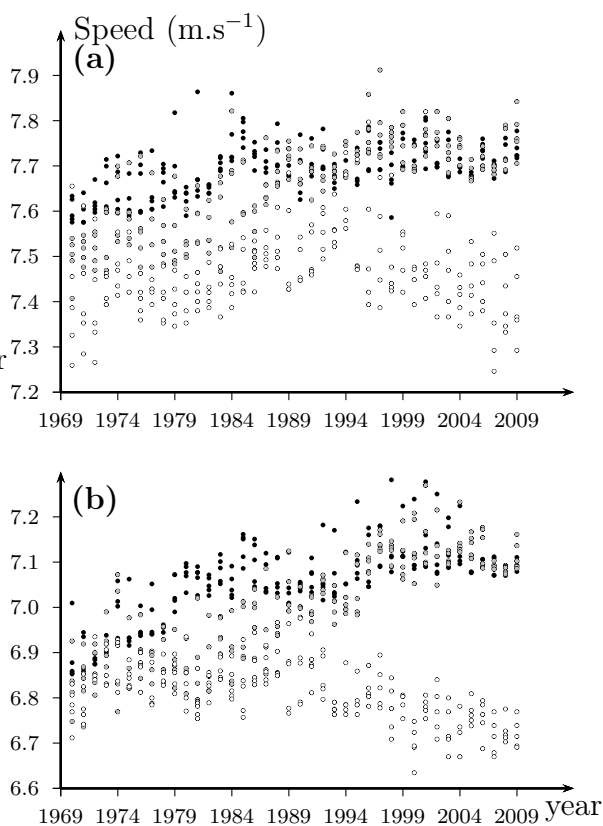
\captionof{figure}[One figure]{\label{Fig. 4.4} {\footnotesize Five best human performances in 800m. \textbf{(a)} and 1500m. \textbf{(b)} races in runners from Nordic countries (open circles), East African countries (grey circles) and rest of World (black circles).\\[.1cm]}}
The similarity in speed patterns (progression, asymptotes, CV) across species is somewhat unexpected since the balance of genetic and environmental forces acting on this pattern is different. Dogs for about 20 generations, and horses for 25 generations have been submitted to intense artificial selection over the period considered (artificial selection was indeed initiated much earlier in both species) \cite{Willet1975}.\\[.3cm]
Quite clearly, artificial selection has reached a phenotypic limit with regard to the maximal speed and minimal variance across the 10 BP, at least under the environmental conditions imposed by humans on dog and horse champions. Interestingly the quantitative genetic analysis indicated some genetic variance and heritability in horses. These genetic effects are larger than those reported recently for lifetime earnings in a larger thoroughbred population \cite{Wilson2008}. This marked genetic influence on speed appears surprising when considering the reduced speed progression over the last 40 years. However the animal model was run on a much more reduced data-set than that used for getting the general trend, and the ancestors of the current 10 BP did not belong to the 10 BP in their generation. It might also that training and raising environments for best horse racing should have been homogenized over time enhancing in return $h^2$ estimates. The heritability in dogs was on the verge of statistical significance ($p = 0.08$). Previous studies reported that selective breeding may have led to high homogeneity of predispositions for running \cite{Gu2009, Hill2010} suggesting reduced chance for the occurrence of genetically gifted individuals. Our results highlight that genetic predispositions for running fast in these particular populations are still good indicators of individual speed.\\[.3cm]
The optimization of running speed in humans probably essentially relies on training and national detection systems, but these might also allow identifying favorable polygenic profiles for elite runners \cite{Yang2003, William2008, Bejan2010}. It remains therefore possible that part of speed progression in humans is due to genetic (evolutionary) processes because there might be some selection to run fast under some conditions. However this effect might only be minor given the very short time period considered. Our data also suggest that estimated speed limits of the 10 BP have already been achieved in 200-400m. and 800-1500m., although exceptional athletes may sometimes set atypical performances \cite{berthelot2010a}. Over the century, the number of individuals engaged in athletic competitions evolved with societies\textquoteright~development, and their emphasis on sport \cite{marion1}. America and Europe largely contributed to the 10 BP with differences according to distance, while Asia and Oceania did not. The contribution of Africa to the 10 BP in middle distance over the most recent decades is noticeable. Initially less engaged, East Africans achieved speed records comparable to those of Nordic athletes on the 1500 m in the 1970\textquoteright s. The striking speed improvement over the last 40 years now placed them in the first ranks, comparable to the other world regions. It is no less striking that running speeds of natives from Nordic countries stagnated over the same period in 800m. races, and even declined in 1500m. Trainers and scientists from Northern Europe were indeed largely involved in the development of modern training methods and scientific results \cite{Astrand1991}, questioning their real impact on current best performances. On the other hand, Nordic athletes are still successful in other sports (i.e., jumps and throws in athletics, speed skating, cross-country skiing, rowing) suggesting that their genetic potential may be more favored under the environmental constraints of other sports.\\[0.3cm]
Over the last decades, the maximal human speed has progressed with the involvement of runners from new geographical origins leading to optimized phenotypes \cite{marion1}. The enlargement of the runner population in humans certainly enlarged the genetic pools upon which elite athletes were detected \cite{William2008}. Furthermore, some studies suggested that successful runners from East Africa originated from distinct ethnic and environmental backgrounds compared to the general population of these countries \cite{Onywera2006, Scott2009}. Quite clearly, improving speed through artificial selection in dogs and horses or detecting better athletes through an increasing population can only be actually efficient under appropriate environmental conditions, including training and competing conditions \cite{fdd1, Davids2007}. The decrease in the 10 BP during the two world wars \cite{marion1}, is an indication of the role of environmental conditions. Speed increase requires prosperous societies with a flourishing economy as described for improved health and body size over the 19$^{th}$ and 20$^{th}$ centuries \cite{fogel}.\\[0.3cm]
Our analysis is based on patterns exhibited in long-term data-sets collected in arguably artificial environments. It would certainly be interesting to use more experimental approaches to test whether other species are not far away from their limit, and whether this limit can be experimentally manipulated and increased. Speed increase has indeed been described in nature when environmental conditions allow for the improvement of maximal capacities \cite{Miles2004, husak2006, berthelot2011}. For example, dietary and thermal conditions experienced both during embryogenesis and early in life may favor the expression of genetic predisposition for running \cite{Elphick1998, LeGalliard2004}. Trade offs are of course expected between running performances and appropriate behavior or the ability to migrate towards more favorable environmental conditions \cite{Irschick1999, HusakFox2006}. In natural conditions, such adaptations in animals have been associated to survivorship and reproductive success \cite{Miles2004}. It is ironical that behavior-training and migration to more favorable environments also occur in competitive sports.\\[0.3cm]
\textbf{\textsc{Conclusion}}
\\[0.3cm]
The parallel progression of maximal running speeds in these three species over the last decades suggests that performances will no longer progress despite genetic selection in animals and best population detection in humans. Regardless of differences between species (biological, environmental, and competition history), human pressure, which has accelerated the biological adaptations allowing to run faster, is a process with limited potential and reduced benefits in the near future.
\end{multicols}
\chapter{The phenotypic expansion}
In the previous chapter we studied the developments of performances with time: the progression rate of sport performances is slowing down and is even halted in a majority of matures disciplines. In the present chapter, we focus on the core aspect of performance development with aging, at the performer scale and at the population scale. We assume that the performance-ageing pattern is redundant in many natural processes and at different scales (human body, organ, portion of an organ (ie. tissue), cell, etc.). Such a philosophy is shared with G. West that promotes an integrated and systemic view in order to better understand the physiological processes in medical research \cite{West2012}. We then define and investigate the concept of \textbf{phenotypic expansion} based on the development of sport performances with aging and time.
\section{Development of performances with age}
\label{sec:PhenotypicExpansion}
In order to investigate the development of sport performances with age, two types of data are gathered: the individual careers of athletes and the best performances per age class. Data approximatively cover the average life span, with missing data mostly distributed at the younger age and after the masters series. We gather the career of elite athletes in 13 events in T\&F,and 20 events in swimming. A total number of 646 and 512 careers are collected in T\&F and swimming respectively. We also investigate the development of chess performances in 96 international grandmasters careers. A model is adjusted to both the single careers and the best performances per age class:
\begin{equation}
  \label{eqmoore0}
  P(t) = a.(1-\exp^{b.t}) + c.(1-\exp^{d.t})
\end{equation}
with $P(t)$ the performance at age $t$. This model was first introduced by Moore in 1975 \cite{moore1975} and adjusted on a few sport events. It is adjusted to the gathered careers with a good accuracy: average $R^2 = 0.997 \pm 1.82\times10^{-3}$ for T\&F careers, $R^2 = 0.998\pm2.29\times10^{-3}$ for swimming careers and $R^2 = 0.755\pm1.99\times10^{-1}$ for chess careers. Beside sport careers, other phenomena in nature show such increasing and decreasing patterns with aging: Kasemsap and Wullschleger demonstrated that the net photosynthesis rate of a leaf with aging exhibits such a biphasic pattern \cite{kasempap1997, wullschleger1990}. This situation is also well known and described in physiology: the bone mineral density \cite{boot2010}, the pulmonary function \cite{schoenberg1978}, the cognitive capacity \cite{salthouse2009} or the human reproduction \cite{kuhnert1} exhibit such a pattern with aging. This is also expected to appear at other scales such as the cell level, although it remains hypothetical for the time being. Different ongoing studies are conducted to identify and quantify the performance of a cell and a population of cells with ageing (personal work).\\[0.3cm]
Technical innovations remain necessary for athletes to reach their peak condition: during the XX\textsuperscript{th} century, the environment of sport is widely impacted by the introduction of new equipments in training, medicine and nutrition. They help the athletes to optimize their potential at each age. By reading the present work, one should have noticed that we made an extensive use of the reference \cite{fogel} in the previous sections of the document. This reference points to the work of Fogel: \textquoteleft\textit{The escape from hunger and premature death, 1700-2100: Europe, America and the Third World}\textquoteright, that is a cornerstone of this work. He introduced the concept of \textquoteleft techno-physiological evolution\textquoteright~as a term for describing the process of ``\textit{the synergism between rapid technological change and the improvement in human physiology}'' \cite{Fogel2004b}. He used height as a proxy for health and general well-being, and observed dramatic improvements in health, body size, and mortality reduction over the past 200 years. In the previous sections, we demonstrated that WR are correlated to historical \cite{berthelot2008, berthelot2010a} and geopolitical \cite{marion1} events. It seems reasonable to think that the sport environment also benefited from such improvements, and also experienced a techno-physio evolution. In section \ref{sec:Technology}, we detailed the development of the technological enhancements dedicated to physiology. We showed that technological innovations were subsequently used for improving sport performances. As an example, one can think of indirect technological innovations such as wind tunnels or swimsuits. Other direct technological improvements are related to innovations in chemistry and physics and their diffusion in the sport environment, such as the polyurethane, first made in 1937 in Germany, as well as doping substances.\\[0.3cm]
Such technological innovations have influenced the development of physiology in sport. The area under Moore\textquoteright s curve may have increased with time, and both processes may have been altered such that the convex envelop may have expanded in all disciplines. A similar function (eq. \ref{eqmoore0}) can be used to describe the development of survival among individuals, with age-related diseases and disorders located at the start and at the end of the lifespan. Again, the same phenomenon occurs with an expansion of the convex envelop, due to technological breakthroughs in nutrition, medicine, biology, pharmacology. The major reduction of infant and elderly mortality extended lifespan to more than 75 years-old on average in developed countries. Such an approach is pictured in figure \ref{Fig.Expansion3}.\\[0.3cm]
The pattern describing the development of performances with aging was described in the article published in Age \cite{berthelot2011} and is presented below:
\\[0.6cm]
\noindent
\textbf{Exponential growth combined with exponential decline explains lifetime performance evolution in individual and human species}\\
G. Berthelot \& S. Len, P. Hellard, M. Tafflet, M. Guillaume, J.-C. Vollmer, B. Gager, L. Quinquis, A. Marc, J.-F. Toussaint
\\[0.6cm]
\noindent
\textbf{\textsc{Abstract}} The physiological parameters characterizing human capacities (the ability to move, reproduce or perform tasks) evolve with aging: performance is limited at birth, increases to a maximum and then decreases back to zero at the day of death. Physical and intellectual skills follow such a pattern. Here, we investigate the development of sport and chess performances during the lifetime at two different scales: the individual athletes\textquoteright~careers and the world record by age class in 25 Olympic sports events and in grandmaster chess players. For all data sets, a biphasic development of growth and decline is described by a simple model that accounts for 91.7\% of the variance at the individual level and 98.5\% of the variance at the species one. The age of performance peak is computed at 26.1 years old for the events studied (26.0 years old for track and field, 21.0 years old for swimming and 31.4 years old for chess). The two processes (growth and decline) are exponential and start at age zero. Both were previously demonstrated to happen in other human and non-human biological functions that evolve with age. They occur at the individual and species levels with a similar pattern, suggesting a scale invariance property.
\begin{multicols}{2}
Sport performances analysis revealed a large progression rate after 1896 \cite{berthelot2008, fdd1, berthelot2010a, berthelot2010b}. These enhancements coincided with major improvements in other human activities: energy production, water supply, transports, knowledge and science. Agriculture yields also dramatically evolved during the same period and enabled a sustainable growth of population. Initially triggered by the industrial revolution, this resulted in a great increase of life expectancy \cite{fogel}. Such a process may be referred to as a large \textit{phenotypic expansion}, as mankind optimized its performance potential over less than ten generations. Despite the strong development of sports performances over the last century, the asymptotic progression pattern demonstrated in several disciplines \cite{berthelot2008, fdd1, berthelot2010a, berthelot2010b} suggests that our progression as a species may soon reach its limits. Nowadays, major efforts are aimed at optimizing the processes that may further improve performance in all human domains. Among them, the process of selecting and hiring future elite performers is partly based on the assessment of individual performances at a young age \cite{hoye1}. The selected athletes are expected to progress and reach a peak under appropriate training programmes. Performance is thus used to determine their ability to later compete with and surpass others.
\\[0.3cm]
Age is also the main determinant in the decline of physical and intellectual capacities. The effect of age has been assessed at the cellular level \cite{kitani1}, in several physiological functions \cite{kuhnert1, aguilaniu1, fotenos1, vandisseldorp1} and can also be observed in various forms of human performance, such as sports \cite{baker2003, balmer2008, fair2007, tanaka2003, tanaka2008, wright2008, baker2010, young2008a, young2008b} and chess \cite{charness1990, fair2007}, which, respectively, illustrate physical and intellectual capabilities with advancing age. Based on a simple inverted U-shaped function, Moore has studied the effect of age on running speed \cite{moore1975}. The present study is based on a large sample of individual data from elite athletes in two major Olympic disciplines (track and field and swimming) and grandmaster chess players. Our aim is to assess and model the physiological development of performances during growth and aging and demonstrate how simple mathematics can shed light on the result of complex natural processes.
\\[0.3cm]
\textbf{\textsc{Material and methods}}
\\[0.3cm]
\textbf{Data}\\
Two types of data were collected: the individual careers of athletes and the top performances or world records established by age class ($WR_a$). The study of these progressions allows for the measurement of the age of peak performance in both sport and chess. We focus on the non-linear function describing these progressions. In the two studied sports, we covered the full range of Olympic distances; 13 events were chosen in track and field: the 100m. (men M and women W), 400m. (M, W), 800m. (M, W), 1500m. (M, W), 5000m. (M), 10000m. (M, W) and marathon (M, W). Data from throwing and hurdle events were not gathered since the rules depend on age. Twelve events were chosen in long course pool freestyle swimming: 50m. (M, W), 100m. (M, W), 200m. (M, W), 400m. (M, W), 800m. (M, W) and 1500m. (M, W). Datapoints of the best 96 all-time chess grandmasters were gathered.
\\[0.3cm]
\textbf{Individual sport careers}\\
The full career data set corresponds to all performances established by a single athlete and their date of achievement. All are recorded in recognized data sources (IAAF \cite{IAAF}; FINA \cite{FINA}). We selected athletes who established at least one performance in the yearly 10 best ranking in the period from 1980 to 2009. Full archived careers were gathered for 1392 subjects in track and field and 815 individuals in swimming. The date of birth of each athlete was recorded (IAAF \cite{IAAF}; T\&F \cite{alltimesTF}). For each year of an athlete\textquoteright s career, when several performances were established, only the best one was kept. The average number of performance points per career was 6.20$\pm$1.36 in track and field and 6.57$\pm$0.44 in swimming. These values represent the average number of marks established by the athletes listed in the official databases. The careers containing six or more points well describe the development of performances, so we decided not to restrict the study to the minimum of four points required for each fit. Thus, all careers containing at least six performance points were used for the modeling. We thereafter use the term \textquoteleft career\textquoteright~to designate the selected performances during the full career of a single athlete.
\\[0.3cm]
\textbf{Individual chess careers}\\
The full career data of chess players were gathered on Jeff Jonas\textquoteright~Chessmetrics (\cite{Chessmetrics, howard2005}). This web site lists ranking of top players back into the nineteenth century. We selected the first 96 chess players that show the greatest average ratings and recovered their whole career. A total of 34481 ratings were recovered, and the yearly best performances were selected for each player. The year of birth of each chess player was also recorded.
\\[0.3cm]
\textbf{World records}\\
In addition to the previously gathered data, the best performances of the cadet series (15.16 years old), junior series (17.18 years old) and master series ($>35$ years old) were also collected for the two sports disciplines (Swimnews \cite{Swimnews}; Swimrankings \cite{SwimRankings}; WMA \cite{WorldMastersAthletics}).
The complete data (junior+elite+master series in the two sports and all the whole careers of the chess players) is used to quantify the best performance established for each available age. We thereafter use the term \textquoteleft age-related world records\textquoteright~or $WR_a$ to designate this data set.
\\[0.3cm]
\textbf{The model}\\
For chronometric sports events, times (second) were converted into speeds (meter per second). Each sport performance is bound to the corresponding age according to:
\begin{equation}
  \label{eq1}
  t = \Delta Y + \frac{\Delta M}{12} + \frac{\Delta D} {365.25}
\end{equation}
where $t$ (in years) is the age, when a given performance is established; $\Delta Y$ (years) is the difference between the year when the performance is established and the performer birth year; $\Delta M$ (months) is the difference between the performer birth month and the month of performance and $\Delta D$ (days) is the difference between the day of birth and the day of performance. Only the year of birth was available for the chess players. The model is adjusted to the progression patterns of 1. all the careers of athletes and chess players and 2. the $WR_a$ for all sports events and chess player using the equation \cite{moore1975}:
\begin{equation}
  \label{eqmoore}
  P(t) = a.(1-\exp^{b.t}) + c.(1-\exp^{d.t})
\end{equation}
The coefficients $a$, $b$, $c$ and $d$ are estimated using a least-square non-linear regression method with:
\begin{equation}
  \label{mooreCstr}
  \left\{
    \begin{array}{ll}
        a,c,d > 0 \\
        b < 0
    \end{array}
  \right.
\end{equation}
For an estimated performance P, the model can be described as the sum of two von Bertalanffy\textquoteright s growth functions (VBGF): $P(t) = A(t) + B(t)$, where $A(t)$ is the increasing exponential process (first VBGF) and $B(t)$ the decreasing exponential process (the second VBGF is modified with $d>0$). The two processes are antagonists, and for appropriate sets of parameters (\ref{mooreCstr}), the resulting curve is always hump-shaped: rising at the beginning, reaching a maximum and falling back to zero: the decreasing $B(t)$ process overwhelm the increasing $A(t)$ process. For each event, the exact peak is computed and corresponds to the age when the performance is maximal. The roots of equation~\ref{eqmoore} are also computed for each series.
\\[0.3cm]
\textbf{\textsc{Results}}
\\[0.3cm]
A total of 646 careers (5,167 performances) were kept for running in track and field, 512 careers (3,129 performances) for swimming and 96 careers (2,969 performances) for chess players (Online Resource 1). The mean number of careers selected per event is $52.73\pm16.26$ (s.d.) for track and field and $42.67\pm6.71$ for swimming. All the careers of chess players were kept, as they all included more than six performances. The mean number of performances by career is $6.20\pm1.36$ for track and field, $6.57\pm0.44$ for swimming and $31.16\pm14.24$ for chess players.
\\[0.3cm]
The model describes the development of careers with age in sports and chess (Online Resource 2 \cite{SupInfA6}). The mean adjusted $R^2$ for the careers is $0.997\pm1.82\times10^{-3}$ for track and field, $0.998\pm2.29\times10^{-3}$ for swimming and $0.755\pm1.99\times10^{-1}$ for chess. It also describes the development of world records with age in the two disciplines: the mean adjusted $R^2$ for the world records is $0.991\pm4.05\times10{-3}$ for track and field, $0.987\pm6.06\times10^{-3}$ for swimming and 0.978 for chess.
\\[0.3cm]
In running, the performance peaks at $25.99\pm2.13$ on average and the gathered peaks in all events range from 23.29 years old (10,000m M) to 31.61 years old (Marathon M). The mean age of peak performance for swimming is younger than the one of track and field ($20.99\pm1.55$) and ranges from 18.36 (1500m M) to 23.14 (50m M). The mean age of peak performance is 31.39 years for chess.
\\[0.3cm]
The average roots value of \ref{eqmoore} of the track and field events is $109.48\pm5.97$ years old. Roots range from 98.72 years old (1,500m W) to 118.91 (100m W).
\\[0.3cm]
The average roots value for swimming events is $110.38\pm3.23$ ($110.44\pm2.14$ for men and $110.31\pm4.28$ for women). Roots range from 104.14 (50m W) to 115.77 (800m W).
\\[0.3cm]
The root for chess is 130.14 years old.
\\[0.3cm]
\textbf{\textsc{Discussion}}
\\[0.3cm]
The analysis of the age-performance relationship for each of these events suggests a biphasic development with two antagonistic processes. This pattern was first described by Moore using a bi-exponential function (eq. \ref{eqmoore}) on five sports events \cite{moore1975}. More recently, the declining process was reanalyzed by Donato and al. \cite{donato2003} and Tanaka and Seals \cite{tanaka2008} and described by Baker et al. \cite{baker2003}, Baker and Tang \cite{baker2010} and Bernard et al. \cite{bernard2010} using an exponential equation:\\
\begin{equation}
  \label{eqbaker}
  Y = 1 - \frac{\exp^{(T-T_0)}}{\tau}
\end{equation}
Stones and Kozma \cite{stones1984} and Bongard et al. \cite{bongard2007} also observed and modeled the declining process in a non-elite population and observed a quadratic decrease of performances. A later model investigated the master athletic world records using third order polynomial functions \cite{rittweger2009}. However, it remained impractical to model the development of performance during the whole lifetime, which could also tend toward positive infinity, making the estimation of life duration unrealistic. Chess ratings were also shown to follow an increasing and decreasing pattern with age \cite{charness1990, roring2007} that was not characterized. The model presented by Moore was used here to describe the development of 1. individual athletes\textquoteright~and chess players\textquoteright~careers and 2. the world records by age class in swimming and running Olympic disciplines and in chess ratings.
\\[0.3cm]
\textbf{Development of careers and world records with aging}\\
More than 11200 performances were used to describe the development of sports and chess careers. While the data used to construct individual careers might remain incomplete (data are not usually recorded at very young ages; many former athletes are still alive, which prevents to record data from their late life), the age span gathered covers a large part of the present human life-span. The model used properly describes the biphasic development of each career (mean $R^2 = 0.92$, Fig. \ref{Fig.Expansion1}).
\\[0.3cm]
This evolution is also observed for the $WR_a$ and successfully described by the same model (mean $R^2=0.99$, Fig. \ref{Fig.Expansion2}a) in both chess and sport (Fig. \ref{Fig.Expansion2}a,b). Using the $WR_a$ series, we computed the roots of eq. \ref{eqmoore} for each event (Online Resource 2 \cite{SupInfA6}). These correspond to a rough estimate of the maximum life-span. The maximal speed measured in track and field (100m) is associated with a value of maximal life-span close to the archived records for both men\textquoteright s and women\textquoteright s longevity \cite{young2009}. Through the analysis of their human maxima, this holistic model establishes a very coherent link between two major parameters of the phenotype: motion speed and life duration.
\\[0.3cm]
The causes and parameters influencing the decline of performances in the process of aging were and still are extensively investigated \cite{hayflick2007,kitani2007,njajou2010}. Donato et al. \cite{donato2003} previously analyzed the performances of American swimmers in US competitions and stated that the magnitude of age-related decline in swimming was smaller than that observed in running. He suggested that it was not due to differences in the training volume with age but rather to a greater reliance on biomechanical technique in swimming.\\[.3cm]
The magnitude of age-related decline in sprint and endurance performances remains a controversial question \cite{rittweger2009}. Previous studies reported that endurance performance was more affected by age than sprint performance \cite{baker2003,donato2003}. Finally, Wright and Perricelli \cite{wright2008} and Donato et al. \cite{donato2003} observed a difference between men and women in track and field and swimming, respectively, and suggested that both the rate and magnitude of decline were greater for women. Fair \cite{fair2007} investigated and compared the rates of decline in athletics and swimming events as well as chess. In athletics and swimming, he found that both men and women generally revealed larger rates of decline at the longer distances, with a larger rate of decline for women, in accordance with Tanaka and Seals \cite{tanaka2003}. We show here that these processes are in fact similar and independent from the considered sport, gender, milieu (ground or water) or principal anatomic-physiological medium (muscle, cardiovascular system, brain) as they are most strongly related to the effect of time upon all living things: aging.\\[0.3cm]
The performances were gathered over a 30-year period starting in 1980. Technological, medical and nutritional innovations appeared in this period and influenced the environment around each performance. Recent studies \cite{berthelot2008,berthelot2010a,nevill1,nevill2} analyzed the evolution of world records/top performers and
revealed that performances reached a reliable stability during this period. However, all the innovations
introduced contributed to widening the sides of the patterns: athletes can compete earlier (at a young age)
or later. There is no possibility to precisely quantify the medical or nutritional improvements, which may
have introduced a bias in the coefficients assessment or in the measurement of the age of peak performance. In the period and disciplines covered, however, only one major technological innovation was introduced: swimsuits, which induced three bursts in the evolution of performances \cite{berthelot2010b}. All athletes took advantage of them. This might have led also to a slight overestimation of the age of peak performance in short distance swimming. Other confounding variables may have impacted the performances, but the model shows a very robust adequacy with each series of performances, suggesting that the shape of the pattern is fixed for the studied period.\\[.3cm]
The progression in performance during the early phase is similar at both careers and $WR_a$ levels for sport events as well as chess: after the age of performance peak, the decline is described by the same equation. Both careers and age-adjusted world records provide a large coherence to illustrate the impacts of these associated mechanisms. The declining process starts from the origin of each career and each $WR_a$, i.e. at day zero. It competes against the development process that also starts from the origin, but the former systematically \textquoteleft wins the battle\textquoteright. Performance inevitably returns to zero: for the individual, this represents death. It is inherent to the model and remains inescapable at the individual and at the species level. The opposition between both biological courses may mathematically represent an anabolic and catabolic dialogue, which drives life from the conception of our first cell up to the reproductive acme then down to the last heart beat.\\[0.3cm]
\textbf{A biphasic development occurring at several scales?}\\
The model was applied at two different scales: the single career (Fig. \ref{Fig.Expansion1}) and the world records (Fig. \ref{Fig.Expansion2}a). It was also applied in two different disciplines: athletics, swimming and in two different fields: sport and chess. This biphasic development is also observed in other sports, such as tennis, basketball, soccer or baseball, where the age for peak performance appears similar \cite{schulz1988,bradbury2009,guillaume2010}. It is shared by a large variety of circumstances \cite{kuhnert1,aguilaniu1,fotenos1} suggesting a widespread common process in physiology.
\\[0.3cm]
The present analysis demonstrates that this pattern is similar in track and field and swimming, despite the
fundamental differences separating the two disciplines. The athletes encounter two fluid mediums, and the magnitudes of the drag forces in air or water are widely different. Running, swimming or flying are techniques developed by organisms to move while minimizing drag and energy losses \cite{alexander2005, bejan2006}. Such abilities play an important role in the theory of senescence and partly determine the capacity of a predator to catch a prey or the optimal response to danger \cite{medawar1952,williams1957,hamilton1966,reznick2004,carlson2007}. In fact, animal performance (i.e. running, swimming or flying speeds) is critically involved in the escape from predators \cite{husak2006,irschick2008}.
\\[0.3cm]
Our results also suggest that this ability to perform may be, to a certain extent, still correlated to life expectancy in the human species. Therefore, the biphasic development described here may also be operant not only in human but in other animal species, such as in insect \cite{dukas1994} and mammals \cite{bronikowski2006}. Several studies also demonstrated that the net photosynthesis rate is related to leaf age \cite{kasempap1997,wullschleger1990} through a biphasic pattern:
\begin{equation}
  \label{eqkasempap}
  Y = a + b.X.\exp^{c.X}
\end{equation}
The similarity of the mathematical models \ref{eqmoore}, \ref{eqbaker} and \ref{eqkasempap}, used to describe processes that fundamentally differ from one another by their nature or scale, suggests that a common law exists for a sizeable set of biological and physiological phenomena, all undergoing the same decaying process. A large number of biological processes in nature are fractals: they exhibit auto-similar shapes at different scales \cite{Mandelbrot1982}. We show here that the biphasic relation between performance and age exhibits a similar pattern at the individual and at the species level, suggesting a scale-invariant process. The growing and declining pattern of performances similarly operates for a living organism (the first studied scale) and a population (the second studied scale). All simultaneous biphasic developments at work determine the life-span of the studied system (cells, organisms, populations, Fig. \ref{Fig.Expansion3}a). This evolution seems to have been optimized through the human phenotypic expansion in the last two centuries (Fig. \ref{Fig.Expansion3}b).

\newpage
\begin{center}
\includegraphics[scale=0.8]{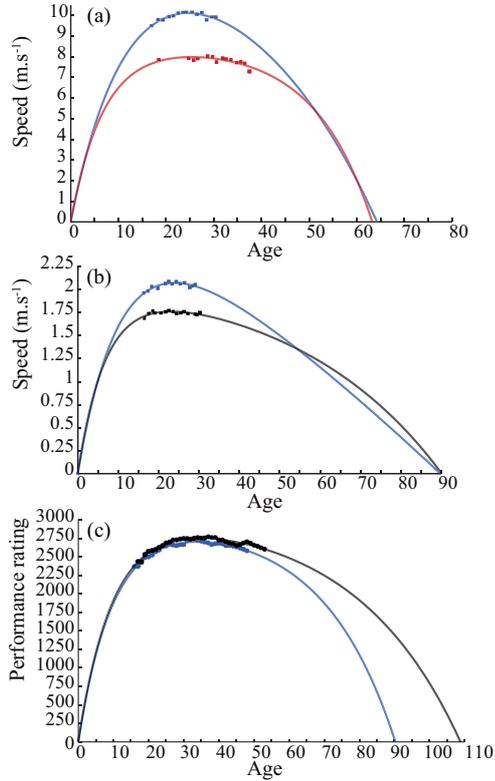}
\captionof{figure}[One figure]{\label{Fig.Expansion1} {\footnotesize The model applied at the individual scale (athletes\textquoteright~and chess players\textquoteright~careers). \textbf{a}. The model is adjusted to two careers in two track and field events: the 100m. straight (blue men career: Ato Boldon; adjusted $R^2$=0.99 and peak=24.63 years old) and the 400 m in track \& field (red women career: Sandie Richards; $R^2$=0.99 and peak=25.37). \textbf{b}. The model is adjusted to two careers in swimming: the 100m. freestyle (blue men career: Peter van den Hoogenband; $R^2$=0.99 and peak=23.92); the 200m. freestyle (black women career: Martina Morcova; $R^2$=0.99 and peak= 23.66). \textbf{c}. The model is adjusted to two careers in chess: Jonathan Simon Speelman (blue): $R^2$=0.97 and peak=33.86 and Jam Timman (black): $R^2$=0.95 and peak=34.67.}}
\end{center}

\begin{center}
\includegraphics[scale=0.81]{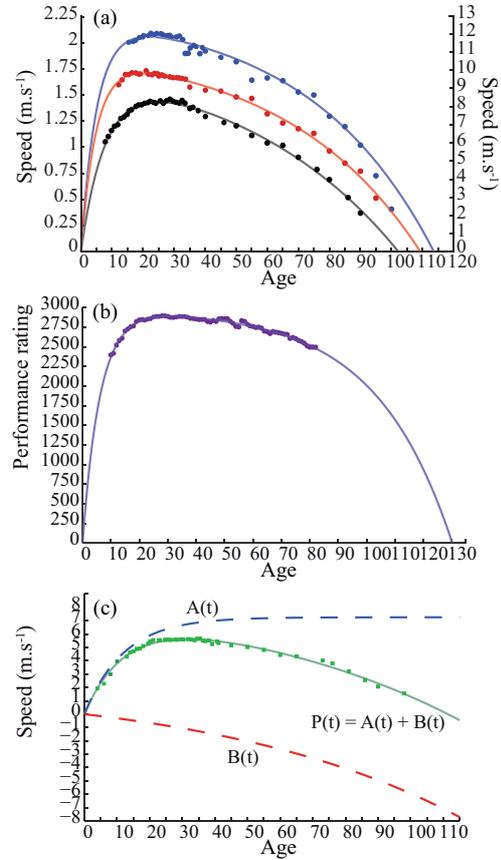}
\captionof{figure}[One figure]{\label{Fig.Expansion2} {\footnotesize The model applied at the species scale. For each age, the maximum performance among the studied careers is gathered. (\textbf{a}) The model is adjusted to two swimming events (left ordinate): 100 m men (blue, $R^2$=0.98 and peak=21.71) and 200 m women (red, $R^2$=0.99 and peak=20.04) and to one track and field event (right ordinate): the 400 m women (black $R^2$=0.99 and peak=24.72). (\textbf{b}) The model is adjusted to the best chess performance by age: (purple fit) $R^2$=0.97 and peak=31.39. (\textbf{c}) The marathon event (men) is fitted ($R^2$=0.99 and peak=31.61). The model used is composed of two antagonists processes: $P(t) = A(t) + B(t)$ (methods) with $A(t)$ the increasing process ($A(t) = 7.17\times(1-e^{-0.084\times age})$) and $B(t)$ the declining process ($B(t) = 1.84\times(1-e^{0.014\times age})$).}}
\end{center}
\newpage
\begin{center}
\includegraphics[scale=0.8]{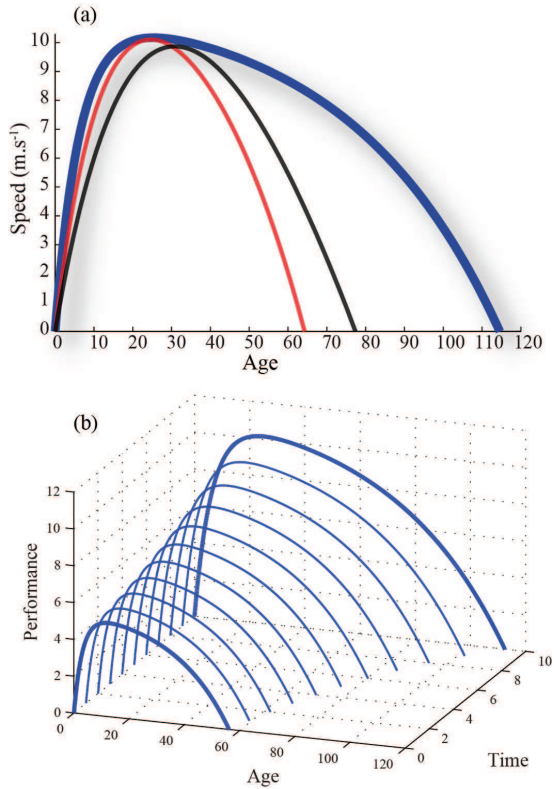}
\captionof{figure}[One figure]{\label{Fig.Expansion3} {\scriptsize Scale invariance and phenotypic expansion. (\textbf{a}) Estimated curves are plotted for two individual careers (red Ato Boldon, $R^2$=0.99, peak=24.63 at 10.10 $ms^{-1}$ and roots=[0, 64.27] years old; black Johnson Patrick, $R^2$=0.99, peak=30.82 at 9.87 $ms^{-1}$ and estimated roots=[0, 77.38] years old) and the $WR_a$ (blue curve) in the 100 m straight in track and field. The model describes how both individual and species scales are related. (\textbf{b}) Conceptual framework of the phenotypic expansion at work through the XX\textsuperscript{th} century (z-axis). Both performance (y-axis) and life expectancy (x-axis) were limited in the early times of industrial revolution. Both increased during time (black arrows) and allowed for optimizing the human performance and population life expectancy. The shape of the phenotype was thus extended. It now culminates to an optimum but at the price of high primary energy consumption and by stressing the pressure on the biomass.}}
\end{center}
In summary, the development of sport and chess performances during the process of aging exhibits two coexisting and conflicting processes that result in a biphasic pattern. It is possible to obtain the expected average performance gain from one age to another by deriving the equation on world records. Thus, characterizing the progression and regression of individual athletes or chess players as they age is mathematically feasible and may lead to nomograms with a physiological pattern and to the identification of \textquoteleft atypical paths\textquoteright~ \cite{berthelot2010a,berthelot2010b} or abnormal trajectories.\\[.3cm]
The estimated age of peak performance of the world records (26.1 years old) is in accordance with other biological and physiological phenomena such as the age of peak bone mineral density \cite{boot2010}, the development of the pulmonary function \cite{schoenberg1978}, the cognitive capacity \cite{salthouse2009} or the human reproduction \cite{kuhnert1}. Inherent to these phenomena, the function includes two mathematical terms associated with developmental growth and decline that closely describe performance development and occur at different levels. From a theoretical point of view, it has the self-similarity characteristics of a scale-invariant process. Technological improvements and the evolution of rules both led to modifications in the evolution pattern of performances. This may have slightly biased the coefficients assessment and the exact age of peak performance. Further, changes in the environment and increased athlete detection efforts might also have influenced the pattern. However, the model was adjusted with a high adequacy to the series of performances gathered here. It may therefore lead to several applications well outside the field of sport. The investigation of the age of peak performances for different phenomena occurring at different scales might lead to a classification of these phenomena, with early or late peaks. Another implication would be to study the interaction of the two phases at different scales and the development of a method to estimate life expectancy for a set of individuals. Injuries and illness and their impact on performances would also help to estimate their influence on the pattern. The use of QALYs and DALYs could help to model and understand this issue.
\\[0.3cm]
The study of the world records progression and top performances revealed a plateau in a majority of studied events. We extended the studied data and the model to a broader context: the development of physiological performance in the process of aging. This questions the upcoming evolution of the biphasic pattern presented here: will the phenotypic expansion continue, plateau or decrease? Do we have the ability to maintain our development in a sustainable way? Current trends and official statements develop multiple scenarios: a better understanding of their different steps and possible regulations will be necessary. Our results might help to define the contours of such questions.
\end{multicols}
\section{In other species}
The previous section demonstrated that the development of sport and cognitive performances with age can be modeled using the Moore equation (eq. \ref{eqmoore0}). The development of speed with aging is investigated in two other species; we study to which extend such a $\rotatebox[origin=c]{180}{U}$-shaped development adequately meets the age-speed relationship in other mammals species. We investigate competitive(greyhounds) and non competitive species (mice), revisit the model of Moore and introduce a new model based on ecological assumptions.
\subsection{Material \& methods}
\subsubsection{Data}
The development of performance with age is collected in three species for both genders: human (200m, 400m and 800m), greyhounds (480m) and mice (distance covered during a day). Dogs data come from the results of international competitions, while mice data are based on the recordings of voluntary physical effort.\\[.3cm]
\textbf{(\textit{i}) Greyhounds}\\
A total of 47,991 performances (39,664 performances for males, 8,327 for females) are collected on several websites \cite{Greyhoundsdata2011}. The entire 100 best times of the 480 meters (typical race distance) are gathered each year on a ten years period (2002-2012). We select the best speed (m.s$^{-1}$) per day of life for a total of 2,821 best speeds (1,552 for males and 1,269 for females).\\[.3cm]
\textbf{(\textit{ii}) Mice}\\
The maximal distance in the wheel activity per day is recorded for Mice (\textit{Mus musculus}) with a previous selective process based on their voluntary behavior to practice wheel activity \cite{Morgan2003}. Each of the 224 selected mice (112 for each gender) performed locomotor activity during their lifespan. A total of 14,241 (7,078 for males and 7,163 for females) wheel revolutions per day are gathered. In order to convert wheel revolution into average running speeds (m.s$^{-1}$) per day of life, we assume that mice ran 16h per day, such that the following transformation is used:
\begin{equation}
  \label{eqMice}
  P(t) = \dfrac{0.7215 \cdot W}{16\cdot3600}
\end{equation}
With $W$ the wheel revolution per day and $P(t)$ the performance at age $t$. The best speed per day of life is then gathered for a total of 1,507 speed recordings (755 for males and 752 for females).\\[.3cm]
\textbf{(\textit{iii}) Humans}\\
In order to compare the two species with man, we gather human performances with age in several websites (\cite{IAAF, alltimesTF, www.mastersathletics.net, www.tilastopaja.org}) in the following distances: 200, 400 and 800m straight. A total of 5,065 speeds (2,683 performances for men and 2,382 for women) are collected for the 200m, 5,013 speeds in the 400m (2,675 for men and 2,338 for women), 5,080 speeds in the 800m (2,754 for men and 2,326 for women). The best speeds per year of life are gathered for a total of 143 top speeds (73 speeds for men, 70 for women) in the 200m, 76 speeds (34 for men and 42 for women) in the 400m and 78 (39 for both genders) speeds in the 800m.
\subsubsection{Revisiting the initial model}
In his equation \ref{eqmoore0}, Moore considered two linearly independent processes. We consider that the two processes that describe performance development are not independent -ie. their effects did not sum one another- but rather interact to lead to senescence. The above equation \ref{eqmoore0} can subsequently be rewritten as the interaction of two Von Bertalanffy growth equations. Considering two specific time references $t_0$, $t_1$ as the respective roots of the equations, it yields:
\begin{equation}
  \label{eqmoore02}
  P(\text{age}) = a (1- \exp^{b \cdot \left(\text{age} - t_0\right)}) \cdot c (1- \exp^{d \cdot \left(\text{age} - t_1\right)})
\end{equation}
with respect to definition \ref{mooreCstr} in order to preserve the physiological increasing and decreasing pattern. We assume that $t_0 = 0$:
\begin{equation}
  \label{eqmoore03}
  P(\text{age}) = (ac)(1- \exp^{b \cdot \text{age}})(1- \exp^{d \cdot \left(\text{age} - t_1\right)})
\end{equation}
setting $A = ac$ and rewriting the variables in a correct order ($d \rightarrow c$), the function can read:
\begin{equation}
  \label{eqmoore04}
  P(\text{age}) = A (1- \exp^{b \cdot \text{age}})(1- \exp^{c \cdot \left(\text{age} - t_1\right)})
\end{equation}
$A$ is the scaling parameter and $b, c$ are two parameters controlling the strength of the growth and de-growth exponential processes respectively. Note that two parameters $t_0$, $t_1$ are related to the point in time where the processes reach 0. We assume that the increasing process starts at age 0 ($t_0=0$, ie the conception moment) and that the decreasing process stops at $t_1$ (ie the moment of death). Biologically, this has no meaning, because the increasing process probably begin somewhere before the birth of the individual (maybe during fertilization). However, in terms of speed of the individual it is convenient to assume that at birth the speed is approximatively 0.
\subsubsection{Introducing a new model}
The model of Moore was applied to the development of performance\textquoteright~maxima with aging and it resulted in a biphasic pattern at the population level \cite{berthelot2011}. This was observed in a range of mammal species and plants \cite{kasempap1997, wullschleger1990}. One leading assumption is the self-similarity feature of the model. It can be applied at both the population scale and at the individual scale, such as the athlete\textquoteright s career. Additionally, Kasemsap and Wullschleger previously demonstrated that the \textquoteleft performance\textquoteright~of leaves also exhibited a biphasic pattern with aging \cite{kasempap1997, wullschleger1990} while the overall yield of the cotton field also vary during time. The eq. (\ref{eqmoore04}) written as the interaction of two Von Bertalanffy growth equations suggests that two distinct processes lead to the observed patterns. In the following text, we use a different approach to explain the model of Moore based on population models. Considering a number of individuals, or \textquoteleft units\textquoteright, we describe \textit{i}) their growth and \textit{ii}) the declining process that starts at the creation of each unit.\\[.3cm]
\textbf{The general issue}\\
Consider a population of biological units $N$ (e.g. molecules, a population of cells, an organ, etc.) at a given scale $S$ that grow during the development phase. The general issue can be written as:
\begin{equation}
  \label{GissueNpop}
  P_S(\text{age}) = \alpha_S(t) \cdot N_S(t)
\end{equation}
where $\alpha(t)$ is a function describing the senescence process and $N(t)$ is a function describing the limited growth of the population. $N_S(t)$ can be a well known function such as the logistic function or Gompertz function.
\newpage
\textbf{A population model}\\
We model the issue \ref{GissueNpop} with two senescence factors: $\alpha_1(t)$, $\alpha_2(t)$. The first senescence parameter $\alpha_1$ is representative of the difficulties of the population of unit to keep reproducing at a constant rate with aging. Instead, units are less and less efficient to reproduce themselves. The second parameter starts at the birth of a unit, and smoothly decreases during aging. Such a population model can be written as:
\begin{equation}
  \label{eq:Model2.1}
   \dot{N_c} = \alpha_1(t) N_c
\end{equation}
where $\alpha_1(t)$ corresponds to:
\begin{equation}
  \label{eq:Model2.2}
   \alpha_1(t) = \alpha_{1_0} \exp^{-\gamma_1.t}
\end{equation}
with $\alpha_{1_0}$, the starting value of $\alpha_1(t)$ and $\alpha_{1_0}, \gamma_1 \in \mathbb{R}^{+*}$. Then eq. \ref{eq:Model2.1} reads:
\begin{equation}
  \label{eq:Model2.3}
   \dfrac{\dot{N_c}}{N_c} = \alpha_{1_0} \exp^{-\gamma_1.t}
\end{equation}
it comes (recall that $ln(u)' = \frac{u'}{u}$):
\begin{equation}
  \label{eq:Model2.4}
   \dfrac{d(\ln N_c)}{d_t} = \alpha_{1_0} \exp^{-\gamma_1.t}
\end{equation}
integrating and considering that at time $t=0$, $N_c = N_{c_0}$:
\begin{equation}
  \label{eq:Model2.5}
   [\ln N_c - \ln N_{c_0}] = -\dfrac{\alpha_{1_0}}{\gamma_1} \left( \exp^{-\gamma_1.t} -1 \right)
\end{equation}
then:
\begin{equation}
  \label{eq:Model2.6}
   \ln N_c = \ln N_{c_0} + \dfrac{\alpha_{1_0}}{\gamma_1} \left( 1 - \exp^{-\gamma_1.t} \right)
\end{equation}
using $\exp$ on both sides:
\begin{equation}
  \label{eq:Model2.7}
   N_c = N_{c_0} . \exp^{\dfrac{\alpha_{1_0}}{\gamma_1} \left( 1 - \exp^{-\gamma_1.t} \right)}
\end{equation}
then:
\begin{equation}
  \label{eq:Model2.8}
   N_c = N_{c_0} . \exp^{\dfrac{\alpha_{1_0}}{\gamma_1}} . \exp^{-\dfrac{\alpha_{1_0}}{\gamma_1} . \exp^{-\gamma_1.t}}
\end{equation}
using $N_{\infty}$ as the value reached when $t \rightarrow \infty$ (ie. the capacity $N_{\infty} = N_{c_0}.\exp^{\dfrac{\alpha_{1_0}}{\gamma_1}}$):
\begin{equation}
  \label{eq:Model2.9}
   N_c(t) = N_{\infty} . \exp^{-\dfrac{\alpha_{1_0}}{\gamma_1}.\exp^{-\gamma_1.t}}
\end{equation}
We apply the second senescence parameter $\alpha_2$, as the time-related decreasing performance parameter of all units. Such that the final equation describing the performance of the system (at a given scale) with time $P(t)$ is:
\begin{equation}
  \label{eq:Fineq}
   P(t) = \alpha_2(t) \cdot N_c(t)
\end{equation}
in line with issue \ref{GissueNpop}. We model the decreasing feature $\alpha_2$ of each unit as a function of age, and use the individual growth or de-growth model of Von Bertalanffy. It is used to express the development of the body length of an organism as a function of time \cite{Bertalanffy1934, Pauly1987}. It has been shown to conform to the observed growth of most fish species and we assume that it can also model the age-related degradation. Thus, we write $\alpha_2(t)$ as a Von Bertalanffy\textquoteright s de-growth equation:
\begin{equation}
  \label{eq:Fineq2}
   \alpha_2(t) = \alpha_{2_0} \cdot \left(1 - \exp^{\gamma_2.(t-t_d)}\right)
\end{equation}
with $\alpha_{2_0}$ the initial value of $\alpha_2(t)$ and $t_d$ the specific death time (when performance reach zero). Note that:
\begin{equation}
  \label{eq:limFineq3}
   \lim_{t \rightarrow \infty} \alpha_2(t) = -\infty
\end{equation}
implying that:
\begin{equation}
  \label{eq:limFineq4}
   \lim_{t \rightarrow \infty} P(t) = -\infty
\end{equation}
In this model, the initial senescence is compensated by the increasing number of units at the given scale (provided $\gamma_1 > \gamma_2$). The two senescence interact to decrease the overall performance of the system (Fig. \ref{FigM1}). Equation \ref{eq:Fineq} can be written as:
\begin{equation}
  \label{eq:Fineq4}
   P(t) = \alpha_2(t) \cdot N_c(t) = A \cdot \left[ \exp^{-\dfrac{\alpha_{1_0}}{\gamma_1}.\exp^{-\gamma_1.t}} \right] \cdot \left[1 - \exp^{\gamma_2.(t-t_d)}\right]
\end{equation}
with $A = \alpha_{2_0} \times N_{\infty}$ such that we make the economy of one coefficient.
\newpage
\begin{center}
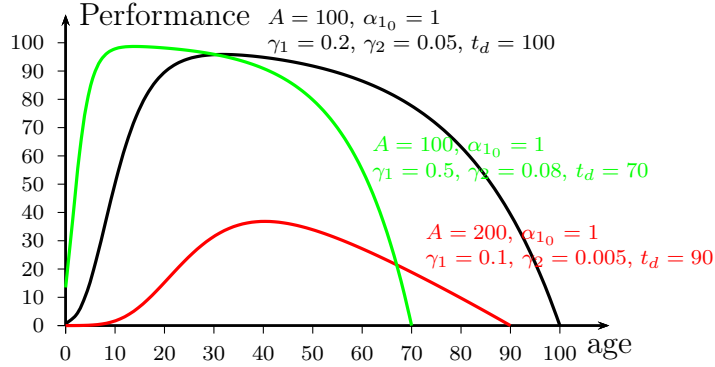

\psset{xunit=.065cm,yunit=0.0375cm}
\begin{pspicture}(0,-5)(110,115)
\psaxes[Dx=10,Dy=10,Ox=0,Oy=0,ticksize=-3pt,labelFontSize=\scriptstyle]{->}(0,0)(110,110)[age,-90][Performance,0]
\psplot[algebraic,linewidth=1.2pt,plotstyle=curve]{0}{100}{
    1 * Euler^(-(1 / 0.2) * Euler^(-0.2*x)) * 100 * (1 - Euler^(0.05*(x-100)))}
\psplot[algebraic,linewidth=1.2pt,plotstyle=curve,linecolor=red]{0}{90}{
    200 * Euler^(-(1 / 0.1) * Euler^(-0.1*x)) * (1 - Euler^(0.005*(x-90)))}
\psplot[algebraic,linewidth=1.2pt,plotstyle=curve,linecolor=green]{0}{70}{
    100 * Euler^(-(1 / 0.5) * Euler^(-0.5*x)) * (1 - Euler^(0.08*(x-70)))}
    \rput[b](70,95){\scriptsize \begin{tabular}[t]{@{}l@{}}$A = 100$, $\alpha_{1_0} = 1$ \\$\gamma_{1}=0.2$, $\gamma_{2}=0.05$, $t_d = 100$\end{tabular}}
    \rput[b](102,19){\scriptsize \textcolor{red}{\begin{tabular}[t]{@{}l@{}}$A = 200$, $\alpha_{1_0} = 1$ \\ $\gamma_{1}=0.1$, $\gamma_{2}=0.005$, $t_d = 90$\end{tabular}}}
    \rput[b](90,50){\scriptsize \textcolor{green}{\begin{tabular}[t]{@{}l@{}}$A = 100$, $\alpha_{1_0} = 1$ \\ $\gamma_{1}=0.5$, $\gamma_{2}=0.08$, $t_d = 70$\end{tabular}}}
\end{pspicture}
\captionof{figure}[One figure]{\label{FigM1} {\footnotesize The model (eq. \ref{eq:Fineq4}) is plotted for different values of the parameters.}}
\end{center}
One would have written $\alpha_2(t)$ as a decreasing Gompertz process, such as the decreasing feature approaches 0 as the age increases: $\lim_{t \rightarrow \infty} \alpha_2(t) = 0$ implying $\lim_{t \rightarrow \infty} P(t) = 0$.
\begin{equation}
  \label{eq:Fineq22}
   \alpha_2(t) = \exp^{-d \exp^{\gamma_2 t}}
\end{equation}
where $d$ sets the $y$ displacement and $\gamma_2$ is the de-growth rate. Then a new formulation of the model is:
\begin{equation}
  \label{eq:Fineq5}
   P(t) = \alpha_2(t) \cdot N_c(t) = A \cdot \left[ \exp^{-\dfrac{\alpha_{1_0}}{\gamma_1}.\exp^{-\gamma_1.t}} \right] \cdot \left[\exp^{-d \exp^{\gamma_2 t}} \right]
\end{equation}
It means that the speed of an individual would progressively approach 0 as $t$ increases. Regarding the biological meaning of this alternate formulation, it suggests that the decreasing process slows down to finally reach 0 as the units dissipate more and more energy. One would state that this energy is distributed to other physiological functions or dissipated into heat, such that the overall yield of the body decreases.
\begin{center}
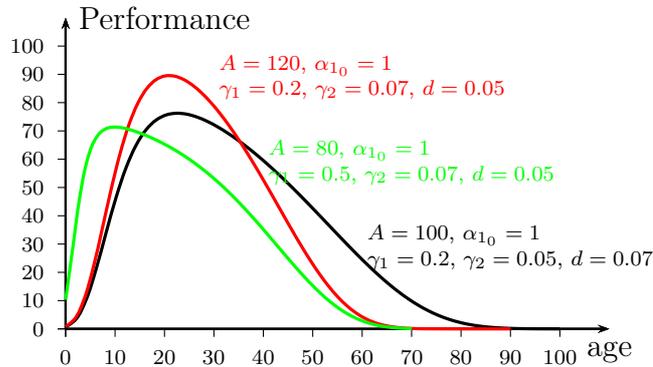

\psset{xunit=.065cm,yunit=0.0375cm}
\begin{pspicture}(0,-5)(110,115)
\psaxes[Dx=10,Dy=10,Ox=0,Oy=0,ticksize=-3pt,labelFontSize=\scriptstyle]{->}(0,0)(110,110)[age,-90][Performance,0]
\psplot[algebraic,linewidth=1.2pt,plotstyle=curve]{0}{100}{
    100 * Euler^(-(1 / 0.2) * Euler^(-0.2*x)) * (Euler^(-0.07 * Euler^(0.05*x)))}
\psplot[algebraic,linewidth=1.2pt,plotstyle=curve,linecolor=red]{0}{90}{
    120 * Euler^(-(1 / 0.2) * Euler^(-0.2*x)) * (Euler^(-0.05 * Euler^(0.07*x)))}
\psplot[algebraic,linewidth=1.2pt,plotstyle=curve,linecolor=green]{0}{70}{
    80 * Euler^(-(1 / 0.5) * Euler^(-0.5*x)) * (Euler^(-0.05 * Euler^(0.07*x)))}
    \rput[b](90,20){\scriptsize \begin{tabular}[t]{@{}l@{}}$A = 100$, $\alpha_{1_0} = 1$ \\$\gamma_{1}=0.2$, $\gamma_{2}=0.05$, $d = 0.07$\end{tabular}}
    \rput[b](60,80){\scriptsize \textcolor{red}{\begin{tabular}[t]{@{}l@{}}$A = 120$, $\alpha_{1_0} = 1$ \\ $\gamma_{1}=0.2$, $\gamma_{2}=0.07$, $d = 0.05$\end{tabular}}}
    \rput[b](70,50){\scriptsize \textcolor{green}{\begin{tabular}[t]{@{}l@{}}$A = 80$, $\alpha_{1_0} = 1$ \\ $\gamma_{1}=0.5$, $\gamma_{2}=0.07$, $d = 0.05$\end{tabular}}}
\end{pspicture}
\captionof{figure}[One figure]{\label{FigM2} {\footnotesize The model (eq. \ref{eq:Fineq5}) is plotted for various values of the parameters.}}
\end{center}
\subsection{Results}
We end up with 4 models: two models with 4 parameters and two models with 5 parameters. We use classical goodness-of-fit indicators (adjusted $R^2$, $\text{RMSE}$) and compare the 4 models using the corrected Akaike information criterion (AICc, eq. \ref{eq:AICc}) \cite{Bozdogan1987} and Schwarz\textquoteright s Bayesian Information Criterion (SBIC, eq. \ref{eq:SBIC}) \cite{Schwarz1978}. SBIC is similar to AICc but penalizes additional parameters more. We note RSS the residual sum of squares:
\begin{equation}
  \label{eq:RSS}
   \text{RSS} = \sum^n_{i=1} \left(y_i - \hat{y_i} \right)^2
\end{equation}
where $\hat{y_i}$ is provided by the model and corresponds to the estimated $y_i$, $n$ is the number of observations. AICc is evaluated with:
\begin{equation}
  \label{eq:AICc}
   \text{AICc} = n \ln \left( \dfrac{\text{RSS}}{n}\right) + 2k + \dfrac{2k(k+1)}{n-k-1}
\end{equation}
where $k$ is the number of parameters of the model. The SBIC is calculated with:
\begin{equation}
  \label{eq:SBIC}
   \text{SBIC} = n \ln \left( \dfrac{\text{RSS}}{n}\right) + k \ln(n)
\end{equation}
Generally, given a set of candidate models for the data, the preferred model is the one with the minimum AICc or SBIC value. The best model is determined by examining their relative distance to the \textquoteleft truth\textquoteright (ie. the model with the lowest AICc or SBIC value). The first step is to calculate the difference between model with the lowest AICc and the others:
\begin{equation}
  \label{eq:disAICc}
   \Delta^{\text{AICc}}_i = \text{AICc}_i - \min\left(\text{AICc}\right)
\end{equation}
$\text{AICc}_i$ is AICc for model $i$, $\min\left(\text{AICc} \right)$ is the minimum AICc value of all models. $\Delta^{AICc}_i$ is the difference between the AICc of the best fitting model and that of model $i$. The same approach is used for the SBIC:
\begin{equation}
  \label{eq:disSBIC}
   \Delta^{\text{SBIC}}_i = \text{SBIC}_i - \min\left(\text{SBIC}\right)
\end{equation}
with $\min\left(\text{SBIC} \right)$ is the minimum SBIC value of all models. The results are provided in tables \ref{TableMooreres1}, \ref{TableMooreres2} and \ref{TableMooreres3}.
\newpage
\subsection{Discussion}
Based on the results provided by the $R^2$ and RMSE (Tab. \ref{TableMooreres1}), the models that best performed are the population model v1 (eq. \ref{eq:Fineq4}), and the revised version of the initial model of Moore (eq. \ref{eqmoore04}). The same conclusion can be drawn when investigating the results of the AICc and the SBIC. Both measures give the same results, except for the 400m straight women where all models fit the data with nearly the same accuracy (apart from the last model, eq. \ref{eq:Fineq5}). Beside the fact that all 4 models fit well on the various data-sets (the minimum $R^2$ in all events is 0.58 and always superior to 0.9 concerning human events), one can withdraw the initial model (eq. \ref{eqmoore0}), based on both its misleading biological approach and its poor performance compared to other models. However, all models perform poorly outside of the data, ie. in the early (0-10) and advanced ages (Fig. \ref{Fig. 33-A1B1}). The models Moore and Moore rev. consider that the speed of an individual is null at age = 0 but starts to increase when age > 0. This assertion is dismissed when considering the two population models, but in some cases the speed is not null at birthdate. Regarding the fact that we initially consider the performance of a population of units that lead to the speed feature of an individual, it indicates that the units start the process of mobility before the birthdate. In all data-set there is no evidence of a convex development of performances when approaching the lifespan of a species. Kasempap et al. observed such a convex trend in the net photosynthesis rate of aged leaves in cotton fields, but we only focus on mammals, and the tail of the curve may differ when considering kingdoms (ie. Animalia, Plantae, Fungi, Protista, Archaea, and Bacteria) \cite{kasempap1997}. We test this approach: the long tailed model (eq. \ref{eq:Fineq5}) generally provides poor estimates of the lifespan of mammal individuals.
\begin{center}
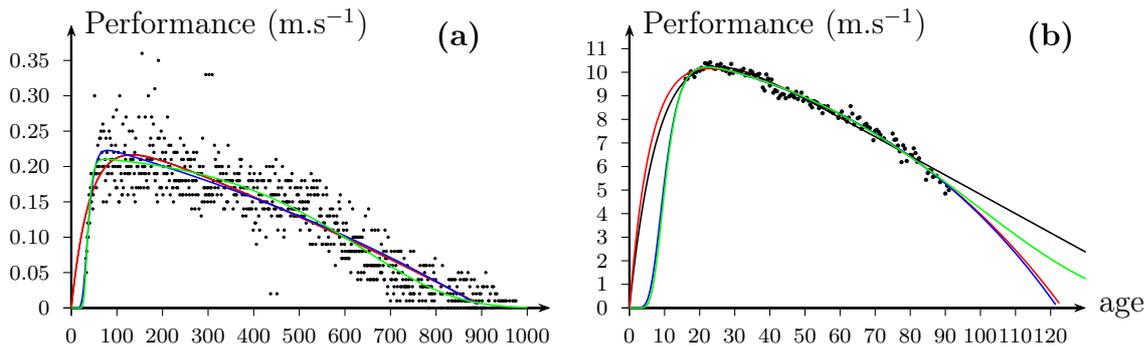

\begin{tabular}{l l l}
\psset{xunit=0.006cm,yunit=9.375cm}
\begin{pspicture}(0,0)(1000,0.40)
    \psaxes[Dx=100,Dy=0.05,Ox=0,Oy=0,ticksize=-3pt,labelFontSize=\scriptstyle]{->}(0,0)(1050,0.40)[,0][Performance (m.s$^{-1}$),0]
    \fileplot[plotstyle=dots, dotscale=0.35]{Figure33-A1.prn}
    \fileplot[plotstyle=line, linecolor=black, linewidth=0.6pt]{Figure33-AM1.prn}
    \fileplot[plotstyle=line, linecolor=red, linewidth=0.6pt]{Figure33-AM2.prn}
    \fileplot[plotstyle=line, linecolor=blue, linewidth=0.6pt]{Figure33-AM3.prn}
    \fileplot[plotstyle=line, linecolor=green, linewidth=0.6pt]{Figure33-AM4.prn}
    \rput[b](850,0.36){\textbf{(a)}}
\end{pspicture} & \hspace{0.5cm} &
\psset{xunit=0.046153846cm,yunit=0.3125cm}
\begin{pspicture}(0,0)(130,12)
    \psaxes[Dx=10,Dy=1,Ox=0,Oy=0,ticksize=-3pt,labelFontSize=\scriptstyle]{->}(0,0)(130,12)[age,0][Performance (m.s$^{-1}$),0]
    \fileplot[plotstyle=dots, dotscale=0.45]{Figure33-B1.prn}
    \fileplot[plotstyle=line, linecolor=black, linewidth=0.6pt]{Figure33-BM1.prn}
    \fileplot[plotstyle=line, linecolor=red, linewidth=0.6pt]{Figure33-BM2.prn}
    \fileplot[plotstyle=line, linecolor=blue, linewidth=0.6pt]{Figure33-BM3.prn}
    \fileplot[plotstyle=line, linecolor=green, linewidth=0.6pt]{Figure33-BM4.prn}
    \rput[b](120,10.8){\textbf{(b)}}
\end{pspicture}
\end{tabular}
\captionof{figure}[One figure]{\label{Fig. 33-A1B1} {\footnotesize The development of speed with aging in (\textbf{a}) mice (females) and (\textbf{b}) human 200m straight women. The age are given in days-old for mice and years-old for humans. The four models are represented (black: Moore, red: Moore revisited, blue: population model 1, green: population model 2 (long-tailed)). Corresponding goodness-of-fit measures (adjusted $R^2$, RMSE, AICc and SBIC) are given in tables \ref{TableMooreres2}, \ref{TableMooreres3} and \ref{TableMooreres1}.}}
\end{center}
\begin{center}
  \small
\begin{tabular}{|l|c|c|c|c|c|c|c|c|c|c|}
  \hline
  $\Delta^{\text{AICc}}_i$ & M (m.) & M (f.) & G (m.) & G (f.) & H200 (m.) \\
  \hline
  Moore & -38.21 & -94.74 & -72.51 & -3.50 & -50.23 \\
  Moore rev. & -35.94 & -94.79 & -12.70 & \textbf{0} & -6.97 \\
  Moore pop1 & \textbf{0} & -83.14 & -0.47 & -2.47 & \textbf{0} \\
  Moore pop2 & -3.38 & \textbf{0} & \textbf{0} & -2.10 & -8.91 \\
\hline
  $\Delta^{\text{AICc}}_i$ & H200 (w.) & H400 (m.) & H400 (w.) & H800 (m.) & H800 (w.)\\
  \hline
  Moore & -0.10 & -26.81 & -0.50 & -29.92 & -0.12 \\
  Moore rev. &  \textbf{0} & -26.68 & -0.2857 & -29.70 & \textbf{0} \\
  Moore pop1 &  -0.73 & \textbf{0} & \textbf{0} & \textbf{0} & -1.26 \\
  Moore pop2 &  -5.08 & -43.63 & -16.92 & -31.70 & -9.10 \\
\hline
\end{tabular}
\captionof{table}{\label{TableMooreres2} {\footnotesize Values of $\Delta^{\text{AICc}}_i$ for each model. Zeros indicate the best models. The higher the distance, the poorer the fit; M denotes mouse G, greyhound and H human while (m.) (f.) (w.) denote males (or men), females and women respectively.}}
\end{center}
\begin{center}
  \small
\begin{tabular}{|l|c|c|c|c|c|c|c|c|c|c|}
  \hline
  $\Delta^{\text{SBIC}}_i$ & M (m.) & M (f.) & G (m.) & G (f.) & H200 (m.) \\
  \hline
  Moore & -33.62 & -90.14 & -67.18 & -3.51 & -47.49  \\
  Moore rev. & -31.35 & -90.19 & -7.36 & \textbf{0} & -4.23 \\
  Moore pop1 & \textbf{0} & -83.14 & -0.47 & -7.61 & \textbf{0} \\
  Moore pop2 & -3.38 & \textbf{0} & \textbf{0} & -7.23 & -8.91 \\
\hline
  $\Delta^{\text{SBIC}}_i$ & H200 (w.) & H400 (m.) & H400 (w.) & H800 (m.) & H800 (w.)\\
  \hline
  Moore & -0.10 & -24.10 & -0.22 & -27.19 & -0.12\\
  Moore rev. & \textbf{0} & -23.97 & \textbf{0} & -26.97 & \textbf{0}\\
  Moore pop1 & -3.38 & \textbf{0} & -2.46 & \textbf{0} & -3.99 \\
  Moore pop2 & -7.73 & -43.63 & -19.38 & -31.70 & -11.84\\
\hline
\end{tabular}
\captionof{table}{\label{TableMooreres3} {\footnotesize Values of $\Delta^{\text{SBIC}}_i$ for each model. Zeros indicate the best models. The higher the distance, the poorer the fit.}}
\end{center}
\newpage
\begin{center}
  \footnotesize
\begin{tabular}{|l|c|c|c|c|}
  \hline
  Model & eq. & data-set & $R^2$ & RMSE\\
  \hline
  Moore & \ref{eqmoore0} & mouse (males) & 0.6309 & 0.0368 \\
  Moore rev. & \ref{eqmoore04} & mouse (males) & 0.6321 & 0.0368 \\
  \textcolor{green}{\textbf{Moore pop1}} & \ref{eq:Fineq4} & mouse (males) & 0.6496 & 0.0359 \\
  Moore pop2 & \ref{eq:Fineq5} & mouse (males) & 0.6481 & 0.0360 \\
  \hline
  Moore & \ref{eqmoore0} & mouse (females) & 0.8221 & 0.0315 \\
  Moore rev. & \ref{eqmoore04} & mouse (females) & 0.8221 & 0.0315 \\
  Moore pop1 & \ref{eq:Fineq4} & mouse (females) & 0.8251 & 0.0313 \\
  \textcolor{green}{\textbf{Moore pop2}} & \ref{eq:Fineq5} & mouse (females) & 0.8434 & 0.0296 \\
  \hline
  Moore & \ref{eqmoore0} & greyhound (males) & 0.7091 & 0.1696 \\
  Moore rev. & \ref{eqmoore04} & greyhound (males) & 0.7201 & 0.1664 \\
  Moore pop1 & \ref{eq:Fineq4} & greyhound (males) & 0.7225 & 0.1656 \\
  \textcolor{green}{\textbf{Moore pop2}} & \ref{eq:Fineq5} & greyhound (males) & 0.7226 & 0.1656 \\
  \hline
  Moore & \ref{eqmoore0} & greyhound (females) & 0.5800 & 0.1983 \\
  \textcolor{green}{\textbf{Moore rev.}} & \ref{eqmoore04} & greyhound (females) & 0.5811 & 0.1980 \\
  Moore pop1 & \ref{eq:Fineq4} & greyhound (females) & 0.5806 & 0.1981 \\
  Moore pop2 & \ref{eq:Fineq5} & greyhound (females) & 0.5808 & 0.1981 \\
  \hline
  Moore & \ref{eqmoore0} & human 200m. (men) & 0.9725 & 0.2420 \\
  Moore rev. & \ref{eqmoore04} & human 200m. (men) & 0.9801 & 0.2059 \\
  \textcolor{green}{\textbf{Moore pop1}} & \ref{eq:Fineq4} & human 200m. (men) & 0.9813 & 0.1998 \\
  Moore pop2 & \ref{eq:Fineq5} & human 200m. (men) & 0.9800 & 0.2065 \\
  \hline
  Moore & \ref{eqmoore0} & human 200m. (women) & 0.9796 & 0.2018 \\
  Moore rev. & \ref{eqmoore04} & human 200m. (women) & 0.9796 & 0.2017 \\
  \textcolor{green}{\textbf{Moore pop1}} & \ref{eq:Fineq4} & human 200m. (women) & 0.9797 & 0.2014 \\
  Moore pop2 & \ref{eq:Fineq5} & human 200m. (women) & 0.9789 & 0.2050 \\
  \hline
  Moore & \ref{eqmoore0} & human 400m. (men) & 0.9805 & 0.2133 \\
  Moore rev. & \ref{eqmoore04} & human 400m. (men) & 0.9805 & 0.2132 \\
  \textcolor{green}{\textbf{Moore pop1}} & \ref{eq:Fineq4} & human 400m. (men) & 0.9842 & 0.1918 \\
  Moore pop2 & \ref{eq:Fineq5} & human 400m. (men) & 0.9780 & 0.2265 \\
  \hline
  Moore & \ref{eqmoore0} & human 400m. (women) & 0.9816 & 0.2012 \\
  Moore rev. & \ref{eqmoore04} & human 400m. (women) & 0.9816 & 0.2011 \\
  \textcolor{green}{\textbf{Moore pop1}} & \ref{eq:Fineq4} & human 400m. (women) & 0.9818 & 0.2000 \\
  Moore pop2 & \ref{eq:Fineq5} & human 400m. (women) & 0.9794 & 0.2129 \\
  \hline
  Moore & \ref{eqmoore0} & human 800m. (men) & 0.9809 & 0.1893 \\ 
  Moore rev. & \ref{eqmoore04} & human 800m. (men) & 0.9809 & 0.1892 \\
  \textcolor{green}{\textbf{Moore pop1}} & \ref{eq:Fineq4} & human 800m. (men) & 0.9849 & 0.1685 \\
  Moore pop2 & \ref{eq:Fineq5} & human 800m. (men) & 0.9808 & 0.1898 \\
  \hline
  Moore & \ref{eqmoore0} & human 800m. (women) & 0.9843 & 0.1652 \\
  \textcolor{green}{\textbf{Moore rev.}} & \ref{eqmoore04} & human 800m. (women) & 0.9843 & 0.1651 \\
  Moore pop1 & \ref{eq:Fineq4} & human 800m. (women) & 0.9843 & 0.1652 \\
  Moore pop2 & \ref{eq:Fineq5} & human 800m. (women) & 0.9833 & 0.1701 \\
  \hline
\end{tabular}
\captionof{table}{\label{TableMooreres1} {\footnotesize Comparison of the 4 models based on the adjusted $R^2$ and the RMSE. Green color indicates which model performs the best. On average, model 2 \& 3 perform better than the initial model (Moore) and the long-tailed version. Note that the initial model of Moore is systematically outperformed by the other models.\\[.3cm]}}
\end{center}
\textbf{Mice}\\
Performances of mice were gathered by T.J. Morgan, T. Garland and P.A. Carter \cite{Morgan2003}. This data set is different from the two others because performance is recorded on the base of voluntary running effort with no time constraints, such that we rather model (\textit{i}) the development and decrease of the average speed during lifespan and (\textit{ii}) appetence for physical exercise than optimized instantaneous speeds. However, the average speed with life is modeled using the model of Moore, revealing that a biological attractor may exist that alters speed as age increases. This attractor may consist in a set of complex interactions between biological elements. It is irreversible.\\[.3cm]
\textbf{Other species}\\
Other athletics species such as horses (thoroughbred) and ostriches have historical records in racing competitions. However, data is very limited. Racing horses are usually retired when they are 6 years old due to health, attitude and performance issues. Assessing the decreasing part of the curve is thus not possible. Ostriches racing is limited to a few countries and data is difficult to gather.\\[.3cm]
Previous studies showed that a $\rotatebox[origin=c]{180}{U}$-shaped development of vital features with aging is observable in non domesticated species: bit force in the mouse lemur \cite{Chazeau2013}, flight performance in honey bee \cite{Vance2009}, cognitive performance in monkey \cite{Herndon199725}, ability of wolves to attack, select and kill elk \cite{MacNulty2009} and photosynthesis yield in cotton leaf \cite{kasempap1997}.\\[.3cm]
\textbf{Conclusion \& perspectives}: We find a common $\rotatebox[origin=c]{180}{U}$-shaped pattern for all studied -athletics and non athletics- species in sprint (humans, greyhounds) and endurance (mice). The 3 new models describe the dynamics of performance with aging with greater accuracy and enhanced biological basis compared with the initial model. We also show that models producing long tailed curves (ie. $\lim_{t \rightarrow \infty} P(t) = 0$) provide poor estimates of lifespan in the studied species and we recommend to use long tailed models with caution when modeling the decrease of performances with aging. However, biological data is still needed to test the self-similarity feature of the model at different scales. The performance records of other species can be used to assess an allometry hypothesis: is body size related to the standardized age of peak performance?
\section{Describing the expansion}
The section \ref{sec:PhenotypicExpansion} introduces the concept of \textbf{Phenotypic expansion}, suggesting that performance developed through age and time until now. The concept, illustrated in Fig. \ref{Fig.Expansion3}, may be expressed as a function of time and aging:
\begin{equation}
  \label{eqMoore2D}
  f(age,t) = f_2(t) \cdot P(age)
\end{equation}
where $P(age)$ is the revisited Moore function (eq. \ref{eqmoore04}), $t$ the time and $f_2(t)$ is the well-known logistic function:
\begin{equation}
  \label{Phi}
  f_2(t) = \frac{1}{1+\phi_1 \cdot exp^{-\phi_2 t}}
\end{equation}
with two parameters: $\phi_1$ and $\phi_2$. It means that both increasing and decreasing von Bertalanffy\textquoteright s processes are influenced by time (Fig. \ref{eqMoore2D}).
\begin{center}
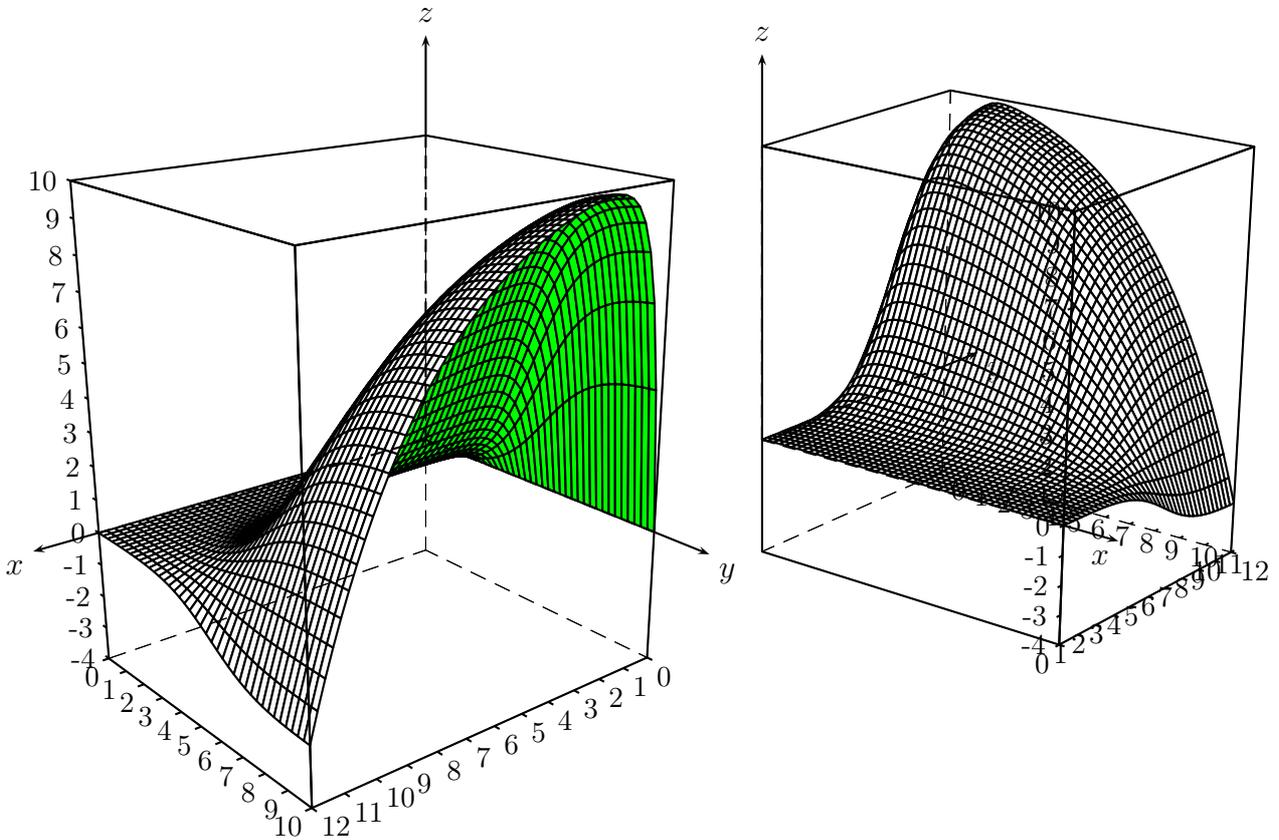

\begin{tabular}{l l}
\psset{viewpoint=50 50 20 rtp2xyz,Decran=20}
\begin{pspicture}(-5,-5)(14,6)
\psSurface[algebraic,ngrid=0.25 0.25,axesboxed,Zmax=10](0,0)(12,10){
     1/(1+300*Euler^(-y))*(11.06*(1 -Euler^(-2.14*x))*(1-Euler^(0.02667*(10*x-113))))}
\end{pspicture} &
\psset{viewpoint=100 -50 20 rtp2xyz,Decran=40}
\begin{pspicture}(10,-5)(24,6)
\psSurface[algebraic,ngrid=0.25 0.25,axesboxed,Zmax=10](0,0)(12,10){
     1/(1+300*Euler^(-y))*(11.06*(1 -Euler^(-2.14*x))*(1-Euler^(0.02667*(10*x-113))))}
\end{pspicture}
\end{tabular}
\captionof{figure}[One figure]{\label{Fig 4.X1} {\footnotesize Equation \ref{eqMoore2D} is represented for different values of $x$ (time), $y$ (age) and $z$ (Performance) from two different perspectives. The equation plotted is eq. \ref{eqMoore2D}: $f_2(y) \times A \left(1- \exp^{b x}\right) \left(1-\exp^{c (10 x - t_1)}\right)$, parameters values are $\phi_1 = 300$, $\phi_2 = 1$, $A = 11.06$, $b = -2.14$, $c = 0.02667$, $t_1 = 113$.}}
\end{center}
In order to investigate the age-time-performance relationship, we gather 3795 unique sport performances from different sources for one popular event: the 100m dash in Track and Field \cite{alltimesTF, MTRanking, ARRS}. For each performance the exact age is calculated (eq. \ref{eq1}). A sample of the data is given (Tab. \ref{Tab.Expansion}).
\begin{center}
     \begin{tabular}{ | l | l | l | p{5cm} |}
     \hline
     Performance (speed m.s$^{-1}$) & Age & Year of performance \\
     \hline
     9.93 & 23.56 &	1968  \\
     9.92 & 22.09 &	1968 \\
     9.93 & 22.86 &	1972  \\
     9.95 & 23.92 &	1975  \\
     9.92 & 21.76 &	1975  \\
     9.94 & 25.93 &	1976  \\
     9.93 & 25.41 &	1976 \\
     \hline
     \end{tabular}
\captionof{table}{\label{Tab.Expansion} {\footnotesize A sample of the data used in the 100m dash to test the age-time-performance relationship. For each performance, the corresponding age of the athlete and the year when the performance occurred are gathered.}}
\end{center}
We round the age values by 5-years classes and make 3 time classes: one containing the best performances values per age class before 1980 (1.74\% of the data-set), one containing the best performances between 1980 and 1990 (6.82\% of the data-set) and the last one containing the best performances after 1990 (90.29\% of the data-set). When plotting the values of the three classes, one can see that the post 1990 period expands: the area under the curve is greater than the areas of the two other periods. These two previous periods are similar. We perform the same analysis on the greyhounds data-set and create three time classes: after 1990 (29.75\% of the data-set), a class containing the best performances values by age band of the period 1970-1990 (24.77\%) and the last one containing the performances anterior to 1970 (45.47\%). Again, the same scheme appears, with an expansion of the performances in the greyhound after 1990.
\begin{center}
\begin{tabular}{l l l}
\psset{xunit=0.054545455cm,yunit=0.340909091cm}
\begin{pspicture}(0,0)(110,11)
    \psaxes[Dx=10,Dy=1,Ox=0,Oy=0,ticksize=-3pt,labelFontSize=\scriptstyle]{->}(0,0)(110,11)[,0][speed (m.s$^{-1}$),0]
    \fileplot[plotstyle=dots, dotstyle=o, fillcolor=black, dotscale=0.75]{Figure50-H1.prn}
    \fileplot[plotstyle=line, linecolor=black, linewidth=0.6pt]{Figure50-H1.prn}
    \fileplot[plotstyle=dots, dotstyle=Bsquare, fillcolor=red, dotscale=0.75]{Figure50-H2.prn}
    \fileplot[plotstyle=line, linecolor=red, linewidth=0.6pt]{Figure50-H2.prn}
    \fileplot[plotstyle=dots, dotstyle=Btriangle, fillcolor=blue, dotscale=0.75]{Figure50-H3.prn}
    \fileplot[plotstyle=line, linecolor=blue, linewidth=0.6pt]{Figure50-H3.prn}
\end{pspicture} & \hspace{0.5cm} &
\psset{xunit=0.003cm,yunit=1.5cm}
\begin{pspicture}(0,15)(2000,17.5)
    \psaxes[Dx=200,Dy=0.5,Ox=0,Oy=15,ticksize=-3pt,labelFontSize=\scriptstyle]{->}(0,15)(2000,17.5)[age,100][speed (m.s$^{-1}$),0]
    \fileplot[plotstyle=dots, dotstyle=o, fillcolor=black, dotscale=0.75]{Figure50-G1.prn}
    \fileplot[plotstyle=line, linecolor=black, linewidth=0.6pt]{Figure50-G1.prn}
    \fileplot[plotstyle=dots, dotstyle=Bsquare, fillcolor=red, dotscale=0.75]{Figure50-G2.prn}
    \fileplot[plotstyle=line, linecolor=red, linewidth=0.6pt]{Figure50-G2.prn}
    \fileplot[plotstyle=dots, dotstyle=Btriangle, fillcolor=blue, dotscale=0.75]{Figure50-G3.prn}
    \fileplot[plotstyle=line, linecolor=blue, linewidth=0.6pt]{Figure50-G3.prn}
\end{pspicture}
\end{tabular}

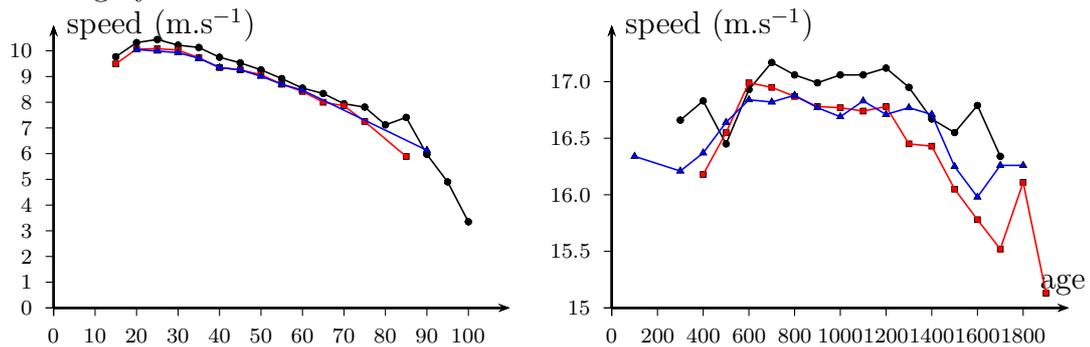
\captionof{figure}[One figure]{\label{Fig 4.X2} {\footnotesize Development of performances (speeds) with aging (years-old for human and days for greyhounds) during three periods in human (left panel) and greyhounds (right panel). Black dots correspond to the post 1990 period, red squares to the 1990-1980 period in humans and to the 1970-1990 period in greyhounds and blue triangles contain the values of performances before to 1980 (humans) and 1970 (greyhounds). An expansion occurs after 1990.}}
\end{center}
The observed phenotypic expansion after 1990 suggests an increase in performances values in all age bands in the 100m straight. However, previous sections reported that top performances are stagnating in this period (see sec. \ref{sec:OlympicSports}, \ref{sec:NonOlympicSports}, \ref{sec:Performance}). Two factors may explain this expansion: repeated data entries for the 20-31 age bands and the massive entry of seniors and the development of master competitions for age bands 31-100. The repeated marks of a few individuals: Usain Bolt, Tyson gay, Justin Gatlin, Yohan Blake and Asafa Powell for age bands 20-31 increase the performance level for the last decade only. But these marks might be modified in the future as all these athletes -to the exception of Usain Bolt so far- are tested positive for a banned substance during their career. This may alter the actual shape of the curve, and may reduce the observed expansion to a narrow margin. The total number of old people ($>65$ years old) is increasing each decade and there have been both an increasing interest in master races and investment from master athletes ($>40$ years old) in competitions \cite{Leyk2007, Leyk2009, Lepers2012}. This may have expanded the observed decreasing pattern. Data is difficult to gather in the 100m straight for age bands 40-100, altering the empirical description of the expansion, since the majority of the data-set is available after 1990. Recovering data on a larger period, prior to 1980 for all age bands, can be convenient to overcome this limitation.\\[.3cm]
The same pattern is visible on greyhounds, with an expansion after the year 1990. This is consistent with the observed pattern in the Fig. \ref{Fig. 4.1} \textbf{b}) (sec. \ref{sec:OtherSpeciesFDD}), where a progression is occurring in the top-10 performances.
\section{Energy as a catalyst in performance development}
\label{sec:Expansion and Energy}
The observed expansion is triggered at the industrial revolution by the massive availability of primary energies. The modern performances are only possible because of the the consumption of primary energy that diffuse through all the elements of the modern society, including sport environment. Maximizing $P(t)$ require huge supplies of energy inputs: a low input of primary energy is likely to imply low performances, whatever the age, while a high input of energy will produce top performances. Today\textquoteright s state of the art technologies require wind tunnels in order to better optimize their initial design. They are used to better understand the fluid dynamics of body suits in ski, cycling and swimming for instance. They are heavy energy consumers, up to 100 MW (Megawatts) for the largest facilities, and energy is a large portion of operational cost \cite{Wagner1}.
\begin{wrapfigure}{l}{0.4\textwidth}
   \begin{center}
    \includegraphics[width=0.4\textwidth]{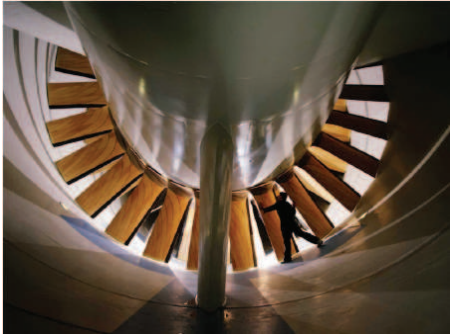}
   \end{center}
   \caption{\label{Fig.X: WindTunnel} {\footnotesize Grady McCoy stands in the Langley Research Center\textquoteright s 16 foot transonic wind tunnel in 1990. Before the tunnel was mothballed in 2004 it supported most major military programs both in their development stage and ongoing propulsion integration research including all fighters since 1960 (F-14, F-15, F-16, F-18 and the Joint Strike Fighter) (source: NASA). Such facilities were used for the design and test of new technologies dedicated to enhance sport performance.}}
\end{wrapfigure}
In the USA, wind tunnels have been decommissioned in the last 20 years -including some historic facilities- and pressure is brought to bear on remaining ones (erratic usage, high electricity costs, etc.) \cite{Goldstein2010} (Fig. \ref{Fig.X: WindTunnel}).\\[0.3cm]
Sport facilities are also high energy consumers: they often offer various services to the champion, ie. health care with cryotherapy or balneotherapy and training places. In particular indoor sport centers with swimming pools have a high expected energy demand. Interest arise in modeling the energy costs in such infrastructures and studies showed that maintenance and running costs significantly contribute to the total cost of actual running facilities \cite{Boussabaine1999, Boussabaine2001, Beuskera2012}.\\[.3cm]
Major sport events such as Olympic games or world championships require high primary energy inputs and the energy / greenhouse gas emissions associated with travel and venue operations are considerable. For a decade, environmental impacts and ecological footprint are being investigated in order to find sustainable solutions in mega-sporting events \cite{Collins2009, Dolles2010}. The economical outcomes of such events are well documented and remain controversial \cite{Pillay2008, Li2013, Kasimati2003}. Beside the constructive impacts there are evidences of obstructive impacts on the hosting city (under-utilized event facilities, preparations and staging of the events, subsequent unemployment, etc.) \cite{Zhao2012}. Moreover, it seems that there are no evidences for a large poverty alleviation. Rather, development benefits in cities are likely to be fairly circumscribed \cite{Pillay2008}, not to mention immediate and long-term environmental impacts that remains difficult to assess \cite{Collins2009}.\\[.3cm]
More generally, the sport performance \textquoteleft chain\textquoteright, that is the whole process of finding, selecting, training and bringing elite athletes to the top comes at the cost of high energy inputs, environmental issues and disputed economical impacts.\\[.3cm]
We previously showed that lifespan -another human performance- increases with time (sec. \ref{sec:Lifespan}). Demographers such as Bruce Carnes and Jay Olshansky \cite{Olshansky23022001, Carnes2003} suggest that we are already approaching this maximum human lifespan and note the difficulty of achieving further gains in human longevity. Although the improvement of individual lifespan remains multifactorial, one important trigger of the observed increase is our ability to exploit and use primary energies. We use them to maintain an adequate temperature during the seasons or to provide high quality supplies including food and water. Early estimates of human longevity suggest that life expectancy at birth was around 30 years old \cite{Wilmoth20001111}. After 1870, the increase of life expectancy become stable and quickly rise from 31 year old in the early XX\textsuperscript{th} to 67.2 years old in 2010. It suggests a stepwise development of longevity with multiples thresholds, or capacities induced by the introduction of a new features powered by energy (medicinal, technological, cultural, etc.). We focus on the lifespan and assume that longevity is a also a finite function of the primary energy input.
\subsubsection{Perspectives on modeling the effects of endogenous and extra-metabolic energy on the observed patterns}
Yukalov et al. implemented such a stepwise varying capacity to portray the change due to the activity of the agents in the system, who can either increase or decrease the carrying capacity by creative or destructive actions \cite{yukalov}. Although the authors aimed at modeling the dynamics at the world scale, we think it can be a starting point to model the dynamics of a population of units leading to the observed $\rotatebox[origin=c]{180}{U}$-shaped patterns (see previous section \ref{sec:PhenotypicExpansion}) that includes:
\begin{itemize}
        \item[] (\textit{i}) (internal) biological stresses, such as biological entropy (ie. the irreversible shift of the molecular structure from organization to anarchy and chaos), leading to the alteration of the increasing and decreasing processes
        \item[] (\textit{ii}) (external) favorable or detrimental environmental stresses leading to the alteration of the increasing and decreasing processes; including the use of extra-metabolic energy \cite{Moses2003} that extends our lifespan \cite{DeLong2010, Burger2011, Smith2013} and lead to the observed phenotypic expansion \cite{berthelot2011}
        \item[] (\textit{iii}) Inter-individual variability
\end{itemize}
Yukalov et al. considered the following delayed logistic equation:
\begin{equation}
  \label{YukalovSornette01}
  \dfrac{dN(t)}{d(t)} = rN(t) \left[ 1 - \dfrac{N(t - \tau)}{K}\right]
\end{equation}
where $r$ is the growth parameter (or reproduction rate), $K$ the carrying capacity, $\tau$ the delay of effective reproduction in the system. In their approach, Yukalov et al. proposed an improved version that included both the development of separate individuals and their collective interactions, such that eq. \ref{YukalovSornette01} can be written as:
\begin{equation}
  \label{YukalovSornette02}
  \dfrac{dN(t)}{d(t)} = \gamma N(t) - \dfrac{CN^2(t)}{K(t)}
\end{equation}
where the growth rate is $\gamma$, the collective effects $C$ and $K(t)$ is a time-varying capacity, corresponding to the maximum possible size of the population:
\begin{equation}
  \label{YukalovSornette03}
  K(t) = A + BN\left(t - \tau\right)
\end{equation}
$A$ is the pre-existing carrying capacity and the second member $BN\left(t - \tau\right)$ embody the delayed creations or destructions related to the environmental stresses (e.g. temperature, oxygen consumption, radiations, etc.). The parameter $B$ is the amplitude of a regenerating ($B \in \mathbb{R}^{+*}$) or decaying ($B \in \mathbb{R}^{-*}$) feedback on the capacity. It is related to the delayed impact of environmental stresses on the carrying capacity. The parameter $\tau$ is the corresponding lag time in which the feedback will take place.\\[.3cm]
The model of Yukalov et al. can be modified to fit the issue pictured in eq. \ref{GissueNpop}. Beside its biological ground, based on the ecological field -where it is designed to model population growth-, it includes a delay $\tau$ that may encapsulate biological heterogeneity and delayed degradation occurring between scales. Constants of eq. \ref{YukalovSornette02} can also be altered to include dependencies on pre-existing internal ($A$) and additional external ($B$) stresses (as exposed in sec. \ref{sec:Expansion and Energy}). Such a model can be convenient to describe the distribution of individual trajectories under the convex envelop (models 1-4 in the previous sec. \ref{sec:PhenotypicExpansion}) where inter-individual variability can be implemented as a random variable. This would be investigated in an upcoming work.
\chapter{Performance in the complex framework}
The previous chapters shed light on the strong interactions between sport performance and its direct or indirect environment. We showed that physiology relies on different parameters including climatic conditions, technology, genetics, medicine, nutrition or the economical and geopolitical context \cite{marion1}. Some of these parameters are related one another with positive or negative feedbacks. The behavior of such systems are generally difficult to assess due to their unpredictability. However in the specific case of sport performances and considering the actual paradigm will not shift in the near future, it remains possible to predict the upper boundaries of physiology (as shown in \cite{Blest1996, nevill1, berthelot2008}). In this chapter, we consider human performance in a wider sense and integrate other metrics of physiology such as lifespan or a population. Such systems with numerous interactions are often defined as \textquoteleft complex\textquoteright~systems.
\section{Complexity and complex systems}
\label{subsec:Complexity}
One can find many definitions of the term \textit{complexity} over the literature. It tends to be used to characterize something with many parts in intricate arrangement. Warren Weaver, one pioneer in this modern field of research, stated that the complexity of a particular system is the degree of difficulty in predicting the properties of the system when the properties of the system\textquoteright s parts are given \cite{Weaver1948}. As an example, one can think of a simple network of connected nodes that share a certain amount of its value with other nodes (Fig. \ref{Fig CS1}). The dynamics of such a network is difficult to predict naturally (ie. without using a computer simulation) in the long frame. The study of the dynamics and intrication of such systems is one of the main goal of complex systems theory. Garnett Williams defined a complex system as a system in which numerous independent elements continuously interact and spontaneously organize and reorganize themselves into more and more elaborate structures over time \cite{Williams1997}. Complex systems have been developed and reported in numerous natural examples: ants colonies, social structures, climate, nervous systems, modern energy or telecommunication infrastructures. Many natural processes that appear simple at first sight result from the complex behavior of a collection of individuals at a smaller scale. The long-term behavior of these systems are sometimes difficult or impossible to forecast (Fig. \ref{Fig CS1}). Complex systems have been studied for a long time but gain considerable attention during the XX\textsuperscript{th} century. They are at the interface of various research fields, such as ecology, social science, mathematics, physics, cybernetics, etc. We will here address a few relevant concepts in complex systems.
~\vspace{0.2cm}
\begin{center}
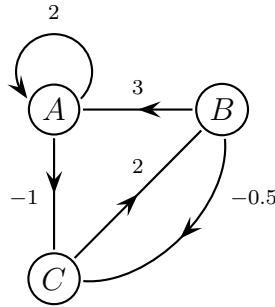

$
\psmatrix[colsep=1.5cm,rowsep=1.5cm,mnode=circle]
A&B\\
C
\psset{arrowscale=2}
\psset{ArrowInside=->,nodesep=1pt}
\everypsbox{\scriptstyle}
\ncline{2,1}{1,2}^{2}
\ncarc[arcangle=50]{1,2}{2,1}>{-0.5}
\ncline{1,2}{1,1}^{3}
\nccircle{->}{1,1}{.5cm}^{2}
\ncline[arcangle=-80]{1,1}{2,1}<{-1}
\endpsmatrix
$
\captionof{figure}[One figure]{\label{Fig CS1} {\footnotesize A graph with 3 connected nodes $A$, $B$ and $C$. At each time step, each node shares some percentage of its value quantity with its neighbors in the network. Specifically, the amount of shared value is divided equally and sent along each of the outgoing links from each node to each other node. If a node has no outgoing links, then it doesn\textquoteleft t share any of it\textquoteleft s value; it just accumulates any that its neighbors have provided via incoming links. As an example of non-trivial forecast in complex systems, one could try to instinctively find the long-term values of the three nodes $A$, $B$ and $C$ in this simple connected graph, given different initial values of the nodes. The values of $A$, $B$ and $C$ converge, in the long term, to fixed values. Such a result is not obvious without simulating the behavior of the system.}}
\end{center}
\subsection{Chaos theory}
Chaos theory investigates the behavior of nonlinear \textquoteleft dynamical\textquoteright~ (ie. that changes over time) systems that are highly sensitive to initial conditions. Nonlinearity means that the output of the system is not proportional to the input and that the system does not conform to the principle of additivity. While linear systems are always exactly solvable, nonlinear systems may or may not be solvable. If it is solvable then it cannot be chaotic because it would be predictable. According to the definition of Stephen Kellert, chaos theory is ``\textit{the qualitative study of unstable aperiodic behavior in deterministic nonlinear dynamical systems}'' \cite{Kellert1993}. S. Kellert defined the behavior of the system as unstable and aperiodic, stressing that the system does not repeat itself. In his definition, he also stated that the system is deterministic, such that chaos is different from random behavior even if its aperiodicity and unpredictability may make it appear to be so (see \cite{Gleick1987}). The idea of iteration (or feedback), in which the output of the system is used as the input in the next calculation, is central to chaos theory. Much of the mathematics involves the repeated iteration of simple mathematical formulas such as the logistic map:
\begin{equation}
  \label{Verhulst}
  X_{t+1} = r X_t (1 - X_t)
\end{equation}
This equation is the discrete version of the logistic model, first published by Pierre-Fran\c{c}ois Verhulst in 1845-1847 as a model of population growth \cite{Verhulst1845, Verhulst1847}. In this equation, $X$ is the population, $t$ is time and $r$ is the rate of population growth (and the control parameter of the equation) also called the Malthusian parameter. The element $(1-X_t)$ in the equation establishes a practical limit to population growth, a constraint related to the existence of famine, disease, and birth control in the real world. When $r$ is less than 3, this system converges towards an equilibrium (or fixed) point, regardless of the initial population level. The concept of equilibrium (or steady state of the system) relates to the absence of changes in a system. In the context of difference equations, the equilibrium point $X^*$ is defined by:
\begin{equation}
  \label{SState}
  X_{t+1} = X_t = X^*
\end{equation}
When $r$ is between 3 and $\approx 3.57$, the system (i.e., the population) converges not on one but on an increasing number of values, which doubles successively from 2 to 4 to 8, and so forth. In the case of the logistic equation these are stable oscillations. A stable oscillation is a periodic behavior that is maintained despite small perturbations. A stable oscillation of period 2 implies that successive generations alternate between two fixed values of $X$: $X_1^*$ and $X_2^*$. Period 2 oscillations (or \textquoteleft two-point cycles\textquoteright) simultaneously satisfy the following two equations:
\begin{equation}
  \label{SO1}
  X_{t+1} = f(X_t)
\end{equation}
\begin{equation}
  \label{SO2}
  X_{t+2} = X_t
\end{equation}
When $r$ falls between 3.57 and 4.0, the system moves into chaos and the population varies erratically. At higher levels of $r$, the system can display either chaotic or non-chaotic behavior (Fig. \ref{Fig. CS1-1-Bifurcation}, \ref{Fig CS1-2}). This equation is a popular example illustrating chaos theory, as it demonstrates that although chaotic behavior is complex, it can arise from simple nonlinear functions.
\begin{center}
    \includegraphics[scale=1]{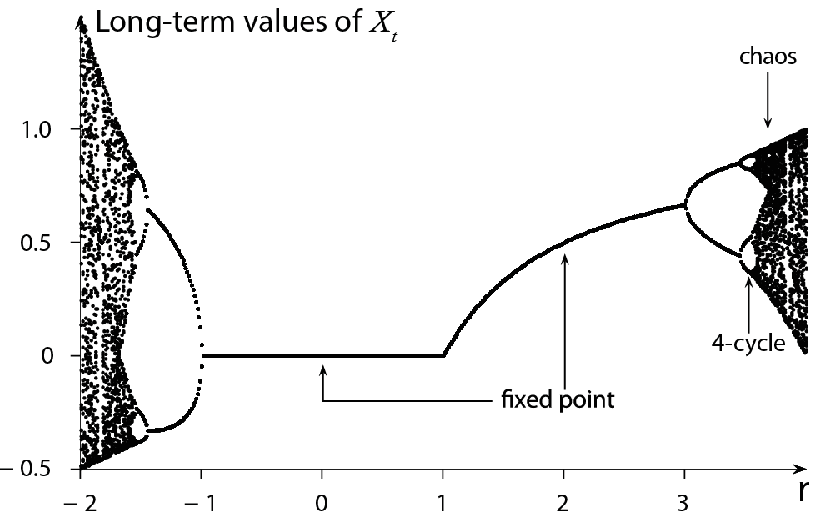}
\caption{\label{Fig. CS1-1-Bifurcation} {\footnotesize \textbf{Bifurcation diagram} of eq. \ref{Verhulst}, where long-term values of eq. \ref{Verhulst} are plotted vs. different values of parameter $r$. A bifurcation diagram shows the possible long-term values (such as fixed points, periodic or chaotic behavior) of a system as a function of a bifurcation parameter in the system. In this diagram, for each value of $r$, the local maximum of values of $X_t$ is reported. The transition from one regime to another is called a bifurcation. One interesting point that can be noticed is that the darkened areas of chaos are interspersed with windows of periodicity, a phenomenon known as \textit{intermittency}.}}
\end{center}
\begin{center}
     \begin{tabular}{ c c c c c }
        \psset{xunit=0.043cm,yunit=20cm}
         \begin{pspicture}(0,0)(100,0.15)
            \psaxes[Dx=10,Dy=0.025,Ox=0,Oy=0,ticksize=-3pt,labelFontSize=\scriptstyle]{->}(0,0)(100,0.15)[$t$,-90][$X_t$,0]
            \fileplot[plotstyle=line, dotscale=0.3]{FigureF-A1-r0.1.prn}
            \fileplot[plotstyle=line, linecolor=blue, dotscale=0.3]{FigureF-A1-r1.12.prn}
          \rput[b](80,0.012){$r=0.1$}
          \rput[b](80,0.12){\textcolor{blue}{$r=1.12$}}
          \rput[b](20,0.12){\scriptsize steady state}
         \end{pspicture} & \hspace{0.3cm} &
            \psset{xunit=0.043cm,yunit=3.333333333cm}
         \begin{pspicture}(0,0)(100,0.90)
            \psaxes[Dx=10,Dy=0.1,Ox=0,Oy=0,ticksize=-3pt,labelFontSize=\scriptstyle]{->}(0,0)(100,0.90)[$t$,-90][$X_t$,0]
            \fileplot[plotstyle=line, dotscale=0.3]{FigureF-A1-r3.4.prn}
          \rput[b](80,0.90){$r=3.4$}
          \rput[b](60,0.08){\scriptsize period 2 oscillations}
         \end{pspicture} & \hspace{0.3cm} &
           \psset{xunit=0.043cm,yunit=3cm}
         \begin{pspicture}(0,0)(100,1)
            \psaxes[Dx=10,Dy=0.1,Ox=0,Oy=0,ticksize=-3pt,labelFontSize=\scriptstyle]{->}(0,0)(100,1)[$t$,-90][$X_t$,0]
            \fileplot[plotstyle=line, dotscale=0.3]{FigureF-A1-r3.81.prn}
          \rput[b](80,1){$r=3.81$}
          \rput[b](80,0.08){\scriptsize chaos}
         \end{pspicture}
     \end{tabular}
\end{center}
\captionof{figure}[One figure]{\label{Fig CS1-2}. {\footnotesize The \textbf{logistic map} for 100 iterations of $X$ as $r$ moves from 0.1 (left panel) to 3.81 (right panel).}}
\subsection{Fractals}
The term \textquoteleft fractal\textquoteright~was coined by Benoit Mandelbrot in 1975. It comes from the Latin term \textit{fractus}, meaning an irregular surface like that of a broken stone. There is still some disagreement amongst scientific authorities about how the concept of a fractal should be formally defined, and many definitions were given over the years. Rather than a formal definition, one could imagine fractals as a seemingly-irregular shape (such as a coastline or cloud) or structure (such as a tree or mountain) formed by repeated subdivisions of a basic form, and having a pattern of regularity underlying its apparent randomness. Every part of a fractal (irrespective of its scale) is essentially a reduced size copy of the whole, and forms an organized hierarchy with its upper level and lower level counterparts. A fractal object contains an infinite amount of detail. They are usually generated using computer programs that repeat (or iterate) the original pattern at multiple scales (Fig. \ref{Fig CS2}). Fractals have the \textquoteleft self-similarity\textquoteright~property, meaning that they are ``\textit{the same from near as from far}'' \cite{gouyet1996}. B. Mandelbrot stated that ``\textit{Fractal Geometry plays two roles. It is the geometry of deterministic chaos and it can also describe the geometry of mountains, clouds and galaxies}''. He was a pioneer in the field and wrote significant literature \cite{Mandelbrot1982}. The concept of fractal has found applications in fields as diverse as astrophysics, chaos theory, and stockmarket analysis.
\begin{center}
\includegraphics[scale=1]{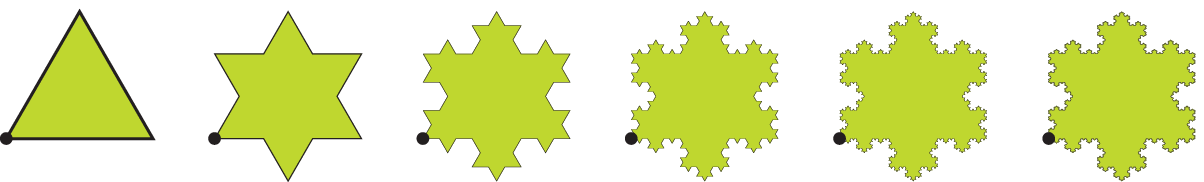}
\end{center}
\captionof{figure}[One figure]{\label{Fig CS2} {\footnotesize The first 6 iterations of the Koch snowflake. It is based on the Koch curve, a popular fractal curve, which first appeared in the 1900\textquoteright s in a publication of Helge von Koch. It can be constructed by starting with an equilateral triangle and recursively altering each line segment by \textit{i}) dividing the line segment into three segments of equal length; \textit{ii}) drawing an equilateral triangle that has the middle segment from step 1 as its base and points outward  and \textit{iii}) remove the line segment that is the base of the triangle from step 2.\\[.1cm]}}
In his introduction to \textit{The Fractal Geometry of Nature}, B. Mandelbrot wrote that ``\textit{Clouds are not spheres, mountains are not cones, coastlines are not circles, and bark is not smooth, nor does lightning travel in a straight line}'' \cite{Mandelbrot1982}, supporting the idea that the irregularities of fractal patterns has to be characterized by a specific indicator \cite{Mandelbrot1967}. In our example, the Koch snowflake (Fig. \ref{Fig CS2}), the length of the curve between any two points is infinite, such that the line is too detailed to be one-dimensional, but too simple to be two-dimensional. Its dimension might best be described not by its usual topological dimension but by its \textquoteleft fractal dimension\textquoteright, which in this case is a real number between 1 and 2. The two mathematicians Felix Hausdorff and Abram Besicovitch proposed to quantify dimensions using non-integer values and demonstrated that though a line has a dimension of 1 and a square a dimension of 2, many curves have an \textquoteleft in-between\textquoteright~dimension related to the varying amounts of information they contain. The Hausdorff is now frequently invoked in defining modern fractals \cite{Mandelbrot1967, Mandelbrot1982}. It can be assessed using:
\begin{equation}
  \label{Hausdorff_d}
  dim_H = \dfrac{\log P}{\log S}
\end{equation}
where $d$ is the Hausdorff dimension, $P$ the number of self-similar copies and $S$ the scaling factor. In the case of the Koch snowflake, the values are $P = 4$ (each line has 4 self-similar copies of the previous level) and $S = 3$ (the size of the segment is reduced by a factor 3) such $dim_H(Koch snowflake) \simeq 1.26185\ldots$. There exists other methods to estimate the fractal dimension of a fractal objet such as the box-counting or Minkowski-Bouligand dimension, but the Hausdorff dimension is the sophisticated standard.\\[0.2cm]
With the increasing capabilities of computers producing high resolution color images, fractal objects started to be used as the basis for digital art and animation in the mid-1980\textquoteright s. This form of algorithmic art is called \textquoteleft Fractal art\textquoteright and represents the iteration in 2 or 3 dimensions of fractal patterns. Some of the pattern designs became famous, such as the 3D Mandelbulb. Some examples of fractals are provided below.
\begin{center}
\begin{tabular}{*{2}{m{0.48\textwidth}}} 
The \textbf{Sierpinski triangle} also called the Sierpinski gasket or the Sierpinski Sieve is a fractal curve. To obtain close approximations to the Sierpinski triangle, one could start with a triangle in a plane, then shrink it and make 3 copies of it. The copies must be placed so that each triangle touches the two others at a corner. Then iterating the procedure leads to the Sierpinski triangle. The Hausdorff dimension of the Sierpinski triangle is $dim_H(\text{Sierpinski triangle}) \simeq 1.5845\ldots$. & \begin{center} \includegraphics[scale=0.8]{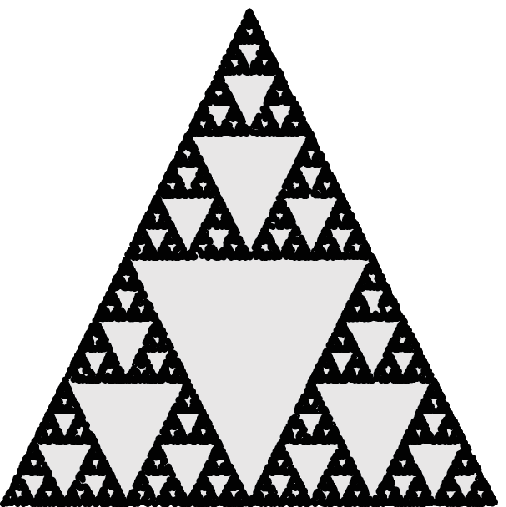} \end{center}\\
\end{tabular}
\end{center}

\begin{center}
\begin{tabular}{*{2}{m{0.48\textwidth}}} 
The \textbf{Pythagoras tree} is a plane fractal constructed with squares placed upon each others. Here are the very first iterations of the construction process: \includegraphics[scale=0.15]{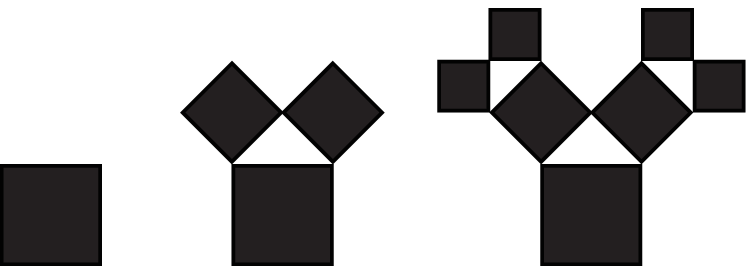}. The Hausdorff dimension of balanced trees is $dim_H(\text{Pythagoras tree}) = 2$. A colored unbalanced pythagoras tree is representated alongside. & \begin{center} \includegraphics[scale=0.6]{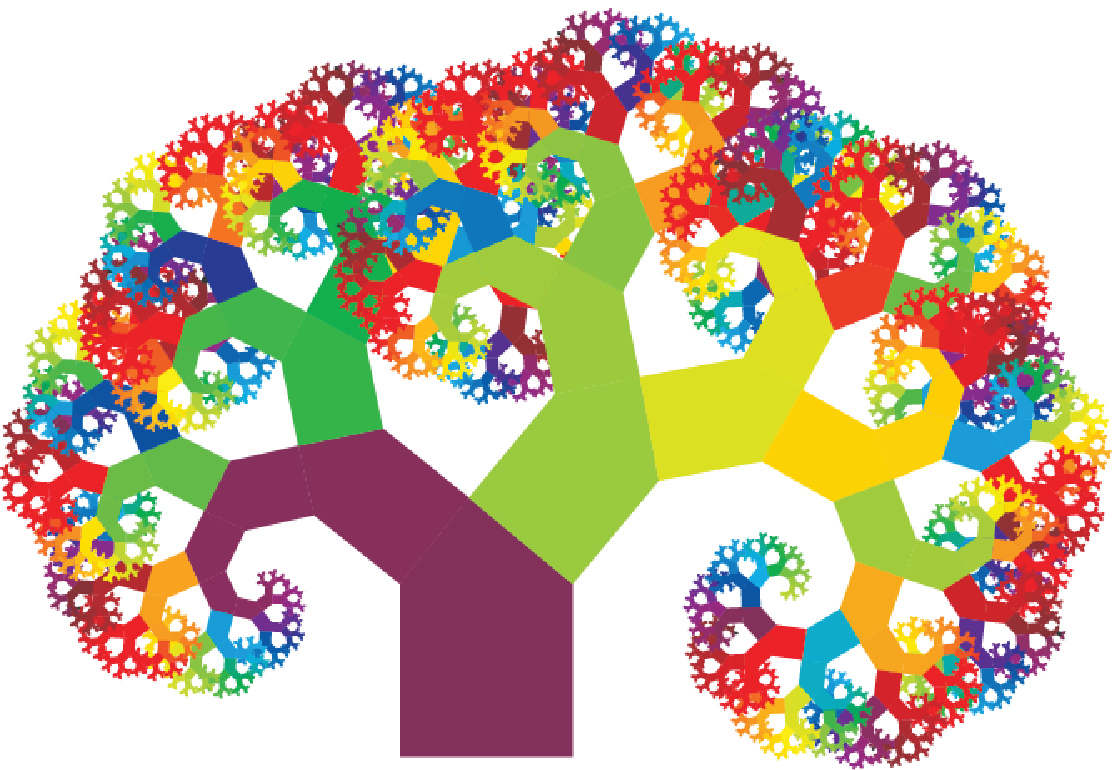} \end{center} \\
\end{tabular}
\end{center}

\begin{center}
\begin{tabular}{*{2}{m{0.48\textwidth}}} 
The \textbf{Menger Sponge} is a fractal curve first described by Karl Menger in 1926. Its approximate Hausdorff dimension is $dim_H(\text{Menger Sponge}) \simeq 2.7268\ldots$. & \begin{center}\includegraphics[scale=0.5]{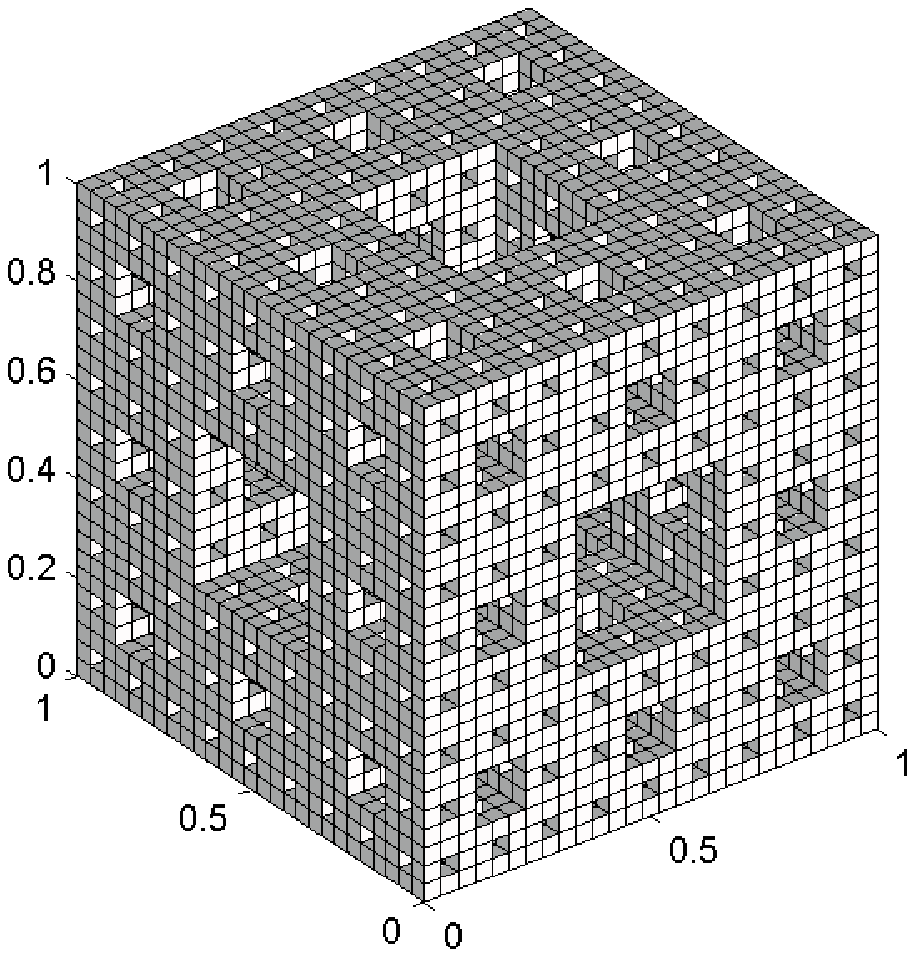}\end{center}\\
\end{tabular}
\end{center}
\subsection{Cellular automata}
Cellular automata (CAs) were first investigated by Stanislaw Ulam and John von Neumann in the 1940\textquoteright s. A cellular automaton (CA) is a discrete model composed of a regular grid of cells of finite or infinite dimensions. Each cell of the grid is in one state and the total possible number of states is finite. A cell usually interacts with its immediate neighborhood at each time step, given a set of simple fixed rules, such as a mathematical function. The rule for updating the state of the cells is the same for each cell and generally does not change over time, and is applied to the whole grid simultaneously. CAs were demonstrated to exhibit complex behaviors with some of them capable of universal computation (see \cite{Wolfram2002}).
\subsubsection{One-dimensional CAs}
Stephen Wolfram intensively studied one-dimensional CAs, and elementary CAs in particular. \textquoteleft Elementary\textquoteright~means that there is a single row of cells, with binary values (ie. there only are two possible states labeled 0 and 1), and update rules that depend only on nearest-neighbor interactions. In elementary CAs, the evolution of the CA is solely determined by its initial state and no other input and the game proceeds entirely as a consequence of the initial rule. It is one of the simplest possible models of computation. Nevertheless, depending on the initial rule, the elementary CAs can exhibit fixed points, oscillators and complex behaviors \cite{Wolfram1983} (Fig. \ref{Fig CS3}). S. Wolfram defined 4 classes of behavior \cite{Wolfram2002}:
\begin{itemize}
   \item Class 1: \textbf{Homogeneous state}. Almost all initial configurations relax after a transient period to the same fixed configuration (limit points in term of dynamical systems theory).
   \item Class 2: \textbf{Simple stable or periodic structures}. Almost all initial configurations relax after a transient period to some fixed point or some periodic cycle of configurations which depends on the initial configuration (limit cycles).
   \item Class 3: \textbf{Chaotic pattern}. Almost all initial configurations relax after a transient period to chaotic behavior (strange attractors).
   \item Class 4: \textbf{Long-lived, complex structures}. Some initial configurations result in complex localized structures, sometimes long-lived (no direct analogue in the field of dynamical systems).
\end{itemize}

\begin{center}
 \includegraphics[scale=0.5]{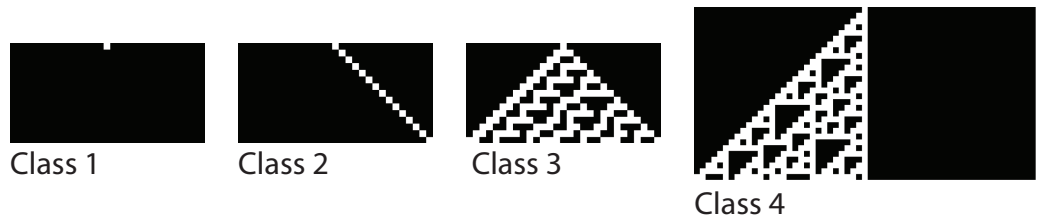}
\end{center}
\captionof{figure}[One figure]{\label{Fig CS3} {\footnotesize Example of patterns in elementary CAs.}}
\subsubsection{Two-dimensional CAs}
S. Wolfram also studied n-dimensional CAs \cite{Wolfram1985}. Among them, the Conway\textquoteright s Game of Life is a popular two-dimensional CA,  in the 1970\textquoteright s, where the survival of every cell is related to its immediate neighbors. The rules are:
\begin{enumerate}
   \item Any live cell with fewer than two live neighbors dies, as if caused by under-population.
   \item Any live cell with two or three live neighbors lives on to the next generation.
   \item Any live cell with more than three live neighbors dies, as if by overcrowding.
   \item Any dead cell with exactly three live neighbors becomes a live cell, as if by reproduction.
\end{enumerate}
In this model, the evolution of the CA is solely determined by its initial state and no other input, ie. the game proceeds entirely as a consequence of Conway\textquoteright s rules. As the system evolve, different kind of patterns were discovered among the years: fixed point patterns, period 2 oscillators and \textquoteleft spaceships\textquoteright, that are patterns that translate themselves across the board (Fig. \ref{Fig CS4}).
\begin{center}
     \begin{tabular}{ c  c  c  c  c }
         \begin{pspicture}(0,0.6)(2,2)
          \put(1,1){\esmallbox}      
          \put(1.5,1){\smallbox}    
          \put(2,1){\smallbox}      
          \put(1,1.5){\esmallbox}    
          \put(1.5,1.5){\smallbox} 
          \put(2,1.5){\smallbox}   
          \put(1,2){\esmallbox}     
          \put(1.5,2){\esmallbox}    
          \put(2,2){\esmallbox}     
          \rput[b](1.5,0.5){Block}
         \end{pspicture} &
          \begin{pspicture}(0,0.6)(2,2)
          \put(1,1){\esmallbox}      
          \put(1.5,1){\esmallbox}    
          \put(2,1){\esmallbox}      
          \put(1,1.5){\smallbox}    
          \put(1.5,1.5){\smallbox} 
          \put(2,1.5){\smallbox}   
          \put(1,2){\esmallbox}     
          \put(1.5,2){\esmallbox}    
          \put(2,2){\esmallbox}     
          \rput[b](1.5,0.5){Blinker}
         \end{pspicture} &
          \begin{pspicture}(0,0.6)(2.5,2.5)
          \put(1,1){\esmallbox}      
          \put(1.5,1){\esmallbox}    
          \put(2,1){\smallbox}      
          \put(2.5,1){\smallbox}      
          \put(1,1.5){\esmallbox}    
          \put(1.5,1.5){\esmallbox} 
          \put(2,1.5){\smallbox}   
          \put(2.5,1.5){\smallbox}   
          \put(1,2){\smallbox}     
          \put(1.5,2){\smallbox}    
          \put(2,2){\esmallbox}     
          \put(2.5,2){\esmallbox}     
          \put(1,2.5){\smallbox}     
          \put(1.5,2.5){\smallbox}    
          \put(2,2.5){\esmallbox}     
          \put(2.5,2.5){\esmallbox}     
          \rput[b](1.5,0.5){Beacon}
         \end{pspicture} &
         \begin{pspicture}(0,0.6)(2,2)
          \put(1,1){\smallbox}      
          \put(1.5,1){\smallbox}    
          \put(2,1){\smallbox}      
          \put(1,1.5){\smallbox}    
          \put(1.5,1.5){\esmallbox} 
          \put(2,1.5){\esmallbox}   
          \put(1,2){\esmallbox}     
          \put(1.5,2){\smallbox}    
          \put(2,2){\esmallbox}     
          \rput[b](1.5,0.5){Glider}
         \end{pspicture}
         \begin{pspicture}(0,0.6)(3,3.5) 
          \put(1,1){\smallbox}      
          \put(1.5,1){\smallbox}    
          \put(2,1){\smallbox}      
          \put(2.5,1){\smallbox}      
          \put(3,1){\esmallbox}      
          \put(1,1.5){\smallbox}    
          \put(1.5,1.5){\esmallbox} 
          \put(2,1.5){\esmallbox}   
          \put(2.5,1.5){\esmallbox}   
          \put(3,1.5){\smallbox}   
          \put(1,2){\smallbox}     
          \put(1.5,2){\esmallbox}    
          \put(2,2){\esmallbox}     
          \put(2.5,2){\esmallbox}     
          \put(3,2){\esmallbox}     
          \put(1,2.5){\esmallbox}     
          \put(1.5,2.5){\smallbox}    
          \put(2,2.5){\esmallbox}     
          \put(2.5,2.5){\esmallbox}     
          \put(3,2.5){\smallbox}     
          \rput[b](1.5,0.5){LWSS}
         \end{pspicture}
     \end{tabular}
\end{center}

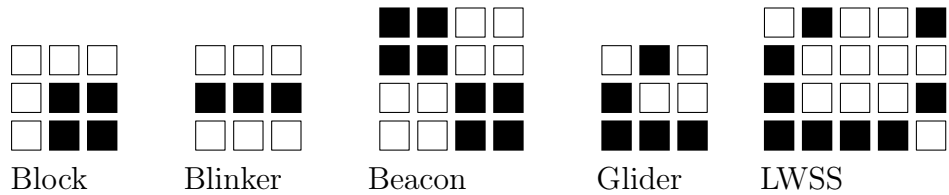
\captionof{figure}[One figure]{\label{Fig CS4} {\footnotesize Example of patterns in the Conway\textquoteright s game of life, with live cells shown in black, and dead cells shown in white. The block is the smallest possible living configuration. The blinker and beacon are period 2 oscillators. Both the glider and the LWSS (lightweight spaceship) are spaceships (ie. a finite pattern that returns to its initial state after a number of generations but in a different location).}}
\subsection{Self-Organization and criticality}
Boccara defined a complex system as a system which consists of large populations of connected agents (or collections of interacting elements) that exhibit an emergent global dynamics resulting from the actions of its parts rather than being imposed by a central controller \cite{Boccara2010}. One could think of ant colonies, where an organization emerges from the collective behavior of social individuals. Former precise definitions of self-organizing systems can be found in the literature \cite{Ashby1962, Mesarovic1962}, but they are embedded in detailed philosophical discussion \cite{Lendaris1964}. It is generally agreed that self-organization is a process where some form of global order or coordination arises out of the local interactions between the components of an initially disordered system (see \cite{Shalizi2004}). According to Scott Camazine, ``\textit{In biological systems self-organization is a process in which pattern at the global level of a system emerges solely from numerous interactions among the lower-level components of the system. Moreover, the rules specifying interactions among the system\textquoteright s components are executed using only local information, without reference to the global pattern}'' \cite{Camazine2003}. Ants colonies are often cited as one of an important type of complex system exhibiting self-organization because they offer an experimental system that is tractable at both the colony (macro-) level and the individual (micro-) level (for an extensive view on the topic see Gordon\textquoteright s work on harvester ants \textit{Pogonomyrmex barbatus} \cite{Gordon1989, Gordon1996, Gordon1999}). Phenomena from mathematics and computer science such as cellular automata, random graphs, and some instances of evolutionary computation and artificial life exhibit features of self-organization (such as Gosper\textquoteright s Glider Gun creating \textit{gliders} (Fig. \ref{Fig CS4}) in the cellular automaton Conway\textquoteright s Game of Life).
\newpage
In 1987, Per Bak, a Danish theoretical physicist coined the term \textit{self-organized criticality} (SOC) \cite{Bak1987}. The definition of self-organized criticality is flexible and SOC is defined for specific systems: there is no all-inclusive definition of this term. In his book, D.L. Turcotte defined SOC this way: ``\textit{The definition of self-organized criticality is that a natural system in a marginally stable state, when perturbed from this state, will evolve back to the state of marginal stability}''. It means that SOC is observed in non-equilibrium, non-linear systems that can undergo massive alterations and get back to a more stable state, without the intervention of external forces. Such systems generally approach a threshold value, defined as a \textquoteleft critical\textquoteright~point, that triggers a collapse in the system to a different state; ie. criticality is a point at which system properties change suddenly. P. Bak showed that simple mathematical representations of natural phenomenon can display SOC, such as models simulating the dynamics of sandpiles, earthquakes, economics, the human brain or traffic jams \cite{Bak1996}. SOC is also observable in the Conway\textquoteright s Game of Life (see previous section and \cite{Bak1989}), a toy model of the formation of organized, complex, societies. In sand-piles models, the pile adjusts itself through avalanches that reduce the criticality, while additional sand builds the criticality back up again, leading to another avalanche. The term avalanche -a sudden release of energy- refers naturally to sand pile behavior, but \textquoteleft avalanche\textquoteright~also refers to any abrupt shift in state or sudden release of energy in any self-organizing critical system. R.J. Wijngaarden demonstrated that the distribution of such avalanches in a rice pile follow a power law and that the maximum avalanche size scales in a particular manner with the size of the system, as predicted by SOC-theory \cite{Wijngaarden2006}. The Ising model \cite{Ising1925} investigates the phase transition between ferromagnetism and paramagnetism and is an example of phase transition criticality. The model consists of discrete variables that represent magnetic dipole moments of atomic spins that can be in one of two states (+1 or -1) (Fig. \ref{Fig CS5}). The spins are arranged in a graph, usually, a rectangular array (or lattice), allowing each spin to interact with its immediate neighbors. Let $J$ be the interaction strength, for low values of $J$, individual domains of cells of aligned spin are randomly distributed. As $J$ increases, there is a critical phase transition from the paramagnetism to the ferromagnetic state, after which the spins become polarized into large domains with aligned spins. At the critical transition value of $J \simeq 0.44$ magnetization domains form a fractal distribution similar to states of self-organized criticality, such as earthquakes and sand-piles, in which the critical state is maintained by fractal avalanches (see \cite{Ising1925} and \cite{Wijngaarden2006}).
\begin{center}
     \begin{tabular}{ c c c c c }
        \includegraphics[scale=0.275]{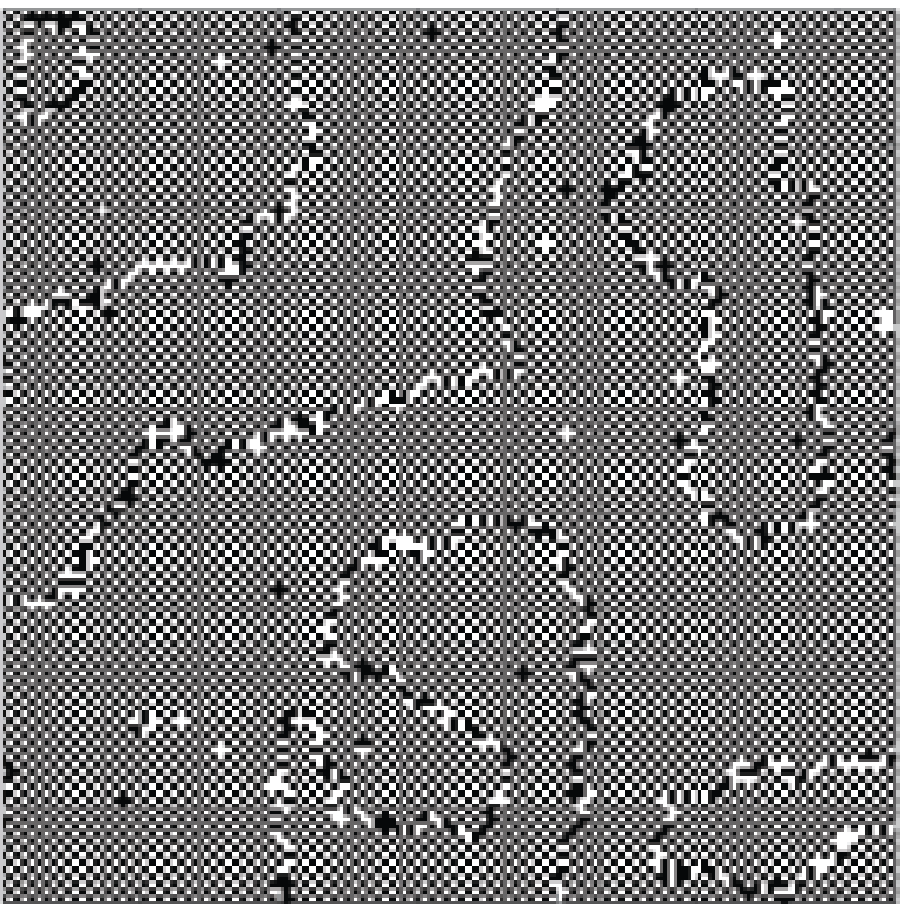} & \includegraphics[scale=0.275]{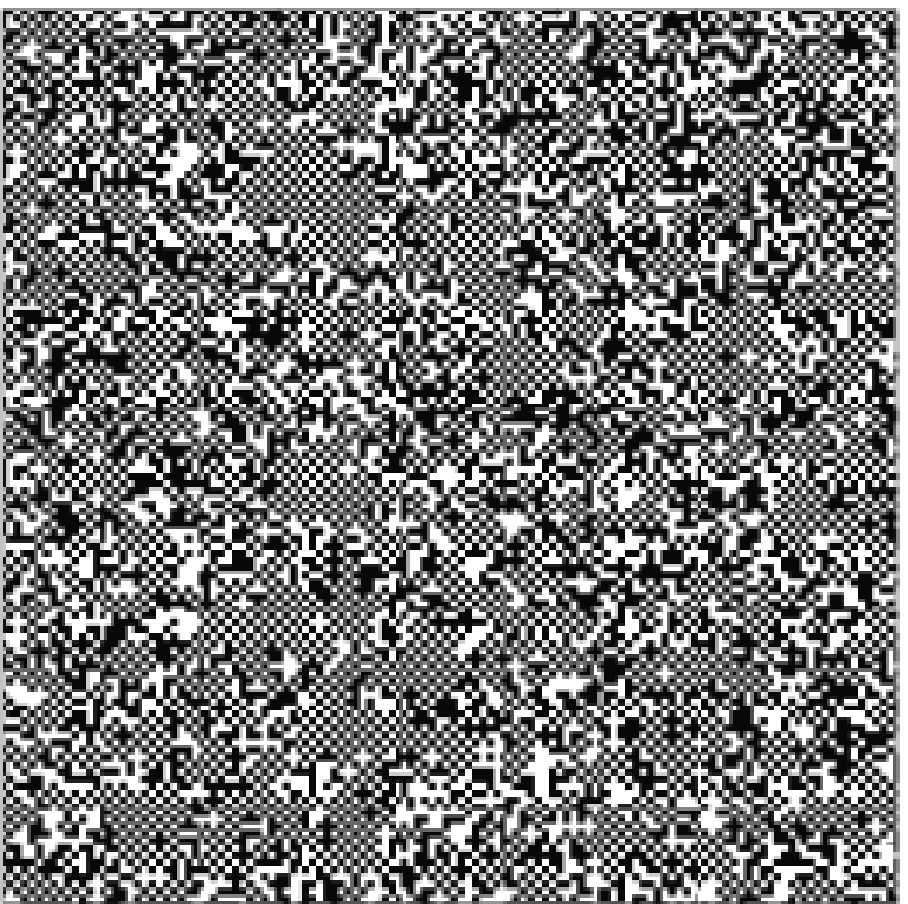} &
        \includegraphics[scale=0.275]{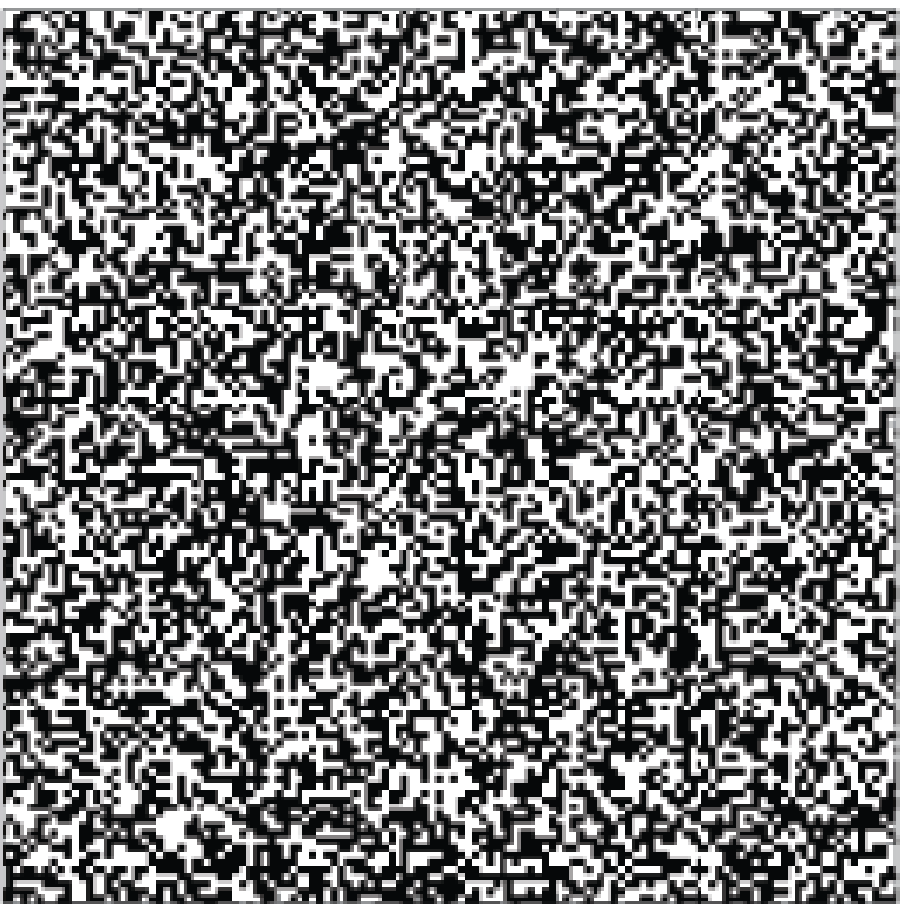} & \includegraphics[scale=0.275]{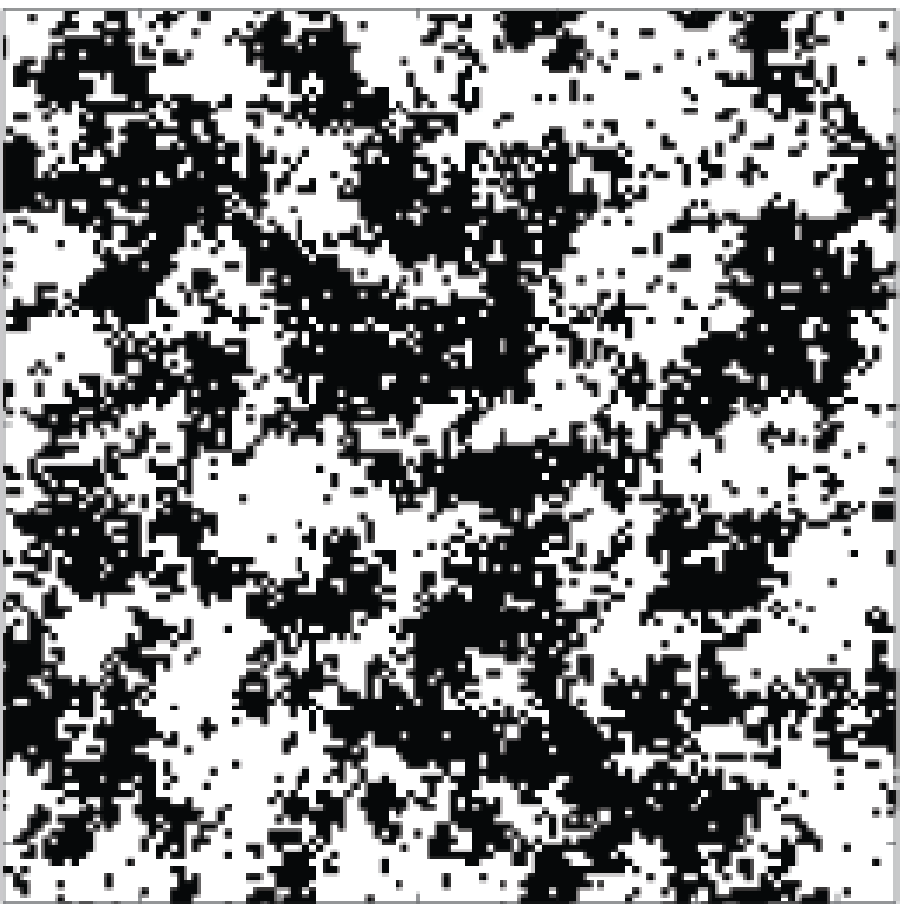} &         \includegraphics[scale=0.275]{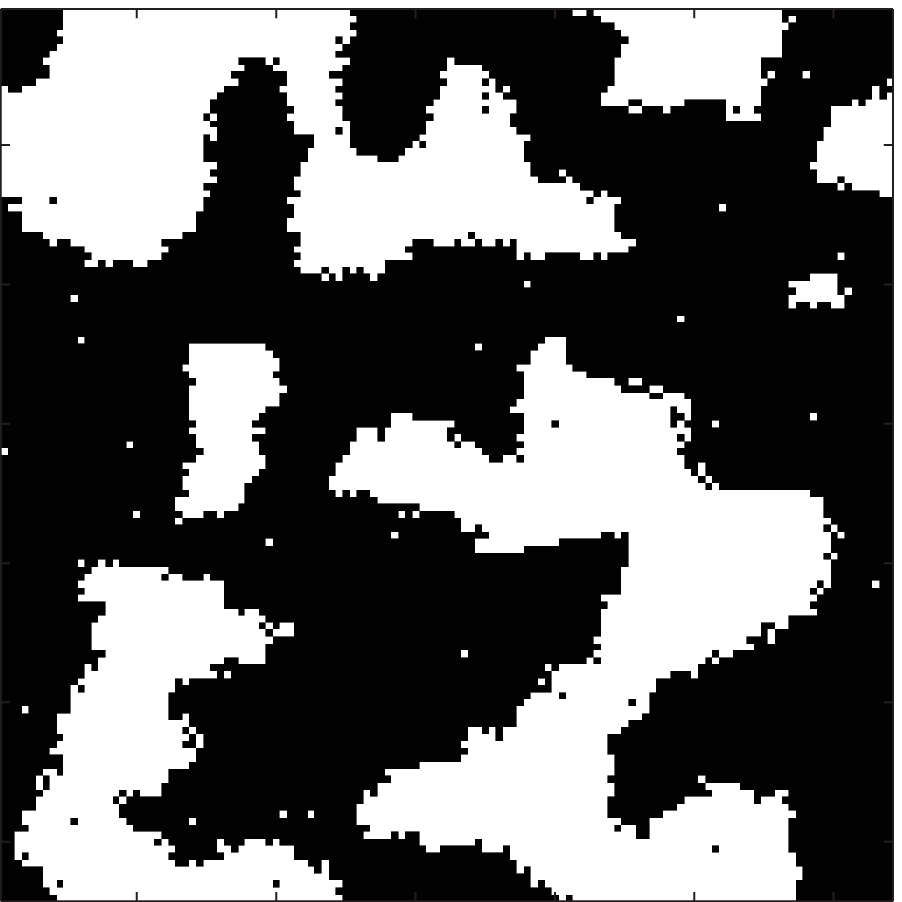}
     \end{tabular}
\end{center}
\captionof{figure}[One figure]{\label{Fig CS5} {\footnotesize The \textbf{Ising model} of ferromagnetism as an example of phase transition criticality. Dynamics of the model with increasing values of the interaction force $J$. From left to right: $J=-0.8$, $J=-0.4406868$, $J=0$, $J=0.4406868$ and $J=0.8$. The model is based on a cellular automaton with each cell element interacting with the four nearest neighbors (above and below and to either side) in a rectangular array (or lattice) containing $128\times128$ cells.\\[.1cm]}}
We previously used several terms relative to these notions of complexity: we used the self-similarity and fractal terms in section \ref{sec:PhenotypicExpansion} to suggest that the observed pattern is similar at different scales. The model introduced by Yukalov and al. in section \ref{sec:Expansion and Energy} may exhibit chaotic patterns with attractors. We also showed that performance is linked to many other factors, and that the system leading to world records in sport is composed of many different variables (technology, genetic predispositions, environment, and so on). In the next section, we use the concept of iteration, by iterating a model in order to determine the possible solutions of a sub-system, defined by several variables and their interactions.
\section{Energy, performance}
In the first chapter we demonstrated that physical performance is limited and we showed that other human features such as cognition and lifespan are limited. All these performances expanded similarly during the last two centuries. R. Fogel described this expansion as a techno-physio evolution with an unique development of health, body size, mass and mortality. These features were optimized with the help of energy (sec. \ref{sec:Expansion and Energy}). We finally replaced human physical abilities into a more global frame, and identified the climatic conditions, technology and energy input as determinants of performance. M. Guillaume also demonstrated that other determinants such as the geopolitical context can also alter the development of performance \cite{marion1}. With the historical view we gathered with these studies, one can perceive sport performance itself as a reflection of a more global trend, as an outcome of the societal infrastructure, synchronized with other features of man such as lifespan. But is such a trend, the mirror of a deeper change?\\[.3cm]
Such a question is part of a larger issue on the ability of man to further adapt upcoming alterations generated by 200 years of human expansion \cite{JFT2012}. J. Aronson et al. suggested that the continual and rapidly growing human population is exhausting Earth\textquoteright s natural capital \cite{Aronson2010}. Of course, one can choose to disconnect the development of physical abilities in international competitions from such a problematic, but we saw that physical performances are tied to external pressures. The alteration of economical conditions already bear additional pressure on the organization of major events such as the Olympic Games. The competitions are dependent on climatic conditions and new pressures are arising from climate change (temperature, growing risks for water scarcity, etc.). Other additional environmental, geopolitical and unknown pressures may arise and lead to a diminished ability to pursue the actual paradigm in sports and more generally in the previous phenotypic expansion. Similarly to the increase of lifespan in the recent years, the actual sport paradigm  \textquoteleft citius, altius, fortius\textquoteright~is based on the starting period of competitions 1890-1910, with the assumption that the surrounding societal infrastructure will remain strong enough to resist international turbulence such as World Wars. This last observation is influenced by the erroneous assumption that natural capital stocks (ie. resources that provide energy or food inputs for instance) are infinite and that technology will always allow for exploiting this stock \cite{Aronson2010}. However, the increasing flow of matter and energy we derive from ecosystems for expanding human features, or generate health benefit for people, returns to ecosystems as waste \cite{Daly2004}. We create a circle that is irresistibly translating to a negative feedback, ultimately leading to the collapse of ecosystems functions, known as \textquoteleft ecological overshoot\textquoteright.\\[.3cm]
One may choose to deny such an observation. In the first chapter we demonstrated that the forecast of sport performances was a subject of debate among experts. After the secular development of elite performances, the audience is still expecting new world records in competition, leading to the collective appreciation of an unlimited physical performance. This is known as the \textquoteleft optimism bias\textquoteright~and is widespread among the population \cite{Sharot2011}. This may lead to undesirable information regarding the future \cite{Sharot2011b} such that expected human features -world record frequency for instance- can be overestimated while others such as physical limits can be underestimated, or more simply, denied. Nevertheless, experts are exploring the variables linked to sport performances since a few years. The ecological footprint, environmental impacts, energy inputs and economical outcomes of sport events are being investigated \cite{Collins2009, Dolles2010, Pillay2008, Li2013, Kasimati2003} and these works can be used for a deeper survey regarding the impact of new policies.\\[.3cm]
Integrating these observations as a whole, we propose an approach to describe a sub-system (ie. a portion) of the structural infrastructure that leads to the phenotypic expansion. In sections \ref{sec:Technology} \ref{sec:Expansion and Energy} we gradually draw the relationship between energy and sport performance. We demonstrated that energy is one of the governing and common determinant, leading to the observed patterns in sport performances and in physiology, provided the environment is stable. Describing the overall system leading to the phenotypic expansion is difficult and may include a huge number of significant variables and interactions. We here focus on describing one of the sub-system -centered on primary energy use- leading to the expansion that includes a reduced number of variables, such as actual lifespan duration, mobility rate and demography features (such as population density with a distinct number of individuals) and the ecosystems capital. Features such as lifespan, mobility or the population are linked to the energy consumption per capita \cite{DeLong2010, Smith2013}.
\subsection{The overall model}
Several models have been proposed to describe and forecast the development of physical performances during the past century \cite{berthelot2008, berthelot2010a, Blest1996, nevill1}. They integrated a finite evolution and exhibited a S-shaped growth pattern in time, revealing a strong improvement of sport performances to finally reach a plateau in the 90\textquoteright s decade. These improvements were triggered by a massive exploitation of fossil fuels energies and observed in other major human feats such as mobility, food production, or reproduction \cite{Wilkinson2007, Smith2013}. The past development of energy consumption at world scale supports this assertion \cite{Grubler2008} and similar shape of growths are observed in all cases, starting with the industrial revolution at the end of the 19\textsuperscript{th} century. Technology changed our living, with breakthroughs in medicine and nutrition. This resulted in an increase in life expectancy during the past century \cite{Oeppen10052002} but the observed trend may not be set in stone \cite{DowD2010}, and the limits of our genes may become more obvious \cite{Heidinger2011} (see sec. \ref{sec:Lifespan}). On the other hand, the activities generated by an increasing number of individuals induced both a climate hazard and biodiversity loss \cite{Tilman1999, Pereira2010, Green2005, Teyssèdre2007, Woodcock2007, Hoffman2011, Barnosky2011, Barnosky2012}. The joint use of primary energies and new technologies led to agriculture expansion, increasing the Earth\textquoteright s carrying capacity, while reducing the abundance and the diversity in species. Nevertheless, it is expected that the rate of deployment of new and future technologies (such as graphene, nuclear batteries, hydrokinetic or low-cost desalinization systems) will likely be determined on the basis of economic criteria rather than their potential effects on biotic and abiotic attributes of the environment \cite{Sutherland2012}. In this context, modeling sustainability has become of interest in various fields and helps to understand the relationships between economical uncertainties, primary energies consumption and environmental issues \cite{Nihoul1998, Todorov2011, Weishbuch2013}. However, proposed models rarely addressed the issues of upcoming resources scarcity and environmental issues: only a few approaches were intended, one of them known as the \textquoteleft World3\textquoteright~model \cite{meadows1, meadows2, meadows3}. Despite the criticism on the modeling or methodological choices \cite{Jahoda1973}, it raised attention and contributed to the debate on the upcoming resources scarcity and population density issues. Projected forecasts revealed a reduction of the demography and other major feats, such as industrial output. In worst scenarios, a great reduction of all variables is observed while a great amount of the pollution is left over in the system \cite{meadows3, Turner2008}. The comparison of these projected forecasts with actual trends is not encouraging \cite{meadows3, Turner2008}, and actual scenarios sketched during the Fourth assessment report of the IPCC supports the idea of a \textquoteleft standard run\textquoteright~scenario (ie. adopted policies remain the same over time \cite{Giampietro2012} and cannot alter the common path called \textquoteleft business as usual\textquoteright). A. Johansen and D. Sornette, suggest that both the world population and economic growth rates can reach a synchronized spontaneous singularity at a critical time: $2052 \pm 10$ \cite{Johansen2001465}. This is due to the super-Malthusian growth of both human population and economic output, leading to recessions \cite{Sornette2003}.\\[0.3cm]
In our approach, we are interested in the third concept introduced by Holling \cite{Holling1987}: the \textquoteleft organizational change\textquoteright, when external events lead to the perturbation of the environmental system, altering the state of a set of entities that: may not change, converge toward attractors, become disordered, or shift from one pattern of organization to another, in an unpredictable way (e.g. the game of life in sec. \ref{subsec:Complexity}). These perturbations may be the results of human interactions with the environment e.g. removing a pest, converting habitats, increasing the number of crops, or introducing new pathogens \cite{Teyssèdre2007}. Therefore, an individual-based approach is used that places the individual (or agent) as the central object of the environment. A numerical simulation is build to explore the complex and global dynamics between energy, lifespan, food input, demography and the ecosystem. We use a discrete time and space approach based on CAs \cite{Wolfram1983, sante2010, Vliet2011} and multi-agents systems (MAS) with local interactions. Agents are arranged on a spatial lattice of finite size and can indefinitely stack up in one site. Although it does mean that the number of agents is theoretically infinite, the carrying capacity of the environment is not: resources provided in the system are limited, thus restricting the number of agents present in the system at a time.
\subsubsection{Agents}
Agents consume energy and food at each turn and could also perform a set of actions during each turn: moving, reproducing, cooperating, etc. They look for an increase of both the energy security in a site and their quality of life. In other words, they search for the best combination in their surrounding area that increases their lifespan. Agent\textquoteright s line of sight corresponds to the spatial area of perception of the agent. Previous studies assumed that perception is mostly homogeneous across individuals \cite{Ritchie1974} while other researches suggested it may differ \cite{Silverson1973, Wright2000}. However, in both studies the predicted heterogeneous perception was not supported and, on the contrary, more similarities in perception were discovered. We choose a line of sight with fixed radius for the agents. Other studies demonstrated that the perception of risk and/or events is related to distance \cite{Giordano2010, Carlos2009, Kellens2011, Pignataro2011}. Thus we decide that distant events (dwindling reserves of oil, alteration of lifespan, etc.) cannot influence agents.\\[.3cm]
Numerous studies reported that human mobility follows an heavy-tail flight distribution often approximated with random walk, L\'{e}vy Flight or truncated L\'{e}vy Flight \cite{Brockmann2006, Havlin2002, Gonzalez2008, Rhee2011}. Such patterns are also observable in the case of animal foraging behaviors \cite{Viswanathan1996, Atkinson2002, Ramos2004}. Other authors pointed out that an important feature of human mobility trajectories is the radius of gyration \cite{Gonzalez2008, Abramowicz1993}. A key result based on the analysis of this last parameter for 100,000 mobile phone users suggests that all individuals seem to follow the same universal probability distribution, which is anisotropic, unimodal and centered on an single position \cite{Gonzalez2008}. We implement a drift in individual motion related to the density of energy and/or food in the surrounding area.\\[.3cm]
Reproduction is a function of numerous biological or non biological factors, one of them being energy supply \cite{DeLong2010, Burger2011, Smith2013}. However, we here consider a fixed growth rate per individual but link the local population density to energy supplies.\\[.3cm]
We previously focused on individual and population lifespan in section \ref{sec:Lifespan}, \ref{sec:Expansion and Energy} and decide to limit the maximum possible lifespan of agents in the system. Ian Roberts, Phil Edwards and other authors pointed out that our dependance on fossil fuels increases the average BMI value: car use, television, and the energy intensity of food production from agriculture itself demonstrate that both the BMI and the primary energy consumption are strongly correlated \cite{Smith2013, iroberts2010, bronwyn2008, owen2010, 1367-2630-5-1-348}. These studies suggest that the maximum lifespan possible can be shortened when consuming excessive primary energy. We do not directly include such a feature in our model as an optimistic approach of the energy-lifespan relationship but include it with other negative effects (population density, ecosystem issues, etc.) \cite{Smith2013}.\\[.3cm]
We focus on the competition and cooperation between agents in the system, and study its evolution according to different initializations (ie. scenarios). The agents are divided into two groups by their consumption profile: fossil fuels friendly or renewable energy friendly. One probability is associated with the cooperation or competition, and at each turn the agent is in one of the two states. Several actions are realized during a turn. If the agent is in a cooperative state, the two actions are performed:
\begin{itemize}
  \item \emph{share}: the agent gives a portion of the energy he consumed to another random agent of the same group in its neighborhood. The share is performed after he consumed it, and regardless of the source of energy it came from. This action is possible only if it remains enough energy or food to consume.
  \item \emph{care}: the agent contributes to a local increase of the ecosystem by reducing its destructive actions. The agent can behave as usual regardless of the ecosystem issues or, in the opposite, nullify its ecological impact. This action is introduced to model the positive interactions of humans with ecosystems, as suggested earlier by Elinor Ostrom \cite{Ostrom1990}. In this particular case, agents adopt a sustainable development and try to maintain long-term resource yields. It implies that agents may increase the food production of their actual site, by improving their agricultural methods or technologies, allowing for an increasing number of individuals, as described in the debated Boserup\textquoteright s theory \cite{Grigg1979, Boserup2003}.
\end{itemize}
The competition occurs between different groups and inside each group of the population. If the agent is competitive, the two following actions are performed:
\begin{itemize}
  \item \emph{convert}: the agent converts another random agent of the opposite group chosen in the site. This action is introduced to model the struggle of behaviors between the two groups.
  \item \emph{push}: the agent pushes a random agent of its own group in an adjacent site: it represents the intra-species habitat competition.
\end{itemize}
A set of probabilities is associated with each agent, and remains constants during the lifetime of the agent, except for the energy profile, that can be updated several times, due to the \textquoteleft convert\textquoteright~action. That the agents are neither able to adapt nor anticipate any rapid change(s) in the system, no matter whether they possessed a wider horizon of perception as an individual or a group, is one leading assumption in the model. The model embed both the zero contribution and collective action theories \cite{Ostrom2000}: competitive actions can be thought of as rational egoism while cooperative actions are collective interest (which may sometimes be a higher order of selfishness).
\subsubsection{Resources}
Resources are consumed by agents at each turn and are necessary for the agents to survive. Because of the different renewal rates occurring for different types of resources, some are considered non-renewable: their production rates are negligible in front of their consumption rate by the population. Thus, resources are split into two categories: the renewable resources (wood, wind, tides, solar radiations, etc.) and the fossil fuels (coal, petroleum, natural gas, etc.). In the simulation, renewable and non-renewable resources represent only the exploitable fraction of the whole renewable energy in the system. M.Z. Jacobson and M.A. Delucchi recently proposed a plan to power 100\% of the world\textquoteright s energy with wind, hydroelectric, and solar power by the year 2030 \cite{Jacobson20111154, Delucchi20111170}. However, they concluded that barriers to such a project maybe more social and political than technological or economical. We assume that much progress in the exploitation of renewable energies can still be achieved with the use of new technologies in a near future. However, in this study we focus on the case of resource scarcity and admit that political, social or technological barriers remained strong enough to not reach such a development.
\subsubsection{Ecosystems and food production}
In this approach, ecosystems are complex set of relationship among the living resources, habitats and residents of an area (including plants, trees, animals, birds, micro-organisms, water, soil, people,\ldots). It could be described spatially as a number, high values representing rich ecosystem whereas low values embodied poor ecosystems. Important alterations of the ecosystems and biodiversity were recently observed, although it was not clear if it was a global reduction of the diversity or a modification in the repartition of species \cite{Mora2011}. On the other hand, it was admitted that the combustion of fossil fuels and the industrial agriculture unmistakably modified our environment \cite{Rockstrom2009}. If more agents stack together in a site, the local ecosystem richness decreases. High density of individuals, such as in urban environment, is regarded as a major threat to biodiversity and ecosystems \cite{Antrop2004, Hansen2005}. Several studies suggested that urban land use profoundly alter all environmental components and that humans are the main drivers of change \cite{Sukopp1979, Gilbert1989, Pickett2001, Alberti2003}. However, the complex mixture of direct (size and level of habitats, structures, etc.) and indirect influences (climate and air pollution, etc.) of urban patterns on ecosystems remains unclear, and some aspects of urban patches may also benefit local ecosystems \cite{Ingo2011, Werner2010}.\\[.3cm]
Several forecasts predicted an increase in food production along with the growing demography \cite{IFPRI, Gilland200247}. The United Nations estimated that an increase of 70\% of the actual agricultural production will be necessary to feed the future population in 2050, despite the soil nutrient depletion, erosion, desertification, depletion of freshwater reserves, etc. \cite{FAO2012}. This late statement strengthen the ecosystems-food relationship, and stress the future trends in required supplies for food production. Ecosystems provides high variety of food (crops, livestock, forestry, fish, etc.) \cite{Diaz2006, fogel, Kroll2012}. In our simulation, food is a local variable related to the richness of ecosystems present in a site and is consumed by the agents at each turn. Previous studies reported that low food supplies can occur in both urban and rural areas \cite{fogel, Kroll2012}. In the model, food supplies are available from everywhere but the amount varies from one site to another. We admit that technological innovations can still increase the production when the demography increases, as suggested in Boserup\textquoteright s theory \cite{Grigg1979, Boserup2003}. We also consider a positive correlation between food production and fossil fuels availability: J. Woods, D.R. Mears, and D. Pimentel pointed out that the production capacity remains heavily dependent on fossil fuels \cite{Woods27092010, UNESCO, Pimentel1973}. M. Harvey and S. Pilgrim noticed that a major effort is necessary to shift to sustainable intensification of cultivation given the food-energy-environment trilemma \cite{Harvey2011}.
\subsection{Nomenclature and range of the simulation}
The simulation rely on:
\begin{itemize}
  \item discrete time and space (flat two-dimensional square \textquoteleft grids\textquoteright~with indexes as discrete representations of space)
  \item active (agents) and passive (resources, richness of ecosystems, etc.) objects
  \item localized and global interactions (resource transactions between agents, exploitation of the surroundings resources, etc.)
\end{itemize}
For a convenient notation, we use:
\begin{itemize}
    \item \textbf{capital letters} $G$, $B$, $R$, $E$ and $F$ for the grids notation, letter $A$ for subsets of agents, letter $K$ for capacity and $L$ for the line of sight
    \item \textbf{lowercase Greek letters} with subscript indices as constants (e.g. $\beta_1$, $\beta_2$, $\gamma_1$, \ldots). Letter $\alpha_{1,\ldots,9}$ was used for the agent\textquoteright~parameters (cooperation value, reproduction probability, \ldots)
    \item \textbf{lowercase letters} $t$ for turn number, $n$ to denote a proportion number (e.g. $n(G,t)$ is the number of agents in grid $G$ at time $t$, $n(F,t)$ the sum of fossil energy at turn $t$), $s$ as a size-related value (such as the size of a grid), $a$ for agents and $q$ for consumed quantities at a given turn
    \item \textbf{subscript letters} as indexes: $_i$,$_j$ for indexes values in grids
\end{itemize}
The following notations are equivalent $\invbreve{G}_{t} = G_{t,i,j}$, $\invbreve{F}_{t} = F_{t,i,j}$. We use $\invbreve{G}_{t}$ to refer to the site $i$, $j$ in grid $G$ at time $t$. The variables are updated at each turn $t$ of the simulation, with $t$ taking all the integer values between $t_0$ and $t_{max}$, $t_0$ being the initial turn and $t_{max}$ the final turn (when the simulation ends):
\begin{equation}
  \label{turns}
  \left\{
  \begin{array}{lll}
      t \in [t_0, t_{max}] \\
      t = t_0, t_0+1, t_0+2, \ldots, t_{max} \\
      t,t_0,t_{max} \in \mathbb{Z}^{+}
  \end{array}
\right.
\end{equation}
where the time step is 1 and also corresponds to the aging rate of the agents (see Aging section).
\subsection{Spatial representation}
The spatial domain is pictured by a square matrix or grid $G$ with a side length of size $s$ and represents the landscape habitable by agents. There is a total of $s\times s$ possible positions (or site) in the grid. Agents are located in each site of the grid at indexes $i, j$ (with origin in the upper left corner) and can move from one site to another. However, agents can stack up in one site without limitation, so that the total number of agents contained in one site can be infinite (if no carrying capacity is considered). The value of an empty site is $0$, while a site occupied by one or more agents is valuated by the number of agents stacked up in this site. We apply the toroidal topology (ie periodic boundary conditions) to avoid edge effects. Four other grids are defined with the same size parameter $s$: a grid containing the local amount of fossil fuels $E$, Renewable energy $R$, ecosystems $B$ and food generated by the landscape $F$. A site at position $i,j$ contains the number of agent in grid $G$, the ecosystems richness in grid $B$, the amount of renewable or fossil energy in $R$, $E$ and the available food in grid $F$ (Fig. \ref{Fig GSIMS1}).
\begin{center}
\begin{pspicture}(-5,-1.2)(5,5.5)
\psset{viewpoint=50 40 30 rtp2xyz,Decran=20}
\psSolid[object=grille,
base=-5 5 -5 5,
linecolor=gray](0,0,0)
\psSolid[object=grille,
base=-5 5 -5 5,
linecolor=gray](0,0,3)
\psSolid[object=grille,
base=-5 5 -5 5,
linecolor=gray](0,0,6)
\psSolid[object=grille,
base=-5 5 -5 5,
linecolor=gray](0,0,9)
\psSolid[object=grille,
base=-5 5 -5 5,
linecolor=gray](0,0,12)

\psSolid[object=grille, base=-4 -3 3 4, linecolor=black, fillcolor=gray](0,0,0)
\psSolid[object=grille, base=-4 -3 3 4, linecolor=black, fillcolor=gray](0,0,3)
\psSolid[object=grille, base=-4 -3 3 4, linecolor=black, fillcolor=gray](0,0,6)
\psSolid[object=grille, base=-4 -3 3 4, linecolor=black, fillcolor=gray](0,0,9)
\psSolid[object=grille, base=-4 -3 3 4, linecolor=black, fillcolor=gray](0,0,12)

\rput(3.3,0){$F$}
\rput(3.3,1.2){$E$}
\rput(3.3,2.4){$R$}
\rput(3.3,3.6){$B$}
\rput(3.3,4.8){$G$}

\psset{linewidth=0.7pt}
\ThreeDput[normal=0 0 1](0,0,0){\psline{->}(2.2,-2)(3.0,2.6)} 
\ThreeDput[normal=0 0 1](0,0,0){\psline{->}(-1.8,2.2)(2.8,2.8)} 
\rput(-2,-1){$i$}
\rput(2,-1){$j$}
\end{pspicture}
\end{center}
\captionof{figure}[One figure]{\label{Fig GSIMS1} {\footnotesize The five grids $G$, $B$, $R$, $E$, $F$ of side size $s =10$ (100 sites). The value of site $i=9$, $j=2$ corresponds to the number of agents in grid $G$, the richness of ecosystems in grid $B$, the amount of renewable or fossil energy (available in each site) in grid $R$, $E$ and the amount of food in grid $F$.}}
\vspace{0.3cm}
The discrete Chebyshev distance is used to measure the distance between two sites $i_1, j_1$ and $i_2,j_2$ in a grid:
\begin{equation}
  \label{Chebyshev}
   dC(i_1,j_1,i_2,j_2) := \max\left(\mid i_2 - i_1\mid, \mid j_2 - j_1\mid\right)
\end{equation}
\subsection{Agents}
\label{AgentsSection}
Agents are randomly sorted before performing any action (such as moving, reproducing, cooperating, etc.). They may realize one of the following actions at each turn: move or reproduce with a given probability. The probabilities to move or reproduce are defined at birth, and remain constant during the lifetime of the agent. At each turn, a real number is drawn in the uniform law and the action is realized if it is lower than the probability. All of these actions take place in the perimeter defined by the Moore neighborhood. We define a set of 9 parameters for each agent (Tab. \ref{Table_Agents}).
\begin{center}
\begin{table*}[htbp!]
\begin{center}
\begin{tabular}{|c|l|c|}
  \hline
  Parameter & description & range of values\\
  \hline
  $\alpha_{1}$ & age at turn $t$ & $[0, \alpha_2]$ \\
  $\alpha_{2}$ & lifespan at turn $t$ & $[0, \xi_2]$ \\
  $\alpha_{3}$ & movement probability & $[0,1]$ \\
  $\alpha_{4}$ & reproduction probability & $[0,1]$ \\
  $\alpha_{5}$ & cooperation probability & $[0,1]$ \\
  $\alpha_{6}$ & energy consumption per turn & $[0,1]$ \\
  $\alpha_{7}$ & food consumption per turn & $[0,1]$ \\
  $\alpha_{8}$ & energy consumption profile & $[0,1]$ \\
  $\alpha_{9}$ & energy group & 0 or 1 \\
  \hline
\end{tabular}
\caption{ \label{Table_Agents} {\footnotesize List of the parameters and range used for the agents.}}
\end{center}
\end{table*}
\end{center}
Parameter of an agent is written as $\alpha_9(a)$ for describing the energy group of agent $a$. At the initialization of the system, agents are distributed in the two groups $\alpha_9 = 1$ (group of fossil fuel) or $\alpha_9 = 0$ (renewable energy) according to their energy consumption profile:
\begin{equation}
  \label{Group_subset}
    \alpha_9(a) =
    \begin{cases}
        0 \text{\, if \,} 0 < \alpha_8(a) \le 0.5\\
        1 \text{\, else}
    \end{cases}
\end{equation}
\subsubsection{Line of sight}
The line of sight $\lambda \in \mathbb{N}^*$ is the minimal discrete Chebyshev distance (eq. \ref{Chebyshev}) in which sites are visible by a given agent and $L(a)$ is the subset of sites in the line of sight of agent $a$:
\begin{equation}
  \label{LOS1}
  L(a) := \{i,j / dC\left(i,j,i_a,j_a \right) \le \lambda\} := \{i,j / i_a-\lambda \le i \le i_a+\lambda, j_a-\lambda \le j \le j_a+\lambda\}
\end{equation}
\subsubsection{Motion}
Between two turns, an agent can move in its immediate neighborhood or stay in the same site. The aim of each agent is to maximize its energy security, given the immediate local knowledge of the surrounding area, defined by the line of sight. When moving, the destination of an agent may only be determined by the resources in its line of sight, aiming to the site were energy and food are the most abundant. The motion of the agent is then drifted to the site with indexes $i,j$ which contains the larger amount of food and/or fossil fuel $c$ in sight:
\begin{equation}
  \label{destinationelection}
   c = \max \left(\invbreve{F}_{t} + \alpha_{8}(a) \cdot \invbreve{E}_{t,i,j}\right)
\end{equation}
with $\{i,j\} \in L(a)$ (eq. \ref{LOS1}) and $\alpha_{8}(a)$ is the consumption profile of agent $a$. If there are no food and no urban patch in sight the motion resumes in a two-dimensional random walk with constant step size and associated diffusion matrix:
\begin{equation}
\label{eq:aMvt1}
\begin{bmatrix}
  1 & 1 & 1\\
  1 & 0 & 1\\
  1 & 1 & 1
\end{bmatrix}
\end{equation}
such as the non-drifted motion is defined as a step in one of the 8 adjacent sites. When there is food or at least one urban patch in sight, the motion is anisotropic -as suggested by Gonzalez \cite{Gonzalez2008}-, otherwise the motion is isotropic.
\subsubsection{Reproduction}
We here consider a fixed growth rate $\alpha_{4}(a)$. Each new children inherit of all parameters $\alpha_{1, \ldots, 9}$ from the initial parents and is placed in the same site.
\subsubsection{Cooperation and competition}
At each turn, an agent tests its cooperation $\alpha_{5}$. If the agent is cooperative, it performs two actions at once in the following order:
\begin{itemize}
  \item \textit{share}: the agent gives half of the quantities $\alpha_{6}$, $\alpha_{7}$ to another agent of the same group, randomly chosen in the site. The share is performed after the agent effectively gathered the quantities and regardless of the source of energy it came from.
  \item \textit{care}: the agent contributes to an increase $\omega$ of the ecosystems richness on the current site by reducing its destructive actions. With $\omega = 0$ the agent behaves as usual, regardless of the ecosystems and with $\omega > 0$ the agent reduces its ecological impact. The parameter $\omega$ is fixed at the start of the simulation.
\end{itemize}
These values are associated and initialized at the birth of each agent and remain constants during their lifetime. Competition occurs between different groups and inside each group of the population. If the agent is competitive, the two following actions are draw in the following order:
\begin{itemize}
  \item \textit{convert}: the agent converts another random agent of the opposite group $\alpha_{9}$ in the site. The profile of consumption $\alpha_{8}$ of the targeted agent is then set to $1 - \alpha_{8}$ and its group to $1 - \alpha_9$. A same agent can be converted several times during the same turn.
  \item \textit{push}: the agent pushes another agent of its own group randomly chosen in the site. The agent is then pushed to a randomly chosen adjacent site in the Moore neighborhood.
\end{itemize}
\subsubsection{Energy consumption}
At each turn, an agent consumes a quantity $\alpha_6(a)$ of energy taken from grid $E$ or $R$. This value can eventually reach 0 in the absence of energy in the site. The grid used is determined by its energy consumption profile $\alpha_8(a)$ such that $E$ is used if $\alpha_8(a) > X$ where $X \sim \mathcal{U}$. Agents are distributed in the subsets $A_1$, $A_2$ regarding their consumption profile at a given turn:
\begin{equation}
  \label{con_subsets}
    \begin{cases}
        a \in A_1 \text{\, if \,} \alpha_8(a) > X \\
        a \in A_2 \text{\, else \,}
    \end{cases}
\end{equation}
The total energy consumed by all agents $q_t$ at turn $t$ can read:
\begin{equation}
  \label{q_t1}
        q_t = q_t(E) + q_t(R)
\end{equation}
where:
\begin{equation}
  \label{q_t2}
     \begin{split}
    q_t(E) = \sum_{a \in A_1} \alpha_6(a) \\[.1cm]
    q_t(R) = \sum_{a \in A_2} \alpha_6(a)
    \end{split}
\end{equation}
and similarly, $\invbreve{q}_t$ is the local energy consumption in site $i,j$:
\begin{equation}
  \label{q_t1b}
        \invbreve{q}_t = \invbreve{q}_t(E) + \invbreve{q}_t(R)
\end{equation}
where:
\begin{equation}
  \label{q_t2b}
     \begin{split}
    \invbreve{q}_t(E) = \sum_{a \in A_1} r(a) \\[.1cm]
    \invbreve{q}_t(R) = \sum_{a \in A_2} r(a)
    \end{split}
\end{equation}
with:
\begin{equation}
  \label{q_t2b2}
    r(a) =
    \begin{cases}
        \alpha_6(a) \text{\, if \,} \{i_a,j_a\} = \{i,j\} \\
        0 \text{\, else \,}
    \end{cases}
\end{equation}
\subsubsection{Food consumption}
At each turn an agent consumes a portion $\alpha_7(a)$ of available food taken from the grid $F$ (that can eventually reach 0 if no food is available). The quantity consumed at each turn is:
\begin{equation}
  \label{q_t31}
    q_t(F) = \sum_{a} \alpha_7(a)
\end{equation}
and:
\begin{equation}
  \label{q_t32}
    \invbreve{q}_t(F) =
    \begin{cases}
        \alpha_7(a) \text{\, if \,} \{i_a,j_a\} = \{i,j\} \\
        0 \text{\, else \,}
    \end{cases}
\end{equation}
\subsubsection{Aging}
At each turn, $\alpha_1$ increase by 1. The lifespan of any agent $\alpha_2$ is bounded by the maximum lifespan $\xi_2$ (sec. \ref{sec:Lifespan}). A minimum lifespan $\xi_1$ is also introduced to model the possible non-zero minimal value when there are no energy, food and external stresses (ie. survival case with low population density). We use two parameters to model the influence of (\textit{i}) consumed food / energy ($\xi_3 \in [0,1]$) and (\textit{ii}) local population density ($\xi_4 \in [0,1]$). At each turn, the lifespan of an agent is computed:
\begin{equation}
  \label{Lifespan}
    \alpha_{2,t+1}(a) = \xi_1 + \left(1 - \xi_3\right) \left(\xi_2 - \xi_1\right) + \xi_3 \alpha_{6,t}(a) \alpha_{7,t}(a) \left( \xi_2 - \xi_1 \right) - \xi_4 \xi_2  s\left(\invbreve{G}_t \right)
\end{equation}
with $s(x)$ the step function:
\begin{equation}
  \label{Lifespan2}
    s(x) =
    \begin{cases}
        0 \text{\, if \,} x < \xi_5\\
        1 \text{\, if \,} x > \xi_6\\[.2cm]
        \dfrac{x - \xi_5}{\xi_6 - \xi_5} \text{\, else \,}
    \end{cases}
\end{equation}
such that between $\xi_6$ and $\xi_5$ the local density of the population $\invbreve{G}_t$ has a linear influence on lifespan. If $\invbreve{G}_t < \xi_5$ the influence is null; if $\invbreve{G}_t > \xi_6$ it is maximal.
\newpage
\begin{center}
\begin{tabular}{c c c}
\psset{xunit=0.1cm,yunit=0.0234375cm}
\begin{pspicture}(0,20)(60,180)
    \psaxes[Dx=10,Dy=20,Ox=0,Oy=20,ticksize=-3pt,labelFontSize=\scriptstyle]{->}(0,20)(60,180)[$t$,-90][Local Population density,0]
    \fileplot[plotstyle=line, linecolor=black]{Figure34-Pop.prn}
    \rput[b](42,120){$\invbreve{G}_t$}
\end{pspicture} & \hspace{0.3cm} &
\psset{xunit=0.1cm,yunit=0.034090909cm}
\begin{pspicture}(0,0)(60,110)
    \psaxes[Dx=10,Dy=10,Ox=0,Oy=0,ticksize=-3pt,labelFontSize=\scriptstyle]{->}(0,0)(60,110)[$t$,-90][$\alpha_{2,t}(a)$,0]
    \fileplot[plotstyle=line, linecolor=red]{Figure34-K1.prn}
    \fileplot[plotstyle=line, linecolor=blue]{Figure34-K2.prn}
    \fileplot[plotstyle=line, linecolor=green]{Figure34-K3.prn}
    \rput[b](60,5){\scriptsize \textcolor{red}{$\xi_4 = 1$, $\xi_5=20$, $\xi_6=100$}}
    \rput[b](60,55){\scriptsize \textcolor{blue}{$\xi_4 = 0.5$, $\xi_5=20$, $\xi_6=100$}}
    \rput[b](60,85){\scriptsize \textcolor{green}{$\xi_4 = 0.25$, $\xi_5=20$, $\xi_6=100$}}
\end{pspicture} \\
\psset{xunit=0.1cm,yunit=3.440366972cm}
\begin{pspicture}(0,0)(60,1.09)
    \psaxes[Dx=10,Dy=0.1,Ox=0,Oy=0,ticksize=-3pt,labelFontSize=\scriptstyle]{->}(0,0)(60,1.09)[$t$,-90][Energy \& Food consumption,0]
    \fileplot[plotstyle=line, linecolor=black]{Figure34-2-pop1.prn}
    \fileplot[plotstyle=line, linecolor=red]{Figure34-2-pop2.prn}
    \rput[b](17,0.25){$\alpha_{7,t}(a)$}
    \rput[b](33,0.25){\textcolor{red}{$\alpha_{6,t}(a)$}}
\end{pspicture} & \hspace{0.3cm} &
\psset{xunit=0.1cm,yunit=0.034090909cm}
\begin{pspicture}(0,0)(60,130) 
    \psaxes[Dx=10,Dy=10,Ox=0,Oy=0,ticksize=-3pt,labelFontSize=\scriptstyle]{->}(0,0)(60,110)[$t$,-90][$\alpha_{2,t}(a)$,0]
    \fileplot[plotstyle=line, linecolor=red]{Figure34-2-K1.prn}
    \fileplot[plotstyle=line, linecolor=blue]{Figure34-2-K2.prn}
    \fileplot[plotstyle=line, linecolor=green]{Figure34-2-K3.prn}
    \rput[b](40,18){\scriptsize \textcolor{red}{$\xi_3 = 1$}}
    \rput[b](40,52){\scriptsize \textcolor{blue}{$\xi_3 = 0.75$}}
    \rput[b](40,85){\scriptsize \textcolor{green}{$\xi_3 = 0.25$}}
\end{pspicture} \\
\end{tabular}

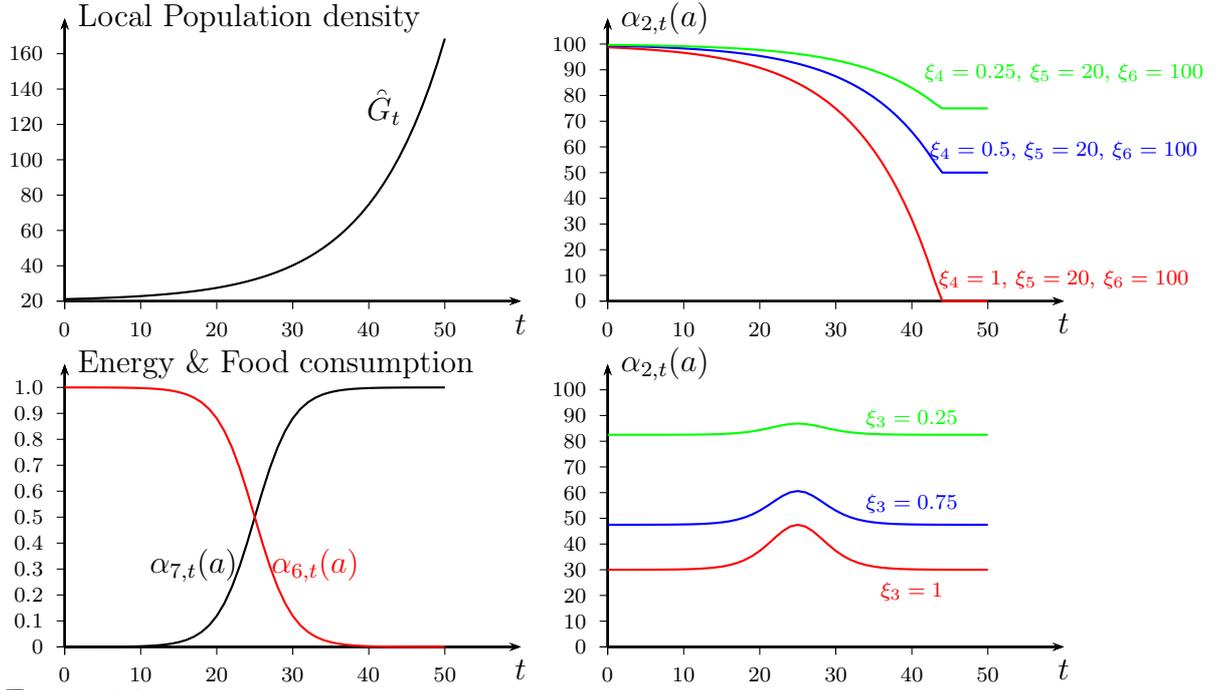
\captionof{figure}[One figure]{\label{Fig.M0} {\footnotesize Dynamics of eq. \ref{Lifespan} with different parameters values. In the two examples, the values of $\xi_1$ and $\xi_2$ are set to 30 and 100 respectively.}}
\end{center}
\subsection{Dynamics and localized interactions}
Here we describe the local dynamics and interactions occurring between agents and their environment.
\subsubsection{Energy production and consumption}
The renewable resources are considered as volatile and cannot be preserved in one site between two turns. However, the proportion $\gamma$ produced at each turn is constant and not related to agents density. We assume no major development or decline in the production of renewable energy over time:
\begin{equation}
  \label{Rt}
    \invbreve{R}_{t+1} = \gamma - \invbreve{q}_t(R)
\end{equation}
The production of fossil energy is assumed to be null:
\begin{equation}
  \label{Ft}
  \invbreve{E}_{t+1} = \invbreve{E}_{t} - \invbreve{q}_t(E)
\end{equation}
where $\invbreve{q}_t(R)$, $\invbreve{q}_t(E)$ respectively are the quantities of renewable and fossil fuels consumed by the agents at turn $t$. Both quantities are related to $\invbreve{G}_{t}$ and depend on the consumption profile of the agents (see sec. \ref{AgentsSection}).
\subsubsection{Ecosystems}
Each site $\invbreve{B}_{t}$ of grid $B$ contains a real number representing the richness of ecosystems in this area at turn $t$. Let $f_1(x)$ be the function describing the renewal (or \textquoteleft production\textquoteright) of local ecosystems. In our approach, a high density of agents negatively alters the richness of a site. The richness available in a site $i,j$ is thus related to the number of agents in the corresponding site in $G$:
\begin{equation}
  \label{Biodivrenew1}
        \invbreve{B}_{t+1} = \invbreve{B}_{t} + f_1(\invbreve{B}_{t}) - \beta_2 \cdot \invbreve{G}_{t}
\end{equation}
with
\begin{equation}
  \label{Biodivrenew2}
    f_1(\invbreve{B}_{t}) =
    \begin{cases}
        \beta_1 \cdot \invbreve{B}_{t} & \text{if \,} 0 < \invbreve{B}_{t} < \beta_K\\
        0 & \text{else \,}
    \end{cases}
\end{equation}
where $\beta_1 > 1$ is the rate of renewal, $\beta_2 \ge 0$ is the parameter related to the destructive actions of agents and $\beta_K$ is the local richness capacity. Ecosystems can be locally destroyed if the value of a site becomes null or negative (ie. $\invbreve{B}_{t} \le 0$). Agents can locally act on ecosystems in order to limit their destructive action (see sec. \ref{AgentsSection}).
\subsubsection{Food production}
Each site $\invbreve{F}_{t}$ contains a real number relative to the available food in this area. $\invbreve{F}_{t+1}$ admits supplies and a capacity of production related to the energy and ecosystems available in the environment:
\begin{equation}
  \label{Food}
    \invbreve{F}_{t+1} = \phi_1 \cdot \invbreve{F}_{t} + \phi_2 \cdot \left( \invbreve{E}_t \invbreve{B}_t \invbreve{R}_t \right) + \phi_3 \cdot \left( \invbreve{B}_t \invbreve{R}_t \right) - \invbreve{q}_t(F)
\end{equation}
where $\phi_1 \in [0,1]$ is the spoilage of food stock with time, $\phi_2$ the proportion of primary energy used to generate a high amount of food output from ecosystems and renewable energy. The parameter $\phi_3$ is the stand-alone local production and is related to the agricultural biodiversity \cite{FAOAB} (sometimes called \textquoteleft Agrobiodiversity\textquoteright). This last term is introduced to model the fact that only a few species with local energy -solar radiation for instance- can still be used to produce reduced amounts of edible food supplies (such as poultry, cattle, etc.), even if the local fossil energy $\invbreve{E}_t$ is depleted. The proportion $ \invbreve{q}_t(F)$ is the proportion consumed by the agents at each turn (see sec. \ref{AgentsSection}).
\subsection{Initialization}
Initialization of the various parameters is a critical step in setting up the model: it may be sensitive to different initial conditions and may exhibit different behaviors. We are interested in resource scarcity scenarios and we set up the parameters accordingly while basing the estimates of the initializations on literature and known estimations. However, some of the parameters are set arbitrarily due to the lack of empirical studies. In such cases the values are set in a very large interval, exceeding the actual forecasts in order to include possible future developments.\\[.3cm]
Three parameters: movement ($\alpha_3(a)$), cooperation / competition ($\alpha_5(a)$) and one parameter $\alpha$ defining energy profile ($\alpha_8(a) \sim \text{Beta}(\alpha,\beta)$) are intensively investigated. We test different initial values of these parameters using a multi-dimensional mesh (see mesh section).
\subsubsection{Grid size and time}
Due to the time needed to perform calculations, we run a restricted number of iterations and reduced grid size. We set $s = 100$ for a total of 10,000 sites in the grids, $t_{max} = 1,500$ and consider this as long term values (it corresponds to approximatively 15 life cycles). For added convenience in parameters initialization, we define the time step as being 1 year.
\subsubsection{Agents}
The number of agents at the start of the simulation is $n(G,t_0) = 100$ and $G$ is populated with this value. As of 2010, about 50.5\% of the world population lived in urban areas \cite{CIAWorld}. The agents are distributed into the grid $G$ with respect to the urban patches: 50\% of the agents are randomly distributed among the urban patches while the others are randomly placed in the lattice.
\subsubsection{Reproduction}
The crude birth rate corresponds to a number of births per 1000 people per year over the total population in the studied period. The crude birth rate in 2005-2010 is 20 \cite{UNSTATS}. We consider that the rate is basically the same and set $\alpha_{4}(a)=0.02$.
\subsubsection{Motion}
Different values of $\alpha_{3}(a)$ are tested over the interval $[0,1]$ (see mesh section).
\subsubsection{Line of sight}
As a proxy for setting the value of $\lambda$, we use measurements of the perception of risk. Although difficult to assess, the perception of risk from hazards such as volcano eruptions or nuclear power plants is quantified in the scientific literature. The investigation of G. Pignataro and G. Prarolo revealed that the area of risk perception in the case of the set up of new nuclear power plants is over 100km \cite{Pignataro2011}. In another case, J.C. Gavilanes-Ruiz et al. demonstrated that the risk perception of being affected by the eruption of a volcano is up to 14.6km \cite{Carlos2009}. On the other hand, people from a vast majority of countries were able to follow the development of the recent financial crisis thanks to the new information technologies and the instant availability of news \cite{Huang2010}. The author stresses that different forms of risk are present in our daily lives: the energy and financial crisis or the global terrorism. Our hypothesis is that such global risks are broadcasted at world scale but may not influence the people as strongly as more direct hazards, such as volcanoes or nuclear dangers that might directly affect the individuals. According to the International Organization for Migration, the financial crisis had slightly slowed down emigrations and did not appear to have stimulated substantial return migration: the risk seemed not strong enough to produce huge migrations \cite{WMR}. The energy crisis is not mentioned in this outlook and does not appear to be a major case of migration for the time being. It supports the idea that agents have a narrow horizon of perception in time and in space.\\[.3cm]
We use the value introduced by Pignataro \cite{Pignataro2011} and set $\lambda=3$, corresponding to an approximative value of 120km in our system (related to world size $s$ and circumference of earth).
\subsubsection{Cooperation / competition}
Different values of $\alpha_{5}(a)$ are tested over the interval $[0,1]$ (see mesh section). See sub-section \textit{Ecosystems} for estimate of $\omega$.
\newpage
\subsubsection{Energy consumption per turn}
The energy use per capita (kg of oil equivalent per capita (kgoe)) per year are gathered for all countries (\cite{WBank} \& IEA) from 1960 to 2010. The distribution of energy use per year for all countries reveals that the shape of the consumption didn\textquoteright t change much over the period containing all the data (after 1970, Fig. \ref{Fig.M1}).
\begin{center}
    \includegraphics[scale=0.8]{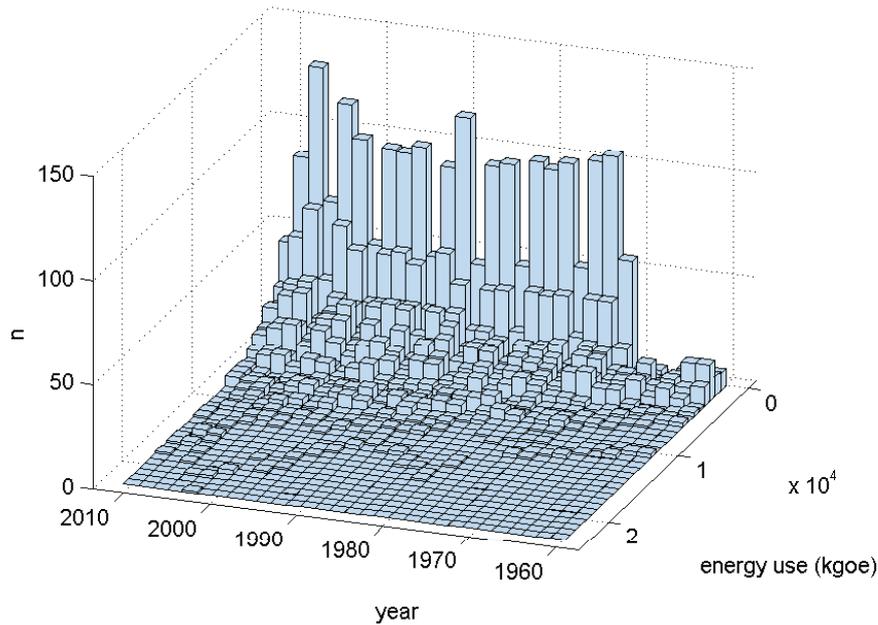}
\end{center}
\captionof{figure}[One figure]{\label{Fig.M1}  {\footnotesize Two dimensional distribution of energy use per capita per year for all countries. The data is not complete during the first decade (1960-1970).\\[.1cm]}}
We focus on the years were data is the most complete: from years 2005 to 2007, 168 countries reported their energy use per capita each year, totalizing 504 values. The distribution of the values reveals that the data seems to follow a log-normal distribution (Fig. \ref{Fig.M2}textbf{a}). Each agent is thus initialized with a random value in the log-normal distribution: $\alpha_{6}(a) \sim \ln \mathcal{N}$ with parameters $\mu=-2.88$, $\sigma=1.22$ estimated from the scaled and truncated ($\alpha_{6}(a) \in [0,1]$) empirical distribution (Fig. \ref{Fig.M2}\textbf{b}):
\begin{equation}
  \label{ScaledTruncated}
    x = \dfrac{x}{\max(x)}
\end{equation}
where $x$ is the energy use per capita per year for the period 2005-2007.
\begin{center}
\begin{tabular}{ c c c }
\psset{xunit=0.0003333cm,yunit=0.03440367cm}
\begin{pspicture}(0,-5)(18000,109)
\psaxes[Dx=4000,Dy=10,Ox=0,Oy=0,ticksize=-3pt,labelFontSize=\scriptstyle]{->}(0,0)(18000,109)[{\scriptsize E. use (kgoe)},100][$n$,0]
    \readdata{\data}{Figure36-pop.prn}
    \listplot[linecolor=blue,plotstyle=bar,barwidth=0.08cm,fillcolor=blue!30,fillstyle=solid,opacity=1]{\data}
    \fileplot[plotstyle=line, linecolor=red, linewidth=1pt]{Figure36-logN.prn}
    \fileplot[plotstyle=line, linecolor=green, linewidth=1pt]{Figure36-Weib.prn}
    \fileplot[plotstyle=line, linecolor=black, linewidth=1pt]{Figure36-Gamma.prn}
\end{pspicture} & \hspace{0.4cm} &
\psset{xunit=6cm,yunit=0.03440367cm}
\begin{pspicture}(0,-5)(1,109)
\psaxes[Dx=0.1,Dy=10,Ox=0,Oy=0,ticksize=-3pt,labelFontSize=\scriptstyle]{->}(0,0)(1,109)[Energy use,100][$n$,0]
    \readdata{\data}{Figure37-pop.prn}
    \listplot[linecolor=blue,plotstyle=bar,barwidth=0.08cm,fillcolor=blue!30,fillstyle=solid,opacity=1]{\data}
    \fileplot[plotstyle=line, linecolor=red, linewidth=1pt]{Figure37-logN.prn}
\end{pspicture}
\end{tabular}

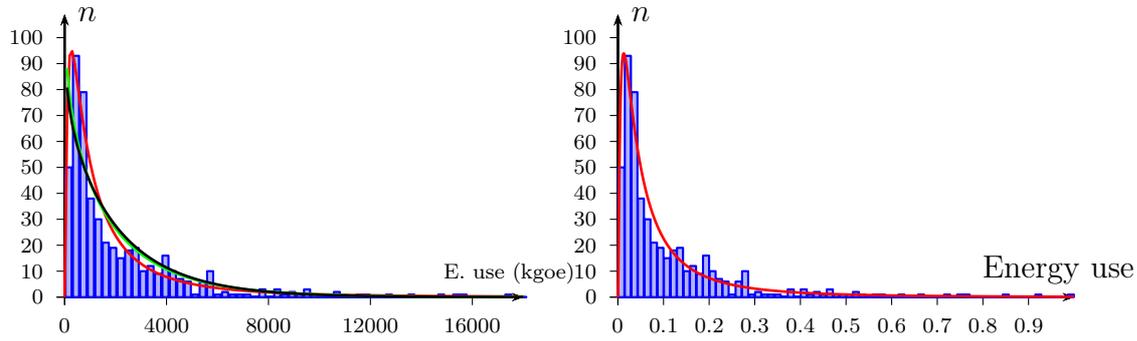
\captionof{figure}[One figure]{\label{Fig.M2}  {\footnotesize \textbf{a}. Distribution of energy use per capita for 168 countries over the period 2005-2007 (for a total of 540 values, blue bars). Three distributions are adjusted to the data: log-normal (red line, parameters: $\mu=7.05$, $\sigma=1.22$), Weibull (green line, $\lambda=2085.82$, $k=0.90$) and the gamma (black line, $k=0.90$, $\theta=2466.10$). \textbf{b}. The same distribution is scaled (eq. \ref{ScaledTruncated}) and truncated between $[0,1]$ (blue bars). The log-normal distribution is adjusted with parameters $\mu=-2.88$, $\sigma=1.22$ and is used to draw the starting values of $\alpha_6(a)$.}}
\end{center}
\subsubsection{Food consumption per turn}
According to several studies, a positive correlation exists between energy and the increase of the average BMI value \cite{iroberts2010, bronwyn2008, owen2010, 1367-2630-5-1-348}, supporting the possible correlation between energy use and food supply. The food supply, in kg per capita per year, is gathered for the year 2007 for a total of 3391 items and 172 countries \cite{FAO}. The sum of all items is computed per country resulting in the total food supply delivered for all items per capita / country in 2007. The relationship between energy use $x$ (see previous section) and food input $y$ in 2007 is non-linear and modeled by an exponential model (Fig. \ref{Fig.M3}). We use a similarly shaped exponential model with values $\in [0,1]$:
\begin{equation}
  \label{ScaledTruncated2}
    \alpha_7(a) = 1-\exp^{-5 \alpha_6(a)}
\end{equation}
\begin{center}
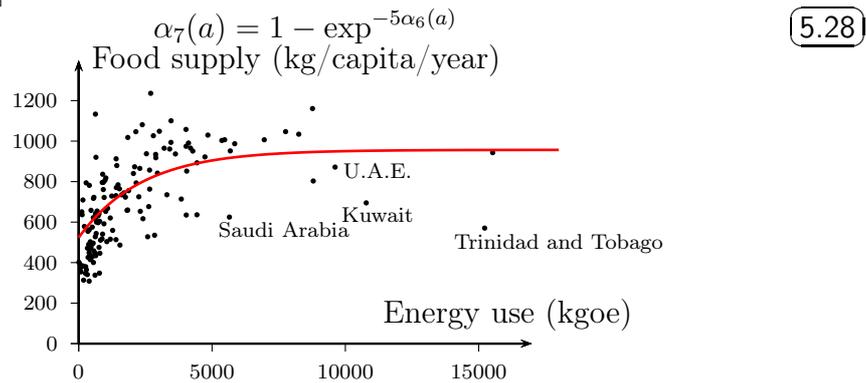

\psset{xunit=0.000352941cm,yunit=0.002678571cm}
\begin{pspicture}(0,-80)(17000,1400)
\psaxes[Dx=5000,Dy=200,Ox=0,Oy=0,ticksize=-3pt,labelFontSize=\scriptstyle]{->}(0,0)(17000,1400)[Energy use (kgoe),100][Food supply (kg/capita/year),0]
    \fileplot[plotstyle=dots, dotscale=0.55]{Figure38-A1.prn}
    \fileplot[plotstyle=line, linecolor=red, linewidth=1pt]{Figure38-A2.prn}
    \rput[b](18000,450){\scriptsize Trinidad and Tobago}
    \rput[b](11200,600){\scriptsize Kuwait}
    \rput[b](7700,525){\scriptsize Saudi Arabia}
    \rput[b](11200,820){\scriptsize U.A.E.}
\end{pspicture}
\captionof{figure}[One figure]{\label{Fig.M3}  {\footnotesize Energy use (kgoe) vs. Food supplies (kg/capita/year) for the year 2007 in 172 countries (U.A.E is United Arab Emirates). The model $y = -a \exp^{-b x} + c$ is applied to the data ($a=433.2$, $b=4.23\cdot 10^{-4}$, $c=956.8$, adjusted $R^2 = 0.36$). Outliers are represented by Middle Eastern and less populated countries. In Middle East countries, fossil energy is abundant and available compared with food supplies.}}
\end{center}
\subsubsection{Energy consumption profile}
In the simulation, we aim at having different profile values distributed among all agents. Some agents frequently consume fossil fuels, while others frequently consume renewable energy. Competition or cooperation occur between the two profiles (see cooperation / competition section). In order to investigate the range of possible $\alpha_8(a)$ distributed in the population at $t_0$, we use the beta distribution that can take a wide range of shapes over a finite interval $[0,1]$: $\alpha_8(a) \sim \text{Beta}(\alpha, \beta)$. We set $\beta=6-\alpha$ and the values of $\alpha$ are explored (see mesh section, Fig. \ref{Fig.M5}). The different values of $\alpha$ used for the initializations allow for a large number of fossil fuels and renewable profiles distributed in the population (Fig. \ref{Fig.M5}).
\subsubsection{Lifespan}
The maximum possible lifespan $\xi_2$ was set to 100. According to Wilmoth, the lifespan in minimal conditions $\xi_1$ is approximatively 24\% of the maximum lifespan and we set $\xi_1 = 24$ \cite{Wilmoth20001111}. We assume that lifespan is heavily dependent on food / energy interaction \cite{Smith2013} and set $\xi_3=0.8$.\\[.3cm]
The dependance on local population density is also considered important and many factors are embedded in this parameter (exposure to stress, local water or air quality, etc.) \cite{Christenfeld1999, Smith2013}. We set $\xi_4=0.6$, assuming $\xi_3 > \xi_4$. Such density effects appear in largest cities (Paris, Mexico, Chongqing, New York, etc). We define the density thresholds $\xi_5, \xi_6$ as the ratios between the number of population in large cities and the actual population with the hypothesis that a mega-city containing $x\%$ of the population is expected to have increased effects compared with a smaller city. The most populated city -Shanghai- is 17,800,000 individuals (0.25\% of the actual population) according to the concept of city proper. The lowest effect threshold is chosen as more than 1 individual and a hypothetical city of 1,000,000,000 individuals as the highest threshold (14.3\% of the actual population).  The lower bound is set to $\xi_5 = 2$ and the upper bound is set to $\xi_6 = 0.143\% \cdot n(G,t_0) = 14.3$. Such that agents being in a site with more than 15 other agents suffers the full effect of population pressure.
\subsubsection{Mesh}
A three-dimensional mesh $M$ is used in order to initialize $\alpha_3(a)$, $\alpha_5(a)$ and $\alpha$ (for $\alpha_8(a))$. There are $m$ linearly spaced nodes in each dimension, and each simulation is initialized with the value corresponding to the coordinate of a node (Fig. \ref{Fig.M4}). Boundaries of the mesh are $\alpha_3(a) \in [0,1]$, $\alpha_5(a) \in [0,1]$ and $\alpha \in [1,5]$ (Fig. \ref{Fig.M5}).
\begin{center}
\psset{xunit=4cm,yunit=4cm,Alpha=160,Beta=15}
\begin{pspicture}(0,-0.1)(1,1.2)
\pstThreeDCoor[xMin=0, xMax = 1.2, yMin = 0, yMax = 1.2, zMin = 0, zMax = 1.2,
    IIIDticks = true, IIIDlabels = true, IIIDzTicksPlane=yz, IIIDzticksep=-1.2,
    IIIDxTicksPlane=xy, IIIDxticksep=-0.01, IIIDyTicksPlane=yz, IIIDyticksep=0.2, Dx=1.25, Dy=-0.25, Dz=-1.25,
    nameX = $\alpha_3(a)$, nameY = $\alpha_5(a)$, nameZ = $\alpha$]

\pstThreeDPlaneGrid[xsubticks=2,ysubticks=2,planeGrid=yz,linecolor=gray,linewidth=.5pt,linestyle=dashed](0,0)(1,1)
\pstThreeDPlaneGrid[xsubticks=2,ysubticks=2,planeGrid=xz,linecolor=gray,linewidth=.5pt,linestyle=dashed](0,0)(1,1)
\pstThreeDPlaneGrid[xsubticks=2,ysubticks=2,planeGrid=xy,linecolor=gray,linewidth=.5pt,linestyle=dashed](0,0)(1,1)
\pstThreeDPlaneGrid[xsubticks=2,ysubticks=2,planeGrid=yz,linecolor=gray,linewidth=.5pt,linestyle=dashed,planeGridOffset=4](0,0)(1,1)
\pstThreeDPlaneGrid[xsubticks=2,ysubticks=2,planeGrid=xz,linecolor=gray,linewidth=.5pt,linestyle=dashed,planeGridOffset=4](0,0)(1,1)
\pstThreeDPlaneGrid[xsubticks=2,ysubticks=2,planeGrid=xy,linecolor=gray,linewidth=.5pt,linestyle=dashed,planeGridOffset=4](0,0)(1,1)
\pstThreeDPlaneGrid[xsubticks=2,ysubticks=2,planeGrid=yz,linecolor=gray,linewidth=.5pt,linestyle=dashed,planeGridOffset=2](0,0)(1,1)
\pstThreeDPlaneGrid[xsubticks=2,ysubticks=2,planeGrid=xz,linecolor=gray,linewidth=.5pt,linestyle=dashed,planeGridOffset=2](0,0)(1,1)

\pstThreeDDot[dotscale=1.2](1,0,0)
\pstThreeDDot[dotscale=1.2](0,1,0)
\pstThreeDDot[dotscale=1.2](0,0,1)
\pstThreeDDot[dotscale=1.2](0,0,0)
\pstThreeDDot[dotscale=1.2](1,1,1)
\pstThreeDDot[dotscale=1.2](1,1,0)
\pstThreeDDot[dotscale=1.2](0,1,1)
\pstThreeDDot[dotscale=1.2](1,0,1)
\pstThreeDDot[dotscale=1.2](0.5,0,0)
\pstThreeDDot[dotscale=1.2](0,0.5,0)
\pstThreeDDot[dotscale=1.2](0,0,0.5)
\pstThreeDDot[dotscale=1.2](0.5,0.5,0)
\pstThreeDDot[dotscale=1.2](0,0.5,0.5)
\pstThreeDDot[dotscale=1.2](0.5,0,0.5)
\pstThreeDDot[dotscale=1.2](0.5,1,1)
\pstThreeDDot[dotscale=1.2](1,0.5,1)
\pstThreeDDot[dotscale=1.2](1,1,0.5)
\pstThreeDDot[dotscale=1.2](0.5,0.5,1)
\pstThreeDDot[dotscale=1.2](1,0.5,0.5)
\pstThreeDDot[dotscale=1.2](0.5,1,0.5)
\pstThreeDDot[dotscale=1.2](0.5,1,0)
\pstThreeDDot[dotscale=1.2](1,0.5,0)
\pstThreeDDot[dotscale=1.2](1,0,0.5)
\pstThreeDDot[dotscale=1.2](0.5,0,1)
\pstThreeDDot[dotscale=1.2](0,0.5,1)
\pstThreeDDot[dotscale=1.2](0,1,0.5)
\pstThreeDDot[dotscale=1.2](0.5,0.5,0.5)

\end{pspicture}

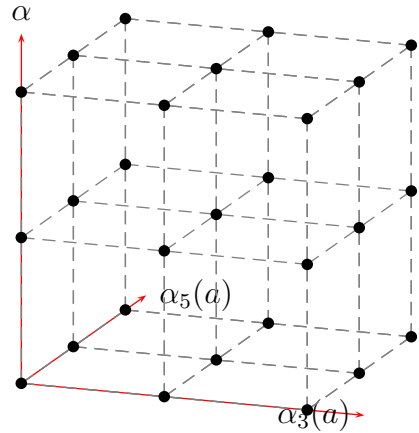
\captionof{figure}[One figure]{\label{Fig.M4} {\footnotesize Example of a 3-dimensional mesh with resolution $m = 3$. Black dots are the nodes and corresponds to the 27 different values of $\alpha_3(a)$, $\alpha_5(a)$ and $\alpha$ to explore.}}
\end{center}

\begin{center}
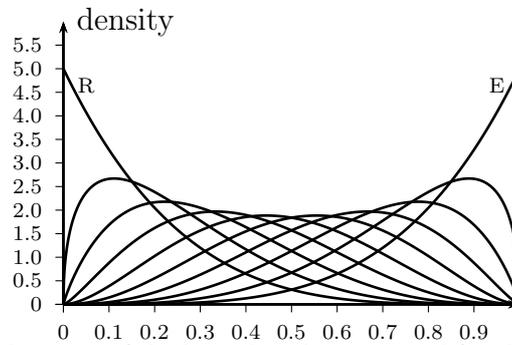

\psset{xunit=6cm,yunit=0.625cm}
\begin{pspicture}(0,-0.13)(1,6)
    \psaxes[Dx=0.1,Dy=0.5,Ox=0,Oy=0,ticksize=-3pt,labelFontSize=\scriptstyle]{->}(0,0)(1,6)[,-90][density,0]
    \fileplot[plotstyle=line, linecolor=black, linewidth=1pt]{Figure39-A1.prn}
    \fileplot[plotstyle=line, linecolor=black, linewidth=1pt]{Figure39-A2.prn}
    \fileplot[plotstyle=line, linecolor=black, linewidth=1pt]{Figure39-A3.prn}
    \fileplot[plotstyle=line, linecolor=black, linewidth=1pt]{Figure39-A4.prn}
    \fileplot[plotstyle=line, linecolor=black, linewidth=1pt]{Figure39-A5.prn}
    \fileplot[plotstyle=line, linecolor=black, linewidth=1pt]{Figure39-A6.prn}
    \fileplot[plotstyle=line, linecolor=black, linewidth=1pt]{Figure39-A7.prn}
    \fileplot[plotstyle=line, linecolor=black, linewidth=1pt]{Figure39-A8.prn}
    \fileplot[plotstyle=line, linecolor=black, linewidth=1pt]{Figure39-A9.prn}
    \fileplot[plotstyle=line, linecolor=black, linewidth=1pt]{Figure39-A10.prn}
    \rput[b](0.05,4.5){\scriptsize R}
    \rput[b](0.95,4.5){\scriptsize E}
\end{pspicture}
\captionof{figure}[One figure]{\label{Fig.M5} {\footnotesize The 10 densities of the beta distributions used for the initializations. The density situated on the utmost left part of the graphic corresponds to the scenario where renewable energy profiles are predominant in the population at $t=0$ (labeled $R$ on the graph). The utmost right density is the scenario where fossil energy profiles are predominant in the population ($E$).}}
\end{center}
For each initialization, we run $n$ simulations in order to study the convergence of population and lifespan values. The total number of simulations is $m^3\times n$ and can rapidly grow if $m$ and $n$ are large. We make a compromise between computation time and results and set $m=n=10$, leading to a total of 10,000 simulations.
\subsubsection{Resources production and supplies}
Current population prospects suggest that the actual population can grow to a maximum of approximatively 143\% of its actual value with assumptions on energy, development, etc \cite{Lutz2001, UN2008}. On the opposite, numerous studies reported that this total number of individuals is not approachable because of the limited capacity of earth (including energy supplies) \cite{meadows1, diamond, fogel}. The debate lead us to arbitrary set the value of $\gamma$ and the initial amount of fossil energy in the system in order to study resource scarcity. We assume that the local renewable energy production is of $\gamma=10$, allowing energy supply for more than 10 individuals in a site. The initial amount of fossil fuel in the system is set to $n(E,t_0) = 10,000$, allowing for a total number of individual in the system of more than 10,000 ($100 n(G,t_0)$). The fossil energy is randomly distributed among all urban patches. The total number of urban patches or areas is estimated to 422, as the number of urban areas that count over 1,000,000 population in 2011 and reported in \cite{Demographia}.
\subsubsection{Ecosystems}
Each site of $B$ contains a real number representing the richness of ecosystems in this area at turn $t$. Spatial heterogeneity in species richness at world scale is generally accepted \cite{Gaston2000}. In order to preserve this spatial feature, all sites are randomly initialized and brutal variations in the number of species between two adjacent sites can occur. The total richness of the ecosystem in the system is not known and we use a large number as the starting richness: $n(B,t_0)=10,000$.\\[.3cm]
Estimating the features of ecosystems (resilience, renewal, etc.) arose as an important field of research in the recent years. Ecological resilience is difficult to measure, and tend to be generally conceptualized or modeled using dynamical systems, state transition or structural models \cite{Petchey2009, Thrush2009}. The lack of experimental studies and empirical measure that could help with assessing the resilience of an ecosystem stressed by a high density of individuals -such as in urban areas- lead us to consider the study of D. Tilman in 1999 \cite{Tilman1999}. He stated that the agricultural intensification leads to a reduction in soil fertility and productivity, suggesting we reached one of the limits of ecological resilience in the recent times.\\[.3cm]
Haberl et al. introduced the Human Appropriation of Net Primary Production (HANPP) as a measure of the aggregate impact of land use on biomass available each year in ecosystems \cite{Haberl2007}. It was estimated as 23.8\% of potential net primary productivity in the year 2000. Other estimates of human impact on ecosystems have been introduced such as the ecological footprint, that is a measure of human demand on the Earth\textquoteright s ecosystems. One result shows that humanity\textquoteright s demand exceeded in 1999 the planet\textquoteright s biocapacity by more than 20\% \cite{Wackernagel2002}. Lester Brown of the Earth Policy Institute states that ``\textit{It would take 1.5 Earths to sustain our present level of consumption}'' \cite{Brown2011}. According to these observations, we set $\beta_2 = 1.2$ and $\omega = 0.5$ such that agents can choose to limit almost half of their destructive actions.\\[.3cm]
Biomass production can be used as a proxy to measure ecosystems production. It is known that richness in plant diversity and overall surrounding biodiversity increase biomass \cite{Cardinale2007}. A rough estimate of annual world biomass production is 146 billion metric tons a year, consisting of mostly uncontrolled plant growth \cite{Cuff1980}. The biomass production per person per year is thus approximatively $1,905.8$ metric tons (using the overall increase in world population growth between 2000 and 2005). However, ecosystems richness is assumed to grow much more slowly, as actual biomass production results from long established ecosystems. We set the rate of production to $\beta_1 =2$, such that the richness production will double from one year to another. The local capacity of ecosystems richness is not known, and we set this value to $\beta_K = 10$.
\subsubsection{Food}
The total amount of food available at start is randomly spread among all sites of $F$. Recall $\alpha_{6}(a) \sim \ln \mathcal{N}$ with parameters $\mu=-2.88$, $\sigma=1.22$ and eq. \ref{ScaledTruncated2}. The mean food consumption per agent is $1-\exp^{-5 \exp^{\mu + \sigma/2}} = 0.40$, leading to an average global initial consumption of $0.4 \times n(G,t_0) = 40$. Assuming all agents consume $\alpha_6(a) = 1 $, the upper boundary is $1 \times n(G,t_0) = 100$. We started the simulation with a large number of food in the system, such as $n(F,t_0) = 10,000$, allowing for more than ten times the upper initial consumption bound.\\[.3cm]
Frozen foods have a shelf life of several months or years while fresh supplies have a shelf life of days or weeks \cite{Kaale2011}. The optimal and recommended conservation of frozen food is up to 2 years and the yearly average consumption of frozen meals is about of 6.5\% of the global food intake for an American and around 5\% for the European countries \cite{Harris1, FAFPAS}. We set $\phi_1 = 0.1$ as the average proportion of food reported from one year to another, in order to take into account frozen foods (0.065) and other possible food (0.035) that can be kept that long.\\[.3cm]
Fossil fuels remain the dominant source of energy for agriculture: more than 50\% of natural gas and coal are used in commercial agriculture and between 30 and 75\% of energy inputs in U.K. agriculture \cite{Woods27092010}. We set $\phi_2 = 0.6$ as a rough average. We set $\phi_3 = 0.001$ in order to encompass the fact that a reduced number of agents can possibly exist using food production with no energy input.
\newpage
\subsection{Initializations (summary)}
We design the system with a high plasticity such that it can approximately sustain more than twice the initial population in terms of energy. We voluntary choose an initial amount of fossil resources $n(E,t_0)$ inferior to $n(R,t_0)$ to capture the actual paradigm: the potential stock of renewable resources is tremendous but still requires several technological leaps to be fully exploited. Initializations can be summed up into one table:
\begin{center}
\scriptsize
\begin{tabular}{|l|l|l|}
   \hline
   Parameter & value & description\\
   \hline
   $t_0$ & 1 & initial time\\
   $t_max$ & 1500 & ending time\\
   $s$ & 100 & side size of the grids\\
   $n(G,t_0)$ & 100 & initial population\\
   $n(B,t_0)$ & $10^4$ & initial amount of ecosystems richness\\
   $n(E,t_0)$ & $10^4$ & initial amount of fossil energy\\
   $n(R,t_0)$ & $\gamma s^2 = 10^5$ & initial amount of renewable energy\\
   $n(F,t_0)$ & $10^4$ & initial amount of food\\
   $n(up,t_0)$ & 422 & initial amount of urban patches\\
   $\gamma$ & 10 & local renewable energy production rate \\
   $\lambda$ & 3 & agents\textquoteright~line of sight \\
   $\omega$ & 0.5 & limitation of destructive actions\\
   $\beta_1$ & 2 & local ecosystems renewal rate\\
   $\beta_2$ & 1.2 & agents\textquoteright~local destruction amount \\
   $\beta_K$ & 10 & local ecosystems capacity\\
   $\phi_1$ & 0.1 & amount of food reported from one year to another\\
   $\phi_2$ & 0.6 & rate of food production from fossil energy\\
   $\phi_3$ & 0.001 & rate of food production from ren. energy / biodiv.\\
   $\xi_1$ & 24 & minimal lifespan (if $\xi_3 = 1$) \\
   $\xi_2$ & 100 & maximal lifespan \\
   $\xi_3$ & 0.8 & energy \& food dependence of lifespan\\
   $\xi_4$ & 0.6 & population density dependence of lifespan\\
   $\xi_5$ & 2 & minimal threshold for density effect\\
   $\xi_6$ & 14.3 & maximal threshold for density effect\\
   \hline
   $\alpha_1(a)$ & 0 & initial age\\
   $\alpha_2(a)$ & 100 & initial lifespan\\
   $\alpha_3(a)$ & $\alpha_3(a) \ in [0,1]$ & \textbf{movement probability}\\
   $\alpha_4(a)$ & 0.02 & reproduction probability\\
   $\alpha_5(a)$ & $\alpha_5(a) \ in [0,1]$ & \textbf{cooperation probability}\\
   $\alpha_6(a)$ & $\alpha_{6}(a) \sim \ln \mathcal{N}(\mu=-2.88,\sigma=1.22)$ & energy consumption per turn\\
   $\alpha_7(a)$ & $1-\exp^{-5 \alpha_{6}(a)}$ & food consumption per turn\\
   $\alpha_8(a)$ & $\alpha_8(a) \sim \text{Beta}(\alpha, \beta=6-\alpha)$ and $\alpha \in [1,5]$ & \textbf{energy consumption profile}\\
   $\alpha_9(a)$ & 0 if $\alpha_8(a) \le 0.5$ 1 else & energy group\\
   \hline
   $d$ & 3 & dimension of the mesh\\
   $m$ & 10 & nb of nodes per dimension\\
   $n$ & 10 & nb of simulation runs\\
   \hline
\end{tabular}
\captionof{table}{\label{Tab.M2} {\footnotesize Table of initializations. Bold descriptions are the 3 parameters set using the mesh (see mesh sec.).}}
\end{center}
\subsection{Results}
In the results we first investigate the complete 10,000 simulations regardless of their ending times, and we then focus on non-sustainable and sustainable results by eventually running the simulation for specific cases of results.
\subsubsection{All initializations}
Simulations stop at turn $t_e$ when there is no agent left in the system or when $t=t_{max}$. The distribution of ending times $t_e$ reveals that sustainable initializations of the system (ie. $n(G,t_{max})>0$) exist for different setups of $\alpha_3(a)$, $\alpha_5(a)$ and $\alpha$ in the mesh (Fig. \ref{FigRS1}). Among the 10,000 simulations performed, most of them end around $t=500$ and 135 (1.35 \%) simulations reach $t_{max}$.
\begin{center}
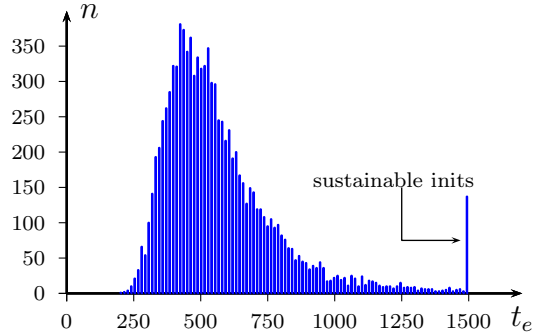

\psset{xunit=0.003529412cm,yunit=0.009375cm}
\begin{pspicture}(0,-20)(1700,400)
\psaxes[Dx=250,Dy=50,Ox=0,Oy=0,ticksize=-3pt,labelFontSize=\scriptstyle]{->}(0,0)(1700,400)[$t_e$,-90][$n$,0]
    \readdata{\data}{Figure40.prn}
    \listplot[linecolor=blue,plotstyle=bar,barwidth=0.005cm,fillcolor=blue!30,fillstyle=solid,opacity=1]{\data}
    \rput[b](1220,150){\scriptsize sustainable inits}
    \psline[linecolor=black, linewidth=0.5pt]{->}(1250,150)(1250,75)(1470,75)
\end{pspicture}
\captionof{figure}[One figure]{\label{FigRS1} {\footnotesize Distribution of the simulations\textquoteright~ending times $t_e$. Sustainable initializations of the simulations are those reaching $t_{max}$.}}
\end{center}
We investigate all the trajectories of the simulations with time, regarding the different initializations. We plot the 10,000 trajectories versus time for each initialization (left panels of the three figures \ref{Fig.RS5}, \ref{Fig.RS4} and \ref{Fig.RS6}) with a yellow color gradient. Dark and bright colors corresponds to the initializations with values 0 and 1 respectively for figures \ref{Fig.RS5}, \ref{Fig.RS4} and 1 and 5 for figure \ref{Fig.RS6}. Such that dark areas in figure \ref{Fig.RS5} are related to non-mobility (initializations with $\alpha_3(a) \approx 0$) and bright areas corresponds to extreme mobility ($\alpha_3(a) \approx 1$). Similarly, dark and bright areas are related to competition and cooperation in figure \ref{Fig.RS4}. Dark areas correspond to a majority of renewable energy profiles distributed in the population at $t_0$ while bright areas correspond to a majority of non-renewable (or fossil) energy profiles in figure \ref{Fig.RS6}.\\[.3cm]
The right panels of the three figures \ref{Fig.RS5}, \ref{Fig.RS4} and \ref{Fig.RS6} correspond to the mean trajectories: for each $10$ initializations we gather the mean value per turn, resulting in 10 curves per studied variable.

\newpage
\begin{center}
    \includegraphics[scale=0.7]{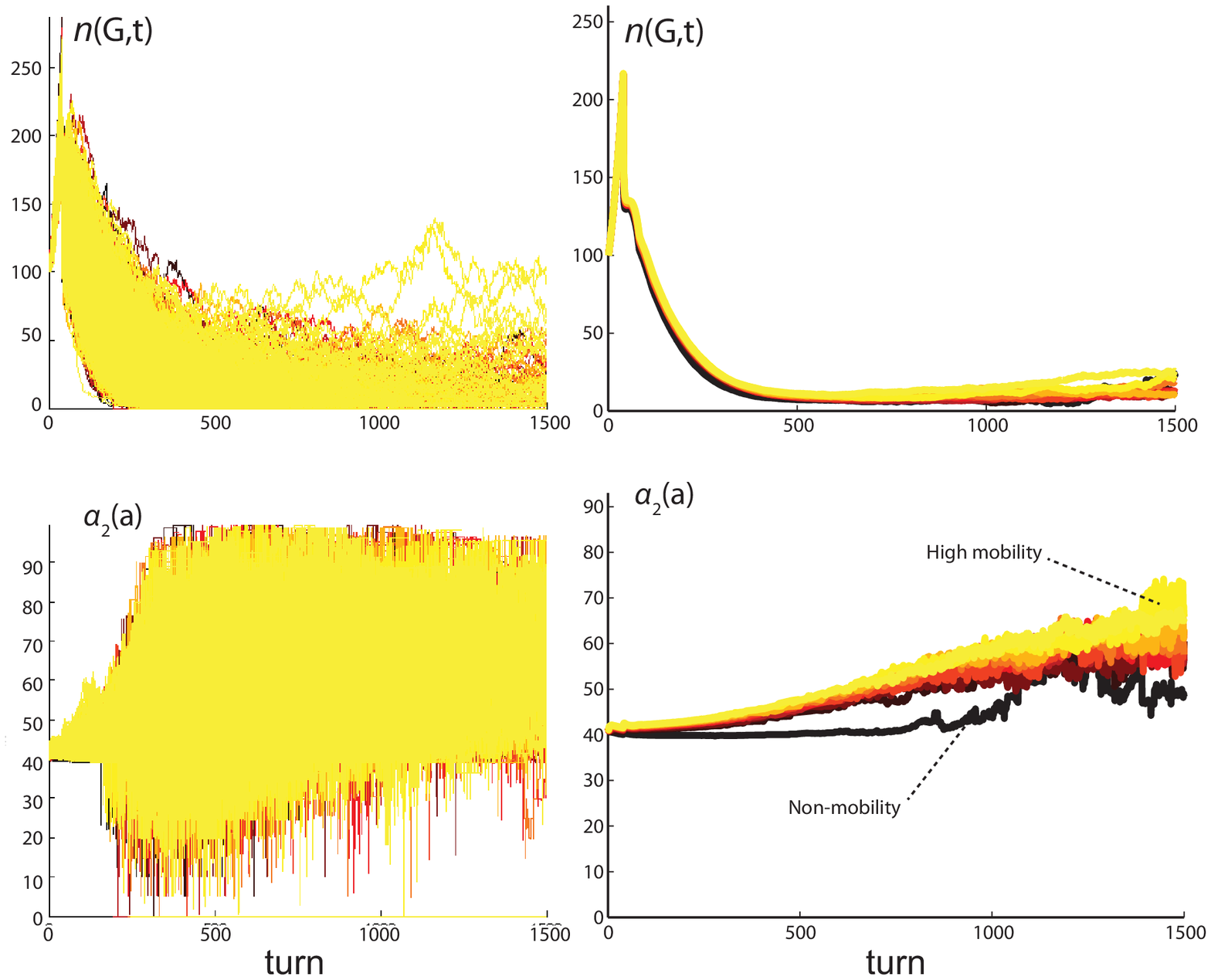}
\end{center}
\captionof{figure}[One figure]{\label{Fig.RS5} {\footnotesize \textbf{Trajectories given various initializations of motion value ($\alpha_3(a)$)}. \textbf{Left panels}: All the trajectories are plotted with a yellow color gradient according to their initializations. \textbf{Right panels} Bright and dark yellow/red areas represent the mean trajectories for the simulations at each $t$, according to different initializations values. \textbf{Upper panels} are the number of individuals per turn, \textbf{lower panels} are the average lifespan value per turn.}}

\newpage
\begin{center}
    \includegraphics[scale=0.7]{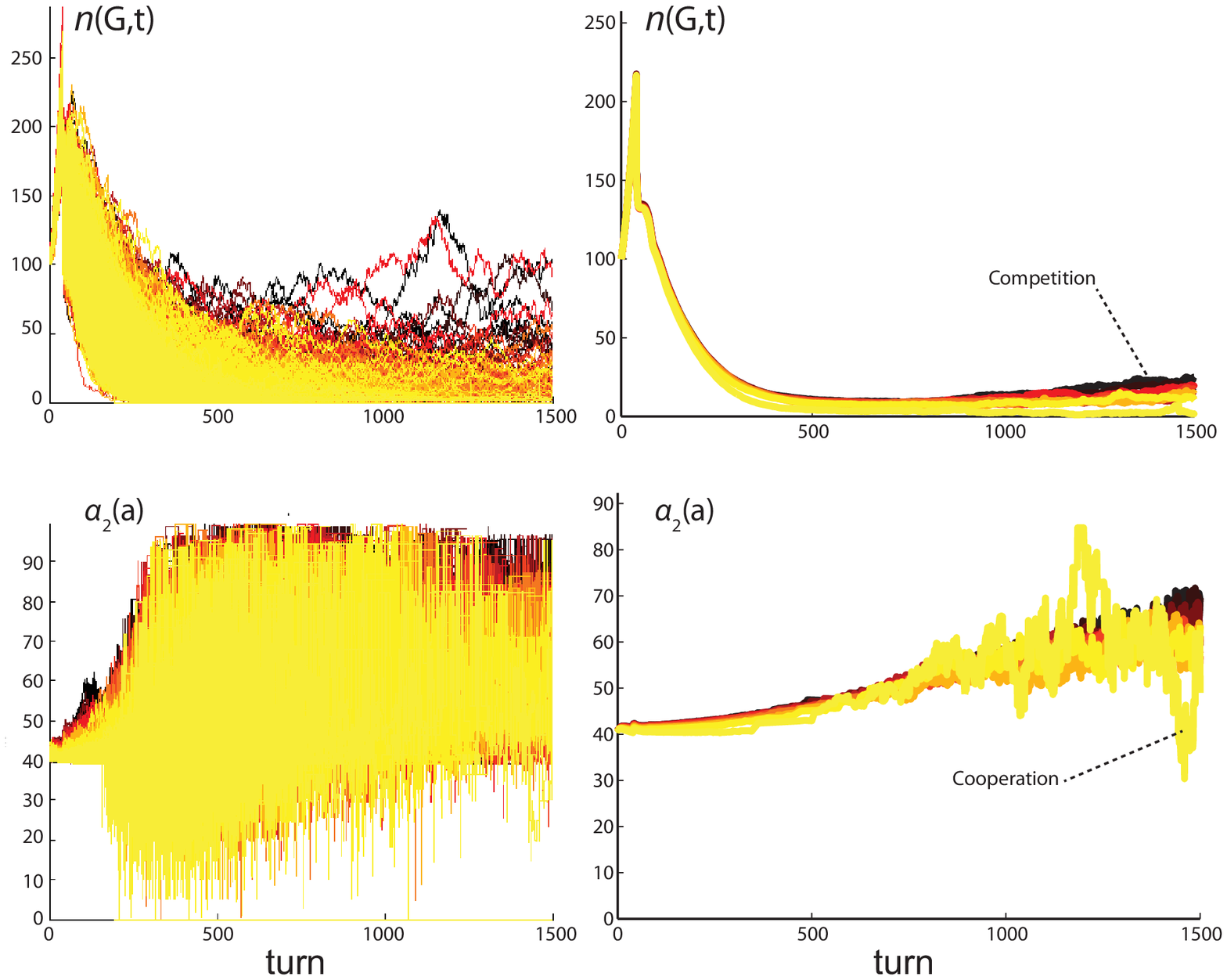}
\end{center}
\captionof{figure}[One figure]{\label{Fig.RS4} {\footnotesize \textbf{Trajectories given various initializations of cooperation/competition value ($\alpha_5(a)$)}. \textbf{Left panels}: All the trajectories are plotted with a yellow color gradient according to their initializations. \textbf{Right panels} Bright and dark yellow/red areas represent the mean trajectories for the simulations at each $t$, according to different initializations values. \textbf{Upper panels} are the number of individuals per turn, \textbf{lower panels} are the average lifespan value per turn.}}

\newpage
\begin{center}
    \includegraphics[scale=0.7]{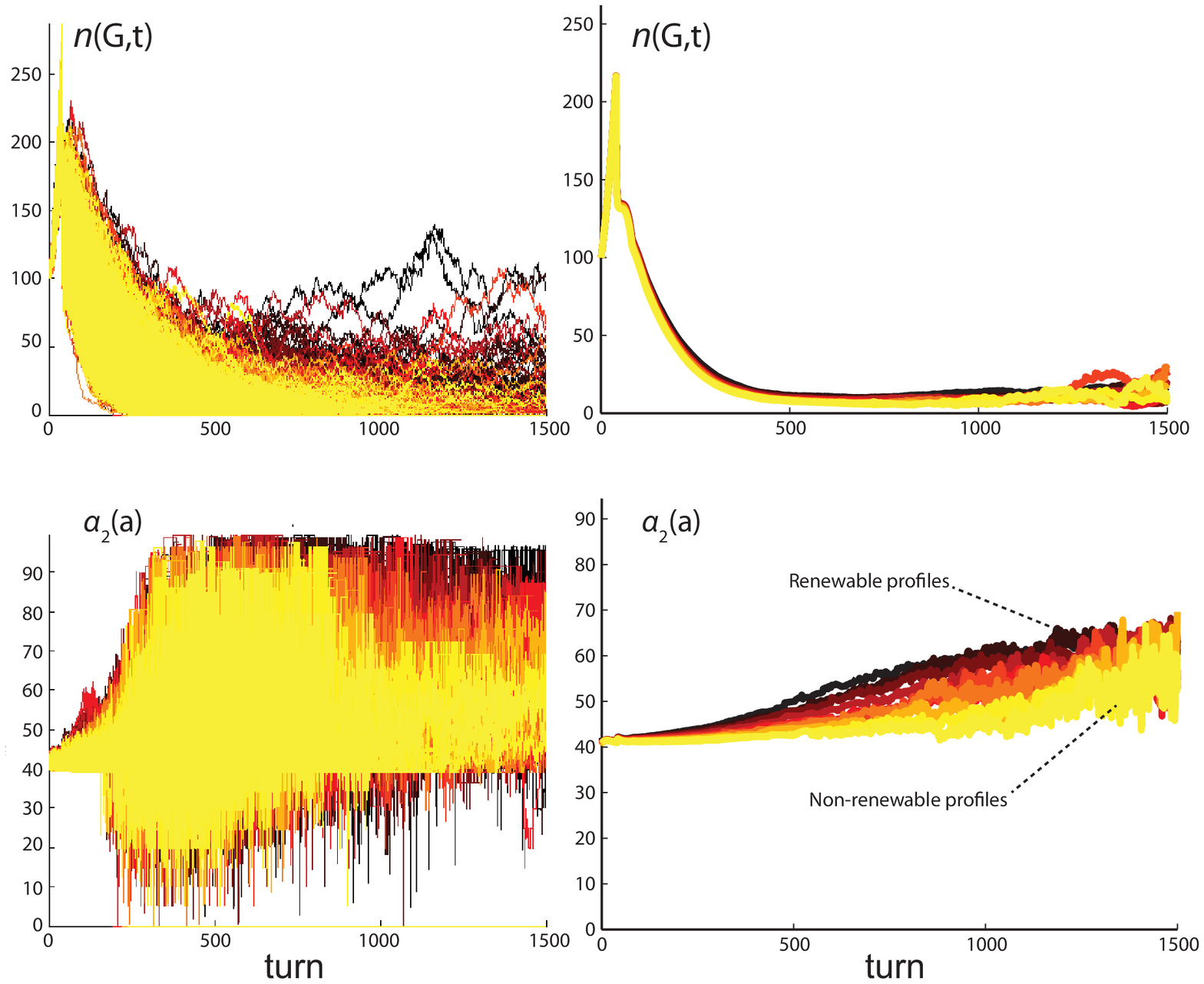}
\end{center}
\captionof{figure}[One figure]{\label{Fig.RS6} {\footnotesize \textbf{Trajectories given various initializations of consumption profile ($\alpha$, one of the two parameters setting the $\text{Beta}$ distribution of $\alpha_8(a)$ values at start)}. \textbf{Left panels}: All the trajectories are plotted with a yellow color gradient according to their initializations. \textbf{Right panels} Bright and dark yellow/red areas represent the mean trajectories for the simulations at each $t$, according to different initializations values.\textbf{Upper panels} are the number of individuals per turn, \textbf{lower panels} are the average lifespan value per turn.}}

\newpage
For each of the 1,000 different simulation setups, we perform 10 simulation runs. We collect the average ending times ($\bar{t_e}$) and the number of times a simulation reached $t_{max}$ $n(t_{max})$ for these 10 runs (Tab. \ref{Tab.R1}, \ref{Tab.R2}), thus corresponding to the \textquoteleft success\textquoteright~of a given initialization (ie. the chance to reach $t_{max}$). Top scores of $n(t_{max})$ (Tab. \ref{Tab.R2}) reveal that only one initialization reaches 4 times $t_{max}$, another one reaches 3 times $t_{max}$, 23 (2.3\%) reach 2 times $t_{max}$ and 82 (8.2\%) reach 1 times $t_{max}$, while the vast majority (89.3\%) do not reach $t_{max}$ at all in the 10 runs.\\[.3cm]
We also plot the distribution of the values of the three parameters $\alpha_3(a)$,$\alpha_5(a)$,$\alpha$ used in each initialization for the 100 best $\bar{t_e}$ and $n(t_{max})$ (Fig. \ref{FigRS2} and \ref{FigRS3}).
\begin{center}
\begin{tabular}{|c|c|c|c|c|c|}
   \hline
   sim id & $\bar{t_e}$ & s.d. & $\alpha_3(a)$ & $\alpha_5(a)$ & $\alpha$\\
   \hline
902	& 1023.1 & 423.69 &	1.00 & 0.00 & 1.44\\
901 & 1018.1 & 444.61 & 1.00 & 0.00 & 1.00\\
821 & 995.1 & 347.44 & 0.89 & 0.22 & 1.00\\
811 & 955.5 & 359.50 & 0.89 & 0.11 & 1.00\\
622 & 946.7 & 443.27 & 0.67 & 0.22 & 1.44\\
612 & 946.6 & 315.82 & 0.67 & 0.11 & 1.44\\
331 & 940.3 & 456.15 & 0.33 & 0.33 & 1.00\\
841 & 930.1 & 350.66 & 0.89 & 0.44 & 1.00\\
122 & 910.7 & 367.22 & 0.11 & 0.22 & 1.44\\
921 & 901.2 & 371.36 & 1.00 & 0.22 & 1.00\\
   \hline
\end{tabular}
\captionof{table}{\label{Tab.R1} {\footnotesize Table of 10 best average ending times, s.d. is the standard deviation.}}
\end{center}

\begin{center}
\begin{tabular}{|c|c|c|c|c|}
   \hline
   sim id & $n(t_{max})$ & $\alpha_3(a)$ & $\alpha_5(a)$ & $\alpha$\\
   \hline
901 & 4 & 1.00 & 0.00 & 1.00\\
761 & 3 & 0.78 & 0.67 & 1.00\\
122 & 2 & 0.11 & 0.22 & 1.44\\
261 & 2 & 0.22 & 0.67 & 1.00\\
282 & 2 & 0.22 & 0.89 & 1.44\\
331 & 2 & 0.33 & 0.33 & 1.00\\
341 & 2 & 0.33 & 0.44 & 1.00\\
602 & 2 & 0.67 & 0.00 & 1.44\\
622 & 2 & 0.67 & 0.22 & 1.44\\
631 & 2 & 0.67 & 0.33 & 1.00\\
   \hline
\end{tabular}
\captionof{table}{\label{Tab.R2} {\footnotesize Table of 10 best $n(t_{max})$.}}
\end{center}

\newpage
\begin{center}
\begin{tabular}{ c c c }
\psset{xunit=6cm,yunit=0.1875cm}
\begin{pspicture}(0,0)(1.09,20)
\psaxes[Dx=0.1,Dy=2,Ox=0,Oy=0,ticksize=-3pt,labelFontSize=\scriptstyle]{->}(0,0)(1.09,20)[$\alpha_3(a)$,100][$n$,0]
    \readdata{\data}{Figure41-d1.prn}
    \listplot[linecolor=blue,plotstyle=bar,barwidth=0.5cm,fillcolor=blue!30,fillstyle=solid,opacity=1]{\data}
    \rput[b](0.05,18){\scriptsize idle}
    \rput[b](0.95,18.5){\scriptsize mobile}
\end{pspicture} & \hspace{0.3cm} &
\psset{xunit=6cm,yunit=0.1875cm}
\begin{pspicture}(0,0)(1.09,20)
\psaxes[Dx=0.1,Dy=2,Ox=0,Oy=0,ticksize=-3pt,labelFontSize=\scriptstyle]{->}(0,0)(1.09,20)[$\alpha_5(a)$,100][$n$,0]
    \readdata{\data}{Figure41-d2.prn}
    \listplot[linecolor=blue,plotstyle=bar,barwidth=0.5cm,fillcolor=blue!30,fillstyle=solid,opacity=1]{\data}
    \rput[b](0.2,18){\scriptsize competitive}
    \rput[b](0.8,18){\scriptsize cooperative}
\end{pspicture} \\
\psset{xunit=1.5cm,yunit=0.083333333cm}
\begin{pspicture}(1,0)(5.3,52)
\psaxes[Dx=0.5,Dy=5,Ox=1,Oy=0,ticksize=-3pt,labelFontSize=\scriptstyle]{->}(1,0)(5.3,45)[$\alpha$,-90][$n$,0]
    \readdata{\data}{Figure41-d3.prn}
    \listplot[linecolor=blue,plotstyle=bar,barwidth=0.6cm,fillcolor=blue!30,fillstyle=solid,opacity=1]{\data}
    \rput[b](1.75,35){\scriptsize Renewable profile}
    \rput[b](4.5,35){\scriptsize Fossil profile}
\end{pspicture}
\end{tabular}
\captionof{figure}[One figure]{\label{FigRS2} {\footnotesize Distributions of $\alpha_3(a)$, $\alpha_5(a)$ and $\alpha$ regarding the top-100 values of $\bar{t_e}$.}}
\end{center}

\begin{center}
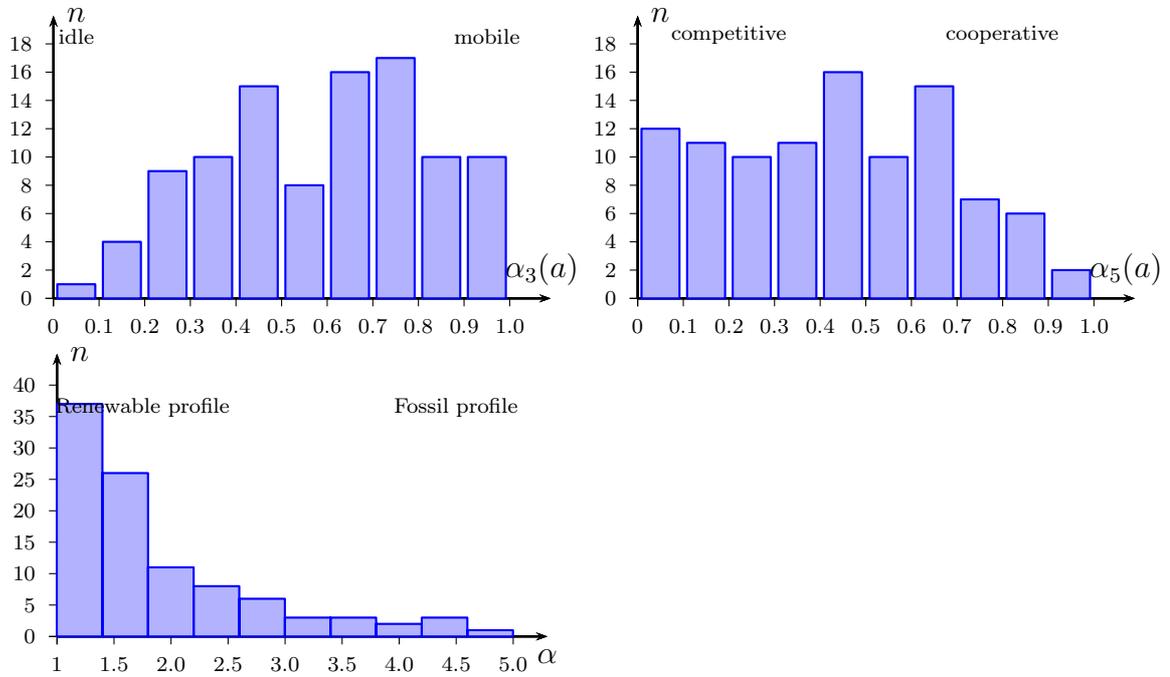

\begin{tabular}{ c c c }
\psset{xunit=6cm,yunit=0.1875cm}
\begin{pspicture}(0,0)(1.09,20)
\psaxes[Dx=0.1,Dy=2,Ox=0,Oy=0,ticksize=-3pt,labelFontSize=\scriptstyle]{->}(0,0)(1.09,20)[$\alpha_3(a)$,100][$n$,0]
    \readdata{\data}{Figure42-d1.prn}
    \listplot[linecolor=blue,plotstyle=bar,barwidth=0.5cm,fillcolor=blue!30,fillstyle=solid,opacity=1]{\data}
    \rput[b](0.05,18){\scriptsize idle}
    \rput[b](0.95,18){\scriptsize mobile}
\end{pspicture} & \hspace{0.3cm} &
\psset{xunit=6cm,yunit=0.1875cm}
\begin{pspicture}(0,0)(1.09,20)
\psaxes[Dx=0.1,Dy=2,Ox=0,Oy=0,ticksize=-3pt,labelFontSize=\scriptstyle]{->}(0,0)(1.09,20)[$\alpha_5(a)$,100][$n$,0]
    \readdata{\data}{Figure42-d2.prn}
    \listplot[linecolor=blue,plotstyle=bar,barwidth=0.5cm,fillcolor=blue!30,fillstyle=solid,opacity=1]{\data}
    \rput[b](0.2,18){\scriptsize competitive}
    \rput[b](0.8,18){\scriptsize cooperative}
\end{pspicture} \\
\psset{xunit=1.5cm,yunit=0.083333333cm}
\begin{pspicture}(1,0)(5.3,52)
\psaxes[Dx=0.5,Dy=5,Ox=1,Oy=0,ticksize=-3pt,labelFontSize=\scriptstyle]{->}(1,0)(5.3,45)[$\alpha$,-90][$n$,0]
    \readdata{\data}{Figure42-d3.prn}
    \listplot[linecolor=blue,plotstyle=bar,barwidth=0.6cm,fillcolor=blue!30,fillstyle=solid,opacity=1]{\data}
    \rput[b](1.75,35){\scriptsize Renewable profile}
    \rput[b](4.5,35){\scriptsize Fossil profile}
\end{pspicture}
\end{tabular}
\captionof{figure}[One figure]{\label{FigRS3} {\footnotesize Distributions of $\alpha_3(a)$, $\alpha_5(a)$ and $\alpha$ regarding the top-100 values of $n(t_{max})$.}}
\end{center}
\subsubsection{Non-sustainable initializations}
We here focus on the causes for not reaching $t_{max}$ in the $8,930$ scenarios that do not reach $t_{max}$. In these scenarios, the average values of $n(E,t)$ decreases to 85\% of the initial stock at around $t = 1,000$. The ecosystems and food values decrease but do not reach 0. However, the number of agents that are likely to consume renewable energies reach 0 between $t=450$ and $t=650$.
\subsubsection{Sustainable initializations}
We investigate the sustainable trajectories associated with extreme mobility and renewable energy profiles distributed in the population at start. In order to perform a deeper analyze of such trajectories, we set $\alpha_3(a)=1$ and $\alpha = 1$ and test $m=200$ different values of cooperation / competition ($\alpha_5(a)$) linearly spaced between $[0,1]$. We also set 100 simulations per value of $m$ for a total of $20,000$ simulations.\\[.3cm]
Among all the simulations performed, a majority of them ends around $t=500$ and 214 (1.07 \%) simulations reach $t_{max}$ (Fig. \ref{FigRS7}).
\begin{center}
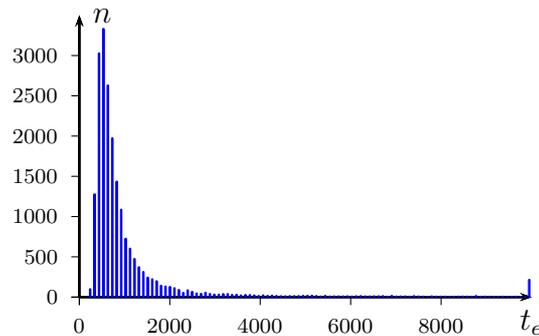

\psset{xunit=0.0006cm,yunit=0.001071429cm}
\begin{pspicture}(0,-100)(10000,3500)
\psaxes[Dx=2000,Dy=500,Ox=0,Oy=0,ticksize=-3pt,labelFontSize=\scriptstyle]{->}(0,0)(10000,3500)[$t_e$,-90][$n$,0]
    \readdata{\data}{Figure51-d1.prn}
    \listplot[linecolor=blue,plotstyle=bar,barwidth=0.005cm,fillcolor=blue!30,fillstyle=solid,opacity=1]{\data}
\end{pspicture}
\captionof{figure}[One figure]{\label{FigRS7} {\footnotesize Simulations\textquoteright~ending times $t_e$. Sustainable initializations of the simulations are those reaching $t_{max}$.}}
\end{center}
As previously, we collect the average ending times and the number of times a simulation reaches $t_{max}$ for these 10 runs (Tab. \ref{Tab.R3}, \ref{Tab.R4}). \begin{center}
\begin{tabular}{|c|c|c|c|}
   \hline
   sim id & $\bar{t_e}$ & s.d. & $\alpha_5(a)$\\
   \hline
2	& 2480.9 &	2989.2 &	0.0101\\
3	& 2448.1 &	3004.9 &	0.0202\\
6	& 2216.3 &	2792.7 &	0.0505\\
1	& 2153.1 &	2884.9 &	0\\
4	& 2109.6 &	2541.1 &	0.0303\\
7	& 1938.7 &	2370.4 &	0.0606\\
12	& 1760.5 &	2221.4 &	0.1111\\
9	& 1701.8 &	2224.8 &	0.0808\\
5	& 1676.2 &	2261 &	0.0404\\
14	& 1580.3 &	1960 &	0.1313\\
   \hline
\end{tabular}
\captionof{table}{\label{Tab.R3} {\footnotesize Table of 10 best average ending times, s.d. is the standard deviation.}}
\end{center}

\begin{center}
\begin{tabular}{|c|c|c|c|}
   \hline
   sim id & $n(t_{max})$ & $s$(\%) & $\alpha_5(a)$\\
   \hline
3 &	17 & 8.5 & 0.0202\\
1 &	16 & 8.0 & 0\\
2 &	15 & 7.5 & 0.0101\\
4 &	11 & 5.5 & 0.0303\\
6 &	10 & 5.0 & 0.0505\\
7 &	8 &	4.0 & 0.0606\\
5 &	6 &	3.0 & 0.0404\\
9 &	6 &	3.0 & 0.0808\\
12 & 6 & 3.0 & 0.1111\\
13 & 6 & 3.0 & 0.1212\\
   \hline
\end{tabular}
\captionof{table}{\label{Tab.R4} {\footnotesize Table of 10 best $n(t_{max})$, $s$ is the success ratio (ie. the percentage of initializations $n(t_{max}) / 200 \times 100$ that reached $t_{max}$).}}
\end{center}
We finally plot the $\bar{t_e}$ plus its associated standard deviation and the success ratio for reaching $t_{max}$ versus $\alpha_5(a)$ (Fig.\ref{FigRS8}).
\begin{center}
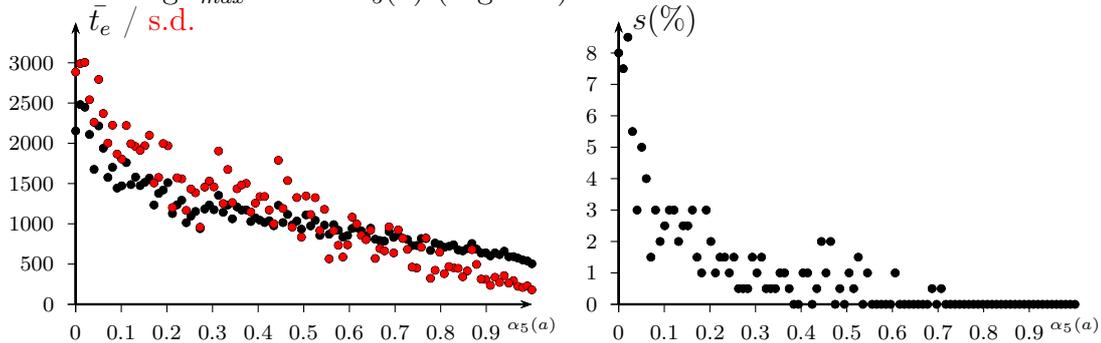

\begin{tabular}{ c c c }
\psset{xunit=6cm,yunit=0.001071429cm}
\begin{pspicture}(0,0)(1,3500)
    \psaxes[Dx=0.1,Dy=500,Ox=0,Oy=0,ticksize=-3pt,labelFontSize=\scriptstyle]{->}(0,0)(1,3500)[\tiny $\alpha_5(a)$,-90][$\bar{t_e}$ / \textcolor{red}{s.d.},0]
    \fileplot[plotstyle=dots, dotstyle=o, fillcolor=black, dotscale=0.90]{Figure52-A1.prn}
    \fileplot[plotstyle=dots, dotstyle=o, fillcolor=red, dotscale=0.90]{Figure52-A2.prn}
\end{pspicture} & \hspace{0.3cm} &
\psset{xunit=6cm,yunit=0.416666667cm}
\begin{pspicture}(0,0)(1,9)
    \psaxes[Dx=0.1,Dy=1,Ox=0,Oy=0,ticksize=-3pt,labelFontSize=\scriptstyle]{->}(0,0)(1,9)[\tiny $\alpha_5(a)$,-90][$s(\%)$,0]
    \fileplot[plotstyle=dots, dotstyle=o, fillcolor=black, dotscale=0.90]{Figure52-B1.prn}
\end{pspicture}
\end{tabular}
\captionof{figure}[One figure]{\label{FigRS8} {\footnotesize The average $t_{max}$ reached per value of $\alpha_5(a)$ (black dots, red dots are the standard deviation, abbreviated s.d.) and the success ratio to hit $t_{max}$ per value of $\alpha_5(a)$ (right panel, black dots).}}
\end{center}
\subsubsection{Patterns}
We here detail three two-dimensional patterns that appear as time increase. These patterns are related to the initial values of $\alpha_3(a)$ and $\alpha_5(a)$.\\[.3cm]
\begin{center}
\begin{tabular}{m{0.3\textwidth} m{0.7\textwidth}}
\begin{center}
    \begin{pspicture}(0,0.6)(2,2)
          \put(1,1){\esmallbox}      
          \put(1.5,1){\esmallbox}    
          \put(2,1){\esmallbox}      
          \put(1,1.5){\esmallbox}    
          \put(1.5,1.5){\smallbox} 
          \put(2,1.5){\esmallbox}   
          \put(1,2){\esmallbox}     
          \put(1.5,2){\esmallbox}    
          \put(2,2){\esmallbox}     
    \end{pspicture}
\end{center} & The \textquoteleft dot\textquoteright~is a pattern appearing when the cooperation is 1. Agents stack into one site, resulting in a rapid increase of the population density and pressure on agents lifespan. Such a pattern is a short-lived structure (ie. it disappears in a few turns).\\
\end{tabular}
\end{center}

\begin{center}
\begin{tabular}{m{0.45\textwidth} m{0.55\textwidth}}
\begin{center}
\begin{tabular}{c c c}
    \begin{pspicture}(0,0.6)(2,2)
          \put(1,1){\esmallbox}      
          \put(1.5,1){\smallbox}    
          \put(2,1){\smallbox}      
          \put(1,1.5){\smallbox}    
          \put(1.5,1.5){\smallbox} 
          \put(2,1.5){\esmallbox}   
          \put(1,2){\esmallbox}     
          \put(1.5,2){\esmallbox}    
          \put(2,2){\esmallbox}     
    \end{pspicture} & \hspace{0.2cm} &
    \begin{pspicture}(0,0.6)(2,2)
          \put(1,1){\esmallbox}      
          \put(1.5,1){\smallbox}    
          \put(2,1){\esmallbox}      
          \put(1,1.5){\smallbox}    
          \put(1.5,1.5){\smallbox} 
          \put(2,1.5){\smallbox}   
          \put(1,2){\esmallbox}     
          \put(1.5,2){\smallbox}    
          \put(2,2){\smallbox}     
    \end{pspicture}
\end{tabular}
\end{center}
& The \textquoteleft cloud\textquoteright~is a pattern appearing when the cooperation is set to 0 (full competition). Agents try to stack into one site, resulting in a rapid increase of the population density and pressure on agents lifespan. Agents are distributed around the site and form a pile with a width > 1 site. Such a pattern is a long-lived structure and can last hundred of turns, if not more. It allows for a limited population density in each site, thus favoring lifespan expansion.\\
\end{tabular}
\end{center}

\begin{center}
\begin{tabular}{*{2}{m{0.5\textwidth}}}
\begin{center}
\includegraphics[scale=0.6]{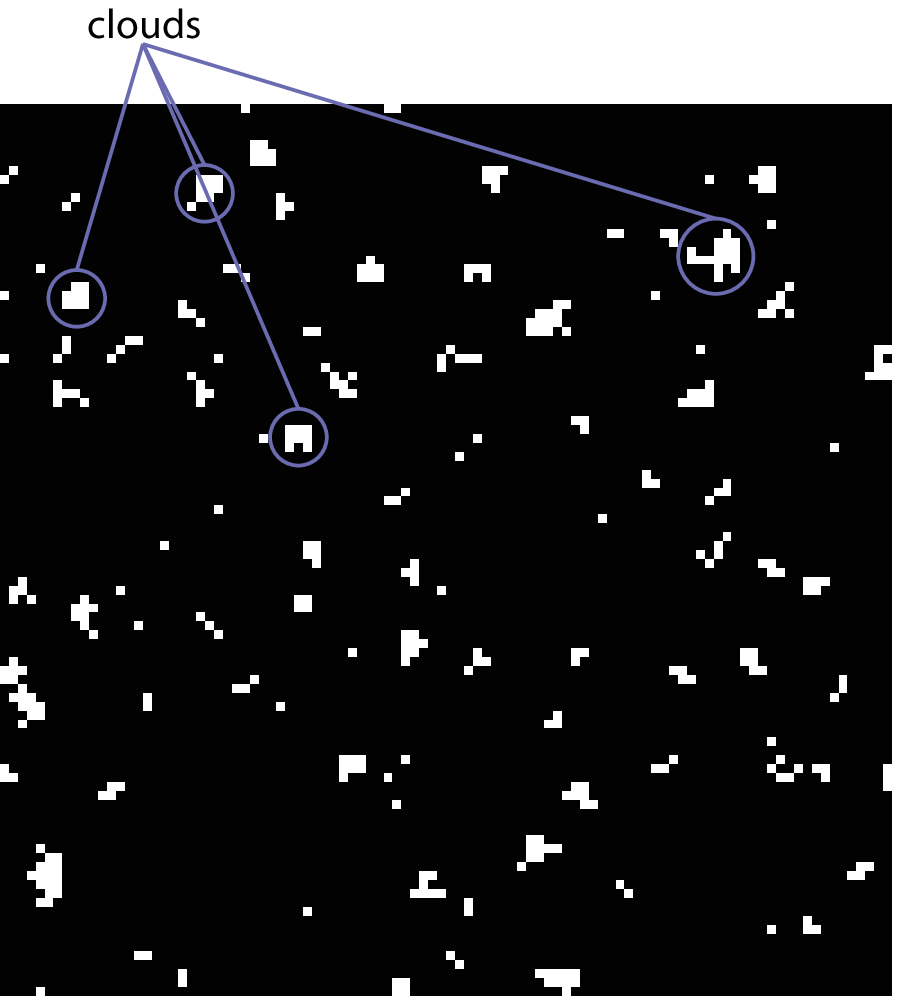}
\end{center} & The \textquoteleft generalized cloud\textquoteright~is a pattern appearing when the mobility rate is 1 and when the cooperation is 0. Agents are forming clouds at regular intervals, with a minimum separation distance that is equal to agent\textquoteright s line of sight, that populate the grid $G$. The structure is self-sustained and durable. Agents travel from clouds to clouds across the grid as they are pushed in adjacent sites due to high competition pools in clouds. Such a pattern is a long-lived structure and can last hundred of turns, if not more, and is robust to local resources depletion (agents are then redistributed to other clouds).\\
\end{tabular}
\end{center}

\newpage
\subsection{Discussion}
Potential scenarios leading to the increase of the population are promoted by high mobility, high inter-agents competition and renewable energy consumption. According to results plotted in Fig. \ref{Fig.RS5} we show that bright trajectories of mobility (ie. mobility $\alpha_3(a) \approx 1$) lead to more sustainable trajectories in terms of population density and lifespan. We do not link mobility to fossil fuels in the model. However, there are evidences that -in the specific case of human- mobility (and the distance covered with time) is positively correlated with primary energy consumption \cite{Smith2013}. We here rather investigate different scenarios of mobility and show that a high mobility favors overall population density in the long run. Reichenbach et al. previously demonstrated that mobility -up to a certain threshold value- promotes ecosystems in the long term \cite{Reichenbach2007}. Competition seemed to be more related with long-term sustainable solutions than cooperation (sim. id 3, $s = 8.5\%$, in Tab \ref{Tab.R4} and Fig. \ref{FigRS8}). In Fig. \ref{Fig.RS4} trajectories initialized with $\alpha_5(a) \approx 1$ promote population density, lifespan. Ecosystem richness is also spared as agents form \textquoteleft clouds\textquoteright~structures (see previous section) and rapidly change from site to site. In Fig. \ref{Fig.RS6} there is a clear tendency for trajectories $\alpha \approx 1$, leading to a distribution of $\alpha_8(a)$ favoring renewable energy profiles (left density of Fig. \ref{Fig.M5}, labeled $R$).\\[.3cm]
In a majority of results, the population decreases as time increases, after a first period of exponential growth of about 100 turns. The initial breakdown at $t=500$ is due to the dramatic fossil reduction in urban patches leading to the extinction of fossil energy profiles. If the system can sustain this initial breakdown, the population is often reduced to clouds of agents with renewable energy profiles. Then the grid $G$ is slowly re-populated with clouds of agents that remain stable in the long term. Clouds can eventually generalize in the grid and lead to a specific configuration of the system that can re-organize itself and endure local resource depletion.\\[.3cm]
One can attempt to link the results to the current societal infrastructure and show that current ending times $t\approx500$ correspond to the duration of approximatively 5 maximum lifespan. J. Diamond \cite{diamond}, P. Ehrlich \cite{ehrlich}, K.R. Smith \cite{Smith2013}, J. \cite{Aronson2010} and many other authors questioned our ability to go through the upcoming energy and environmental crisis. One of the main findings of The Millennium Ecosystem Assessment (MA) -released in 2005- was ``\textit{The challenge of reversing the degradation of ecosystem while meeting increasing demands for services can be partially met under some scenarios considered by the MA, but will involve significant changes in policies, institutions and practices that are not currently under way}'' \cite{MilleniumEA}. Such sustainable scenarios exist but are unlikely to produce given the actual societal inertia. Futures energy sources, such as nuclear fusion (that is not set to be technologically practicable until 2050 \cite{Lior2010}), may change the actual development. However, in our model, we make the assumption that such technological breakthrough are not possible in the future, but the results underline that renewable energy consumption is one leading parameter to promote sustainable trajectories. In eq. \ref{Food} we authorize the production of food without the use of fossil energy. If an agricultural transition occur, one can write $\phi_3 > \phi_2$, leading to a sustainable scenario, as suggested in the MA. We do not deeply test long-term breakdowns, related to non-sustainable agriculture (ie. $\phi_2 > \phi_3$ with $\phi_3$ a small value). We test a few new initializations by reducing the food and ecosystem inputs at the start of the simulation and see their utmost influence on simulations ending times, dramatically reducing $t_e$.\\[.3cm]
Sustainable trajectories are characterized by two-dimensional patterns in the long term and we are investigating the formation of new ones and their behavior by extending the range of the time studied (ie. $t_{max} > 10,000$). We assume that, provided there is still food and ecosystem in the system, other two-dimensional patterns can form and live or disappear. Such cases may be very rare, such as only $1.35\%$ simulation runs reach $t_{max}$ in the general case (Fig. \ref{FigRS7}) and only $1.07\%$ runs reach $t_{max}$ when choosing favorable initializations (sec. Sustainable initializations). The total chance of achieving long term trajectories may therefore be much less when starting the system with low mobility and non-renewable energy profiles. Not to mention that we use optimistic initializations in our approach considering the initialization of the ecosystem richness / food inputs and production.\\[.3cm]
We limit the model to a predefined set of cooperative and competitive actions. Including / mixing some cooperative features inside the competitive set of actions may lead to a better and sustainable combination: introducing cooperative actions to promote ecosystems while keeping the intra-species habitat competition (\textit{push} action) for instance. Moreover we use a low level mesh resolution and other solutions may lead to sustainable trajectories and new two-dimensional patterns. However, such solutions do not come to light when sketching the scene of possible results in the initial phase of initializations.\\[.3cm]
We plan to test the long-term outcome of each action, separately, on a larger grid, with a larger population and an extended time frame. However, we need to overcome computation issues with large meshes. We also plan to introduce an adaptation feature among the agents of the system, by linking their birth rate to their fitness at a given time. We also plan to link mobility with energy and study if extreme mobility is possible, given resource scarcity. Finally, we look forward linking population density with food production, in order to further investigate Boserup\textquoteright s theory.\\[.3cm]
The population-energy-environment relationship involve a great number of variables (food production, local conflicts for food security, economical transactions, access to resources, energy storage capacity, etc.) with strong or weak interactions, positives or negatives feedbacks and delays. The present model captures the possible behaviors of a simplified representation of the problem in the current technological paradigm. Including all the variables and modeling the overall behavior of the problematic is too complicated. Holistic approaches such as the one intended in this work -and the current IGSM and World3 models- offer practical alternatives for studying the complex interactions and their impact on the overall behavior. However, they are not numerous despite their relevance in anticipating the upcoming difficulties \cite{JFT2012, Aronson2010, Johansen2001465}.
\begin{savequote}[10pc]
\sffamily
A pessimist sees the difficulty in every opportunity; an optimist sees the opportunity in every difficulty.
\qauthor{Winston Churchill (1874-1965)}
\end{savequote}
\chapter{Discussion} 
``\textit{Citius, altius, fortius}'', the Olympic motto, proposed in 1894 by Pierre de Coubertin and introduced in 1924 at the Olympic Games of Paris, is now facing upheaval. The analysis of the development of sports performances through the modern era, from 1896 to 2010, reveals that athletes reach a ceiling in terms of physical ability \cite{berthelot2008, fdd1} (see section \ref{sec:OlympicSports}). The most part of athletes are stagnating for mature sport events, in particular in track and field, even if exceptionable or atypical athletes still arise \cite{berthelot2010a}. This is partly due to the population increase during the past century; more rare genotypes and phenotypes have been produced. Thus, the chance that an athlete holds exceptional genes allowing for the combination of speed and height is increasing. It would be common sense to assume that we would then see an increase in performance. However, as demonstrated in section \ref{sec:Performance}, this is no longer the case.
\\[0.3cm]
\noindent
At the same time, the development of technology allowed for the improvement of the detection and selection of these unique phenotypes by advancing the detection methodology. It also enabled the introduction of new equipment such as swimsuits in swimming \cite{berthelot2010b}, new bikes in cycling \cite{nour1}, klap skates in speed skating, and many others innovations in almost all events \cite{neptune} (sec. \ref{sec:Technology}). These strongly influenced the evolution of performances \cite{nourTHESIS}. One typical example is the evolution of the hour record in cycling (Fig. \ref{FigFDD4}): the Union Cycliste Internationale (UCI) chose to strictly enforce the use of upright bicycles, while the International Human Powered Vehicle Association (IHPVA) used streamlined recumbent bicycles. The world record for men is 49.7 kms using Eddy Merckx\textquoteright s upright bike and was set in 2005 by Ondrej Sosenka. The world record for IHPVA is 91.6 kms. The performance increased by about 90.6\%. Technology also led to the introduction of doping, so that athletes can reach and optimize their peak condition and defeat the competition, while being less constrained by their genotype. Several studies demonstrate that doping protocols have existed in a number of countries \cite{franke} and were exacerbated in the Cold War \cite{marion1}. However, despite all the means given to athletes to establish new world records and top marks, the progression of performances remains finite. No man or woman will run the 100m in track and field in 1 second or less, although recently suggested \cite{tatem}. Both legal and illegal innovations allowed for the optimization of performances in sports. One would question the interest of an increased use of technology in the establishment of new records. But it\textquoteright s part of the inner spirit of man to always try to repel and push his limit, sometimes regardless of the cost and downside. In a larger view, we can add that human performance, in a broad sense, has only increased through technology.
\\[0.3cm]
\noindent
The trilogy published by Donella H. Meadows, Dennis L. Meadows, J\o rgen Randers, and William W. Behrens III in 1972 \cite{meadows1}, 1992 \cite{meadows2} and 2004 \cite{meadows3} was related to the identification and modeling of important parameters that drive the performances of mankind (such as life expectancy, demography, etc.) in the recent times. A large panel of human evolution sub-systems were browsed (ie. food, industrial, population, non-renewable resources and pollution sub-systems). This resulted in the discovery that the relations between such sub-systems are composite, with both positive and negative retro-action loops and complex dynamics (sec. \ref{subsec:Complexity}). The projections of their model lead to different scenarios. Some of them are pessimistic, painting a world where the size of the human population will become so high, that it will head for an \textquoteleft overshoot\textquoteright~and collapse. This term is used to describe a signal that exceeds its steady state value, such as an imbalanced population of individuals outweighing the environment\textquoteright s carrying capacity. In this case, the authors depict a world with a high welfare loss (ie. when there is no replacement for a resources that starts to deplete), one where the biosphere is under siege. They also stated that ``\textit{once the collapse was well under way, no one could stop the fall}'' \cite{meadows3}. We do agree that these types of scenarios, although very dark for humanity, might happen if nothing is done to limit the overshoot.
\\[0.3cm]
\noindent
Some other scenarios are more optimistic; they lead to a moment in time where equilibrium would arise, at the cost of a reduced population and welfare and provided additional technologies are added to abate pollution, conserve resources, maintain land yield and protect agricultural land. Such transitions to a sustainable system are only feasible if mankind rapidly reacts and makes dramatic changes among its habits. To our knowledge such mass reactions are rare, not spontaneous and may not be conceivable in a democratic regime. Moreover, during the past century, economy has become the heart of matter and given how considerable the world inertia is, it seems unlikely to change in the near future. Other authors have also pointed out the frontiers of our development, such as Jared Diamond \cite{diamond}, Robert William Fogel \cite{fogel}, Paul Ehrlich \cite{ehrlich}, James Lovelock \cite{lovelock}, Richard Heinberg \cite{heinberg} and show that the problem is more and more prominent nowadays among the scientific community.
\\[0.3cm]
\noindent
In such a context, would it be possible to head toward a continuous development of physical, cognitive, physiological and societal performances in a sustainable way? The initial part of this work was based on the analysis of sport performances, as a practical measurement of physiological development and limitations. We addressed the issue of performance development with aging and time in human and athletic species under human influence, leading to the concept of \textquoteleft Phenotypic Expansion\textquoteright. A further step consisted in extending this analysis to a more general scheme, with a focus on primary energy use and the resulting human ecological footprint. In this last section we specifically studied the mobility of individuals in several species, and the relationship between primary energy consumption, ecosystem loss and performance increase. We built a simple model to investigate the extent of primary energy use in the observed patterns and in a specific sub-system without including economical, geopolitical, geographical and cultural constraints. We considered a strong societal inertia and introduced a simple model in that direction by maintaining the actual paradigm: the evolution of the studied system is solely determined by its initial context and no other input such that the model entirely proceeds as a consequence of the initial conditions. Among the limitations of the model, one should have considered linking the birth rate and motion speed of the agents to primary energy \cite{DeLong2010, Smith2013}. We decided to focus on the investigation of a range of possible scenarios for motion. Birth would be a common way to integrate an adaptation trait in the model, allowing for the agents to sustain various constraints.
\subsubsection{Perspectives}
This work lead to several pending issues:
\begin{itemize}
  \item \textbf{Which scenario for future performances?}. A few events still show progression, despite the observed stagnation in a majority of sport events. As an example, men world records in the 100m straight in track \& field have improved until 2009; however outstanding athletes tested positive to banned substances. Anti-doping agencies are facing turbulent economical conditions and new technologies are experimented with, uncontrolled, hazardous new molecules (and, perhaps, gene doping in the future). A possible scenario would be the reduction of allocated resources to fight against doping while a new technology allows for increasing performance beyond human present capabilities.
  \item \textbf{Identification of atypical performances}. This subject recently gained attention. Nabbing atypical performance is a methodological challenge considering the bias of unknown doped performances in the data-set. However, we demonstrate that it is possible to historically spot atypical developments (sec. \ref{sec:Performance}). Investigating different methodologies may be useful and complementary to traditional biological tests.
  \item \textbf{Measuring the phenotypic expansion}. Although the empirical development of performances with age and time is not precisely known, there may exist historical data-files containing additional information for more precisely assessing the phenotypic expansion. We also investigate the expansion in other fields, measuring scientific prizes and publications or political positions with age.
  \item \textbf{Ageing, performance and statistical entropy}. We introduced 4 models in the present work (sec. \ref{sec:PhenotypicExpansion}). We consider new models, based on a possible relation between statistical entropy and aging. We focus on the field of cellular biology and consider other simple animal models, such as the nematode (\textit{Caenorhabditis elegans}), in order to investigate the mechanisms responsible for limiting the development of common biological processes, such as body-size, and more generally growth \cite{Lui01062011} and include them in our model.
  \item \textbf{Sophistication in modeling sub-systems}. We intend to develop the model recently introduced, by linking the birth rate of an agent to \textit{i}) the local population density and \textit{ii}) lifespan of nearby agents, ie. maximizing $\alpha_2(a) \times n(G,t)$ in the Moore neighborhood. The introduction of a genetic algorithm will also be considered in order to evolve the agents toward a sustainable behavior. Finally we plan to test a wider range of values for motion, cooperation/competition and energy profiles.
\end{itemize}
\chapter{Appendix}
\section{Programming influence and environmental conditions in performance}
In the section \ref{sec:NonOlympicSports}, we detailed the influence of environmental conditions on performances such as temperature. In the following article we focus on both the programming and temperature influence on athletic performance. We aim at measuring the weight of organization calendar of track competitions in the sprint and middle distance running events.
\\[0.6cm]
\noindent
\textbf{Programming influence and environmental conditions in performance athletics}\\
A. Ha\"{\i}da, F. Dor, M. Guillaume, L. Quinquis, A. Marc, L.-A. Marquet, J. Antero-Jacquemin, C. Tourny-Chollet, F. Desgorces, G. Berthelot, J.-F. Toussaint
\\[0.6cm]
\noindent
\textbf{\textsc{Abstract}}
\textbf{Purpose}: Achievement of athletes\textquoteright~performances is related to several factors including physiological, environmental and institutional cycles where physical characteristics are involved. The objective of this study is to analyze the performance achieved in professional sprint and middle-distance running events (100m to 1500m) depending on the organization of the annual calendar of track events and their environmental conditions. \textbf{Methods}: From 2002 to 2008, all performances of the Top 50 international athletes in the 100m to 1500m races (men and women) are collected. The historical series of world records and the 10 best annual performances in these events, amounted to a total of $26,544$ performances, are also included in the study. \textbf{Results}: Two periods with a higher frequency of peak performances are observed. The first peak occurs around the 27.15\textsuperscript{th} $\pm$ 0.21 week (first week of July) and the second peak around $34.75$\textsuperscript{th} $\pm$ 0.14 week (fourth week of August). The second peak tends to be the time of major international competitions (Olympic Games, World Championships, and European Championships) and could be characterized as an institutional moment. The first one, however, corresponds to an environmental optimum as measured by the narrowing of the temperature range at the highest performance around 23.25 $\pm$ 3.26$\,^{\circ}\mathrm{C}$. \textbf{Conclusions}: Both institutional and eco-physiological aspects contribute to performance in the 100m to 1500m best performances and define the contours of human possibilities. Sport institutions may take this into account in order to provide ideal conditions to improve the next records.
\begin{multicols}{2}
Since the beginning of the modern Olympic era (1896), the best performance (BP) are in a process of exponential growth which now seems to have reached its limits \cite{berthelot2008}. Performance is often understood as a very broad term which involves many components such as: psychomotor abilities, flexibility and joint stiffness, muscle strength and power \cite{Atkinson2005}. Athletes, like any living organism, depend on physiological regulations that respect the biological nycthemeral seasonal or vital cycles \cite{Magnanou2009}. There are variations in physiological factors such as maximum oxygen consumption (VO$_2$ max) or concentrations of melatonin on the basis of seasonal rhythms \cite{Lincoln2005}. This seasonal rhythmicity has been demonstrated for certain factors such as mood \cite{Boivin1997}, lung function \cite{Spengler2000} and the core body temperature. It is also observed in the physical activity of the general population, which tends to be higher during summer \cite{McCormack2010, Hamilton2008, Chan2006}. Chan and colleagues (2006) found a significant change in physical activity in the general population, as number of steps walked per day being related to temperature, precipitation and wind speed\cite{Chan2006}.\\[0.3cm]
There is a limited amount of research that has investigated effects of seasonality in sports on sprint athletes\textquoteright~performances. Yet the annual schedule of events seems to be a contributing factor to performance. Comparison of track and field world records (WR) shows that performance prevails in summertime. The influence of environmental parameters on physiology (eco-physiology) partly determines the evolution of human performance \cite{fdd1, Marino2004}. Marathon optimal performances are set at a temperature around 10$\,^{\circ}\mathrm{C}$. This performance dependency on temperature occurs not only for elite-standard athletes but for all participants also \cite{Nour2012, Cheuvront2001, Kenefick2007}.\\[0.3cm]
The objective of this study is to compare the date and temperature of the BP in sprint and middle distance races (100m to 1500m) for men and women during the annual calendar of international competitions, and observe their evolution over the Olympic era in order to assess the environment and scheduling effects on sprint and middle distance running performances.\\[0.3cm]
\textbf{\textsc{Material and methods}}\\[0.3cm]
\textbf{Data collection}\\
From 2002 to 2008, all performances of the Top 50 international athletes in running events ranging from 100m to 1500m races for men and women are collected from the official web-site of the International Association of Athletics Federations (IAAF) \cite{IAAF}. For each event, data collection includes: full name of the athlete, the completion date and place of the competition: 23,746 performances are collected, 11,813 from males and 11,933 from females. Performances are divided into five categories defining the performance as a percentage in relation to the BP obtained at the event. The percent categories (PC) used were: [95-96\%[, [96-97\%[, [97-98\%[, [98-99\%[, [99-100\%]. Within each group, performance is collected according to the competition type: (\textit{i}) major competitions (the Olympic Games (OG), World Championships (WC), European Championships (EC) and American selections (US)), (\textit{ii}) the international circuit represented by the Golden League and (\textit{iii}) other meetings.\\[0.3cm]
Date, place and name of the athlete when WR were set for the same distances between 1952 and 2010 are collected (181 WR) and the performances of the top 10 male and female 100m race are gathered from 1891 to 2008, representing 2,617 performances.\\[0.3cm]
Temperatures for each city, at the time of the competition, are recorded from 97 to 100 PC in 100m, 200m, 400m, 800m and 1500m. In order to improve resolution, half PC are defined to study temperature density: [97-97.5\%[, [97.5-98\%[, [98-98.5\%[, [98.5-99\%[, [99-99.5\%[, [99.5-100\%]. Temperature data are collected from the weather underground website \cite{WUW2011}. The total number of performances collected for this study is 26,544.\\[0.3cm]
\end{multicols}
\begin{center}
    \includegraphics[scale=1.2]{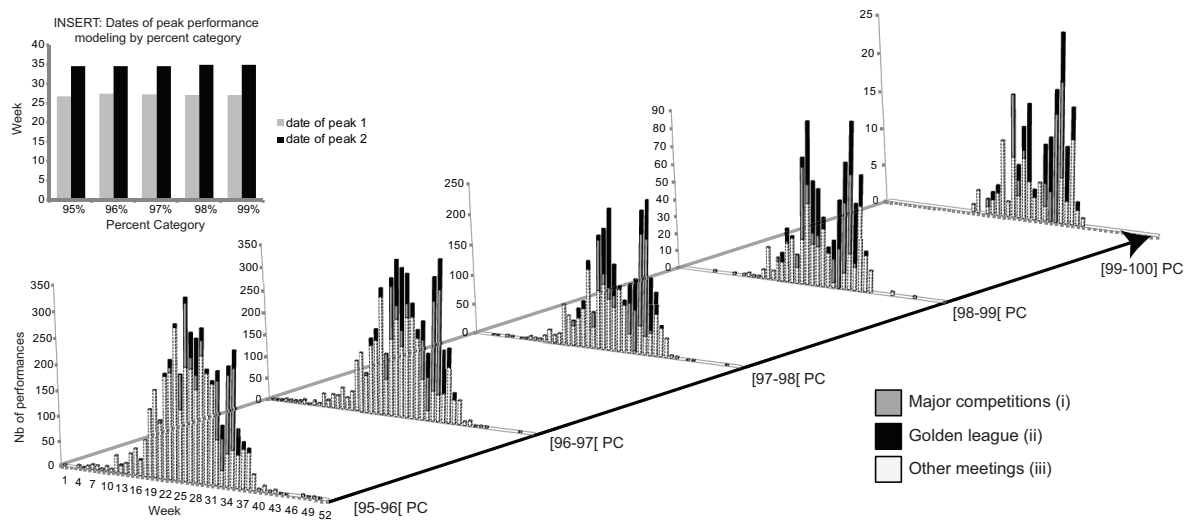}
\end{center}
\captionof{figure}[One figure]{\label{Fig.ANN1} {\footnotesize Number of performances per week and per percent category (PC) by (\textit{i}) major competitions (Olympic Games (OG), World Championships (WC), American selections (US), European championships (EC)), (\textit{ii}) Golden League and the others meetings (\textit{iii}). INSERT: Dates (week) of peaks performance modeling by PC.}}
\begin{multicols}{2}
\noindent
\textbf{Modeling the distribution}\\
The performance data from 2002 to 2008 is based on the distribution of performance per weeks of the year depending on the PC and the type of competition: mathematical analysis and modeling were done using Matlab. In order to estimate the two dates where the greatest number of performances occur, two functions were adjusted using the least squares method (eq. \ref{DoubleGauss1} and \ref{DoubleLorentz1}). For each PC the best model was chosen, based on the adjusted $R^2$ and the root mean squared error (RMSE). The date of the thermal ($x_{01}$) and cultural ($x_{02}$) peak are estimated for each category, along with the proportion of elite performances included in the two peaks ($p_1$) and ($p_2$).
\end{multicols}
\textbf{1) The double Gaussian and double Lorentz}\\
In order to estimate the two dates where the highest number of performances occurs for both thermal and cultural peak, we investigated two functions:\\[.3cm]
(\textit{i}) The double Gaussian function $f_1(x)$ that is the sum of two Gaussian functions:
\begin{equation}
  \label{DoubleGauss0}
  f_1(x) = f_{1,1}(x) + f_{1,2}(x)
\end{equation}
where
\begin{equation}
  \label{DoubleGauss1}
  f_1(x) = a_1 \cdot \exp^{-\left( \dfrac{x-b_1}{c_1} \right)^2} + a_2 \cdot \exp^{-\left( \dfrac{x-b_2}{c_2} \right)^2}
\end{equation}
(\textit{ii}) The double Lorentz function $f_2(x)$ that is the sum of two Lorentz functions:
\begin{equation}
  \label{DoubleLorentz0}
  f_1(x) = f_{2,1}(x) + f_{2,2}(x)
\end{equation}
where
\begin{equation}
  \label{DoubleLorentz1}
  f_2(x) = c_1 \dfrac{\dfrac{2}{a_1 \pi}}{1+\left( \dfrac{x-b_1}{a_1 / 2} \right)^2} + c_2 \dfrac{\dfrac{2}{a_2 \pi}}{1+\left( \dfrac{x-b_2}{a_2 / 2} \right)^2}
\end{equation}
The functions $f_1$ and $f_2$ are two-peak functions and $x$ is the number of elite performances in a week. Resulting adjusted $R^2$ and RMSE were gathered for both functions and the elected function for a given percent category was the function that presented the best statistics.\\[.3cm]
\textbf{2) Estimates of $x_{01}$, $x_{02}$, $p_1$, $p_2$}\\
For each elected function in each percent category, the two peaks $x_{01}$, $x_{02}$ were estimated, and the proportions $p_1$, $p_2$ were given by estimating the area under the curve in the interval [0, 52] for the functions $f_{1,1}$, $f_{1,2}$, $f_{2,1}$, $f_{2,2}$. For notation convenience, we denoted $i$ and $j$ as the indexes of the functions, such as when $i = 1$ and $j=1$, $f_{i,j}$ referred to $f_{1,1}$. Integration of the functions was given by:
\begin{equation}
  \label{GenInt0}
  \int_0^{52} f_{i,j}(x)dx = F_{i,j}(52) - F_{i,j}(0)
\end{equation}
where
\begin{equation}
  \label{GenInt1}
  F_{1,j}(x) = -\sqrt{\pi} \times a_j \times c_j \times \text{erf}\left( \dfrac{b_j-x}{c_j} \right)
\end{equation}
and
\begin{equation}
  \label{GenInt2}
  F_{2,j}(x) = -2 \dfrac{c_j \times \tan^{-1} \left( \dfrac{2 \left( b_j - x \right)}{a_j} \right)}{\pi}
\end{equation}
where $\text{erf}(x)$ is the error function and $tan^{-1}(x)$ the inverse tangent function. The proportion of performances in the two peaks was estimated by computing the area under the curve (proportion of Performances) of each elected model and for each PC:
\begin{equation}
  \label{GenInt3}
  \int_0^{52} f_{1}(x) = \int_0^{52} f_{1,1}(x) + \int_0^{52} f_{1,2}(x)
\end{equation}
\begin{equation}
  \label{GenInt4}
  \int_0^{52} f_{2}(x)dx = \int_0^{52} f_{2,1}(x)dx + \int_0^{52} f_{2,2}(x)dx
\end{equation}
And the proportions were calculated in percentages using:
\begin{equation}
  \label{GenInt5}
  p_1 = \dfrac{\int_0^{52} f_{i,1}(x)dx}{\int_0^{52} f_{i}(x)dx} \times 100
\end{equation}
\begin{equation}
  \label{GenInt6}
  p_2 = \dfrac{\int_0^{52} f_{i,2}(x)dx}{\int_0^{52} f_{i}(x)dx} \times 100
\end{equation}
Where $i=1$ or 2, depending on the elected function $f_1(x)$ or $f_2(x)$ at each PC.
\begin{multicols}{2}
\textbf{Statistical analysis}\\
The performance data from 2002 to 2008 is based on the distribution of performance per week of the year depending on the PC and the type of competition: mathematical analysis and modeling are done using Matlab. To estimate the two dates when the greatest numbers of performances occur, two functions are adjusted using the least squares method: the double Gaussian and double Lorentzian functions. For each PC, the best-fitted function is selected on the basis of adjusted $R^2$ and the mean square error (RMSE) (Fig. \ref{Fig.ANNS1}, Tab. \ref{Tab.ANN1} \ref{Tab.ANN2}). The two dates of peak performance are estimated using the elected model for each PC (Insert Fig. \ref{Fig.ANN1}). The proportion of performances in the two peaks is estimated by computing the area under the curve (proportion of performances) of each elected model and for each PC (Fig. \ref{Fig.ANNS2}, Tab. \ref{Tab.ANN2}).
\begin{center}
    \includegraphics[scale=0.7]{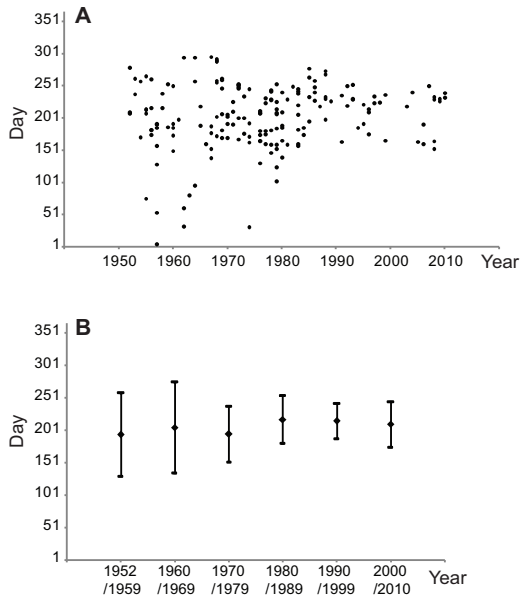}
\end{center}
\captionof{figure}[One figure]{\label{Fig.ANN2} {\footnotesize \textbf{A}. Distribution of world records (WR) date (day) in 100m, 200m, 400m, 800m and 1500 m running events from 1952 to 2010. \textbf{B}. Mean distribution of world records (WR) date (day) in 100m, 200m, 400m, 800m and 1500m running events by decade from 1952 to 2010. The mean date is: $206,09$\textsuperscript{th} day.\\[0.1cm]}}
\textbf{Distribution of WR}\\
Effect size for One-Way ANOVA is Cohen\textquoteright s $d$ and is evaluated with Cohen\textquoteright s conventional criteria \cite{Field2009}. It is used to study the stability of the WR mean date by decades (1952-1959, 1960-1969, 1970-1979, 1980-1989, 1990-1999, 2000-2010). Statistical significance is considered at $p<0.05$.\\[.3cm]
\textbf{Temperature}\\
For the temperature, the statistical analyzes are done on R, Version 3.0.0 (R Core Team,Vienna, Austria, 2013) and results are expressed as a mean $\pm$ standard deviation. Fisher test is used to compare the dispersions between the different PC with a value of $p<0.05$ considered significant. We estimate the density of temperature degrees for each of the PC over a homogeneous mesh of $5\times6$ nodes. The resolution used is of 7.5$\,^{\circ}\mathrm{C}$ in the x-axis (temperature) and 0.59 percent in the y-axis PC.\\[.3cm]
\textbf{\textsc{Results}}\\[0.3cm]
\noindent
\textbf{Distribution of performance by PC and competition}\\
The distribution of weekly performances for each PC (95 to 100 PC) over the competition calendar shows two high frequency periods (Fig. \ref{Fig.ANN1}). The estimated dates of the two peak performances are constant within all PC (on mean 27.15\textsuperscript{th} $\pm$ 0.21 week for peak 1 and 34.75\textsuperscript{th} $\pm$ 0.14 for peak 2) (Fig. \ref{Fig.ANN1}, Insert). The number of performances during major competitions (OG/WC/US/EC) increases from 16.7\% for the 95 PC to 25.7\% for the 99 PC. The performances recorded during the Golden League increase from 7.7\% for the 95 PC to 29.1\% for the 99 PC. Conversely, the number of performances in the other competitions decreases from 75.6\% to 45.1\% (Fig. \ref{Fig.ANN1}), Tab. \ref{Tab.ANN3}).\\[0.3cm]
\noindent
\textbf{Distribution of WR}\\
The mean distribution of WR date by decade from 1952 to 2010 is concentrated at the 206.09\textsuperscript{th} day $\pm$ 46.17. The variability of WR date decreases considerably. In the first period (1952-1959), SD is 64.34, in the last period (2000-2010), SD is 35.09. However the mean day remains stable throughout the period ($p=0.29$) (Fig. \ref{Fig.ANN2}), with a large effect size ($d=0.98$).\\[.3cm]
\noindent
\textbf{Influence of temperature on performance}\\
The analysis of the distribution of PC according to temperature shows a restriction in the thermal interval when reaching the highest performance level. This interval narrows from 10-32$\,^{\circ}\mathrm{C}$ at 97 PC of the BP to 20-27$\,^{\circ}\mathrm{C}$ for the 100 PC with a mean temperature of 23.23 $\pm$ 4.75$\,^{\circ}\mathrm{C}$. Subdividing the data into PC, the mean temperature is 23.13$\pm$4.80$\,^{\circ}\mathrm{C}$ for the PC [97 to 97.5[, 23.49$\pm$4.88$\,^{\circ}\mathrm{C}$ for the PC [97.5 to 98[, 23.23$\pm$4.92$\,^{\circ}\mathrm{C}$ for the PC [98 to 98.5[, 22.89$\pm$4.56$\,^{\circ}\mathrm{C}$ for the PC [98.5 to 99[, 22.63$\pm$3.72$\,^{\circ}\mathrm{C}$ for the PC [99 to 99.5[ and 23.25$\pm$3.26$\,^{\circ}\mathrm{C}$ for the BP [99.5 to 100]. Figure \ref{Fig.ANN3} highlights the narrowing of the temperature range at the BP interval. The peak value of the density mesh is 362 temperature values at 23$\,^{\circ}\mathrm{C}$ and at 97.59\%. The density decreases in both dimensions (temperature and PC) from this point confirming the mean temperature value stated above (Insert, Fig. \ref{Fig.ANN3}).\\[.3cm]
\noindent
\textbf{Top 10 sprinters from 1891 to 2008}\\
There is no evolutionary trend in the completion date on the 100m performances throughout the modern Olympic era. Since 1891, men accomplish their best performance around July 10\textsuperscript{th} ($\pm$50 days) \textit{id est} during the 28\textsuperscript{th} week and since 1921, women perform best around July 20\textsuperscript{th} ($\pm$44 days) \textit{id est} during the 29\textsuperscript{th} week (Fig. \ref{Fig.ANN4}).\\[.3cm]
\textbf{\textsc{Discussion}}\\[0.3cm]
Previous studies have mostly analyzed seasonality in the rhythms of daily life \cite{Atkinson2005, Foster2008} or in marathon runners \cite{Nour2012} but no studies have demonstrated effects of seasonality through environmental or institutional conditions on performance in sprint.\\[.3cm]
Two yearly performance peaks are observed for all levels in this study. The first peak corresponds to the 27\textsuperscript{th} week of the year (first week of July) suggesting an environmental optimum for sprint events. The second peak occurs at the 34\textsuperscript{th} week (fourth week of August), which is related to the main sporting events such as: Olympic Games, European and World Championships. As seen in Fig. \ref{Fig.ANN1}, both peak dates are stable throughout all performance categories.\\[.3cm]
\end{multicols}
\newpage
\begin{center}
    \includegraphics[scale=0.8]{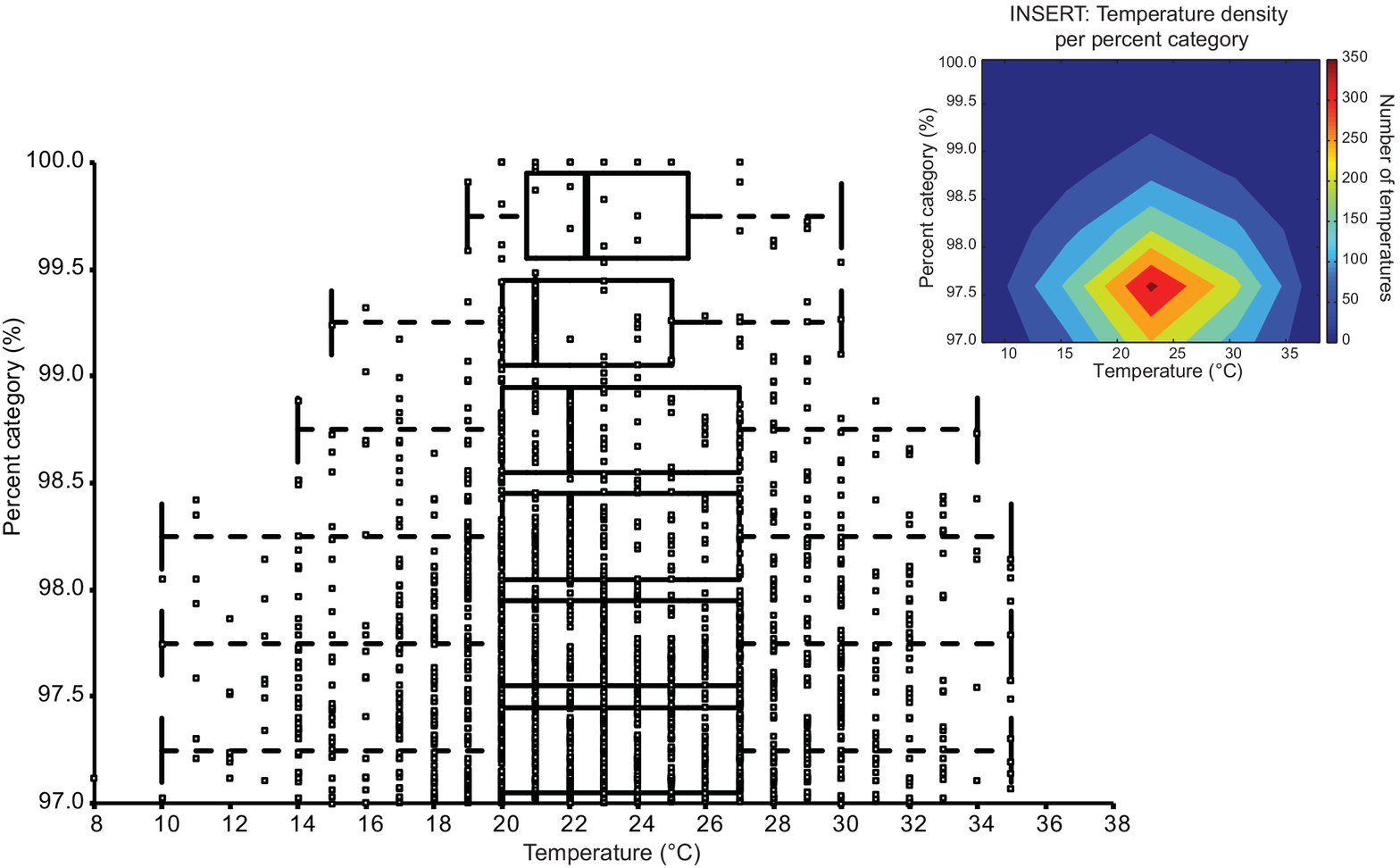}
\end{center}
\captionof{figure}[One figure]{\label{Fig.ANN3} {\footnotesize Percent category (PC) depending on temperature: comparison of temperature at different level from 97 PC to 100 PC in 100m, 200m, 400m, 800m and 1500m. Respectively, in each half PC the means are  23.12$\,^{\circ}\mathrm{C}$, 23.49$\,^{\circ}\mathrm{C}$, 23.23$\,^{\circ}\mathrm{C}$, 22.89$\,^{\circ}\mathrm{C}$, 22.63$\,^{\circ}\mathrm{C}$ and 23.25$\,^{\circ}\mathrm{C}$ and the medians are 23.00$\,^{\circ}\mathrm{C}$, 23.00$\,^{\circ}\mathrm{C}$, 22.00$\,^{\circ}\mathrm{C}$, 22.00$\,^{\circ}\mathrm{C}$, 21.00$\,^{\circ}\mathrm{C}$ and 22.50$\,^{\circ}\mathrm{C}$. INSERT: Temperature density (ie. number of recorded temperatures) per PC computed over a mesh. The maximal density is computed at 23$\,^{\circ}\mathrm{C}$ and 97.59\% and progressively decreases as PC increase (due to the decrease in performance number). The density decreases as temperature increases or decreases from the maximal density (due to the effect of temperature on performance).}}
\begin{multicols}{2}
\textbf{Peak at the 34\textsuperscript{th} week}\\
The impact of major international competitions corresponds to the performance peak in August. The calendar scheduling for world championships or Olympic Games can be considered as an institutional attractor. IAAF hosts competitions taking place outdoor between February and October. Major competitions such as the World Championships, European championships and Olympic Games are usually scheduled in August whereas the international circuit of the Golden League covers the whole period between June and September. National federations plan their own schedules proposing competitions that allow their athletes to qualify for the major competitions.
\end{multicols}
\begin{center}
    \includegraphics[scale=0.6]{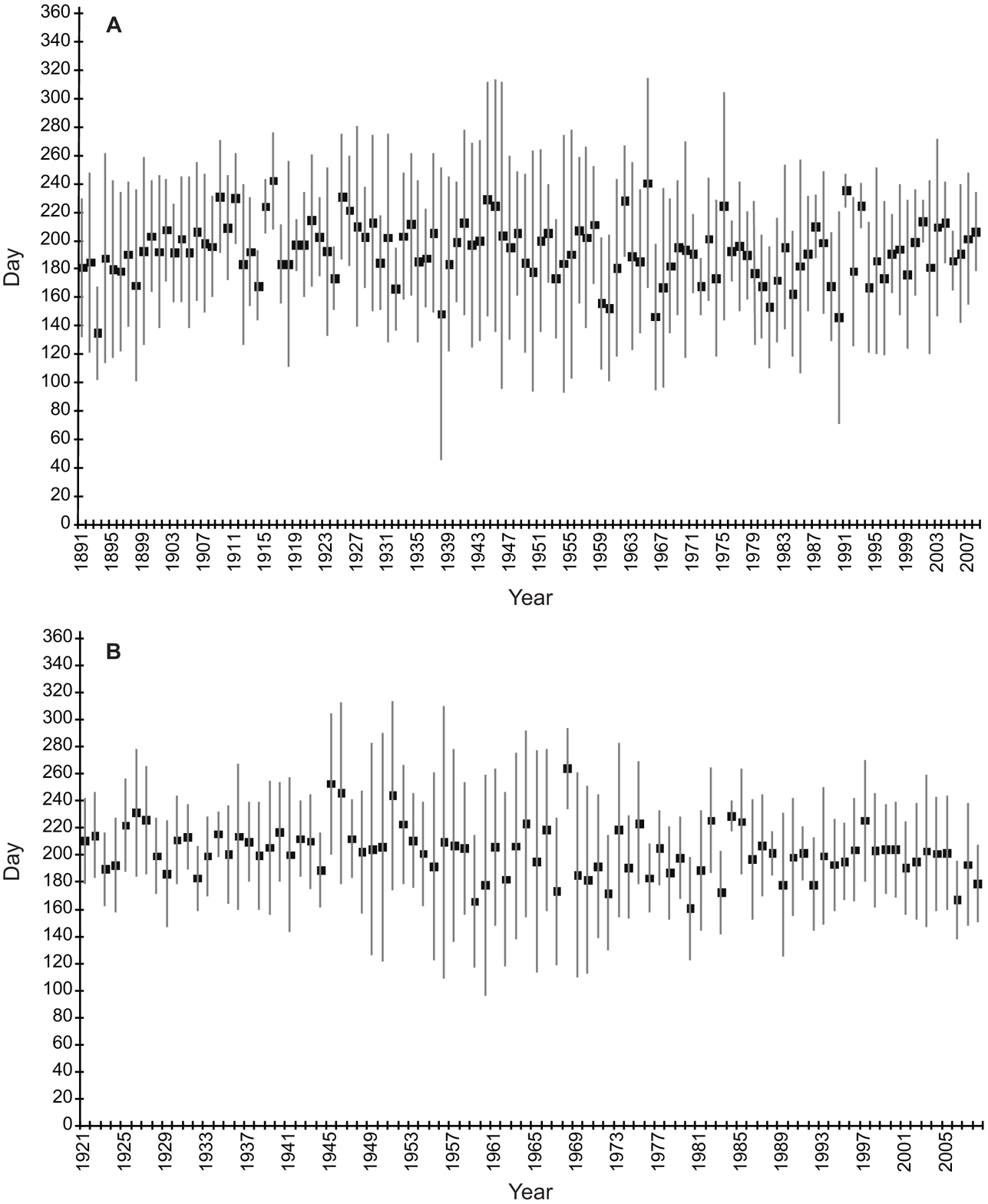}
\end{center}
\captionof{figure}[One figure]{\label{Fig.ANN4} {\footnotesize Relation between day of the performance and year in men and women. \textbf{A}. Average day of the achievement of the performance in the top 10 at the 100m men since 1891. For all years combined, the average day is the 192.76\textsuperscript{th} $\pm$ 49.77. \textbf{B}. Average day of the achievement of the performance in the top 10 at the 100m women since 1921. For all years combined, the average day is the 202.52\textsuperscript{th} $\pm$ 44.0.}}
\begin{multicols}{2}
This study highlights the existence of a cultural peak (second peak) occurring at the same times as the major international events. Globally, top athletes focus on the same goal: to be the most physically and mentally fit for this time of the year (Fig. \ref{Fig.ANN1}). This second peak corresponds to the athlete\textquoteright s own planning for major competitions, which is a result of long term training, technical analysis, strategic choice, awareness of physical and psychological limits [19-21].\\[.3cm]
Although, training and preparation are essential to reach a BP at a specific moment, environmental factors will allow the achievement of the highest level of performance.\\[.3cm]
\textbf{Peak at the 27th week}\\
The analysis of WR (Fig. \ref{Fig.ANN2}) and the top 10 BP in 100m sprint (men and women) illustrates this first peak (Fig. \ref{Fig.ANN4}). Numerous studies have demonstrated effects of environmental conditions on the performance of marathon runners \cite{Ely2007b, Montain2007}. Marathon requires a number of appropriate environmental conditions for thermoregulation of any runner, elite or amateur. The humidity, barometric pressure, dew point, and temperature are all essential in the quest of achieving optimal performance \cite{Ely2007a}. A recent study analyzed the impact of environmental parameters on the performance of marathon running. It established a distribution of performances depending on temperature, observed regardless of the athlete\textquoteright s level. This distribution function defines the field limits of the human possibilities \cite{Nour2012}. The impact of temperature and season on biological parameters is largely documented in the literature \cite{Ely2007a, Galloway1997, Atkinson1996}.\\[.3cm]
In this present study, the results show a distribution for top performance in sprint and middle-sprint where the effective temperature range decreases with performance level (10-32$\,^{\circ}\mathrm{C}$ at the bottom (97 PC of the BP); 20-27$\,^{\circ}\mathrm{C}$ at the top (100 PC of the BP)) (Fig. \ref{Fig.ANN3}). Competitions are mainly organized in the northern hemisphere. The range of temperatures collected from the different host cities was large: ranging from 10 to 38$\,^{\circ}\mathrm{C}$ but the mean temperature when achieving the BP is 23.23 $\pm$ 4.75$\,^{\circ}\mathrm{C}$. The standard deviations decrease progressively with increasing level, but all categories remain centered on the 23.23$\,^{\circ}\mathrm{C}$ value. This suggests a very regulated process at all performance levels.\\[.3cm]
\textbf{The effects of temperature on biological parameters}\\
All biological structures and processes (human or not) are affected by temperature \cite{Somero2002}. Performance depends on physiological responses to exercise performance in an interaction between body temperature and environmental temperature \cite{Atkinson1996}. Performance decreases progressively as the environmental heat stress increases \cite{Galloway1997}. As with other biological rate processes, muscle function is strongly influenced by temperature. Specifically, muscle contraction rates (the rates of both force development and relaxation) are accelerated by an increase in temperature in both invertebrates and vertebrates \cite{Bennett1985, Josephson1984}. Fundamental biological functions like metabolic activity synchronize with the rhythmic phases of environmental change such as temperature. For gradually intensity increasing aerobic exercise the plasma concentration of certain ions (K, Ca) and lactic acids appear differently when muscular exercise takes place at thermal neutrality (21$\,^{\circ}\mathrm{C}$) in comparison to exercise performed at 0$\,^{\circ}\mathrm{C}$ \cite{Therminarias1984}.\\[.3cm]
At the favorable season, body temperature and metabolic rates increase and so does growth rate. Mammal growth depends on seasonal variation even for their bones structure \cite{Koehler2012}. Climates and seasons have a marked influence on human biology \cite{Steinacker2000} including mental abilities \cite{Magnusson2000}, sexual activity \cite{Regnier-Loilier2011} or territorial conflicts \cite{Hsiang2011}.\\[.3cm]
\textbf{Temperature and mortality}\\
Many chronobiological health aspects depend on season and temperature cycles. Affective disorders show a predictable onset in the fall/winter months and, reversely, a reduction in the spring/summer period \cite{Magnusson2000}. Large-scale population studies have shown seasonal variations in mortality rates in different parts of the world peaking during the cold winter months \cite{Healy2003, Donaldson1997}. Relations between mortality and cardiovascular disease (CVD) in the winter months have been reported for many countries and might be partly explained by seasonal changes of risk factors. Cardiac death also depend on the season even after adjustment for age, cholesterol, blood pressure, and body mass index \cite{Donaldson1997, Ghebre2012}. Several studies have reported the existence of optimal ranges of air temperatures \cite{Ballester1997, Rocklov2008}. Specifically, cold weather has been reported to be associated with increased risk of death from cardiovascular causes and respiratory infections \cite{Rocklov2008, Eccles2002, Barnett2007}. The mortality rate is lower on days in which the maximum temperatures range between 20-25$\,^{\circ}\mathrm{C}$ \cite{Ballester1997}. This means that survival rate is highest at this temperature range.\\[.3cm]
\end{multicols}
\begin{center}
    \includegraphics[scale=0.7]{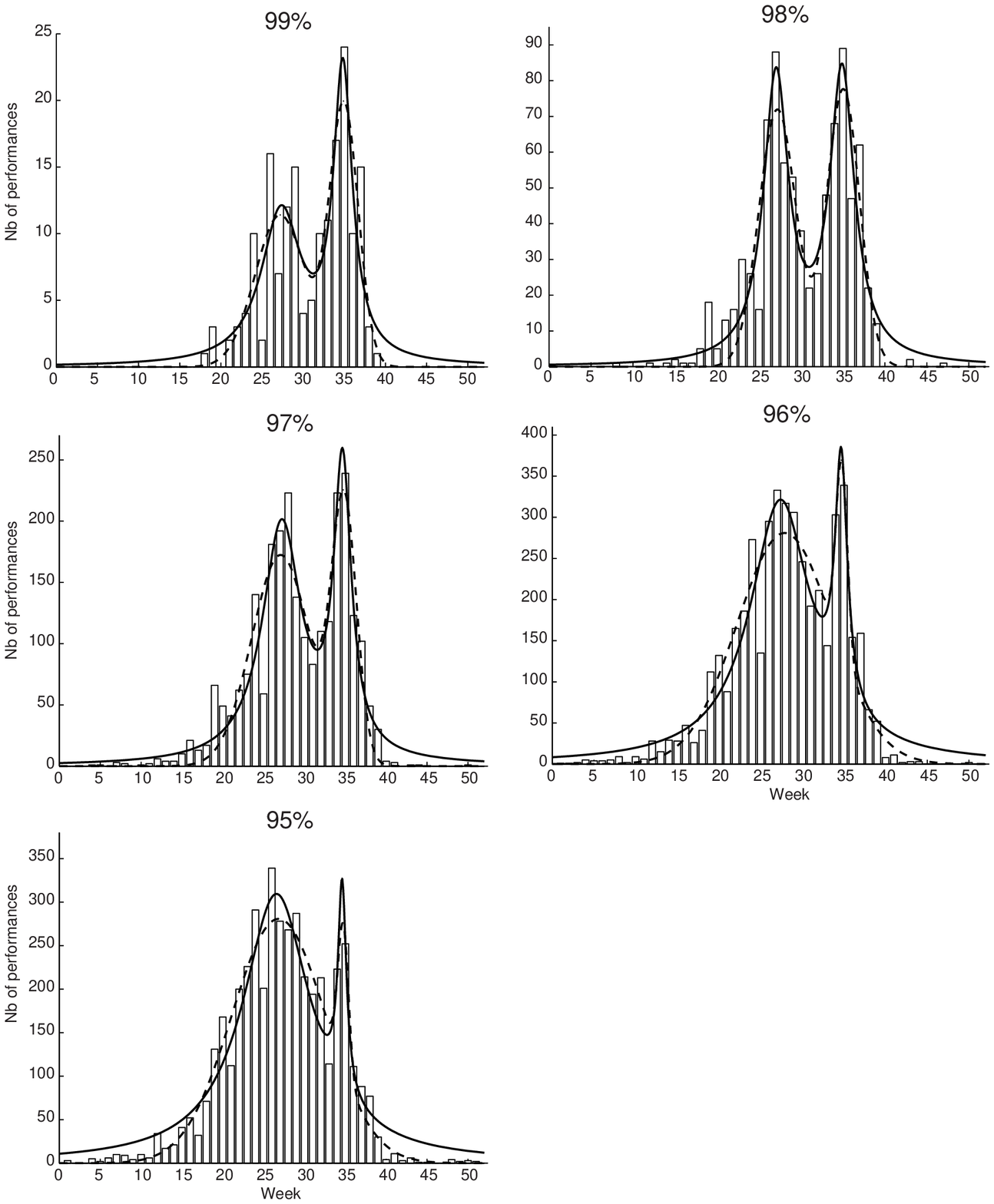}
\end{center}
\captionof{figure}[One figure]{\label{Fig.ANNS1} {\footnotesize Date of peak performance modeling for 95\% to 99\% categories. The Double Lorentzian (continuous line) and Double Gaussian functions (broken line) are adjusted to each percent category. Although the two models differ in the estimate of the tails, they roughly provide the same estimates of $x_{01}$ and $x_{02}$ (maximum difference is around 0.5 week).}}
\begin{multicols}{2}
Our results show a mean temperature of 23.23$\pm$4.75$\,^{\circ}\mathrm{C}$ for the BP which is converging with the temperature of the lowest mortality rates. Therefore, both survival capability and physiological capacities of the human are optimal at 20-25$\,^{\circ}\mathrm{C}$. The two peaks of performance change its distribution in function of the performance level (Fig. \ref{Fig.ANNS1} and \ref{Fig.ANNS2}). However, the first peak persists even at the highest level of competition. This demonstrates that despite the presence of an institutional attractor, represented by the major competitions, the environmental attractor remains omnipresent with an ideal temperature period for maximal performance.\\[.3cm]
\textbf{Conclusion}\\
The range of possible combination of these environmental and institutional components is narrowed for the top performers. On sprint and middle-sprint races, when progressing toward the highest levels of performance, the importance of the institutional component regularly increases with a balanced effect for the top performers. Calendars for major competitions should take this into account in order to increase the probability of breaking the next records.
\end{multicols}
\begin{center}
    \includegraphics[scale=0.8]{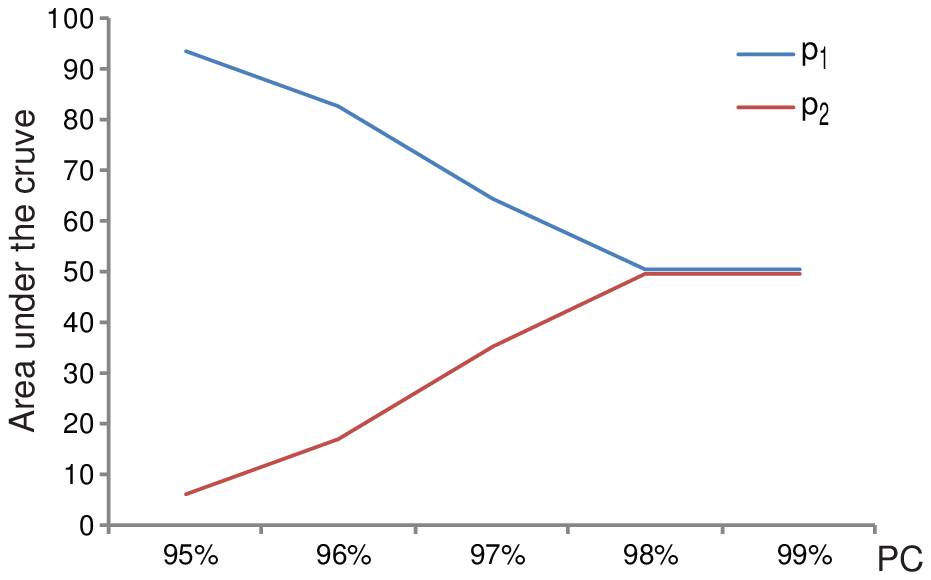}
\end{center}
\captionof{figure}[One figure]{\label{Fig.ANNS2} {\footnotesize Area under the curve for the elected function and both peaks in each PC. Area under the curve (AUC) for the elected functions and both peaks in each PC. Equations \ref{GenInt5} and \ref{GenInt6} are used to estimate $p_1$ and $p_2$. The AUC of both peaks converge toward a unique value as the PC increase (50\%). The proportion of performances: The estimation of the area under the curve for the two peaks shows that, when increasing the PC, the proportion of performances in each peak progressively converge to the same value, from 93.67\% (peak 1) vs. 6.33\% (peak 2) for PC = 95\% to 50.65\% (peak 1) vs. 49.35\% (peak 2) with PC=99\%.}}

\begin{center}
\footnotesize
\begin{tabular}{|l|l|l|l|l|l|l|l|l|l|l|l|}
   \hline
   Percent category & \multicolumn{2}{c}{95\%} & \multicolumn{2}{c}{96\%} & \multicolumn{2}{c}{97\%} & \multicolumn{2}{c}{98\%} & \multicolumn{2}{c|}{99\%} \\
   Type of competition & n & \% & n & \% & n & \% & n & \% & n & \% \\
   \hline
   Major competitions (OG, WC, US, EC) & 712 & 16.7 & 826 & 18.4 & 576 & 23 & 188 & 22.4 & 45 & 25.7 \\
   Major competitions (Golden League) & 461 & 7.7 & 743 & 16.5 & 591 & 23.6 & 247 & 29.4 & 51 & 29.1 \\
   Other meetings & 3309 & 75.6 & 2930 & 65.1 & 1336 & 53.4 & 405 & 48.2 & 79 & 45.1 \\
   \hline
\end{tabular}
\captionof{table}{\label{Tab.ANN1} {\footnotesize \textbf{Number of performances per depending on the type of competition and the percent category.} N is the number for different percent category (PC) and its equivalent percentage. On the overall performance analyzed, 2,347 were conducted during major competitions and 2,093 during the Golden League. The other 8,079 performances were done in other competitions (OG: Olympic Games; WC: World championships; US: American selections; EC: European Championships).}}
\end{center}

\begin{center}
\footnotesize
\begin{tabular}{|l|l|l|l|}
   \hline
   PC & Stats & Double Gaussian & Double Lorentzian \\
   \hline
   99\% & $R^2$ adjusted & \textbf{0.832} & 0.824 \\
     & rMSE & \textbf{2.360} & 2.414 \\
     & sse & \textbf{256.141} & 268.000 \\
   \hline
   98\% & $R^2$ adjusted & 0.893 & \textbf{0.913} \\
     & rMSE & 8.166 & \textbf{7.348} \\
     & sse & 3067.119 & \textbf{2483.700} \\
   \hline
   97\% & $R^2$ adjusted & 0.900 & \textbf{0.933} \\
     & rMSE & 21.680 & \textbf{17.703} \\
     & sse & 21621.279 & \textbf{14415.400} \\
   \hline
   96\% & $R^2$ adjusted & 0.929 & \textbf{0.930} \\
     & rMSE & 29.559 & \textbf{29.439} \\
     & sse & 40192.703 & \textbf{39866.000} \\
   \hline
   95\% & $R^2$ adjusted & \textbf{0.952} & 0.931 \\
     & rMSE & \textbf{22.789} & 27.304 \\
     & sse & \textbf{23889.801} & 34294.200 \\
   \hline
\end{tabular}
\captionof{table}{\label{Tab.ANN2} {\footnotesize \textbf{Statistics of the two models}. For each percent category (PC), the adjusted $R^2$, rMSE and sse are given. Statistics of the elected function are mentioned in bold.}}
\end{center}

\begin{center}
\footnotesize
\begin{tabular}{|l|l|l|}
   \hline
   \textbf{95}\% & & \\
   Estimates & Double Gaussian & Double Lorentzian\\
   $x_{01}$ & 26.835 & 26.575 \\
   $x_{02}$ & 34.675 & 34.600 \\
   $f(x_{01})$ & 280.725 & 309.427 \\
   $f(x_{02})$ & 276.114 & 327.423 \\
   Area under the curve & 4302.785 & 4737.448 \\
   $p_1$ (\%) & 93.666 & 90.326 \\
   $p_2$ (\%) & 6.334 & 9.674 \\
   \hline
   \textbf{96}\%  & & \\
   $x_{01}$ & 27.885 & 27.400 \\
   $x_{02}$ & 34.678 & 34.644 \\
   $f(x_{01})$ & 280.938 & 321.513 \\
   $f(x_{02})$ & 369.480 & 385.732 \\
   Area under the curve & 4430.167 & 4738.177 \\
   $p_1$ (\%) & 90.845 & 82.787 \\
   $p_2$ (\%) & 9.155 & 17.213 \\
 \hline
   \textbf{97}\% & & \\
   $x_{01}$ & 27.085 & 27.249 \\
   $x_{02}$ & 34.748 & 34.628 \\
   $f(x_{01})$ & 172.411 & 201.690 \\
   $f(x_{02})$ & 225.526 & 259.893 \\
   Area under the curve & 2332.208 & 2551.242 \\
   $p_1$ (\%) & 65.756 & 64.519 \\
   $p_2$ (\%) & 34.245 & 35.481 \\
 \hline
   \textbf{98}\% & & \\
    $x_{01}$ & 27.208 & 27.065 \\
   $x_{02}$ & 35.037 & 34.891 \\
   $f(x_{01})$ & 71.955 & 83.803 \\
   $f(x_{02})$ & 77.671 & 84.819 \\
   Area under the curve & 778.209 & 858.340 \\
   $p_1$ (\%) & 51.441 & 50.481 \\
   $p_2$ (\%) & 48.559 & 49.519 \\
 \hline
   \textbf{99}\% & & \\
    $x_{01}$ & 27.212 & 27.391 \\
   $x_{02}$ & 34.898 & 34.804 \\
   $f(x_{01})$ & 11.405 & 12.139 \\
   $f(x_{02})$ & 19.952 & 23.203 \\
   Area under the curve & 172.262 & 188.397 \\
   $p_1$ (\%) & 50.650 & 51.967 \\
   $p_2$ (\%) & 49.350 & 48.033 \\
   \hline
\end{tabular}
\captionof{table}{\label{Tab.ANN3} {\footnotesize \textbf{Results of the two models.} For percent category and model, the estimated date of peak (x01, x02), value of peak (f(x01), f(x02)), the total proportion of performance (area under the curve) , and p1, p2 are given. Results of elected function are given in bold.}}
\end{center}

\bibliographystyle{unsrt}
\bibliography{bibfile}
\end{document}